\documentclass[apj,twocolumn,floatfix]{openjournal}

\usepackage{placeins} 

\usepackage[dvipsnames]{xcolor}

\usepackage{booktabs}
\usepackage{graphicx}
\usepackage{multirow}
\usepackage{longtable}

\usepackage{makecell}

\usepackage{siunitx}
\usepackage{url}

\usepackage{amsmath}

\renewcommand{\ion}[2]{#1\,\textsc{#2}}

\makeatletter
\renewcommand\@makecaption[2]{%
  \par
  \vskip\abovecaptionskip
  \begingroup
    \footnotesize\rmfamily
    \begingroup
      \samepage
      \flushing
      \let\footnote\@footnotemark@gobble
      \ifnum\pdfstrcmp{\@captype}{table}=0
        \@make@capt@title{\textsc{Table \thetable}}{#2}%
      \else
        \ifnum\pdfstrcmp{\@captype}{figure}=0
          \@make@capt@title{\textsc{Figure \thefigure}}{#2}%
        \else
          \@make@capt@title{#1}{#2}%
        \fi
      \fi\par
    \endgroup
  \endgroup
  \vskip\belowcaptionskip
}
\makeatother

\usepackage{textgreek}
\usepackage[utf8]{inputenc}
\usepackage[english]{babel}

\definecolor{linkcolor}{rgb}{0.0,0.3,0.5}
\usepackage{xurl}

\usepackage{colortbl}

\DeclareGraphicsExtensions{.bmp,.png,.jpg,.pdf}
\usepackage{verbatim}
\usepackage[normalem]{ulem}
\usepackage{orcidlink}
\usepackage{soul}

\usepackage{hyperref}
\hypersetup{
    unicode, 
    colorlinks=true,
    linkcolor=linkcolor,
    citecolor=linkcolor,
    filecolor=linkcolor,
    urlcolor=linkcolor,
}

\graphicspath{{./}{figures/}}

\begin{document}
\title{ATLAS22kjn (AT 2022fpx): A Coronal Line Emitter with an Early Light Curve Bump and Mid-Infrared Dust Echo}

\author{\vspace{-1.3cm}
Athena C. Engholm\,\orcidlink{0000-0001-6970-7782}$^{1, 2, \star}$,
Jason T. Hinkle\,\orcidlink{0000-0001-9668-2920}$^{1, 3, 2, \dagger}$,
Benjamin J. Shappee\,\orcidlink{0000-0003-4631-1149}$^{2}$, 
Katie Auchettl\,\orcidlink{0000-0002-4449-9152}$^{4}$,\\
Dhvanil D. Desai\,\orcidlink{0000-0002-2164-859X}$^{2}$,
Willem B. Hoogendam\,\orcidlink{0000-0003-3953-9532}$^{2, \ddagger}$,
Christopher S. Kochanek\,\orcidlink{0000-0001-6017-2961}$^{5,6}$,
Nicholas Earl\,\orcidlink{0000-0003-1714-7415}$^{1}$,\\
K. Decker French\,\orcidlink{0000-0002-4235-7337}$^{1}$,
Michael A. Tucker\,\orcidlink{0000-0002-2471-8442}$^{5, 6, \mathsection}$,
Chris Ashall\,\orcidlink{0000-0002-5221-7557}$^{2}$,
Aaron Do\,\orcidlink{0000-0003-3429-7845}$^{7}$,
Allison Blum\,\orcidlink{0009-0000-9959-5216}$^{2}$, \\
Thomas deJaeger\,\orcidlink{0000-0001-6069-1139}$^{15}$,
Mark E. Huber\,\orcidlink{0000-0003-1059-9603}$^{2}$,
Anna Payne\,\orcidlink{0000-0003-3490-3243}$^{16}$,
Jose L. Prieto\,\orcidlink{0000-0003-1072-2712}$^{12, 13}$,
}

\affiliation{
$^{1}$Department of Astronomy, University of Illinois, 1002 W. Green St., Urbana, IL 61801, USA\\
$^{2}$Institute for Astronomy, University of Hawai`i, 2680 Woodlawn Dr., Honolulu, HI 96822, USA\\
$^{3}$NSF – Simons AI Institute for the Sky (SkAI), 172 E. Chestnut St., Chicago, IL 60611, USA\\
$^{4}$School of Physics, University of Melbourne, VIC 3010, Australia\\
$^{5}$Centre for Cosmology and Astroparticle Physics, The Ohio State University, 191 W. Woodruff Av., Columbus, OH 43210, USA\\
$^{6}$Department of Astronomy, The Ohio State University, 140 West 18th Av., Columbus, OH 43210, USA\\
$^{7}$Institute of Astronomy and Kavli Institute for Cosmology, Madingley Road, Cambridge, CB3 0HA, UK\\
$^{8}$Department of Physics and Astronomy, Texas Tech University, 2500 Broadway, Lubbock, TX 79409, USA\\
$^{9}$Department of Astronomy, School of Physics, Peking University, Yiheyuan Rd. 5, Haidian District, Beijing 100871, China\\
$^{10}$Kavli Institute for Astronomy and Astrophysics, Peking University, Yi He Yuan Road 5, Hai Dian District, Beijing 100871, PR China\\
$^{11}$National Astronomical Observatories, Chinese Academy of Science, 20A Datun Road, Chaoyang District, Beijing 100101, China\\
$^{12}$ Instituto de Estudios Astrofísicos, Facultad de Ingeniería y Ciencias, Universidad Diego Portales, Av. Ejército Libertador 441, Santiago, Chile\\
$^{13}$Millennium Institute of Astrophysics MAS, Nuncio Monsenor Sotero Sanz 100, Off. 104, Providencia, Santiago, Chile\\
$^{14}$Department of Physics, The Ohio State University, 191 W. Woodruff Ave., Columbus, OH 43210, USA\\
$^{15}$LPNHE (CNRS/IN2P3, Sorbonne Université, Université Paris Cité), Laboratoire de Physique Nucléaire et de Hautes Énergies, 75005, Paris, France\\
$^{16}$Space Telescope Science Institute, 3700 San Martin Drive, Baltimore, MD 21218, USA
}
\thanks{$^{\star}$e-mail: engholm2@illinois.edu}
\thanks{$^{\dagger}$NHFP Einstein Fellow}
\thanks{$^{\ddagger}$NSF Graduate Research Fellow}
\thanks{$^{\mathsection}$CCAPP Fellow}

\begin{abstract}
    We present an analysis of the TDE ATLAS22kjn (AT 2022fpx), whose high-ionisation coronal lines (CLs) and pre-peak light curve bump provide distinctive opportunities to investigate the physical mechanisms powering tidal disruption events (TDEs). In addition to CLs, the optical spectra show common TDE features, including a strong, blue continuum and broad Balmer and \ion{He}{ii} lines. The CLs appear before UV/optical light curve peak, preceding the detection of X-rays by $\sim 300$ days and persisting after X-rays are no longer detected, suggesting the X-ray emission is obscured at both early and late times. 
    Using the CL luminosities, we constrain the temperature evolution of the ionising source, finding a decrease of $\lesssim 10 \%$ over 500 days. 
    In the UV/optical light curve, we observe a $9 \substack{+4 \\ -2}$\,day bump that peaks $125 \substack{+5 \\ -3}$ rest-frame days before the peak of the main flare. 
    Although we cannot definitively determine the origins of the bump, we find that its timescale and luminosity are most consistent with theoretical predictions for a precursor feature produced by a stream-stream collision or a wind-stream collision. ATLAS22kjn also shows a prominent dust echo in its mid-infrared (MIR) light curves, indicating a high dust covering fraction $f_c \simeq 0.40 \pm 0.03$, similar to the covering fractions of other CL-emitting TDEs. From the multi-wavelength observations of ATLAS22kjn, we estimate the size and relative radii of the emission regions in its nuclear environment and determine that the CL region lies between the broad line region and the MIR-emitting dust.
    ATLAS22kjn demonstrates the importance of multi-wavelength and early-time observations, and the utility of CLEs in characterising the otherwise unobservable EUV/ultrasoft X-ray emission of TDEs.
\end{abstract}


\maketitle

\section{Introduction}

The centre of nearly every massive galaxy hosts a supermassive black hole \cite[SMBH; e.g.,][]{magorrian1998, kormendy2013}. As evidenced by the tight correlation between black hole mass and host-galaxy bulge velocity dispersion \cite[e.g.,][]{kormendy2013}, SMBHs play pivotal roles in galaxy evolution. About 1-5\% of the SMBHs in the local Universe are strongly accreting \citep{Kauffmann2003, Haggard2010} active galactic nuclei \citep[AGNs;][]{Salpeter1964, ZeldovichNovikov1964}. While the accretion powering these AGNs allows us to study them relatively easily, direct detection of quiescent SMBHs is limited by the ability to resolve the small SMBH sphere of influence \cite[e.g., ][]{ford94, atkinson05, gebhardt11}. Additional probes of SMBHs are necessary to develop a complete picture of SMBH properties and accretion behaviours.

Tidal disruption events (TDEs) are one such probe. A TDE occurs when a star passes within the tidal radius of a SMBH and is torn apart when the tidal forces from the SMBH exceed the star’s self-gravity. In a complete TDE, roughly half the material of the disrupted star remains bound to the SMBH, while the other half is unbound. The fallback of bound stellar material onto the SMBH results in a luminous flare that serves as an observational signpost for the otherwise-quiescent SMBH \citep[e.g.,][]{Rees1988, phinney89, evans89, ulmer99}. 
Although TDEs provide real-time monitoring of accretion disk formation, the details are not well understood 
for TDEs \citep{kochanek94, metzger16, kochanek16b, Gezari2021, Mockler_2026}. 
Consequently, the emission mechanisms of TDEs remain an active area of investigation,  particularly at early times.

UV/optical emission dominates the observed spectral energy distributions (SEDs) of optically-selected TDEs. These SEDs are well-fit by a 
near-constant blackbody temperature of $\sim$$20,000$--$50,000 \ \textrm{K}$ 
\citep[e.g.,][]{gezari12b, holoien14b}. Conversely, AGN SEDs are better modelled by a power law \citep[e.g.,][]{vandenberk01}, and have harder
X-ray emission than TDEs \citep{ricci17, auchettl18}. 
The X-ray emission of most TDEs remains soft throughout their evolution \citep[e.g.,][]{auchettl18}, with hardness ratios between $-1$ and 0.3 \citep[][]{auchettl17}. This emission is also more diverse, with X-ray-to-optical ratios of $L_X/L_{\textrm{opt}}\sim 1$ to $\gg 1$ \citep[e.g.,][]{auchettl17}, with an estimated $\ge 40 \%$ of optically-selected TDEs exhibiting X-ray luminosities $L_X \ge 10^{42} \ \textrm{erg s}^{-1}$ \citep[e.g.,][]{hammerstein23, Guolo_2024}.
While these X-ray properties reveal details about accretion processes, the source of UV/optical emission in TDEs is still debated.

UV/optical light curves provide insight into the origins of UV/optical emission, and can be used to distinguish between AGN and TDE flares. Unlike AGN light curves, which exhibit stochastic variability \citep[e.g.,][]{bianchi05, macleod12, Tarrant_2025}, the UV/optical light curves of TDEs typically exhibit no short-term variability, remaining smooth throughout the flare. Instead, TDEs generally show an approximately month-long rise in brightness to a single peak, followed by a gradual power-law decline \citep[e.g.,][]{Arcavi2014, Gezari2021}. However, we have now observed a handful of TDEs with non-monotonic rises to the UV/optical peak. Pre-peak ``bumps" like those observed for ASASSN-19dj \citep{hinkle21a, Faris2024}, AT 2020wey \citep{Charalampopoulos_2023}, ASASSN-18ap \citep{Wang2024}, AT 2019mha \citep{Wang2024}, AT 2019qiz \citep{Wang2024}, and AT 2023lli \citep{Huang_2024} have been predicted by a range of physical mechanisms, including vertical shock compression during the pericentre passage \cite[also ``nozzle shocks''; e.g.,][]{evans89, kochanek94, yalinewich19, Bonnerot2022, Steinberg_2024, Hu_2026}, the cooling of stellar debris not captured by the SMBH \cite[e.g.,][]{KasenRamirez2010}, disk-stream collisions in the case of a repeating partial TDE \cite[rpTDE; e.g.,][]{Huang2023b, Sun_2025}, the 
collision of two streams formed in a double TDE \citep[e.g.,][]{Mandel2015, Wu_Yuan_2018, Bonnerot_Rossi_2019}, 
collision of the tidal debris stream with itself \cite[e.g.,][]{piran15, Jiang_2016, Huang_2023_stream_stream, huang_2024_pre_peak_emission}, and wind-stream collision \citep[e.g.][]{Calderon2024}. Early-time features like these bumps can provide crucial insight into the physics driving the early optical emission of TDEs. While distinguishing between emission models remains challenging, resolved bumps enable measurements of key observables---including luminosities, temperatures, and timescales---that provide additional constraints not obtainable from the otherwise smooth rise of TDE light curves. With the increasing cadence and depth of all-sky surveys, detection of these bumps is becoming increasingly common \citep{Wang2024}.

Additional insight into the accretion behaviours of TDEs can be gained from their spectra.
Both TDE and AGN spectra exhibit strong UV/optical continuum emission and broad emission lines. Commonly observed lines for both include broad hydrogen and helium lines, as well as nitrogen and oxygen lines \citep[e.g.,][]{antonucci93, ho08, Koss_2017, vanvelzen21, Gezari2021, Charalampopoulos_2022}. AGNs typically exhibit broad-line widths in the range of $\sim10^3 \ \textrm{km s}^{-1}$ to $\sim10^4 \ \textrm{km s}^{-1}$ \citep[e.g.,][]{peterson93, ho08}, while TDE broad lines are typically of order $\sim10^4\ \textrm{km s}^{-1}$ \citep[e.g.,][]{arcavi14, vanvelzen21}.   
The broad lines of TDEs often demonstrate a positive correlation between line width and luminosity \citep{holoien16a, holoien19b, Hinkle2021b} in contrast with AGNs, which typically show an inverse correlation between line width and luminosity \citep{Peterson2004, denney09}. 
High-ionisation ($\gtrsim 100 \ \textrm{eV}$) emission lines---known as coronal lines (CLs) since they were first observed in the solar corona \citep[e.g.,][]{Oliva1997, Ferguson1997, Clark2024}---are also observed for AGNs \citep[e.g.,][]{lamperti17, Negus_2023}.
AGN CLs may be produced at the inner edge of the dusty torus \citep[e.g.][]{Rodriguez_Ardila_2002, Gelbord_2009, Mullaney_2009, Glidden_2016, Riffel_2021, Negus_2023}, while others are associated with jets and outflows \citep[e.g.][]{Rodriguez_Ardila_2002, Almudena_Prieto_2005, Gelbord_2009, Mazzalay_2010, MullerSanchez_2011, Riffel_2021, Rodriguez_Ardila_2020, TrindadeFalcao_2022, Negus_2023}.

Recent studies have also identified ``coronal line emitters'' \citep[CLEs; e.g.,][]{komossa08, wang11, yang13, palaversa16, onori22, Short2023, callow2024, Hinkle2024}, whose CL luminosities are similar to that of the [\ion{O}{iii}] $\lambda$5007 line \citep[e.g.,][]{komossa08, wang11, wang12, vanvelzen21b, Clark2024}. These luminosities are inconsistent with AGN CLs, which are typically only a few percent of the [\ion{O}{iii}] $\lambda$5007 line \citep[e.g.,][]{wang12, Frederick_2019}. Instead, CLEs are proposed to be TDEs in gas-rich environments, as the EUV/soft X-ray emission from these events is sufficient to photoionise gas in the nucleus of the host galaxy to produce these lines \citep[e.g.,][]{Ferguson1997, Mullaney_2009, Mazzalay_2010, yang13, Hinkle2024}.
The distinction in CL strengths reflects timescale differences between TDEs and AGNs. In a TDE, the short-lived flare drives the emission lines. On the timescale of the flare, light travel times mean that only central, high-density \citep[$n_{H} \sim 10^6 - 10^8 \ \textrm{cm}^{-3}$, e.g.,][]{Mummery_2025} regions are illuminated, and in such regions, [\ion{O}{iii}] $\lambda$5007 is collisionally de-excited, making the CL emission comparatively strong for TDEs. In contrast, AGNs are luminous for far longer and so can drive [\ion{O}{iii}] $\lambda$5007 emission from much larger volumes at much lower densities, making the [\ion{O}{iii}] emission stronger than CL emission.
For similar reasons, the low-ionisation [\ion{Ne}{v}] lines are among the most reliable tracers of AGN activity \citep{Negus_2023}. 
In both AGNs and TDEs, commonly observed UV/optical CLs include [\ion{Fe}{vii}] through [\ion{Fe}{xiv}], [\ion{Ar}{xiv}] $\lambda$4414, and [\ion{S}{xii}] $\lambda$7612 \citep{komossa08, wang12}. 

In this paper, we present a detailed analysis of the object ATLAS22kjn, which shows both CLs and a pre-peak light curve bump.
Optical spectra, optical polarization, UV/optical photometry, and X-ray observations of ATLAS22kjn were first studied by \cite{Koljonen2024}. The X-ray emission, late-time plateau, mid-infrared (MIR) flare, colour, dust properties, and aspects of optical spectra were then studied by \cite{Lin_2025}. Here, we expand upon these studies by analysing additional UV/optical photometry---including TESS coverage of the prominent early-time bump---mid-infrared (MIR) photometry, X-ray observations, and further spectral coverage in the optical and NIR. This includes a spectroscopic time series, which we use to characterise the evolution of the CLs.
We describe the discovery of ATLAS22kjn and the photometric and spectroscopic data used in this study in Section \ref{sec:Discovery and Observations}. We analyse this multi-wavelength dataset in Section \ref{sec:Analysis}, including UV/optical photometry and spectra, MIR photometry, and X-ray observations. We analyse the EUV/soft X-ray emission from ATLAS22kjn using CLs in Section \ref{sec:CLs_lum_trend_analysis}. We discuss potential mechanisms for powering the early UV/optical bump in Section \ref{sec:bump discussion}. In Section \ref{sec:CLE discussion}, we compare ATLAS22kjn with other CL-emitting TDEs. 
In Section 
\ref{sec:implications for SMBH environments}, we use the multi-wavelength data of ATLAS22kjn to construct a picture of emission regions surrounding the central SMBH. Section \ref{sec:Conclusions} concludes this paper and summarises our findings. We assume a cosmology of $H_0 = 69.6 \ \textrm{km s}^{-1} \textrm{Mpc}^{-1}$, $\Omega_M = 0.29$, and $\Omega_{\Lambda} = 0.71$ \citep{wright06, bennett14} throughout.

\section{Discovery and Observations} \label{sec:Discovery and Observations}

In this section, we detail the discovery of ATLAS22kjn (Section \ref{sec:discovery}), the characteristics of its host galaxy (Section \ref{sec:host}), survey photometry (Section \ref{sec:photometry}), X-ray observations (Section \ref{sec:xray}), and spectroscopic data (Section \ref{sec:spectroscopic observations}) used in this study.

\subsection{Discovery of ATLAS22kjn}
\label{sec:discovery}

ATLAS22kjn ($\alpha_{\textrm{J2000}},\delta_{\textrm{J2000}}$ = 15:31:03.700, +53:24:19.26) was discovered in $o$-band data from the Asteroid Terrestrial-impact Last Alert System \citep[ATLAS;][]{tonry18} on the 31st of March, 2022 ($\textrm{MJD}=59669.5$), and assigned the name AT 2022fpx\footnote{\url{https://www.wis-tns.org/object/2022fpx}} by the Transient Name Server (TNS). ATLAS22kjn was then classified as a possible TDE by \cite{Perez-Fournon2022} based on the presence of strong Balmer and helium lines in its classification spectrum, and a light curve that was still rising $\sim 90$ days post-discovery. ATLAS22kjn is located in the nucleus of the galaxy SDSS J153103.70+532419.3, at a redshift of $z=0.073$ \citep{Perez-Fournon2022}, corresponding to a luminosity distance of 332.0 Mpc. 

\subsection{Host Galaxy Archival Data}
\label{sec:host}

\begin{table}[t]
    \centering
        \caption{\normalfont The AB magnitudes used to fit the photometry for SDSS J153103.70+532419.3. The $NUV$ magnitude is computed from GALEX \citep{martin05} images using a 5" radius aperture. The $ugriz$ photometry are the measured SDSS cModel mags. The $W1$, $W2$, $W3$, and $W4$ magnitudes come from the AllWISE catalog. All magnitudes are presented in the AB system.}
        \vspace{1mm}
    \begin{tabular}{SSS} \toprule
        {Filter} & {AB Magnitude (mag)} & {Magnitude Uncertainty (mag)} \\ \midrule
        {NUV} & 22.67 & 0.45 \\
        {u} & 20.58 & 0.10 \\
        {g} & 18.66 & 0.01 \\
        {r} & 17.90 & 0.01 \\
        {i} & 17.55 & 0.01 \\
        {z} & 17.40 & 0.02 \\
        {W1} & 17.77 & 0.03 \\
        {W2} & 18.24 & 0.05 \\ 
        {W3} &  17.5 & 0.2 \\
        {W4} & > 16.19 & N/A \\ \bottomrule
    \end{tabular}
    \label{tab:host magnitudes}
\end{table}

SDSS J153103.70+532419.3 was observed by the Sloan Digital Sky Survey \citep[SDSS;][]{Kollmeier2019}, the Galaxy Evolution Explorer \citep[GALEX;][]{martin05}, and the Wide-field Infrared Survey Explorer \citep[WISE;][]{wright10}. No archival spectra were available on the SDSS, MaNGA, or NED archives. We obtained $ugriz$ magnitudes from the SDSS database, and $W1$, $W2$, $W3$, and $W4$ magnitudes from the AllWISE catalog \citep{Wright2019, Cutri2021}, listed in Table \ref{tab:host magnitudes}. Using gPhoton \citep{Million2016}, we measured a 5" radius aperture magnitude from GALEX images. We then fit the $NUV$, $ugriz$, and $W1$/$W2$ magnitudes using Code Investigating GALaxy Emission \citep[CIGALE;][]{burgarella05, Noll2009, boquien19}.

\begin{table}[t]
    \centering
    \caption{\normalfont Host-galaxy properties resulting from CIGALE fits.}
    \vspace{1mm}
    \begin{tabular}{SS} \toprule
        {Property} & {Value} \\ \midrule
        {Age (Gyr)} & {$3.8 \pm 0.1$} \\
        {$M_{\textrm{stellar}} (10^{10}M_{\odot})$} & {$1.1 \pm 0.1$} \\
        {SFR ($M_{\odot}/$yr)} & {$0.12 \pm 0.02$} \\
        {$L_{\textrm{AGN}}$ (erg/s)} & {$(4.2 \pm 0.7) \times 10^{41}$} \\
        {$L_{\textrm{stellar}}$ (erg/s)} &  {$(3.7 \pm 0.2) \times 10^{43}$} \\
        {$L_{\textrm{AGN}}/L_{\textrm{stellar}}$} & {$(1.1 \pm 0.2) \times 10^{-2}$} \\ \bottomrule
    \end{tabular}
    \label{tab:host galaxy results}
\end{table}

\begin{figure}[ht]
\centering
 \includegraphics[width=0.47\textwidth]{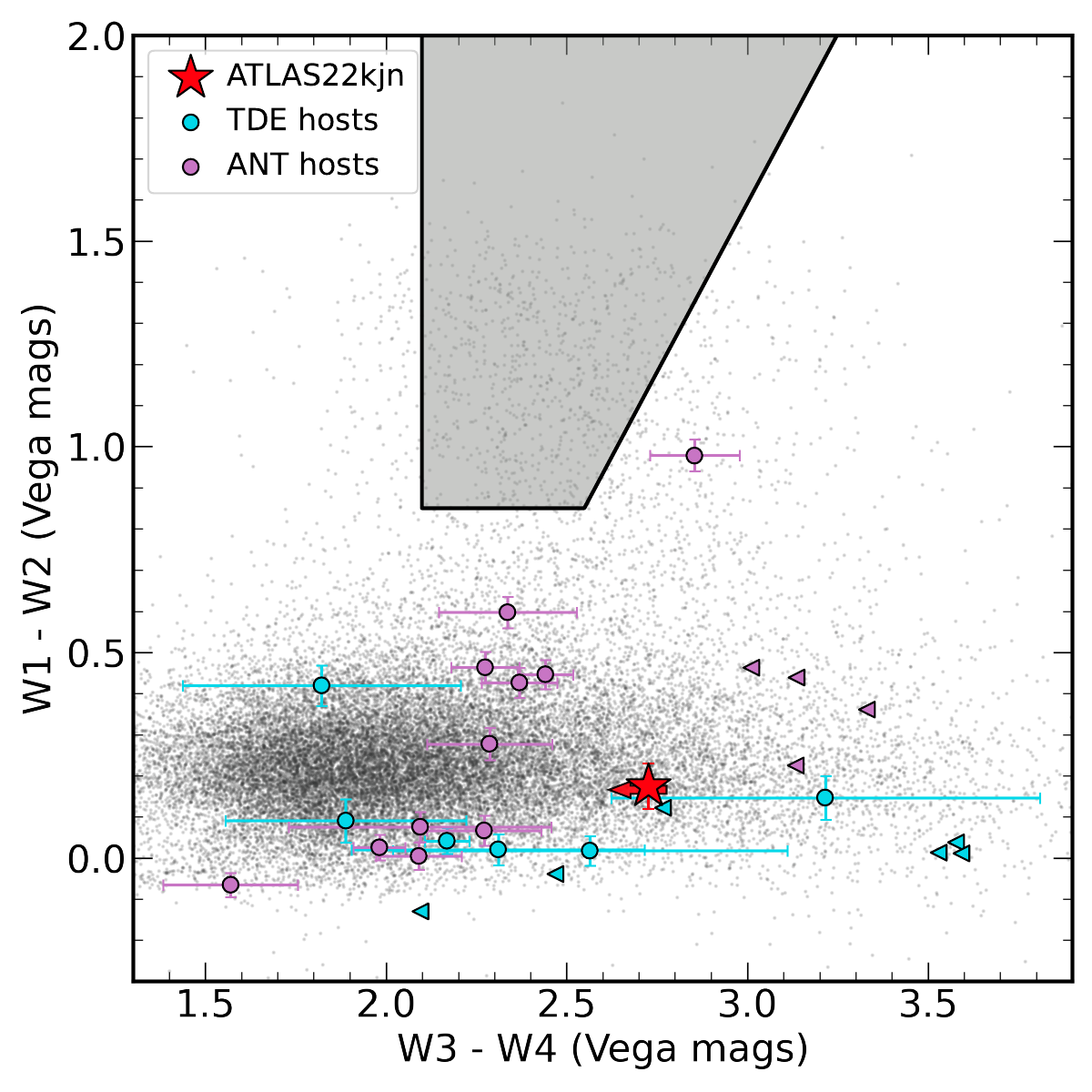}
 \caption{WISE colour-colour diagram used to distinguish between highly obscured AGNs and star formation. The grey-shaded box is the AGN region from \protect\cite{assef13}. The TDE hosts (blue) and ANT hosts (purple) are from \protect\cite{hinkle21b}. ATLAS22kjn is plotted as the red star. The red arrow shows the upper limit $(W3-W4) < 2.73$ mag. Sample background galaxies from HyperLeda \citep{makarov14, Treiber_2023} are plotted as small grey points behind the identified hosts. Upper limits for hosts are plotted as left-pointing triangles. None of the TDE and ANT hosts included here lie in the AGN region.}
 \label{fig:wise_hosts}
\end{figure}

The resulting stellar mass, star formation rate (SFR), and AGN fraction $L_{\textrm{AGN}}/L_{\textrm{stellar}}$ for the host galaxy are presented in Table \ref{tab:host galaxy results}. This AGN fraction of $L_{\textrm{AGN}}/L_{\textrm{stellar}}=(1.1 \pm 0.2) \times 10^{-2}$ indicates the galaxy does not host a strong AGN, but is consistent with a low-luminosity AGN \citep[LLAGN; e.g.,][]{tozzi06, ho08, marchesi16, liu17, ricci17, Yuan_2014}. Similarly, the AllWISE colour $(W1-W2) = 0.17 \pm 0.06$ mag is too blue for there to be significant AGN emission. This result is also shown in Figure \ref{fig:wise_hosts}, where the host galaxy of ATLAS22kjn lies outside of the strong AGN region described in \cite{assef13}. As the galaxy is not detected in $W4$, its location on the WISE colour-colour diagram suggests the galaxy is dominated by stellar light without a strong AGN component. \citep{wright10}.

Using the $M_{\textrm{BH}}$--$M_{\textrm{stellar}}$ scaling relation from \cite{reines15}, we estimate a SMBH mass of $\log (M_{\textrm{BH}}/M_{\odot}) = 6.44 \pm 0.13$, with additional intrinsic scatter of 0.24 dex. This agrees with the estimates by \cite{Koljonen2024} and \cite{Lin_2025}, who used the same scaling relation with different stellar masses. 
Our first estimate also agrees with $\log (M_{\textrm{BH}}/M_{\odot}) = 6.82 \pm 0.58$ using the $M_{BH}-M_{\star}$ relation derived by \citep{Greene_2020}. This relation includes dynamically-measured stellar masses of $M_{\star} < 10^{10} M_{\odot}$ and was determined by \cite{Mummery_2023} to agree with their scaling between TDE plateau-luminosity and SMBH mass.

\subsection{Photometry}
\label{sec:photometry}

\begin{figure*}[t]
\centering
 \includegraphics[width=1\textwidth]{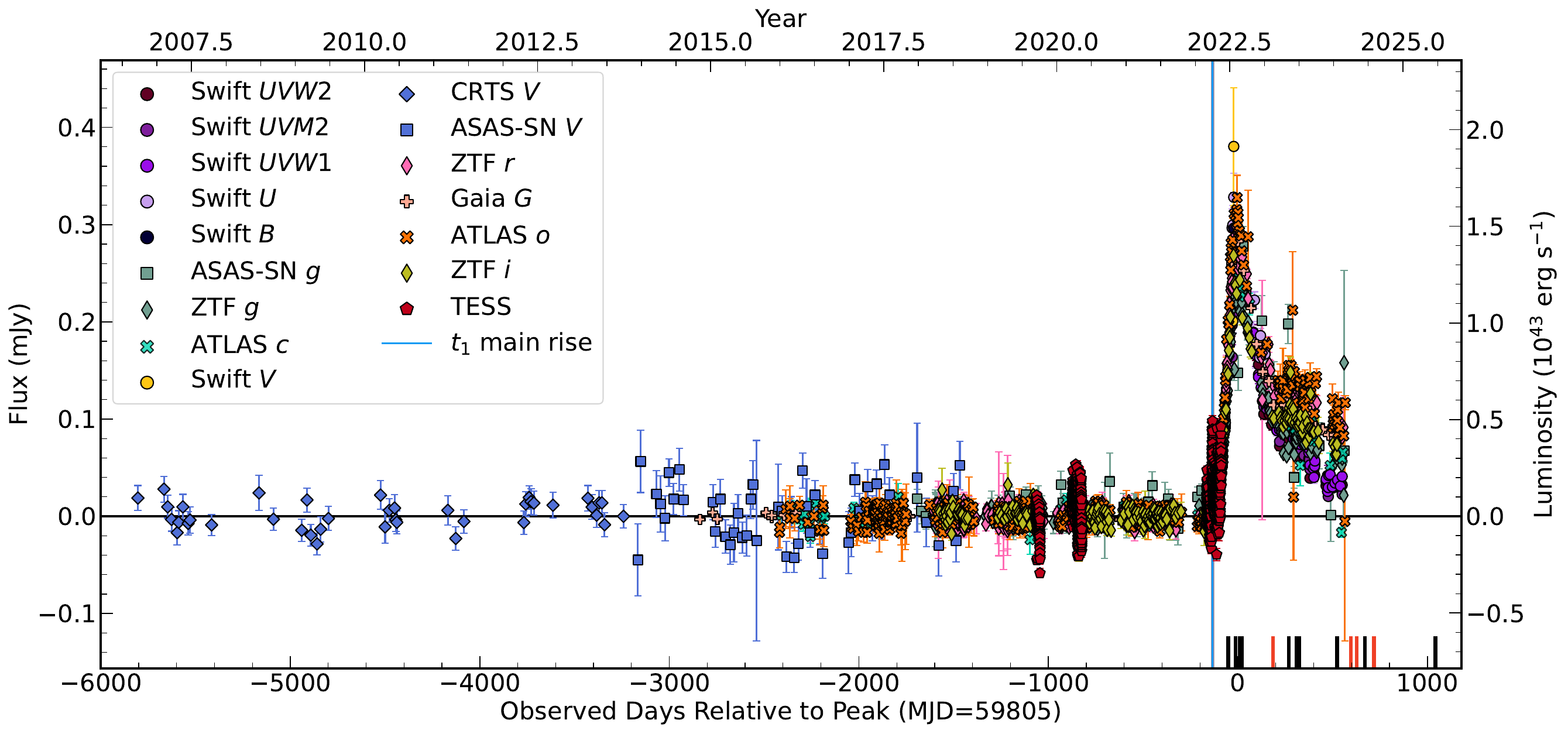}\hfill
 \caption{Long-term light curve of ATLAS22kjn.
 All data have been host-subtracted and corrected for Milky Way extinction. The CRTS $V$-band data have been stacked in 2-day bins, the ATLAS data in 4-hr bins, the ASAS-SN data in 500-hr bins, and the TESS data in 2-hr bins. 
 Black tick marks at the bottom of the plot indicate optical spectral observations, while red tick marks indicate NIR spectral observations. The blue vertical line is the estimated rise start time $t_1$ determined in Section \protect\ref{sec:Early-Time Rise Modelling}. The bump in the light curve prior to the main flare occurs around this time (see Figure \ref{fig:bump lightcurve} for a zoom-in). The luminosity axis is $\nu L_{\nu}$, computed for the TESS band.
 }
 \label{fig:overall light curve}
\end{figure*}

Below, we describe the ground- and space-based photometry used in this paper. UV/optical photometric data in this section are shown in Figure \ref{fig:overall light curve}, up to $\textrm{MJD}=60370.9$. We corrected all data for Galactic foreground extinction \citep{schlafly11}.

We obtained ATLAS \citep{tonry18, Smith_2020} light curves in the $c$- and $o$-bands from the ATLAS forced point-spread function (PSF) photometry service\footnote{\url{https://fallingstar-data.com/forcedphot/}} \citep{Shingles_2021}. The forced photometry service computes forced PSF photometry on difference images, with zero points calculated from nearby reference stars. For both bands, we fit a flat line to the pre-flare flux and subtracted this from the light curves to remove any host galaxy contribution. In Figure \ref{fig:overall light curve}, we show the data stacked in 4-hr bins, and exclude pre-flare data that is more than one standard deviation from the mean flux.

We reduced ASAS-SN images using the automated ASAS-SN pipeline, which uses the ISIS package \citep{alard98, Alard2000}. We created a reference image using ASAS-SN $g$-band images of high-quality taken with different cameras over a number of years, excluding those taken during the flare (beginning $\sim 180$ days before UV/optical peak). The completed reference image was then subtracted from all the $g$-band data. As the ASAS-SN $V$-band light curves ended in 2018 before the rise of ATLAS22kjn, we used the default ASAS-SN references to reduce this data. To produce differential light curves from this data, we employed the IRAF \texttt{apphot} package with an aperture radius of 2 pixels, corresponding to $\sim$16", and performed aperture photometry on each of the subtracted images. We calibrated photometry from this using the AAVSO Photometric All-Sky Survey \citep{Henden2015}. Finally, we stacked all pre-flare photometry in 20-day bins, excluding any measurements with FWHM $\ge 1.67$ pixels. In Figure \ref{fig:overall light curve}, we show the data stacked in 500-hr bins. Pre-flare, we exclude data that is more than one standard deviation from the mean flux.

We retrieved observations of ATLAS22kjn in the $g$-, $r$-, and $i$-bands from the ZTF forced-photometry service \citep{masci19, Masci_2023}. For each band, we fit a flat line to the pre-flare flux and subtracted this from the light curves to remove any host galaxy contribution. In Figure \ref{fig:overall light curve}, for the pre-flare data, we exclude data that is more than one standard deviation from the mean flux.

Data for the host of ATLAS22kjn was taken in the CSS $V$-band and covers $\textrm{MJD} \sim 54000$ to $\textrm{MJD} \sim 56600$, before the transient. We retrieved these observations from the CSS positional cone search\footnote{\url{http://nunuku.caltech.edu/cgi-bin/getcssconedb_release_img.cgi}}. We subtracted the average flux from this data to account for host galaxy contamination.

Between 2022 July 2 and 2024 July 15, seventy Neil Gehrels Swift Gamma-ray Burst Mission \citep[\emph{Swift};][]{gehrels04} target of opportunity (ToO) observations were executed for ATLAS22kjn (Swift Target ID 15260; PIs: Jiang, Guolo, Lin). These observations were taken using the UltraViolet and Optical Telescope \citep[UVOT;][]{roming05}. Observations were obtained in the Swift $UVW2$, $UVM2$, $UVW1$, $U$, $B$, and $V$ bands. UVOT data contains multiple images per epoch in each filter. We combined these into a single image per filter, using the HEASoft \texttt{uvotimsum} package. We extracted source counts with a 5\farcs0 radius region centered on ATLAS22kjn's position. We extracted background counts using a source-free region with a radius of $\sim$ 50\farcs0. Using calibrations from \cite{poole08} and \cite{breeveld10}, we then translated these count rates into flux and magnitude measurements. To eliminate any host-galaxy contamination of the photometry, we computed host galaxy fluxes by performing synthetic photometry using the best-fit host galaxy SED, then subtracted the host flux from each filter.

The Transiting Exoplanet Survey Satellite \citep[TESS;][]{ricker15} has observed the location of ATLAS22kjn in a total of 6 sectors to date. ATLAS22kjn was observed in sectors 16, 23, and 24 (Cycle 2, prime mission) prior to its flare, and sectors 49, 50, and 51 (Cycle 4, extended mission 1) during its rise. We calibrated the light curves for sectors 49, 50, and 51 on a sector-by-sector basis. For each sector, we applied a Savitzky-Golay filter \citep{Savitzky_Golay_1964} with a 5-day window to smooth the TESS light curve. We then linearly interpolated the smoothed data to the epochs of ZTF $i$-band observations. Since no contemporaneous ZTF $i$-band data were available for sector 49, we interpolated the smoothed sector 49 data to the ATLAS $o$-band instead. We then fit a model to estimate the flux offset between the interpolated TESS points and the ZTF $i$-band points. For sector 50, we fit a model with both an additive and multiplicative offset, while for sectors 49 and 51, the best-fit multiplicative offset artificially compressed the scatter in the data; we therefore re-fit sectors 49 and 51 with only an additive offset.

We retrieved additional space-based photometry for ATLAS22kjn from Gaia Photometric Science Alerts (``Gaia Alerts''), in the Gaia $G$-band.
After retrieving Gaia photometry in average magnitudes, we discarded all `null' and `untrusted' measurements. We estimated uncertainties on these measurements using the ESA-provided Python script for Gaia EDR3 photometry\footnote{\url{https://www.cosmos.esa.int/web/gaia/fitted-dr3-photometric-uncertainties-tool}}. We then converted our magnitudes and uncertainties to flux in mJy, as well as AB magnitudes, using the equations specified by the external calibration section of the Gaia Data Release Documentation\footnote{\url{https://gea.esac.esa.int/archive/documentation/GDR2/Data_processing/chap_cu5pho/sec_cu5pho_calibr/ssec_cu5pho_calibr_extern.html}}. We fit a flat line to the pre-flare flux and subtracted this from the light curves to remove any host galaxy contribution.

We obtained $W1$ and $W2$ magnitudes for ATLAS22kjn from the single exposure catalogue in the NEOWISE 2023 Data Release \citep{mainzer23}, and combined the per-visit exposures using weighted averages. Many nuclear transients, particularly those in gas-rich environments like CLEs, show MIR flares consistent with transient-heated dust. To search for this reprocessing ``echo'' and constrain the nuclear dust content of ATLAS22kjn's host, we followed \cite{hinkle22b}, subtracting transient flux from post-flare light curves to isolate dust emission. We estimated the host galaxy flux by fitting a flat line to 16 WISE epochs before $\textrm{MJD}=59500$, estimating uncertainties by summing the standard error on the flux and the median single epoch flux uncertainty in quadrature. This yielded host magnitudes of $W1 = (17.76 \pm 0.02) \ \textrm{mag}$ and $W2 = (18.14 \pm 0.07) \ \textrm{mag}$, consistent with the AllWISE values (Table \ref{tab:host magnitudes}). We subtracted these values from the light curves to subtract the host-galaxy contribution, and used our bolometric light curves (Section \ref{sec:BB modelling}) to estimate and subtract the transient contribution at each epoch.

\subsection{X-ray Observations}
\label{sec:xray}

In addition to the UVOT observations, \textit{Swift} obtained simultaneous X-ray observation using the \textit{Swift} X-ray Telescope \citep[XRT;][]{burrows05} in photon-counting mode. All observations were reprocessed with the standard filter and screening criteria\footnote{\url{https://swift.gsfc.nasa.gov/analysis/xrt_swguide_v1_2.pdf}} as well as the most recent calibration files from level one XRT data using \textsc{XRTPIPELINE} version 0.13.7. To increase the signal-to-noise (S/N) of our observations, we combined individual \textit{Swift} observations into 23 time bins using \textsc{XSELECT} version 2.5b. To extract both background-subtract count rates and spectra, we used \textsc{XRTPRODUCTS} and a region with a radius of 30\farcs0 centered on the position of ATLAS22kjn and a source-free background region centered at ($\alpha,\delta)$=(15:30:18.7036,+53:24:31.351). Extracted count rates were aperture corrected (see e.g., \citealt{moretti04}). We merged our \textit{Swift} observations into two time bins consisting of observations 00015260024-00015260046 and 00015260047-00015260071 so that we can extract two X-ray spectra at different phases of the X-ray evolution. Ancillary response files were generated using \textsc{XRTMKARF} and a merged exposure map that was produced by merging individual exposure maps using \textsc{XIMAGE} version 4.5.1. We also used the ready-made response matrix files from the \textit{Swift} calibration files, while each spectrum was grouped using the \textsc{FTOOLS} command \textsc{GRPPHA} to have a minimum of 10 counts per energy bin.

ATLAS22kjn was observed twice by the \textit{XMM-Newton Observatory}, once on 2023-02-10 (ObsID:0894200301, PI:Liodakis) and once on 2023-05-07 (ObsID:0894200401, PI:Liodakis). We reduced these observations using the \textit{XMM-Newton} science system (SAS) version 20.0.0, standard event screening\footnote{See \url{https://www.cosmos.esa.int/web/xmm-newton/sas-threads} and \url{https://xmm-tools.cosmos.esa.int/external/xmm_user_support/documentation/uhb/XMM_UHB.pdf} for more details} and with the most up-to-date calibration files. Due to the increased sensitivity of the PN detector in comparison to the MOS detectors, we only use the PN data for our analysis. As \textit{XMM-Newton} suffers from periods of high background and/or proton flares, we generated count rate histograms in the 10-12 keV energy range to identify and remove these events. We find that these observations are partially affected, given an effective exposure time of 9.4\,ks, and 8.9\,ks, respectively. All files were corrected for vignetting using \textsc{EVIGWEIGHT}. We find significant X-ray emission from ATLAS22kjn in the observations taken on 2023-05-07 and no X-ray emission in the first \textit{XMM-Newton} observation, consistent with the \textit{Swift} XRT. Using a 20\farcs0 radius region centered on ATLAS22kjn and a 80\farcs0 radius background region centered at $(\alpha, \delta)$=(15:31:14.8930,+53:25:25.954), we extracted a spectrum from the PN observation of 0894200401 using the SAS task \textsc{EVSELECT} and similarly grouped this spectrum using \textsc{GRPPHA} and a minimum of 10 counts per energy bin. Both the \textit{Swift} XRT and \textit{XMM-Newton} spectra were analysed using the X-ray spectral fitting package (XSPEC) version 12.13.1 and $\chi^2$ statistics.

\subsection{Spectroscopic Observations}
\label{sec:spectroscopic observations}

We retrieved the classification spectrum for ATLAS22kjn from TNS. This spectrum was taken approximately 50 days prior to peak by the SPectrograph for the Rapid Acquisition of Transients \citep[SPRAT;][]{piascik14} on the 2-m Liverpool Telescope \citep{steele04}. With this spectrum, \cite{ATLAS22kjn_classification} classified ATLAS22kjn as a TDE, reporting strong Balmer and helium emission as well as rising flux in the light curves. We initiated spectroscopic follow-up observations and obtained a second spectrum of ATLAS22kjn with the SuperNova Integral Field Spectrograph \citep[SNIFS;][]{lantz04} on the 2.2m University of Hawai`i telescope (UH2.2m) as part of the SCAT survey \citep{Tucker2022, Tucker_2026} about 40 days later. All additional spectra were taken after the peak, including four more spectra with SNIFS, one with the Low-Resolution Imaging Spectrometer \citep[LRIS;][]{oke95} on the 10-m Keck II telescope, three with the Keck Cosmic Web Imager \citep[KCWI;][]{Morrissey2018} on the 10-m Keck II telescope, and one with the Gemini-North Multi-Object Spectrograph \citep[GMOS;][]{hook04} on the 8.1-m Gemini North telescope. All spectra were reduced following the standard steps of bias-subtraction, flat-fielding correction, wavelength calibration using arc lamps, and flux calibration. This reduction process was done using the Spectroscopic Classification of Astronomical Transients reduction pipeline \citep[SCAT;][]{Tucker2022} for the SNIFS data, using the Python Spectroscopic Data Reduction Pipeline \citep[\textsc{pypeit};][]{pypeit:zenodo, pypeit:joss_arXiv, pypeit:joss_pub} for the LRIS data, using Data Reduction for Astronomy from Gemini Observatory North and South \citep[DRAGONS;][]{Labrie2023} for the GMOS data, and using the KCWI data reduction pipeline\footnote{\url{https://kcwi-drp.readthedocs.io/en/latest/}}. To improve our initial flux calibration computed from standard star observations, we mangled all our spectra to match multiple bands of follow-up photometry at the same phase, with a linear fit to the colour different, and then corrected for Milky Way extinction with $R_V = 3.1$ \citep{schlafly11}. All spectroscopic observations are described in Table \ref{spectra table} of the Appendix.

In addition to the optical spectra, we obtained four near-infrared (NIR) spectra with SpeX \citep{Rayner2003} on the NASA Infrared Telescope Facility (IRTF). These spectra were reduced using SpeXtool \citep{Cushing2004}, with flux and telluric calibrations done using an A0 standard taken on the same night and at a similar airmass. We corrected these spectra for Milky Way extinction as was done for the optical spectra. 

\section{Analysis} 
\label{sec:Analysis}

Here we analyse the early-time rise to UV/optical peak in Section \ref{sec:Early-Time Rise Modelling}, our UV/optical blackbody SED model in Section \ref{sec:BB modelling}, the early-time bump in the UV/optical light curves in Section \ref{sec:bump_analysis}, the MIR light curves in Section \ref{sec:mir light curve}, our spectral fits in Section \ref{sec:opt spectra}, the evolution of the CLs in Section \ref{sec:CL evolution}, and our X-ray blackbody model in Section \ref{sec:Analysis-Xray}. 

\subsection{Early-Time Rise Modelling}
\label{sec:Early-Time Rise Modelling}

\begin{figure}[t]
\centering \includegraphics[width=0.48\textwidth]{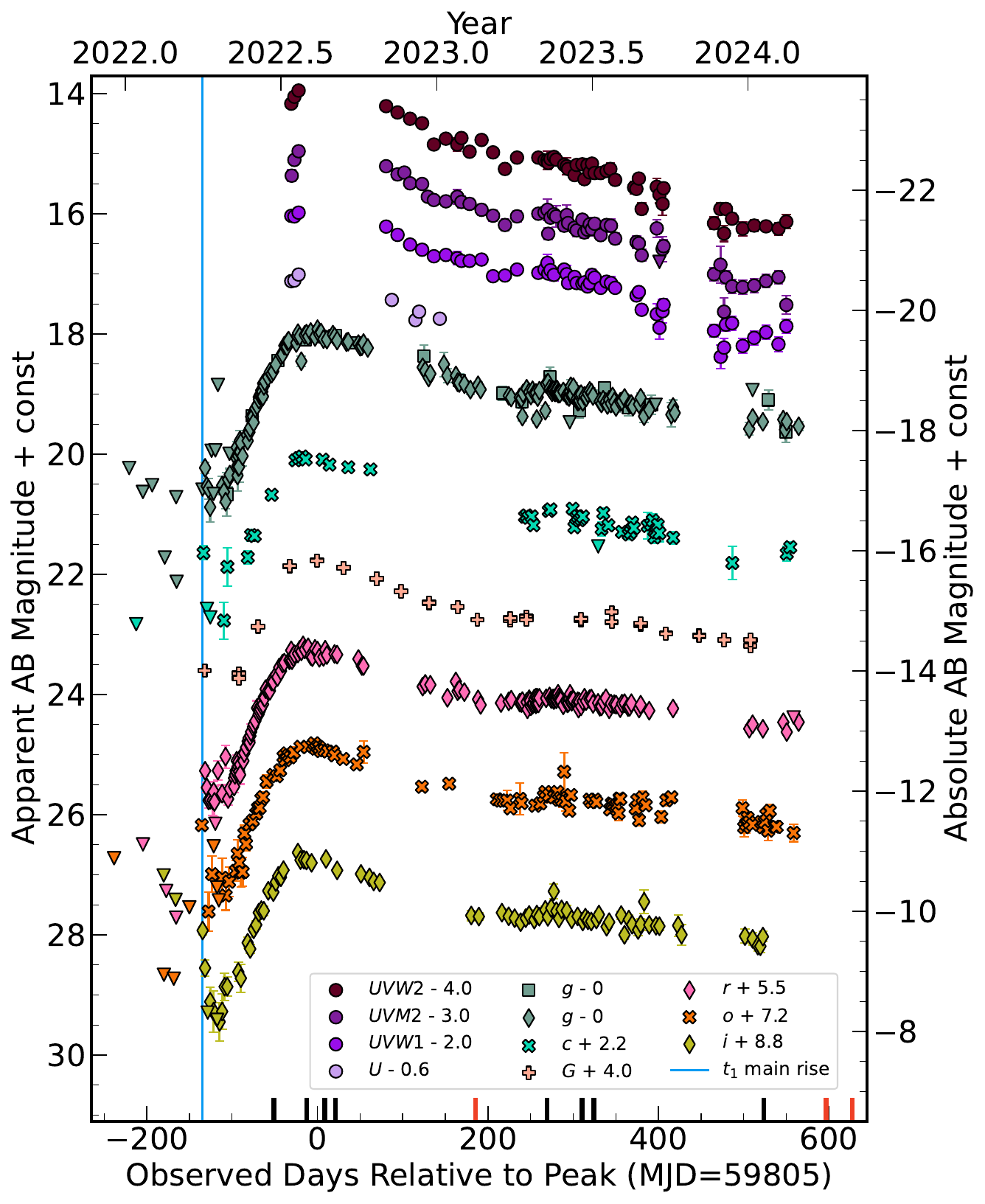}
 \caption{Host-galaxy subtracted, Milky Way extinction-corrected UV and optical light curves of ATLAS22kjn. Photometry from the ASAS-SN $g$-band is shown as green squares, ATLAS $c$ as cyan x's and $o$ as orange x's, ZTF $g$ as green diamonds, ZTF $r$ as pink diamonds, and ZTF $i$ as olive diamonds, Gaia $G$ as plus-signs, and Swift $UV$+$U$ as circles in various shades of purple. and ASASSN $g$-band is stacked in 30-day bins. Prior to the flare, ATLAS and ZTF data are stacked in 10-day bins. Some error bars are smaller than the points. During flare, ATLAS and ZTF are stacked in 4-hour bins.  Black (red) tick at the bottom of the plot mark where optical (NIR) spectra were obtained. The blue vertical line is the rise start time $t_1$, estimated in Section \protect\ref{sec:Early-Time Rise Modelling}; these are the same as shown in Figure \ref{fig:overall light curve}. The bump in the light curve prior is also around this time (see Figure \ref{fig:bump lightcurve} for a zoom-in).}
 \label{fig:flare lightcurve}
\end{figure}

Figure \ref{fig:overall light curve} shows the long-term UV/optical light curve of ATLAS22kjn from $\sim$5700 days before discovery, until $\sim$810 days post-discovery. 
The light curve shows a long-lived flare of $\sim900$ days which began at $\textrm{MJD}\sim59670$ and has yet to return to continuum levels at the end of the data used here ($\textrm{MJD}=60370.9$).
We isolate the flare emission in magnitude space in Figure \ref{fig:flare lightcurve}. We highlight the bands with the most complete coverage in this plot, excluding the Swift $B$- and $V$- band points due to their lower temporal sampling.

\begin{figure*}[t]
\centering 
\includegraphics[width=1\textwidth]{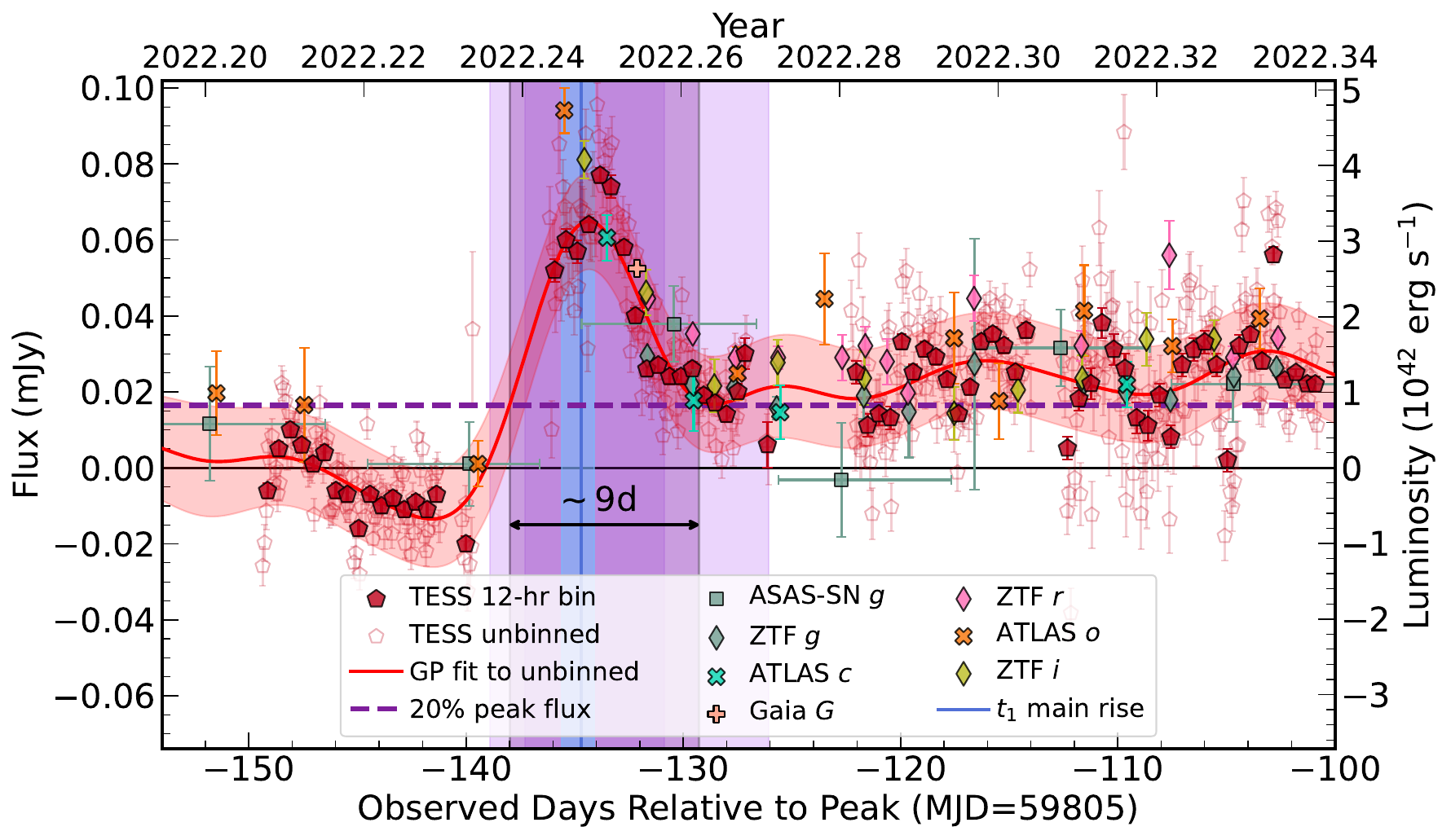} 
 \caption{
 A zoom-in of the bump, including our GP fit to the unbinned 10-minute cadence TESS light curve during the bump (Section \protect{\ref{sec:bump_analysis}}). Though we fit to the TESS data with the main rise model subtracted (see Section \ref{sec:Early-Time Rise Modelling}), for visual consistency between the different bands, we plot all bands without main-rise subtraction. In addition to the unbinned data (empty red pentagons), we plot the data stacked in 12-hr bins (solid red pentagons) to highlight the shape of the light curve. The purple dashed line is 20\% of the peak flux in the GP fit, and corresponds to the $\sim 9$ day duration shown with the central purple shaded region. The purple shaded region is where the GP model is brighter than 20\% of its peak. We plot purple shading for the shortest and longest reasonable bump durations, determined in Section \protect{\ref{sec:bump_analysis}}.
 The ASAS-SN $g$-band data is stacked in 200-hr bins, ZTF $g,r$ -band and ATLAS $c,o$ -band in 6-hr, and TESS in 12-hr bins. There is one Gaia $G$-band data point during this time. The blue line and shading are the calculated start of the transient $t_1$ and $1\sigma$ uncertainty (Section \protect\ref{sec:Early-Time Rise Modelling}), and is the same as shown in Figures \ref{fig:overall light curve} and \ref{fig:flare lightcurve}. The luminosity axis is defined as in \ref{fig:overall light curve}.}
 \label{fig:bump lightcurve}
\end{figure*}

As shown in Figure \ref{fig:bump lightcurve}, there is a bump in the light curves starting at $\textrm{MJD}\sim59667$ and declining by $\textrm{MJD}\sim59690$, prior to the main UV/optical peak. 
\cite{Koljonen2024} first identified this feature as a `precursor' from the ZTF $g$- and $r$- band light curves, which only captured the decline. With the addition of the TESS data, we resolve both the rise and decline of the feature, and therefore refer to it as the pre-peak ``bump". We analyse the bump further in the following section.

To constrain the rise to the peak of the main flare, we fit each UV/optical light curve from before the flare until the peak of the event.  
We used Markov Chain Monte Carlo (MCMC) methods in \texttt{emcee} \citep{ForemanMackey2013}, and the \cite{Vallely2020} curved power-law model
\begin{equation} \label{eq:power law}
   f(t) = f_0 + \frac{h}{(1+z)^2} \left(\frac{t-t_1}{1+z}\right)^{a_1 \big(1+a_2 \cdot (t-t_1)/(1+z)\big)} \ .
\end{equation}
The factor $1+z$ accounts for time dilation, $h$ is a flux scale, $f_0$ is any residual background flux, $t_1$ is the time at which the rise begins, and $a_1$ and $a_2$ are the power-law indices. The $a_2$ parameter allows for the model to follow the curvature of the light curve peak, and is related to the rise time by
\begin{equation}
    a_2^{-1} = -t_{\textrm{rise}} [1+\ln(t_{\textrm{rise}})] \ .
\end{equation}
Because this model does not account for the bump, we exclude the flux measurements between $\textrm{MJD}=59667-59690$ from our fits.

\begin{figure}[t]
\centering
 \includegraphics[width=0.47\textwidth]{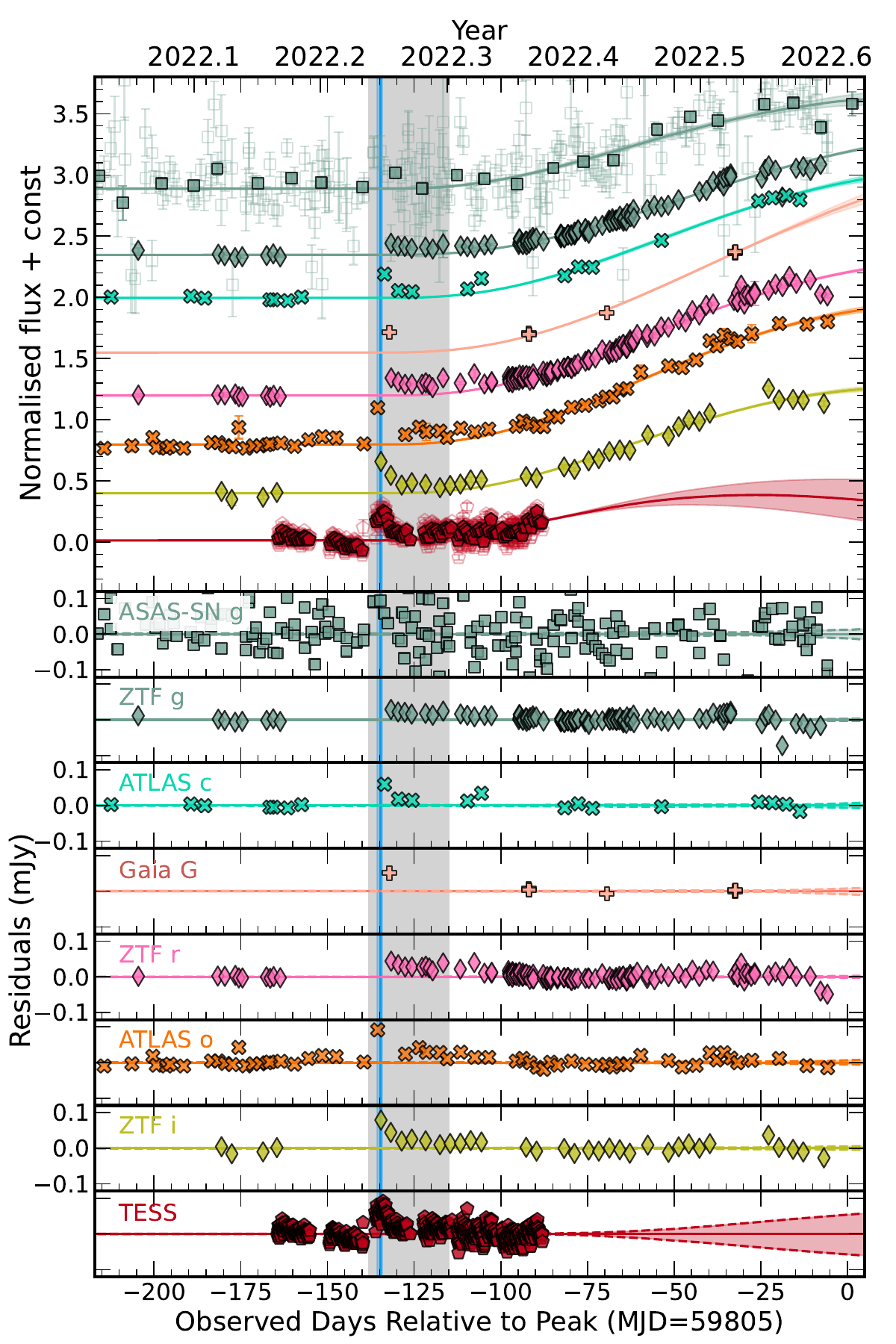}\hfill
 \caption{Early-time power-law fits (top) and residuals (bottom), where the colours and symbols are the same as those in Figures \ref{fig:overall light curve}, \ref{fig:flare lightcurve}, and \ref{fig:bump lightcurve}. The solid lines show the median models, while the shaded regions show the 1$\sigma$ spread in the models. The vertical grey shading indicates the approximate duration of the bump, which we exclude from these fits. The blue vertical line and shading is the calculated rise start time $t_1$ and is the same as in Figures \ref{fig:overall light curve}, \ref{fig:flare lightcurve}
 and \ref{fig:bump lightcurve}. For the ASAS-SN $g$- and TESS bands, we show the data used in fitting as empty points, and show the binned data as solid points to highlight the shape of the light curve. These use 200-hr (12-hr) bins for the ASAS-SN $g$-band (TESS) data. The bottom panels show residuals for each band in mJy.}
 \label{fig:rise_modelling}
\end{figure}

We simultaneously fit the eight pre-peak light curves with the best sampling---ASAS-SN $g$, ZTF $g$, ZTF $r$, ZTF $i$, ATLAS $c$ and $o$, Gaia $G$, and TESS.  We treat $f_0$, $h$, $a_1$, and $a_2$ as free parameters for each light curve, while $t_1$ is treated as a shared free parameter across all light curves. 
We use flat priors of $-0.1 \ \textrm{mJy} \le f_0 \le 0.01 \ \textrm{mJy}$, $10^{-7} \ \textrm{mJy} \le h \le 10^{-5} \ \textrm{mJy}$, $0.005 \le a_1 \le 5.0$, and $-0.05 \ \textrm{day}^{-1} \le a_2 \le -0.0005 \ \textrm{day}^{-1}$ for each set of survey data, and $59500 \le t_1 \le 59700$. 
Figure \ref{fig:rise_modelling} shows the results of these fits, and the median model parameters for each band can be found in Appendix \ref{ap:rise_model_params}. The rise start time of $t_1=59670.3 \substack{+0.3 \\ -0.9}$ (MJD) implies that ATLAS22kjn was discovered 1 day before it began to rise, and that the early bump begins before $t_1$. The initial rise slope $a_1 \sim 3$ is consistent across the different bands, suggesting a relatively constant temperature during the rise to UV/optical peak. This slope at constant temperature implies the emission region is accelerating in size.

\subsection{UV/Optical Blackbody Modelling}
\label{sec:BB modelling}

\begin{figure*}[t]
\centering
\includegraphics[width=1\textwidth]{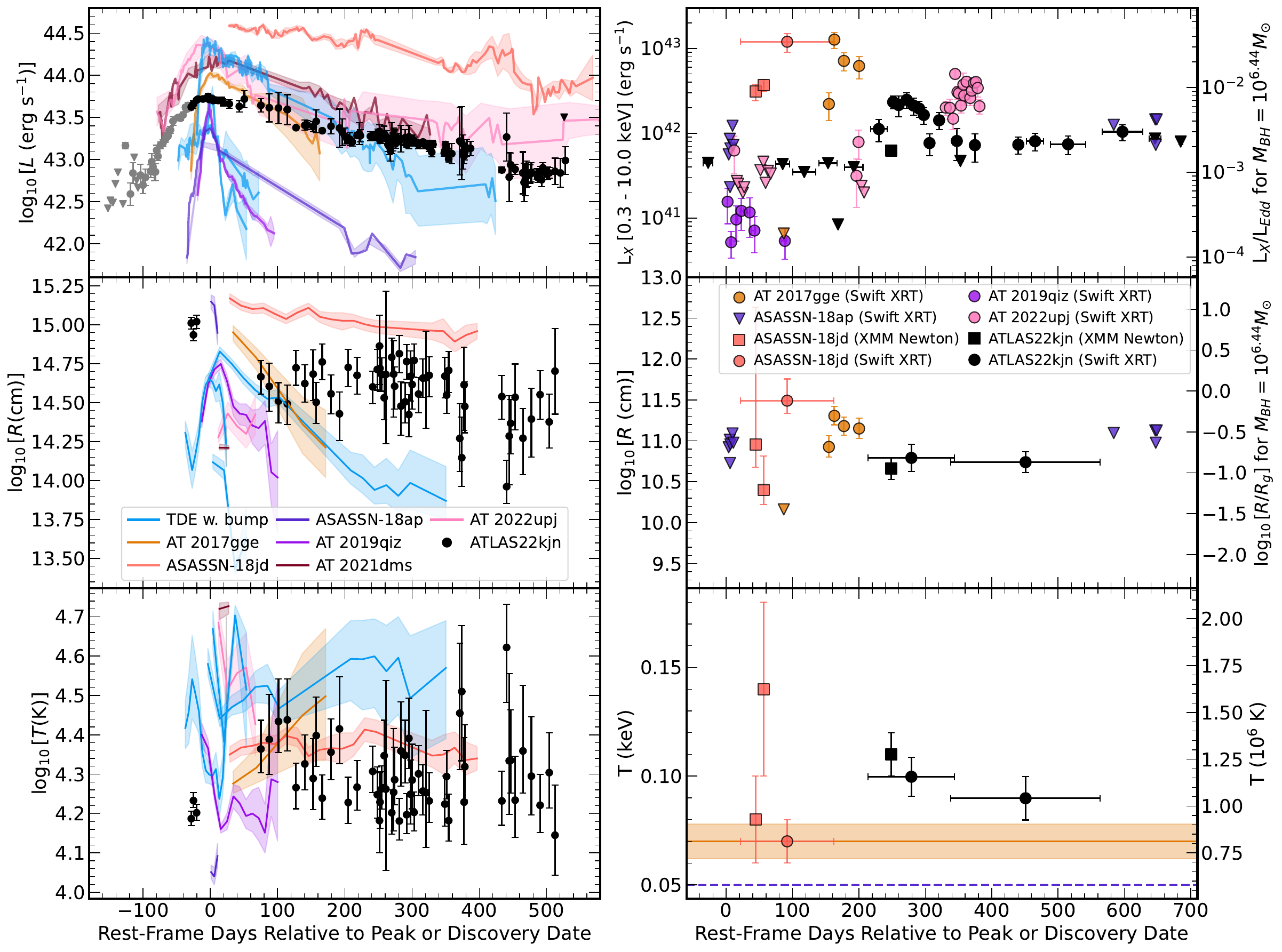}\hfill
 \caption{\textit{Left:} The UV/optical bolometric luminosity (top), blackbody radius (middle), and effective temperature (bottom) evolution of ATLAS22kjn (black points). Grey points in the top panel are where a constant temperature bolometric correction was assumed due to the lack of \textit{Swift} points. The blackbody radius and temperature show a gap due to the lack of \textit{Swift} coverage at this phase; the bolometric luminosity is computed using both \textit{Swift} and scaled ATLAS $o$-band photometry and so does not show this gap.
 The blue lines are other TDEs with a UV/optical bump prior to the main peak. The remaining lines represent CLEs. Objects with purple lines (ASASSN-18ap and AT 2019qiz) have both been classified as CLEs and show a pre-peak bump. Each object is depicted relative to its own peak.
\textit{Right:} The X-ray blackbody luminosity (top), radius (middle)---with a secondary axis with radius as a fraction of the gravitational radius $R_g = GM_{BH}/ c^2 = (4.1 \pm 1.2) \times 10^9 \ \textrm{cm}$ for our SMBH mass estimate of $\log_{10}\left(M_{BH}/M_{\odot}\right) \sim 6.44$ (Section \ref{sec:host})---and temperature (bottom) evolution of ATLAS22kjn (black points). Circles are \textit{Swift} measurements, squares are XMM Newton measurements, and downward-facing triangles show upper limits. Where available, X-ray blackbody measurements are shown for the other CLEs. We do not include blackbody temperature and radius for AT 2022upj \protect\citep{Newsome_2024} or AT 2019qiz \protect\citep{nicholl20} as the X-ray emission was found to be better modelled by a power law. Solid lines with shading are measured blackbody temperatures and $1\sigma$ uncertainties, and dashed lines are assumed temperatures. Each object is depicted relative to its own peak.}
 \label{fig:uvot_xray_bb}
\end{figure*}

The SEDs of optically-selected TDEs are dominated by UV/optical emission that is well-fit by a blackbody, and this is also true of ATLAS22kjn (mean reduced $\chi^2 \sim 1.7$). We use MCMC methods and a flat temperature prior of $10,000 \ \textrm{K} \le T \le 55,000 \ \textrm{K}$ to fit the \textit{Swift} UVOT photometry, which spans $\sim$25 days before UV/optical peak to $\sim$530 days after peak. To construct the bolometric light curve, we scaled the optical light curve to match the \textit{Swift} UVOT bolometric luminosity from our fits \citep[as in e.g.,][]{holoien20, hinkle20a, hinkle21b}. Where optical and UVOT photometry were simultaneous, we interpolated the bolometric correction between \textit{Swift} points. For points before the first \textit{Swift} epoch, we applied a constant bolometric correction by fixing the temperature---as aligns with the constant rise slope $a_1$ determined in Section \ref{sec:Early-Time Rise Modelling}---to the value found for the first \textit{Swift} epoch. Figure \ref{fig:uvot_xray_bb} shows the results compared to other TDEs.
Our comparison sample includes the spectroscopically-selected CLEs ASASSN-18jd \citep{neustadt20}, AT 2021dms \citep{forster21, Hinkle2024}, AT 2022upj \citep{newsome22, fulton22, Newsome_2024}, and AT 2017gge/ATLAS17jrp \citep{wang22b, onori22}. To make comparisons between ATLAS22kjn and other TDEs with pre-peak bumps, we include ASASSN-19dj \citep{hinkle21b,hinkle21a, Faris2024}, AT 2020wey \citep{Charalampopoulos_2023}, AT 2019mha \citep{Wang2024}, and AT 2023lli \citep{Huang_2024}. We also include ASASSN-18ap \citep{Wang2024} and AT 2019qiz \citep{nicholl20, Short2023, Wang2024}, which show both CLs and a bump.

From a quadratic fit to 50 days on either side of the peak bolometric luminosity, we measure a peak luminosity of $(5.4 \pm 0.2) \times 10^{43}$ erg s$^{-1}$ at 
$\textrm{MJD}=59805 \substack{+5 \\ -3}$.
This luminosity lies at the low end of TDE peak luminosities, particularly compared to our comparison sample of CLEs. 
The rise time is therefore $125.4 \substack{+0.3 \\ -0.9}$ rest-frame days.
The fitted blackbody radius peaks at the bolometric luminosity peak and decreases with luminosity, typical for objects classified as TDEs, including the comparison sample shown. The blackbody radius of ATLAS22kjn is larger than others our comparison sample, except for the CLEs ASASSN-18jd and ASASSN-18ap---which also has an early-time bump. At peak, we find the effective temperature to be $\sim 16,000 \ \textrm{K}$, which remains relatively constant over time, in agreement with the findings of \cite{Koljonen2024} and \cite{Lin_2025}. 
The temperature is low compared to those commonly observed for TDEs, but similar to the $18,000 \ \textrm{K}$ temperature found for the CLE AT 2017gge by \cite{onori22}. 

\begin{figure}[t]
    \centering 
    \includegraphics[width=1\linewidth]{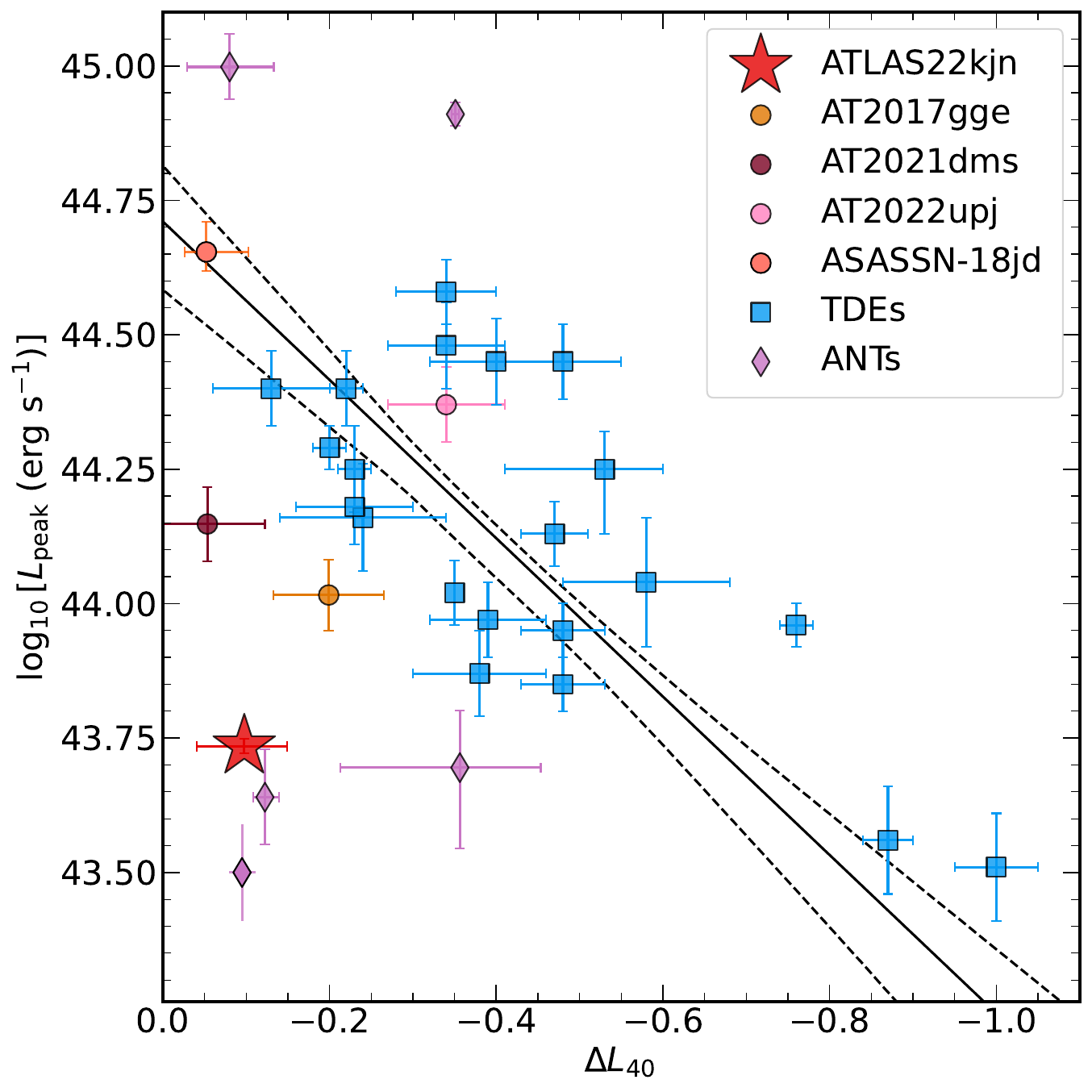}
    \caption{Peak bolometric luminosity vs. decline rate, from \protect\cite{hinkle21b}. $\Delta L_{40}$ is the difference between the log of the peak luminosity and the log of the luminosity at 40 days after peak. ATLAS22kjn is marked by the red star, with a vertical error bar smaller than the point size. The circles with red and orange shades are the CLEs from \protect\cite{Hinkle2024} and use the same colours as in Figure \protect\ref{fig:uvot_xray_bb}. The blue squares are TDEs from \protect\cite{hinkle21b}, and the purple diamonds are ambiguous nuclear transients (ANTs) from \protect\cite{hinkle22a}. The black solid line shows the best-fit relation for TDEs, while the dashed lines represent the 1$\sigma$ uncertainty range. ATLAS22kjn’s decline rate lies well within the CLE range, but its low peak luminosity places it far from the best-fit line for typical TDEs and instead amongst ANTs.}
    \label{fig:d_l40}
\end{figure}

In addition to its slow rise, the decline of ATLAS22kjn is slow compared to most TDEs, as seen in Figure \ref{fig:d_l40} using the luminosity decline rate and peak luminosity relation of \cite{hinkle21b}. Our comparison sample here comprises the TDEs in \cite{hinkle21b} and the CLEs from \cite{Hinkle2024} (also used in Figure \ref{fig:uvot_xray_bb}). ATLAS22kjn is less luminous than most objects in our comparison sample, including the other CLEs. 
Since the \textit{Swift} UV photometry is well-fit by a blackbody, this is unlikely to be an effect of high dust extinction relative to the other TDEs in our comparison sample.

\subsection{Early UV/Optical Bump}
\label{sec:bump_analysis}

We characterise the bump by modelling the unbinned TESS light curve with a Gaussian Process (GP) model using \texttt{scikit-learn} \citep{Pedregosa11}, after subtracting the curved power-law model for the main rise to peak (Section \ref{sec:Early-Time Rise Modelling}). We use the Radial Basis Function kernel \citep{RasmussenWilliams06} for this process. This kernel depends only on the distance between data points, not their absolute positions \citep{RasmussenWilliams06, Duvenaud14}, with $\ell = 4~\text{days}$ to best capture the variability in the data. We used a white noise model and included the baseline flux as a parameter.

This fit is shown in Figure \ref{fig:bump lightcurve}.
The bump peaks $134 \substack{+5 \\ -3}$
observed days
($125 \substack{+5 \\ -3}$ rest-frame days) before the main UV/optical peak with a luminosity $\lambda L_{\lambda} = (3.0 \substack{+ 0.2 \\ -0.4}) \times 10^{42} \ \textrm{erg s}^{-1}$ after subtracting the rise model.
We set a threshold of 20\% of the peak flux in the GP fit to define the overall duration,
finding a bump duration of $9 \substack{+4 \\ -2} \ \textrm{days}$, where the uncertainties span the range of threshold crossings. Integrated over this period, the radiated energy in the TESS band is
$E = (1.5  \substack{+0.9 \\ -0.5})\times 10^{48} \ \textrm{erg}$, with uncertainties corresponding to the range of possible durations.
Using the ZTF bands closest to the peak of the bump, the peak bump colour is $g-r = 0.5 \pm 0.1 \ \textrm{mag}$.
A blackbody produces this colour for a temperature of $7800 \pm 500 \ \textrm{K}$. This temperature is consistent with the pre-peak temperature of $\sim 7500 \ \textrm{K}$ determined by \cite{Lin_2025} at $-21$ days relative to our estimate of UV/optical peak.

Using our estimated $7800 \pm 500$ K with the peak bump luminosity $L = (3.0 \substack{+0.2 \\ -0.4})\times 10^{42} \ \textrm{erg s}^{-1}$ from the bolometrically-corrected TESS light curve implies a blackbody radius of $(1.06 \pm 0.12) \times 10^{15} \ \textrm{cm}$. If we take the rise time of the bump as the time between when the GP model crosses the 20\% threshold and the peak, the rise time is $3.7 \substack{+0.9 \\ -0.7}$ days.
Viewed as an outflow, the blackbody radius and this timescale imply a velocity of $(3.3 \substack{+0.7 \\ -0.9}) \times 10^4 \ \textrm{km s}^{-1}$ ($\sim 0.1c$). 
This velocity is higher than outflow velocities inferred from the optical emission of other TDEs, including $\gtrsim 2000 \ \textrm{km s}^{-1}$ measured from the expansion in the blackbody radius of AT 2019qiz \citep{nicholl20}, and higher than the $\sim 200-9000 \ \textrm{km s}^{-1}$ H$\alpha$ line offsets measured for a sample of TDEs by \cite{Charalampopoulos_2022}. In contrast, outflow velocities $\gtrsim 10^4 \ \textrm{km s}^{-1}$ have been measured from blueshifted absorption lines in the X-ray spectra of some AGNs \citep[e.g.,][]{Tombesi_2011, Gofford_2013, King_Pounds_2015, Igo_2020}. Nevertheless, our inferred velocity is consistent with velocities measured from radio observations of TDE outflows \citep[e.g.,][]{alexander16, alexander17, Anderson_2020, cendes21}, although we note that ATLAS22kjn was not detected in radio observations \citep{Franz_2026}.
Outflow velocities $\sim 10^4 \ \textrm{km s}^{-1}$ are also predicted by theoretical studies. For the case where only a fraction of the stellar debris is accreted while the majority is unbound in an outflow \cite{metzger16} predicted outflow velocities $\ge 10^4 \ \textrm{km s}^{-1}$. Additionally, studying line formation in outflowing gas, \cite{roth18} found if the broad line widths in TDEs are set by electron scattering, lines may be broadened by $\sim 10^4 \ \textrm{km s}^{-1}$, narrowing over time due to decreasing optical depth. We lack spectra at the time of the bump, but since Gaia observed the bump, the future release of low resolution XP (BP/RP) spectra from this time can be used to investigate any outflow-like properties in broad line emission. Furthermore, it is possible that the temperature we determined from the $g-r$ colour is an underestimate, and therefore the bump blackbody radius and velocity are overestimated.

The colour of ATLAS22kjn remains red at the main UV/optical peak. From a simple quadratic fit to each the ZTF $g$- and $r$-band fluxes at $\textrm{MJD}=59805$, using Monte Carlo processes to estimate uncertainties, we find that $g-r = 0.213 \substack{+0.009 \\ -0.008} \ \textrm{mag}$, corresponding to a temperature of $8920 \pm 40 \ \textrm{K}$. This is significantly lower than the $\sim 16,000 \ \textrm{K}$ blackbody temperature we determined in Section \ref{sec:BB modelling}, however, the $g-r$ estimate does not include UV emission as our blackbody fit does.
\cite{Lin_2025} suggested the consistently red colour of ATLAS22kjn may be attributed to high levels of dust extinction from the host galaxy, using the Balmer decrement to estimate reddening under the assumption of Case B recombination \citep[e.g.,][]{osterbrock89}. However, the measured Balmer decrement shows variation and is mostly inconsistent with the Case B recombination value \citep[][see also Section \ref{sec:opt spectra}]{Lin_2025}. The absence of any \ion{Na}{i} D $\lambda \lambda 5890, 5896$ absorption, a reliable tracer of dust along the line of sight \citep[e.g.,][]{Hobbs_1974, Poznanski_2012}, also suggests the line of sight to ATLAS22kjn is not heavily extincted. 
We therefore assume any extinction from the host galaxy to be negligible.

\subsection{MIR Dust Echo}
\label{sec:mir light curve}

\begin{figure}[t]
    \centering
    \includegraphics[width=1\linewidth]{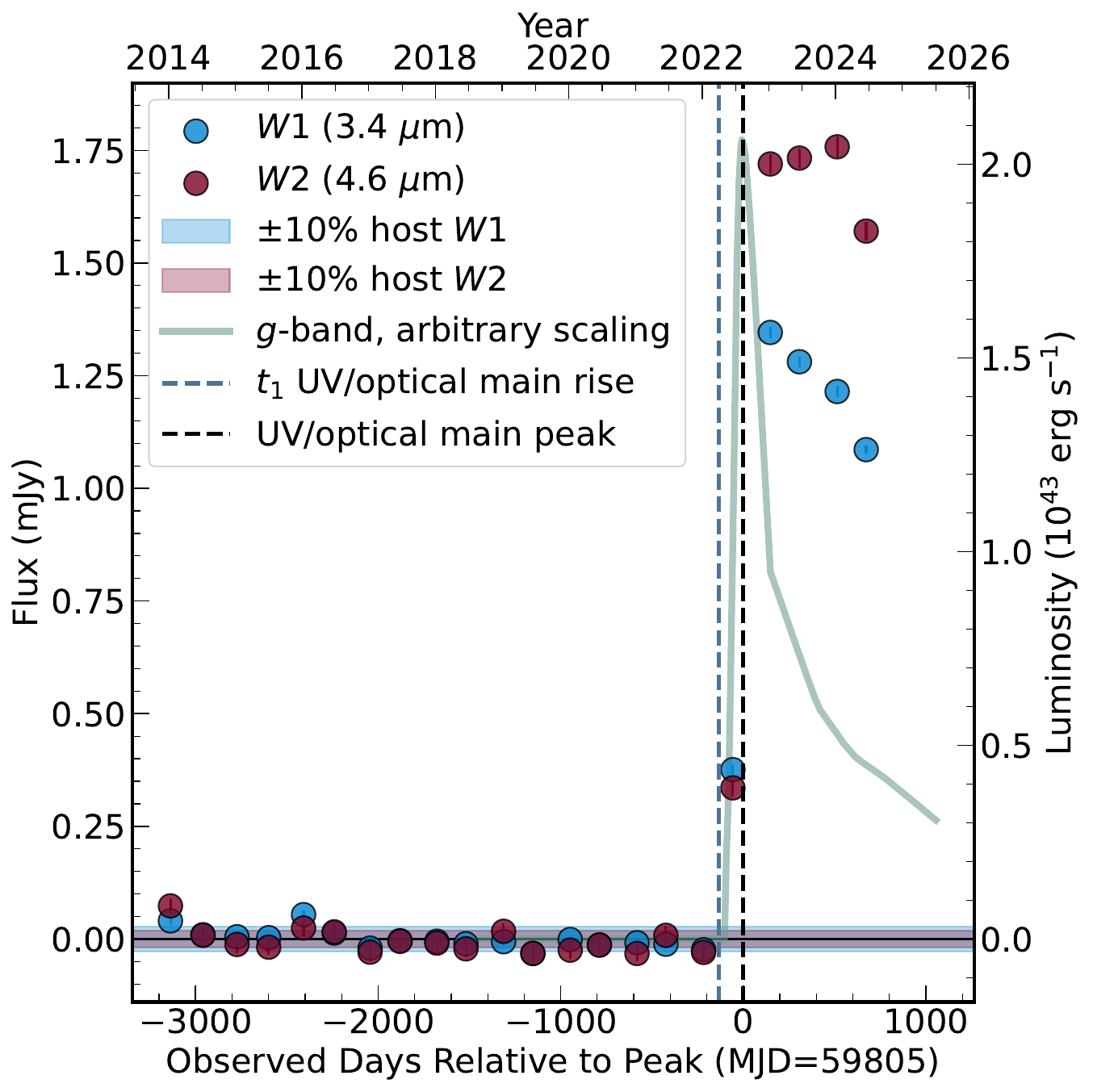}
    \caption{NEOWISE $W1$ (blue points) and $W2$ (maroon points) light curves for ATLAS22kjn. The error bars are smaller than the point sizes. The light curves have had both the host and estimated transient continuum contributions subtracted, leaving the dust-reprocessed emission, and are corrected for Milky Way foreground extinction. The horizontal blue (maroon) shading shows 10\% of the host $W1$ ($W2$) flux around the zero baseline (see Table \protect\ref{tab:host magnitudes}). The green line shows the combined ZTF and ASAS-SN $g$-band light curves, smoothed and scaled arbitrarily to match the peak $W2$ flux. The grey-blue vertical dashed line is the time of UV/optical peak. The black vertical dashed line is the calculated rise start time $t_1$ and is the same as in Figures \ref{fig:overall light curve}, \ref{fig:flare lightcurve}, \ref{fig:bump lightcurve}, and \ref{fig:rise_modelling}. The luminosity axis is $\nu L_{\nu}$, computed for the $W1$ band.}
    \label{fig:mir_light_curve}
\end{figure}

\begin{table}[t]
    \centering
        \caption{\normalfont MIR blackbody parameters from fitting the NEOWISE light curves in Figure \ref{fig:mir_light_curve}.}
    \vspace{1mm}
    \begin{tabular}{cccc} \toprule
        {MJD} & {$\log_{10}[L (\textrm{erg s}^{-1})]$} & {$\log_{10}[T (\textrm{K})]$} & {$\log_{10}[R (\textrm{cm})]$} \\ \midrule
        {59748} & {$42.86 \substack{+0.08 \\ -0.05}$} & {$3.22 \substack{+0.06 \\ -0.05}$} & {$16.56 \substack{+0.07 \\ -0.08}$} \\[1.0ex]
      {59952} & {$43.342 \pm 0.007$} & {$	3.028 \substack{+0.013 \\ -0.012}$} & {$17.19 \pm 0.02$} \\[1.0ex]
      {60112} & {$43.330 \substack{+0.007 \\ -0.006}$} & {$3.006 \substack{+0.012 \\ -0.011}$} & {$17.23 \pm 0.02$} \\[1.0ex]
      {60319} & {$43.324 \substack{+0.008 \\ -0.007}$} & {$2.981 \substack{+0.012 \\ -0.011}$} & {$17.27 \pm 0.02$} \\[1.0ex]
      {60478} & {$43.276 \pm 0.007$} & {$2.981 \substack{+0.010 \\ -0.012}$} & {$17.25 \pm 0.02$} \\ \bottomrule
    \end{tabular}
    \label{tab:MIR_BB}
    \vspace{1mm}
\end{table}

Figure \ref{fig:mir_light_curve} shows the NEOWISE MIR light curves, which show no significant variability prior to the UV/optical flare. We fit the emission as a blackbody, using MCMC methods in a forward modelling approach with a uniform temperature prior of $100 \ \textrm{K} \le T \le 5000 \ \textrm{K}$. The results of these fits are shown in Table \ref{tab:MIR_BB}.
From the ratio of the peak MIR luminosity to the peak optical luminosity, we estimate a dust covering fraction \citep{jiang21a, hinkle22b} of $f_{c} = 0.40 \pm 0.03$. This lies within the range expected for ANTs, $0.05 < f_c < 0.91$  \citep{hinkle22b}, for which the average is $f_c = 0.38 \pm 0.04$ when only detections are considered, or $f_c = 0.29$ when non-detections are included \citep{hinkle22b}.
Compared to the range $0.001 \lesssim f_c \lesssim 0.02$ for TDEs \citep{jiang21b},
our dust covering fraction is $1-2$ orders of magnitude higher, similar to expectations for hot dust in AGNs. This suggests the presence of a dusty torus in the host of ATLAS22kjn \citep{hinkle22b}.

The lag-time between the MIR and UV/optical peaks can be used to obtain a zeroth-order estimate of the distance of the dust from the central SMBH. 
For ATLAS22kjn, we observe a rest-frame lag of $200 \pm 34$ days. Assuming this lag-time corresponds to the light-travel time, the MIR-emitting dust for ATLAS22kjn is a radius of $0.17 \pm 0.03$ pc from the SMBH. 
Luminous AGNs with MIR lags of $\sim200$ days typically have luminosities $\sim 10^{45} \ \textrm{erg s}^{-1}$ \citep[e.g.,][]{Lyu_2019, Li_Shen_2023}. Given our SMBH mass estimate, a $\sim 200$-day light-travel time roughly corresponds to the dust sublimation radius for the SMBH accreting at its Eddington limit. The peak bolometric luminosity determined in Section \ref{sec:BB modelling} is approximately an order of magnitude below this, however, this luminosity does not include the EUV flux component that must necessarily be present to produce the observed CLs (see Section \ref{sec:CL evolution}).

The MIR lag observed for ATLAS22kjn is similar to the $\sim177$-day lag-time estimated by \cite{Hinkle2024} for the CLE AT 2017gge \citep{wang22b, onori22} and the $\sim200$-day lag-time of the TDE PTF-09ge \citep{arcavi14, jiang21b}, both of which have estimated SMBH masses close to that of ATLAS22kjn. In contrast, the $\sim17-21$-day lags estimated by \cite{jiang21b} for the TDEs ASASSN-14li \citep{holoien16a}, PS18kh/AT 2018zr \citep{tucker18_scat, holoien18b, vanvelzen19b}, and ASASSN-18zj/AT 2018hyz \citep{Cendes_2022, Cendes_2026}, which also have comparable SMBH masses to ATLAS22kjn and PTF-09ge, are much shorter. Meanwhile, ASASSN-18pg/AT 2018dyb \citep{leloudas19, Leloudas_2022} shows no lag at all \cite{jiang21b}. These shorter or absent lags suggest the MIR-emitting dust is at a smaller radius than the dust in the environment of ATLAS22kjn.

\subsection{Optical Spectra Fitting and Analysis}
\label{sec:opt spectra}

\begin{figure*}[t]
    \centering
    \includegraphics[width=1\textwidth]{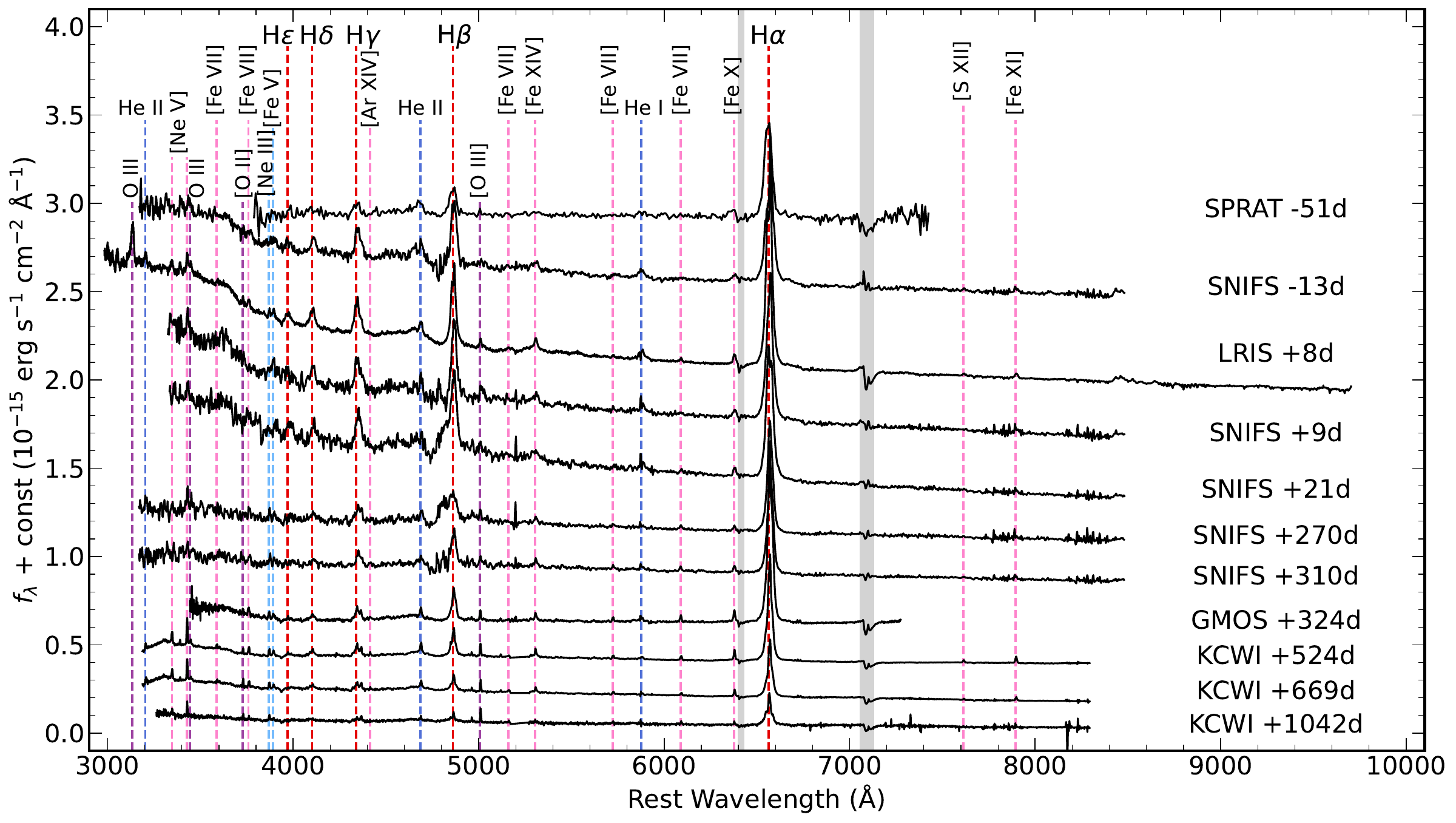}
    \caption{Optical spectra of ATLAS22kjn, corrected for Milky Way extinction and offset by a constant. The spectra are plotted from earliest to latest and labelled with the phase relative to UV/optical peak ($\textrm{MJD} = 59805$) shown. Balmer series lines are indicated by red dashed lines, He lines by blue dashed lines, coronal lines by pink dashed lines, and oxygen features, including [\ion{O}{iii}] $\lambda$5007, by purple dashed lines. Other features that are not coronal are indicated by light blue dashed lines. The grey bands identify prominent telluric features.}
    \label{fig:uv opt spectrum}
\end{figure*}

Figure \ref{fig:uv opt spectrum} shows our optical spectra, corrected for Milky Way extinction and calibrated to the survey photometry of ATLAS22kjn. Our flux calibration for the SPRAT spectrum is more uncertain since the limited wavelength range of the spectrum prevents us from using multiple bands of survey photometry to calibrate the spectrum. 
The spectra range from $-51$ to $+669$ observed days, relative to the UV/optical peak (MJD$=59805$; see Section \ref{sec:BB modelling}).

ATLAS22kjn exhibits the blue continuum emission characteristic of TDEs as well as prominent broad emission lines H$\alpha$, H$\beta$, H$\gamma$, H$\delta$, \ion{He}{i} $\lambda$5876, and \ion{He}{ii} $\lambda$4686. 
Including H$\alpha$ and H$\beta$, most of the broad lines in have two distinct profiles: a broad ($\sim 3000 \ \textrm{km s}^{-1}$) and a narrow ($\sim 600 \ \textrm{km s}^{-1}$) component.
We assume the broad Gaussian component to be due to the transient and the narrow Gaussian to be the host galaxy and LLAGN contribution.

We fit the H$\alpha$ and H$\beta$ lines with the sum of a narrow and a broad Gaussian after subtracting a linear continuum emission model.
Under the assumption that the narrow component is due to the host galaxy and LLAGN, we tightened the priors for the $\sigma$ of the narrow Gaussian to match the 99.7\% CI of the $\sigma$ values for the $\textrm{KCWI} \ + 1042$d spectrum, with an additional margin of $\pm 3 \textrm{Å}$ for H$\alpha$, and $\pm 1.5 \textrm{Å}$ for H$\beta$.
We determined the FWHM and line luminosity of both components for each line and report these measurements and their $1\sigma$ uncertainties in Appendix \ref{ap:spectroscopy_tables}. 
We then fit the luminosity evolution of the broad components using
\begin{equation}
\label{eq:log_luminosity_fit}
    \log(L)= a \cdot \left(\frac{t - t_{\textrm{pk}}}{1+z}\right) \ + \ b \ ,
\end{equation}
where $t_{\textrm{pk}} = 59805$ (MJD), the time of UV/optical peak (see Section \ref{sec:BB modelling}). 
To improve our fits, we inflated our luminosity uncertainties by adding the median uncertainty in quadrature. We used `minimize' in SciPy \citep{scipy}, with a negative log-likelihood function consisting of a log PDF.

\begin{figure}[t]
    \centering    
    \includegraphics[width=1\linewidth]{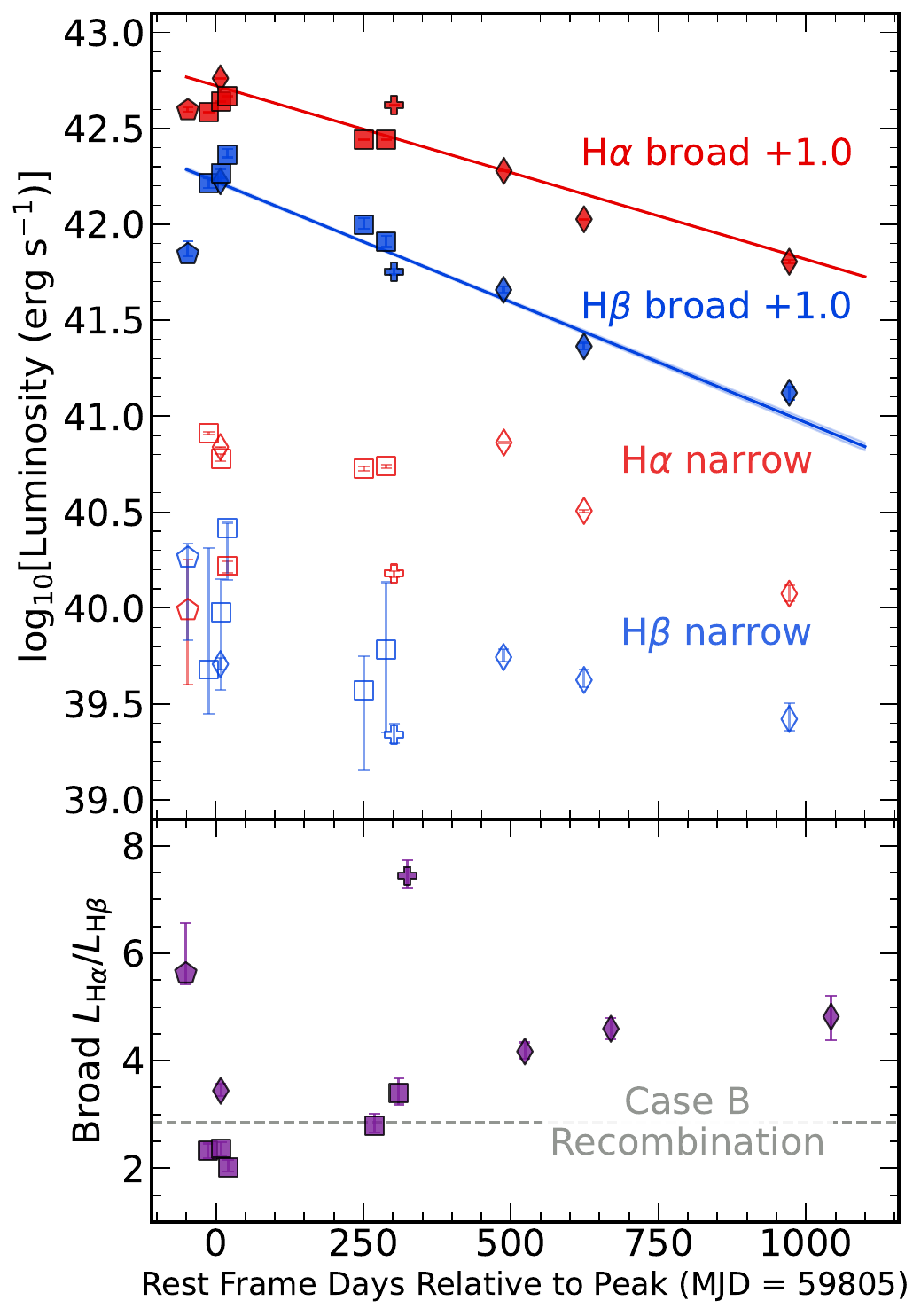}
    \caption{\textit{Top:} Logarithmic luminosity evolution of the broad components of H$\alpha$ and H$\beta$. A constant offset has been added to the broad components for visual purposes. Pentagons are measurements from the Liverpool Telescope spectrum, squares from UH2.2m spectra, diamonds from Keck spectra, and plus signs from the Gemini spectrum. In most cases, the error bars are smaller than the point sizes. We plot shading for the $1\sigma$ spreads around the line fits, but for H$\alpha$ this spread is narrower than the line thickness. \textit{Bottom:} The linear ratio of broad line luminosities $L_{\textrm{H}\alpha}/L_{\textrm{H}\beta}$. The horizontal line is the Case B recombination line ratio \protect\citep{osterbrock89}.}
    \label{fig:Halpha_Hbeta}
\end{figure}

Figure \ref{fig:Halpha_Hbeta} shows our H$\alpha$ and H$\beta$ luminosity measurements and line fits for their luminosity evolutions. The broad component luminosities decrease with time, with H$\alpha$ showing a decrease of $(0.452 \pm 0.002) \ \textrm{dex/500 days}$ and H$\beta$ showing a decrease of $(0.676 \pm 0.003)\ \textrm{dex/500 days}$ (rest-frame days). Accounting for conversions between the 500 and 40-day timescales, this luminosity decline rate of broad H$\alpha$ (H$\beta$) is a factor of $\sim 2.5$ ($\sim 2$) greater than the $\Delta L_{40}$ value measured from the bolometric light curve in Section \ref{sec:BB modelling} and shown in Figure \ref{fig:d_l40}. As shown in the bottom panel of Figure \ref{fig:Halpha_Hbeta}, the broad component Balmer decrement $L_{\textrm{H}\alpha}/L_{\textrm{H}\beta}$ deviates from Case B recombination \citep{osterbrock89}, typically assumed for \ion{H}{ii} regions, varying over a range $\sim 1.8$ to $\sim 7.3$. Similar variations have been observed for several other TDEs \cite{Charalampopoulos_2022}. We note that the SPRAT point is not wholly reliable, as our limited calibration for this spectrum may affect line luminosities. We measure broad H$\alpha$ and H$\beta$ FWHMs of $\lesssim 3000 \ \textrm{km s}^{-1}$, as is low compared to typical broad line FWHMs of TDEs \citep[$\sim 10^4 \ \textrm{km s}^{-1}$; e.g.,][]{arcavi14, vanvelzen21, Charalampopoulos_2022} but common for AGN broad lines \citep[e.g.,][]{peterson93, ho08, Koss_2017, Oh_2022}. Additionally, while TDEs typically show a decrease in broad line width as the luminosity declines, and the reverse is seen for normal AGN variability \citep[e.g.,][]{holoien16a, holoien19b, Hinkle2021b}, we find the line widths of the broad components are flat in time.  

For the narrow components, both the line luminosity and widths are approximately flat in time. For H$\alpha$, we measure a median narrow component luminosity of $\log_{10}[L_{\textrm{H}\alpha}] \sim 40.7$, similar to the median luminosities $\log_{10}[L_{\textrm{H}\alpha}] \simeq 40.2-40.3$ measured for LLAGNs in Seyfert 2 and 1 galaxies \citep{ho08}. LLAGN typically show weak or no H$\beta$ emission \citep[e.g.,][]{ho08}, and the narrow component Balmer decrement we measure ranges from $\sim 6.9$ to $\sim 13.5$ in our high-S/N spectra, similar to values measured for weakly-accreting AGNs \citep[][]{Wu_2023}. Comparing instead with \ion{H}{ii} regions, the median luminosity of H$\alpha$, $\log_{10}[L_{\textrm{H}\alpha}] \simeq 39.2$ \citep{Ho_1997, Ho_2003}, is lower than our measurements. Like the broad component, the narrow component Balmer decrement is not consistent with Case B recombination. To estimate a SFR, we extracted 2\farcs-aperture spectrum from the KCWI data cube at $+1042\textrm{d}$, to include the entire host galaxy flux, and scaled this spectrum to match the host galaxy flux in SDSS $i$-band (magnitude in Table \ref{tab:host magnitudes}). We then re-fit H$\alpha$ in this spectrum, and used the narrow-component luminosity in the H$\alpha$ calibration from \cite{SFR_Kennicutt_2012}, finding a SFR of $(0.19 \pm 0.03) \ M_{\odot} \ \textrm{yr}^{-1}$. This SFR is consistent, within $1.5 \sigma$, with the SFR estimated from our CIGALE fits to host-galaxy photometry (Table \ref{tab:host galaxy results}), suggesting modest star formation. We note, however, that the Balmer decrement values for the narrow H$\alpha$ emission are more consistent with originating from a LLAGN.

\subsection{Coronal Line Evolution} 
\label{sec:CL evolution}

In addition to the Balmer and helium lines, we see the CLs
[\ion{Ne}{v}] $\lambda$3347, 
[\ion{Ne}{v}] $\lambda$3427, 
[\ion{Fe}{vii}] $\lambda$3586, 
[\ion{Fe}{vii}] $\lambda$3760, 
[\ion{Ar}{xiv}] $\lambda$4414, 
[\ion{Fe}{vii}] $\lambda$5160, 
[\ion{Fe}{xiv}] $\lambda$5304, 
[\ion{Fe}{vii}] $\lambda$5722, 
[\ion{Fe}{vii}] $\lambda$6088, 
[\ion{Fe}{x}] $\lambda$6376, 
[\ion{S}{xii}] $\lambda$7612, and 
[\ion{Fe}{xi}] $\lambda$7894 in the optical spectra. The detection of CLs implies the presence of high-energy photons, because these states have $\gtrsim 100$ eV ionisation potentials \citep[e.g.,][]{Mazzalay_2010, yang13}.

\begin{figure*}[t]
    \centering
    \includegraphics[width=1\textwidth]{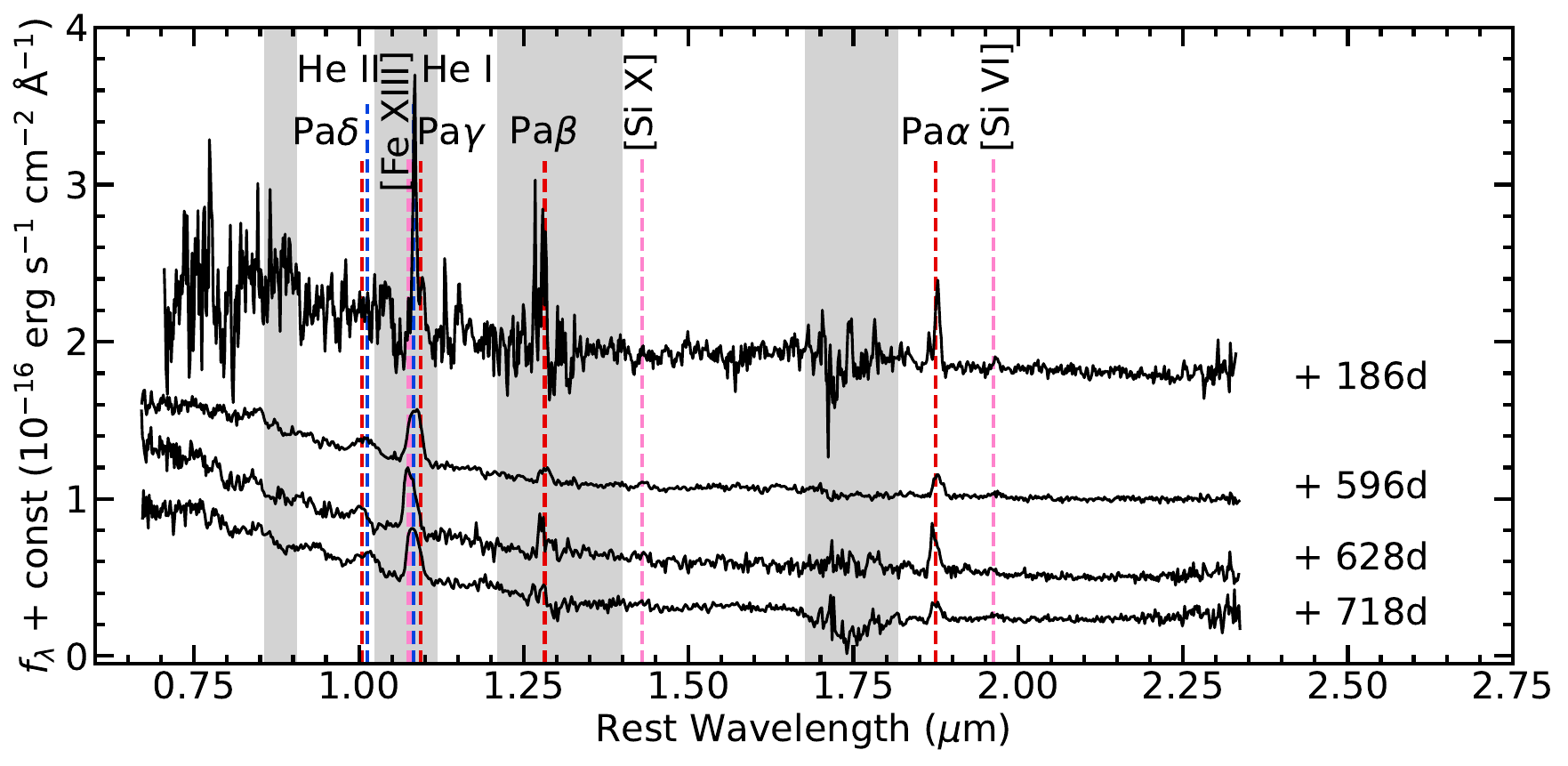}
    \caption{NIR spectra for ATLAS22kjn, with SpeX on IRTF. The phases are labelled relative to the time of UV/optical peak ($\textrm{MJD}=59805$). The blue dashed lines indicate helium features, the red dashed lines indicate hydrogen features, and the pink dashed lines indicate CLs. The top spectrum has been binned to 10 Å bins. Grey bands indicate prominent telluric bands.
    }
    \label{fig:NIR spectra}
\end{figure*}

We also detect CLs in the NIR, as shown in Figure \ref{fig:NIR spectra}. In the earliest NIR spectrum, $+ \ 186\textrm{d}$, we detect the CL [\ion{Si}{vi}] $\lambda$19620. At $+ \ 596\textrm{d}$, we also detect [\ion{Si}{x}] $\lambda$14300. The CLs [\ion{Fe}{XIII}] $\lambda$10747 and [\ion{Fe}{XIII}] $\lambda$107979 may also be present, as indicated, but are blended with each other, He I, and Pa$\gamma$.
We do not detect the NIR CLs [\ion{S}{xi}] $\lambda$19196, [\ion{Si}{xi}] $\lambda$19320, or [\ion{Al}{ix}] $\lambda$20400 in any of the spectra, and their absence is consistent with what has been observed for AGNs \citep{lamperti17}. 
We are unable to make sure the X-ray to NIR  line ratio dignostics in \cite{lamperti17} due to the lack of substantial X-ray emission above $\sim2$\, keV. The lack of hard X-ray emission is commonly seen in TDEs, but is inconsistent with what is seen in AGN \citep[e.g.,][]{auchettl17}.

To determine the luminosity and FWHM of the most prominent CLs, we fit a single Gaussian to the
[\ion{Ne}{v}] $\lambda$3347, 
[\ion{Fe}{vii}] $\lambda$3760, 
[\ion{Ar}{xiv}] $\lambda$4414, 
[\ion{Fe}{x}] $\lambda$6376, 
[\ion{S}{xii}] $\lambda$7612, and 
[\ion{Fe}{xi}] $\lambda$7894 lines. We also fit [\ion{Fe}{xiv}] $\lambda$5304, but since this line appears to be partly blended with [\ion{Fe}{vii}] $\lambda$5722, especially at early times, we fit two Gaussians for this feature and report measurements for the component centered on [\ion{Fe}{xiv}] $\lambda$5304. We omit the $\textrm{SPRAT} \ - 51\textrm{d}$ spectrum due to the limited wavelength coverage of this spectrum. We note that we fit [\ion{Ne}{v}] $\lambda$3347, rather than the stronger [\ion{Ne}{v}] $\lambda$3427 line since the latter is heavily blended with \ion{O}{iii} $\lambda$3444.
For non-detections, we computed 3$\sigma$ upper limits by fitting a Gaussian with a fixed width equal to the average FWHM measured from detections of that CL. Particularly at bluer wavelengths, the noise in the spectra is highly structured, causing some of our upper limits to be lower than detections at nearby epochs. We report our line measurements in Table \ref{tab:CL_measurements} in Appendix \ref{ap:spectroscopy_tables}.

To determine whether the CL emission is due to the transient, we quantify this relative to [\ion{O}{iii}] $\lambda 5007$. We fit [\ion{O}{iii}] $\lambda 5007$ with a single Gaussian and report our measurements in Table \ref{tab:oiii} in Appendix \ref{ap:spectroscopy_tables}). 
We first detect [\ion{O}{iii}] $\lambda$5007 in the early-time high S/N LRIS spectrum at $+ 8 \ \textrm{d}$. The luminosity of [\ion{O}{iii}] $\lambda$5007 shows no strong evolution, with a luminosity of $(8.9 \pm 0.2) \times 10^{39} \ \textrm{erg s}^{-1}$ in this spectrum, and $(7.5 \substack{+ 0.3 \\ -0.4}) \times 10^{39} \ \textrm{erg s}^{-1}$ in the latest spectrum ($\textrm{KCWI}+ 1042 \ \textrm{d}$).
In the $+ 8\textrm{d}$ spectrum, the ratio of the strongest CL ([\ion{Fe}{xiv}]) to [\ion{O}{iii}] is $L_{\textrm{[\ion{Fe}{xiv}]}}/L_{\textrm{[\ion{O}{iii}]}} = 1.38 \substack{+ 0.06 \\ -0.04}$. The last detection of [\ion{Fe}{xiv}] is at $+ 669 \ \textrm{d}$, where $L_{\textrm{[\ion{Fe}{xiv}]}}/L_{\textrm{[\ion{O}{iii}]}} = 0.61 \substack{+ 0.09 \\ -0.05}$. In the last spectrum, at $+ 1042 \ \textrm{d}$, the few CLs that are still detected have $L_{\textrm{CL}}/L_{\textrm{[\ion{O}{iii}]}}$ values between $\sim 0.2$ and $\sim 0.3$.
In typical AGNs, CL luminosities are only a few percent of the [\ion{O}{iii}] $\lambda$5007 line luminosity \citep[e.g.,][]{wang12, Frederick_2019}, so our results suggest that ATLAS22kjn is a CLE.

\begin{figure}[t]
    \centering    \includegraphics[width=1\linewidth]{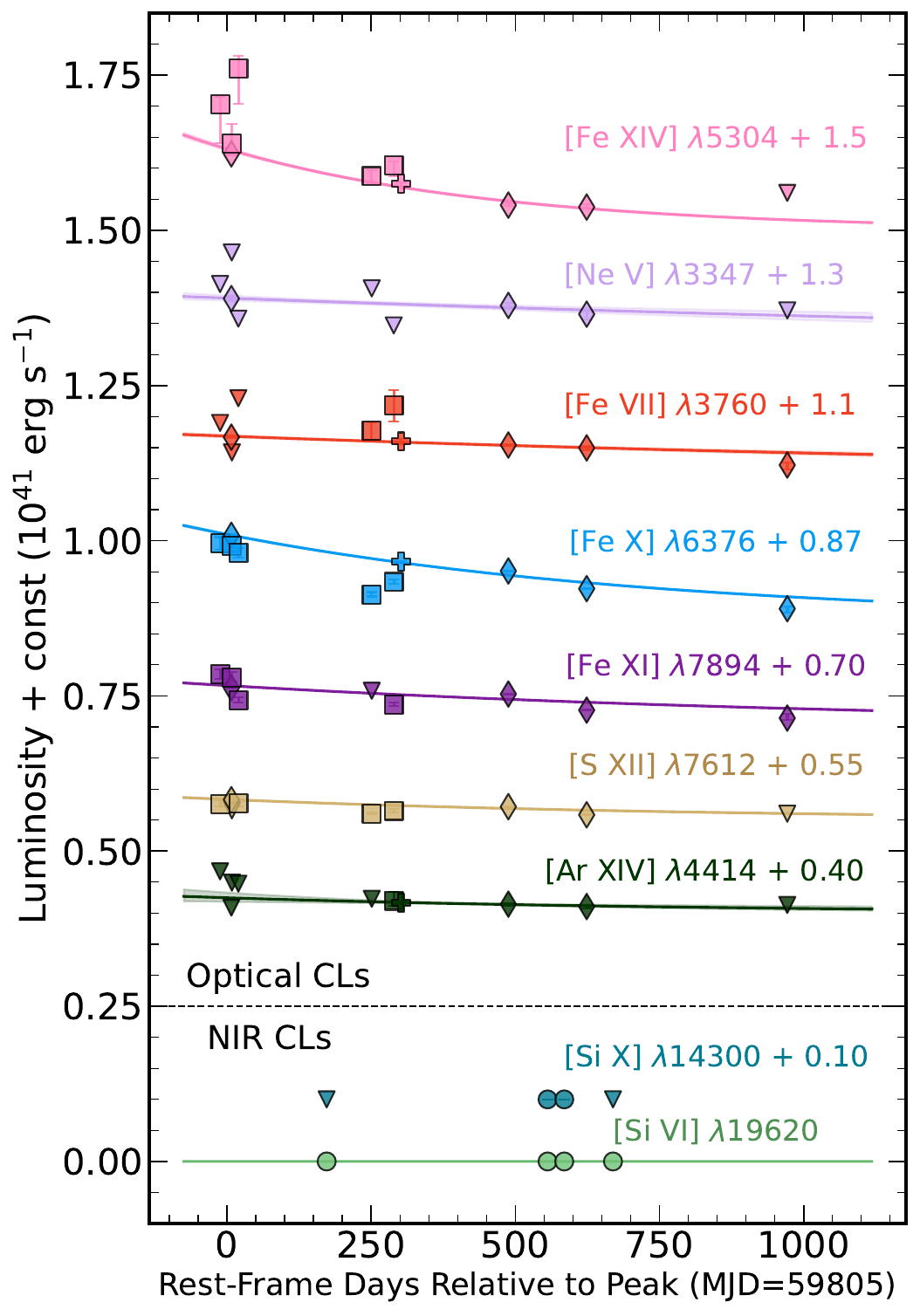}
    \caption{Luminosity evolution of the strongest CLs, offset by a constant. Squares are UH2.2m spectra, diamonds are Keck spectra, and plus signs are Gemini spectra. NIR spectra from IRTF are represented by circles. The horizontal dashed black line separates the optical CLs (above) from the NIR CLs (below). The GMOS wavelength range does include all the lines. Shading around each line corresponds to the 1$\sigma$ spread in the fit. Many of the 1$\sigma$ spreads are narrower than the thickness of the fitted line. We include both data points for detected CLs and upper limits in our fitted data. We did not fit for [\ion{Si}{x}].
    }
    \label{fig:lum_CLs}
\end{figure}

\begin{figure}[t]
    \centering    \includegraphics[width=1\linewidth]{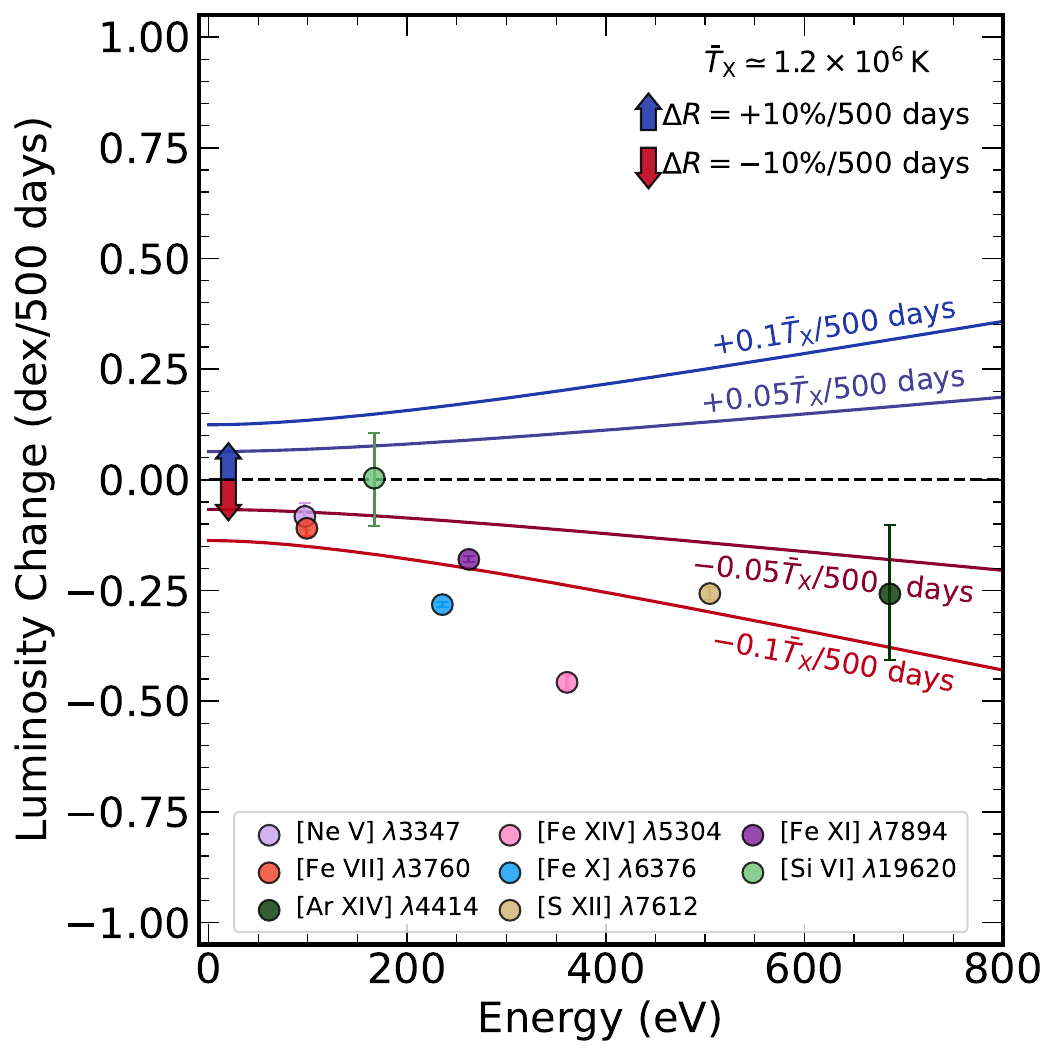}
    \caption{The slope $a$ of the luminosity evolution for each CL, fitted in Figure \protect{\ref{fig:lum_CLs}}, as a function of ionisation potential. Curves are a prediction for the change in the observed line luminosity due to 5\% and 10\% changes in the temperature of the ionising source relative to the weighted-average X-ray temperature $\bar{T}_X$. Blue (red) curves show an increase (decrease) in temperature. The upward-pointing blue and downward-pointing red arrows show the effect of changing the radius of the ionising source by 10\% over 500 days at fixed temperature.}
    \label{fig:lum_IP_CLs_w_lines}
\end{figure}

We fit the luminosity evolution for each CL with the same model used to fit the Balmer line luminosities in Section \ref{sec:opt spectra}, with modifications to leverage both detections and $3 \sigma$ upper limits in our CL fits. 
We minimised a negative log-likelihood function that sums a log PDF for the detections and a log CDF for the $3\sigma$ upper limits.
We used Monte Carlo processes to estimate uncertainties, generating synthetic data from the detections only by drawing from a normal distribution centered on a value, with the width of the distribution set by the uncertainty on the value. 
We do not fit for the luminosity evolution of [\ion{Si}{x}] since the only two data points are not well-separated in time.
In Figure \ref{fig:lum_CLs}, we plot the resulting line fits and corresponding 1$\sigma$ spreads alongside our line luminosity measurements.
We show the slopes $a$ as a function of ionisation potential in Figure \ref{fig:lum_IP_CLs_w_lines}. We recast the luminosity changes $a$ in dex per 500 rest-frame days to illustrate these changes on the timescale of TDE evolution. We used Monte Carlo simulations to estimate $1 \sigma$ uncertainties on the $a$ values.

The evolution of the NIR CL [\ion{Si}{vi}] is approximately flat, while the luminosities of the optical CLs [\ion{Ne}{v}] $\lambda$3347 [\ion{Fe}{vii}], [\ion{Ar}{xiv}], [\ion{Fe}{xiv}], [\ion{Fe}{x}], [\ion{S}{xii}], and [\ion{Fe}{xi}] are declining. 
In Figure \ref{fig:lum_IP_CLs_w_lines}, we show these slopes as a function of ionisation potential.
There appears to be a trend in which lines with higher ionisation potentials decline more rapidly, as would be expected if the ionising source is becoming cooler. We discuss this further in Section \ref{sec:CLs_lum_trend_analysis}.

\subsection{X-ray Blackbody Modelling}
\label{sec:Analysis-Xray}

\begin{figure}[t]
    \centering    \includegraphics[width=1\linewidth]{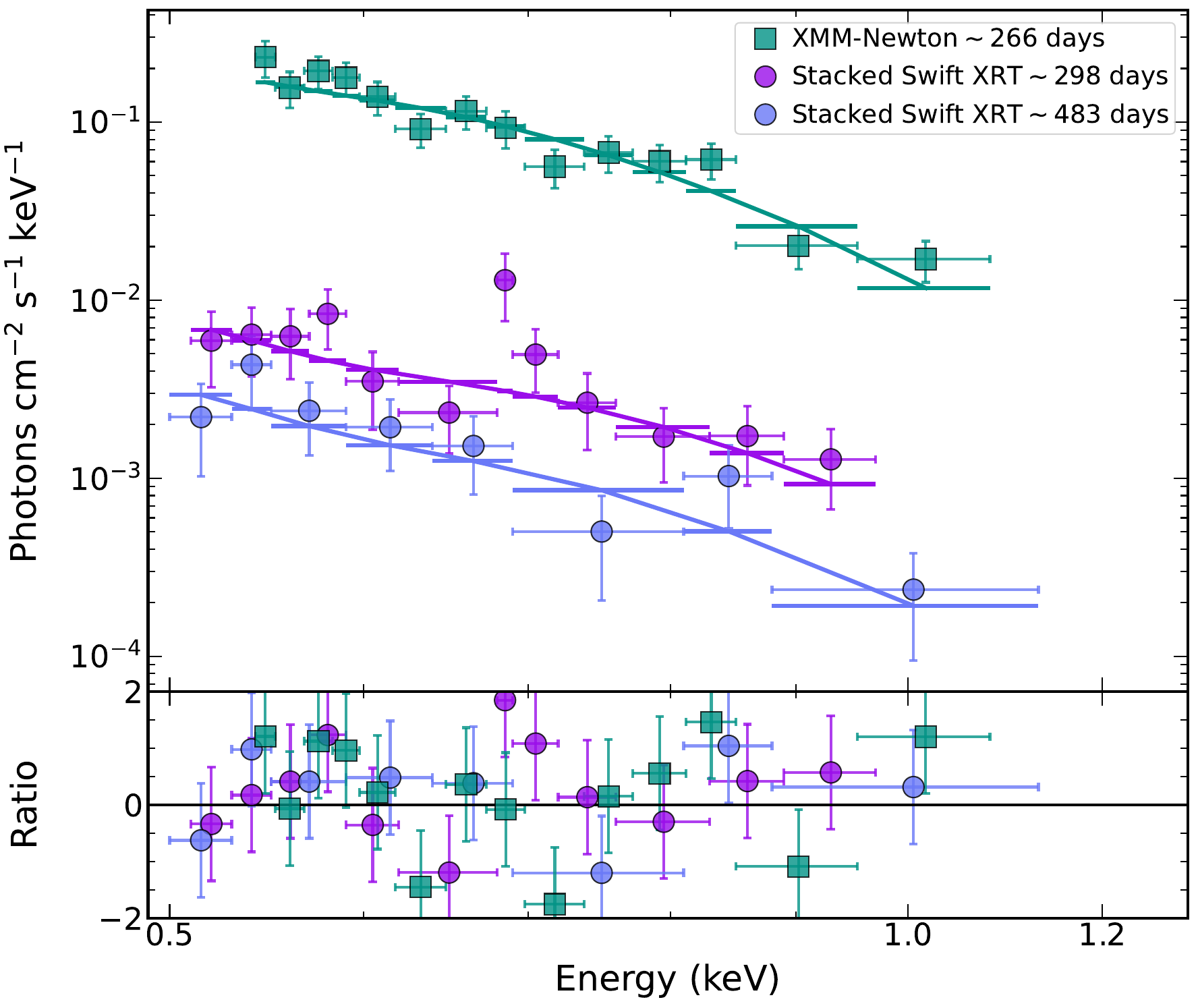}
    \caption{The X-ray spectra for ATLAS22kjn, consisting of \textit{XMM-Newton} and the stacks of the earlier and later \textit{Swift} XRT data, with dates given in observed days relative to UV/optical peak ($\textrm{MJD}=59805$; see Section \protect\ref{sec:BB modelling}). Each spectrum is fit with an absorbed blackbody model with the best fit and residuals shown, while the best fit blackbody values are shown in Figure \ref{fig:uvot_xray_bb}.}
    \label{fig:xray_spectra}
\end{figure}

In addition to luminous UV/optical emission, ATLAS22kjn also shows evolving X-ray emission. Figure \ref{fig:xray_spectra} shows the X-ray spectra of ATLAS22kjn. 
The X-ray emission of ATLAS22kjn is soft, with little to no significant emission detected above $\sim 1 \ \textrm{keV}$, as is commonly observed for TDEs. 
The hardness ratio (HR) of ATLAS22kjn, as presented in Table \ref{tab:xray}, is soft \citep[HR $\lesssim 0.3$; e.g.,][]{auchettl18}. 
The HR we measure from the \textit{Swift} observations is negative throughout, with some variation between $\textrm{HR} = -0.8$ and $-0.4$ for detections, but no clear evolution. From the XMM Newton observations, we measure $\textrm{HR} = -0.3$ around $+270$ days. The HR and HR variation of ATLAS22kjn is similar to that of the CLE ASASSN-18jd \citep{neustadt20} and other TDEs \citep{auchettl17, Guolo_2024, Mondal_2026}. In contrast, AGNs typically show harder X-rays, with emission in the $2-10 \ \textrm{keV}$ range and commonly above $\sim 10 \ \textrm{keV}$ \citep[e.g.,][]{ricci17, auchettl18}. This emission tends to become softer with time and may show TDE-like HRs $<-0.75$ \citep[e.g.,][]{auchettl18, Mondal_2026}. The soft X-ray emission and the lack of evolution in the HR of ATLAS22kjn suggests its X-ray emission is more TDE-like.

We fit the X-ray spectra as an absorbed blackbody (\texttt{tbabs*zashift*bbodyrad} in XSPEC), assuming a host column density of $1.25 \times 10^{20} \ \textrm{cm}^{-2}$, to obtain X-ray luminosity, radius, and temperature measurements. We present these quantities in Table \ref{tab:xray} and show their evolution in the right panel of Figure \ref{fig:uvot_xray_bb}, alongside the X-ray blackbody parameters available for the CLEs AT 2017gge \citep{wang22a}, ASASSN-18jd \citep{neustadt20}, ASASSN-18ap \citep{Wang2024}, AT 2019qiz \citep{nicholl20}, and AT 2022upj \citep{Newsome_2024}. We discuss comparisons with this sample in Section \ref{sec:CLE discussion}.
We note that the X-ray radii we obtain for ATLAS22kjn are less than the gravitational radius and therefore unphysical. This suggests that our choice of a blackbody is a simplification and that detailed disk modelling would provide a better fit.
We find that the X-ray emission begins $\sim 250$ observed days after the UV/optical peak, peaking $\sim 20$ days later. Notably, this peak aligns with a small rise in the UV/optical light curves, as may be evidence for reprocessing. \cite{Koljonen2024} noted this coincident UV/optical and X-ray increase as well, and that the rise in the X-rays coincides with 2.6$\sigma$ increase in the polarization degree, but also that the variability seen in the polarization degree and angle is not easily explained within this scenario.

\begin{table*}
    \centering
    \caption{\normalfont X-ray measurements for detections. From left to right: the instrument, the MJD of the luminosity, luminosity in the $0.3-10 \ \textrm{keV}$ range, the hardness ratio (HR), the MJD spanned by the blackbody parameters, the blackbody temperature, and the logarithmic blackbody radius.}
    \vspace{1mm}
    \begin{tabular}{c c c c c c c} \toprule
        Instrument & $\textrm{MJD}_L$ & \makecell{$L_{0.3-10\ \textrm{keV}}$ \\[1mm] $(10^{42}\ \textrm{erg s}^{-1})$} & HR & $\textrm{MJD}_{BB}$ & $kT_{BB} \ (\textrm{keV})$ & $\log_{10}\left[R_{BB} (\textrm{cm})\right]$ \\ \midrule \\[-1.0ex] 
        XMM Newton & $60072$ & $0.63 \pm 0.04$ & $-0.33$ & $60072$ & $0.110 \pm 0.010$ & $10.66 \pm 0.13$ \\ [1.0ex] \hline \\[-1.0ex] 
        \textit{Swift} XRT & $60052 \pm 12$ & $1.1 \pm 0.3$ & $-0.67$ & \multirow{14}{*}{$60100 \pm 70$} & \multirow{14}{*}{$0.100 \pm 0.009$} & \multirow{14}{*}{$10.79 \pm 0.17$}  \\[1.0ex] 
        \textit{Swift} XRT & $60075 \pm 4$ & $2.3 \pm 0.4$ & $-0.82$  \\ [1.0ex] 
        \textit{Swift} XRT & $60083.7 \pm 1.3$ & $2.2 \pm 0.6$ & $-0.80$ \\[1.0ex]
        \textit{Swift} XRT & $60097 \pm 2$ & $2.5 \pm 0.5$ & $-0.83$  \\[1.0ex] 
        \textit{Swift} XRT & $60107.9 \pm 1.2$ & $2.0 \pm 0.4$ & $-0.83$ \\[1.0ex]
        \textit{Swift} XRT & $60117.0 \pm 1.2$ & $1.6 \pm 0.4$ & $-0.78$ \\[1.0ex]
        \textit{Swift} XRT & $60124 \pm 3$ & $0.8 \pm 0.2$ & $-0.84$ \\[1.0ex]
        \textit{Swift} XRT & $60134 \pm 4$ & $1.4 \pm 0.3$ & $-0.64$ \\[1.0ex]
        \textit{Swift} XRT & $60149 \pm 5$ & $0.8 \pm 0.3$ & $-0.72$ \\[1.0ex]
        \textit{Swift} XRT & $60177.8 \pm 1.3$ & $0.7 \pm 0.3$ & $-0.51$ \\[1.0ex] \hline \\[-1.0ex]
        \textit{Swift} XRT & $60207 \pm 4$ & $0.7 \pm 0.2$ & $-0.44$ & \multirow{7}{*}{$60290 \pm 110$} & \multirow{7}{*}{$0.090 \pm 0.010$} & \multirow{7}{*}{$10.74 \pm 0.13$}  \\[1.0ex]
        \textit{Swift} XRT & $60277 \pm 7$ & $0.8 \pm 0.2$ & $-0.67$ \\[1.0ex] 
        \textit{Swift} XRT & $60304 \pm 13$ & $0.80 \pm 0.19$ & $-0.62$ \\[1.0ex]
        \textit{Swift} XRT & $60358 \pm 27$ & $0.74 \pm 0.18$ & $-0.62$ \\[1.0ex]
        \textit{Swift} XRT & $60446 \pm 31$ & $1.0 \pm 0.2$ & $-0.75$ \\[1.0ex] \bottomrule
    \end{tabular}
    \vspace{3mm}
    \label{tab:xray}
\end{table*}

\subsection{Spectral Energy Distribution}

\begin{figure*}[t]
    \centering
    \includegraphics[width=1\linewidth]{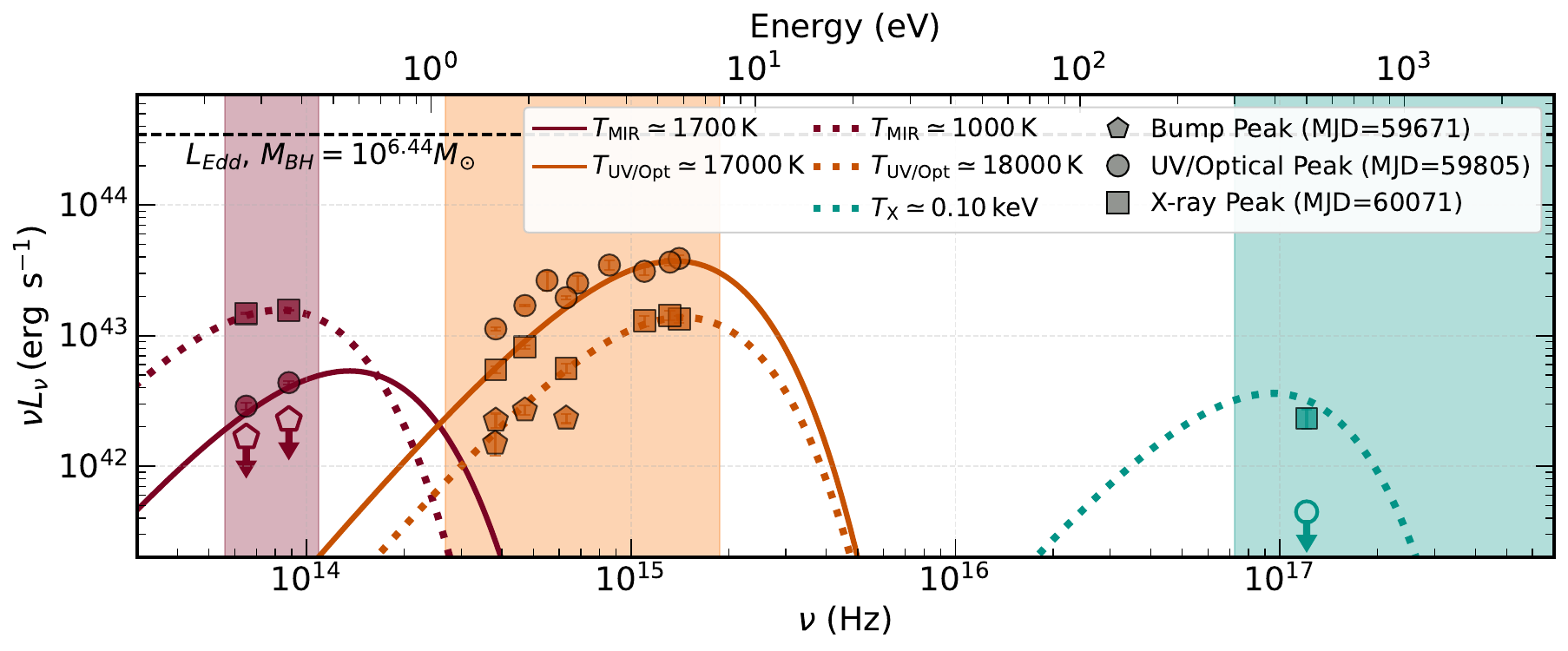}
    \caption{Spectral energy distribution of ATLAS22kjn. The solid lines show the blackbody curves for the epoch of UV/optical peak ($\textrm{MJD} = 59805$), while the dotted lines show the blackbody for the epoch of X-ray peak luminosity  ($\textrm{MJD} = 60071$). The shaded bands show the range covered by the MIR photometric bands (maroon), UV/optical photometric bands (orange), and the $0.3-10 \ \textrm{keV}$ energy range for X-ray observations (teal). The MIR and UV/optical points are placed at the frequency corresponding to the central wavelength of each filter. The MIR points are from the $W1$ and $W2$ light curves (Figure \ref{fig:mir_light_curve}). As no single $W1$/$W2$ epoch matched the bump epoch, for each band, we averaged the two nearest epochs and multiplied the averaged $1 \sigma$ uncertainty to obtain a 3$\sigma$ upper limit (maroon pentagons with downward-facing arrows). The UV/optical points are from the \textit{Swift} $UVW1$-, $UVW2$-, $UVM2$-, $U$-, $V$-, and $B$-band, ZTF $g$-, $r$, and $i$-band (Figures \ref{fig:overall light curve}, \ref{fig:flare lightcurve}, and \ref{fig:bump lightcurve}). For the bump epoch, the TESS luminosity is also included (Figure \ref{fig:bump lightcurve}). There is no \textit{Swift} $U$-, $V$-, or $B$-band coverage at late times. The X-ray points are \textit{Swift} XRT, and are arbitrarily placed at the frequency corresponding to $0.5 \ \textrm{keV}$. No X-ray coverage exists for the bump epoch. The empty teal circle and downward-facing arrow show the \textit{Swift} XRT  3$\sigma$ upper limit near the epoch of UV/optical peak. The black dashed line shows the Eddington luminosity for our SMBH mass estimate ($\log_{10}\left(M_{BH}/M_{\odot}\right) \sim 6.44$; Section \ref{sec:host})}
    \label{fig:SED}
\end{figure*}

We summarise the evolution of the different emission components with the spectral energy distribution (SED) of ATLAS22kjn in Figure \ref{fig:SED}. For the MIR and UV/optical, we show blackbody curves from the fits in Section \ref{sec:BB modelling} and Section \ref{sec:mir light curve} at the epochs nearest to the time of UV/optical peak ($\textrm{MJD} = 59905$) and the time of peak X-ray luminosity ($\textrm{MJD} = 60075$; Section \ref{sec:Analysis-Xray}), with points in the $W1$-, $W2$-, \textit{Swift}, and ZTF bands. We also show points for the the peak time of the bump ($\textrm{MJD} = 59671$; Section \ref{sec:bump_analysis}), including the TESS band. Because no single $W1$/$W2$ epoch is a good match to the bump epoch, we averaged the measurements of the two nearest epochs ($\sim 160$ days apart; see Figure \ref{fig:mir_light_curve}) and determined a 3$\sigma$ upper limit by scaling the $1 \sigma$ uncertainty on the averaged value. We show the X-ray blackbody for the epoch of peak X-ray luminosity, and the available \textit{Swift} XRT measurements.

During the early-time bump, the UV/optical measurements and MIR upper limits are comparable and may be consistent with a MIR excess. The UV/optical emission peaks $\sim 134$ observed days later and dominates over both the rising MIR emission and the X-ray emission. This peak has $L_{\textrm{UV/Optical}}/L_{\textrm{Edd}} \sim 0.1$ for our estimated SMBH mass of $\log_{10}\left(M_{BH}/M_{\odot}\right) \sim 6.44$ (Section \ref{sec:host}), similar to values observed for other TDEs \citep[e.g.,][]{wevers17, hammerstein23}. By the time the X-ray emission peaks $\sim 270$ days later, the UV/optical emission has faded and is comparable to the fading MIR emission, which peaked between these epochs.

\section{Coronal Line Constraints on Extreme UV/Soft X-ray Emission}
\label{sec:CLs_lum_trend_analysis}

Under the assumption that the CL emission is driven by photoionisation, we can use their luminosity evolution as a simplified probe to constrain the temperature evolution of the ionising source. We must also assume that the source of the ionisation and the X-rays is the same. We then determine the change in flux as a function of ionisation potential for $\pm 5 \%$ and $\pm 10 \%$ changes in temperature relative to the X-ray temperature weighted average of $\bar{T}_X \simeq 1.2 \times 10^6 \ \textrm{K}$. The curves are shown in Figure \ref{fig:lum_IP_CLs_w_lines} with the effects of a $\pm 10 \%$ change in the size of the source. Under these assumptions, we see that the change in luminosity with ionisation potential could be explained by a $\lesssim 10\%$ decrease in the temperature of the ionising source.

We attempted an alternative method to constrain the temperature evolution of the CLR more precisely, using MCMC methods to model CL emission with a toy model. However, results from this modelling were inconclusive due to unbreakable degeneracies between our temperature and emission line strength parameters. We believe this method failed precisely because the ionising source is so stable. Our toy model and method can be found in Appendix \ref{ap:alt_CL_analysis}. We strongly encourage others to attempt similar analyses for other well-observed CL-emitting TDEs, or to follow ATLAS22kjn to late times.

\section{Origins of the Early Bump}
\label{sec:bump discussion}

The light curves of ATLAS22kjn are consistent with those of a TDE, showing strong UV/optical emission, with no major variability prior to the main flare, and a slow rise followed by a power-law decay. The light curves in the ZTF $g$-, $r$-, and $i$-band, ATLAS $c$- and $o$-band, the TESS band, and the Gaia $G$-band, also exhibit a pre-peak bump. Similar pre-peak bumps have been observed in a handful of other nuclear transients, including
ASASSN-19dj \citep{hinkle21b, hinkle21a, Faris2024}, AT 2020wey \citep{Charalampopoulos_2023}, ASASSN-18ap \citep{Wang2024}, AT 2019mha \citep{Wang2024}, AT 2019qiz \citep{Wang2024}, AT 2023lli \citep{Huang_2024}, AT 2024lhc \citep{Yao_2026}, and AT 2024kmq \citep{Yao_2026}, all of which have been interpreted as TDEs. These bumps lasted $\sim 10 - 30$ days, peaking $\sim 30-10$ rest-frame days before the main UV/optical peak, except for the ATLAS22kjn bump, which peaked $125 \substack{+5 \\ -3}$ rest-frame days before the main peak. For the bumps of AT 2023lli, AT 2024lhc, and AT 2024kmq---the best-studied bumps---\cite{Huang_2024} and \cite{Yao_2026} both suggested shocks resulting from stream intersection as the most likely mechanism for powering the bump. 

Several physical mechanisms have been proposed to produce pre-peak bumps. For ATLAS22kjn, we discuss six possibilities:

\begin{itemize}
    \item \textbf{Section \ref{sec:nozzle shock}}: A nozzle shock \citep[e.g.,][]{evans89, kochanek94, Bonnerot2022} induced by vertical compression and re-expansion of the debris streams near pericentre \citep{yalinewich19, Steinberg_2024}, which dissipates energy and produces a pre-peak feature.
    
    \item \textbf{Section \ref{sec:cooling of unbound stellar debris}}: The cooling of unbound stellar debris \citep{KasenRamirez2010} to produce a
    short, recombination-powered light curve feature prior to the main TDE flare;
    
    \item \textbf{Section \ref{sec:stream-disk collision in rptde}}: A repeating partial TDE (rpTDE), where a star on an eccentric orbit is repeatedly stripped, and subsequent disk–stream collisions produce recurring bumps in the light curve \citep[e.g.,][]{Huang2023b, Sun_2025};

    \item \textbf{Section \ref{sec:stream-stream collision}}: Stream intersections result in a shock that dissipates energy \citep[e.g.,][]{piran15, Jiang_2016, Huang_2023_stream_stream, huang_2024_pre_peak_emission} and produces a pre-peak feature \citep{Wang2024, Huang_2024};
    
    \item \textbf{Section \ref{sec:dtde}}: A double TDE, in which both members of a stellar binary are disrupted, produces two streams that then collide, resulting in a precursor feature \citep{Bonnerot_Rossi_2019};
    
    \item \textbf{Section \ref{sec:wind-stream collision}}: The spin of the SMBH causes relativistic precession of the tidal debris stream, and when the disk wind interacts with the inner edge of the stream, a bright pre-peak feature is produced \citep{Calderon2024}.
\end{itemize}

In the following sections, we consider how consistent the properties listed in Table \ref{tab:bump_explanations} are with predictions for each of these six mechanisms. Since some of these mechanisms have consequences for the main TDE flare, we leverage properties of both the bump and the ATLAS22kjn main flare to help discriminate between possible bump explanations.
  
\subsection{Nozzle Shock} 
\label{sec:nozzle shock}

Following tidal disruption, a debris stream is formed, part of which falls back toward the SMBH. Returning to pericentre for the first time, this stream can undergo strong vertical compression, resulting in substantial shock dissipation known as ``nozzle shock'' \citep{evans89, kochanek94, Bonnerot2022, Steinberg_2024}. In their investigation into the changing polarization angle of ATLAS22kjn, \cite{Koljonen2024} explored a scenario in which nozzle shocks are the dominant source of energy dissipation during the rising phase of the TDE light curve \citep{Ryu_2023, Steinberg_2024} and are strong enough to begin the circularisation process. This mechanism is also suggested to produce a pre-peak bump \citep{Wang2024, Huang_2024}.

Both \cite{Bonnerot2022} and \cite{Steinberg_2024} assume $M_{\star} = 1M_{\odot}$, $\beta = 1$, and $M_{BH} = 10^6 M_{\odot}$. \cite{Bonnerot2022} predict that the debris stream gas will begin to emit observable radiation only after it passes pericentre a second time. 
The time of this second passage corresponds approximately to the first epoch of synthetic spectra presented by \cite{Steinberg_2024}.
From this first epoch of synthetic spectra, we determined the luminosity in the TESS band, accounting for conversions from $\nu L_{\nu}$. We find that the TESS band luminosity using nozzle shock predictions is $L_{\textrm{TESS}} \simeq 1.6 \times 10^{41} \ \textrm{erg s}^{-1}$, an order of magnitude less than our observed bolometrically-corrected TESS luminosity. This suggests that nozzle shock alone is not energetic enough to produce a bump of the luminosity we observe, aligning with recent theoretical results that suggest nozzle shock is unlikely to contribute significantly to the first light emission of TDEs \citep{Hu_2026}. 

\subsection{Cooling of Unbound Stellar Debris}
\label{sec:cooling of unbound stellar debris}

In a TDE, approximately half the stellar material is bound to the SMBH, and the other half is unbound. As the unbound stellar material expands, its initial thermal energy is rapidly lost through adiabatic expansion. However, ionisation energy---which can amount to $\sim 10^{46}$ ergs per solar mass of ionised hydrogen---is not depleted by adiabatic expansion \citep{KasenRamirez2010}. 
Once the material has cooled to $\sim 10^4 \ \textrm{K}$, recombination sets in \citep{kochanek94}.
The release of this ionisation energy could produce a pre-peak bump. \cite{KasenRamirez2010} produce a model for this emission, predicting the bump would last a few days, beginning of order 1 week after disruption, with a luminosity of
\begin{equation}
\label{eq:KR_luminosity}
    L \sim 2 \times 10^{40} (M_{\star}/M_{\odot})^{5/2} \ \textrm{erg s}^{-1} \ ,
\end{equation}
radiated as a near blackbody with a temperature of $\sim 5,000 \ \textrm{K}$.

The observed $\sim 9$-day duration of the ATLAS22kjn bump is consistent with the $\gtrsim 7$-day durations modelled by \cite{KasenRamirez2010} for higher mass stars ($M_{\star} \ge 5 \ M_{\odot}$).
The temperature predicted by \cite{KasenRamirez2010} is comparable to our estimate from the $g-r$ colours ($7800 \pm 500$ K, see Section \ref{sec:bump_analysis}) Equation \ref{eq:KR_luminosity}, however, matching our measured bump luminosity requires a fairly massive star ($M_{\star}\sim 7 M_{\odot}$), as may be problematic given the low star formation rate we found for the host of ATLAS22kjn in Section \ref{sec:discovery}. Furthermore, even with this high stellar mass, the energy we measure for the bump exceeds the $\sim 10^{46} \ \textrm{erg}$ available from the ionisation energy of hydrogen per solar mass by roughly two order of magnitudes.
The light curves presented by \cite{KasenRamirez2010} also assume a deep encounter with $\beta \gtrsim 3$, as is unlikely to be the case for ATLAS22kjn, given its slow rise to the main peak. Our discussion here would be improved by having predictions for $\beta = 1$ to compare with. 

More problematically, recent studies by \cite{Coughlin_2023} and \cite{Steinberg_2024} found that the optical depths of stream material drop much more slowly than assumed by \cite{KasenRamirez2010}. The peak luminosity at the time of recombination would therefore be much lower, for a maximum $\sim 3 \times 10^{40} \ \textrm{erg s}^{-1}$ in the simulations of \cite{Coughlin_2023}. With these limitations, the cooling of unbound stellar debris is unlikely to explain the ATLAS22kjn bump.

\vspace{10mm}

\subsection{Stream-Disk Collision in Repeating Partial TDE}
\label{sec:stream-disk collision in rptde}

In a rpTDE, a star on an eccentric bound orbit is not fully disrupted at its initial pericentre passage, instead undergoing repeated mass loss at each return to pericentre \citep[e.g.,][]{Huang2023b, Sun_2025, Sharma_2026}. Material stripped from the star may be accreted in stages, allowing for the collision of the stellar debris stream with the disk for each subsequent pericentre passage. An early bump in the light curves is produced when the front stream from the stripped star collides with the disk. Accretion of the stellar material powers the main flare, and the collision of the trailing stream with the disk is observed as rebrightening after the main flare \citep{Huang2023b}. 

Potential observable effects of this model remain relatively unexplored in rpTDE theory. The repeated stream-disk collision scenario has been explored for X-ray quasi-periodic eruptions \citep[QPEs; e.g.,][]{Miniutti_2019, Giustini_2020}, however. First discovered in archival data in 2019 \citep{Miniutti_2019}, QPEs, like rpTDEs, are a class of repeating nuclear transients, characterised by soft X-ray flares that repeat on timescales of hours to days \citep{Arcodia_2021, Arcodia_2024, Nicholl_2024, Chakraborty_2025, Hernandez_Garcia_2025}. A recent study by \cite{Linial_2025} analysed the dynamics of repeated collisions between stellar debris stretched into an elongated stream and an accretion disk. Since no dedicated stream-disk collision models currently exist for the rpTDE scenario, we compare our observations with the predictions of \cite{Linial_2025}. We emphasise, however, that this model makes predictions for X-ray flares rather than optical flares, and would require a reprocessing layer to produce the bump emission we observe.

\cite{Linial_2025} consider a star on a circular orbit that is highly inclined relative to the accretion disk, and two different disk scenarios: a steadily accreting disk, and a viscously spreading TDE disk. \cite{Linial_2025} assume that in the post-shock stream material produced by the stream–disk collision, the production and thermalisation of photons is rapid. 
The authors predict that the stream-disk collision is likely to produce one visible flare per orbit, unless the viewing angle to the disk is edge-on, or the stream is massive enough to punch through the disk.
In both accretion disk scenarios, \cite{Linial_2025} predict that a stream-disk collision results in an X-ray flare with luminosity 
\begin{equation}
    L_{\textrm{bol}} = 2 \times 10^{42} \ m_{*}^{1/3} M_6^{2/3} \ \textrm{erg s}^{-1} \ ,
\end{equation}
and flare duration of 
\begin{equation} \label{eq:porb8}
    \Delta t = 0.9 \ P_{\textrm{orb},8} \left(m_{*}/M_6 \right)^{1/3} \ \textrm{hr} \ ,
\end{equation}
where $P_{\textrm{orb},8}=P_{\textrm{orb}}/8 \ \textrm{hr}$.  
\cite{Linial_2025} do not make an explicit estimate of the energy radiated, as the energy is distributed along the arc of the stream, making this estimate nontrivial. Generally, however, the energy radiated in a QPE flare is observed to scale with the orbital period $P_{\textrm{orb}}$ and can be up to $E_{\textrm{flare}} \approx 10^{48} \ \textrm{erg}$ \citep{Hernandez_Garcia_2025, Chakraborty_2025}. \cite{Linial_2025} predict that for orbital periods $\gtrsim 12 \ \textrm{hr}$, the ejecta stream is significantly less massive than the disk mass interacting with the stellar stream. This means the stream undergoes deceleration and is deflected by the disk instead of punching through it.

For $m_{*}=1$ and our SMBH mass estimate (see Section \ref{sec:host}), the model of \cite{Linial_2025} predicts $L_{\textrm{bol}} \simeq 4 \times 10^{42} \ \textrm{erg s}^{-1}$, close to our measured peak luminosity for the ATLAS22kjn bump. Using $\sim 9$ days in Equation \ref{eq:porb8}, we obtain an orbital period of $\sim 112$ days. 
This orbital period is similar to the separation time between the bump and main peak of ATLAS22kjn, but in the QPE stream-disk model, this orbital period would suggest the stream is deflected from the disk, rather than colliding with it, and no flare is produced. 
\cite{Linial_2025} note that in TDEs, however, the debris stream is not short relative to the distance to where stream-disk collision occurs, as it is in the QPE scenario. This means a TDE stream is not limited by this orbital period and can instead repeatedly penetrate the disk, so stream-disk collision may still be possible for the $\sim 112$-day orbital period we estimate.
The energy radiated in the ATLAS22kjn bump is also similar to the energy expected for a QPE flare---but we caution that this comparison is with observed QPE flares, and not with predictions from a stream-disk collision specifically.

The comparisons we have made with predictions for rpTDEs are limited by the mismatch between the X-ray QPE scenario in the model and the optical rpTDE flare we assume. This includes that \cite{Linial_2025} assume a circular orbit, while an rpTDE is characterised by a star on an eccentric orbit. More problematically, we additionally require a reprocessing layer to produce the observed optical emission from the predicted X-ray flares.
Furthermore, in the rpTDE scenario, if the bump is produced from a stream-disk interaction, the disrupted star must previously have made a pericentre passage to form the disk. The lack of an earlier flare does not rule out this scenario out for the ATLAS22kjn bump, but nonetheless reduces its likelihood. If there had been an earlier encounter, it may either have been too faint to see in CRTS or other archival data, or have had a longer orbit. 
For our SMBH mass estimate and orbital periods exceeding $\sim 25$ years, however, it becomes difficult to keep the star bound \citep[e.g.,][]{Hinkle_2025}.

\subsection{Stream-Stream Collision}
\label{sec:stream-stream collision}

When apsidal precession changes the orbit of the debris stream, it can cause self-intersection of the stream \citep[e.g.,][]{dai15, piran15, lu20, Huang_2023_stream_stream}. The collision of the outgoing and incoming streams will convert some of the stream kinetic energy into radiation. The effects should be most visible in the rising phase of the light curve, or as a precursor event \citep[][]{Wang2024, Huang_2024}.

\cite{Huang_2023_stream_stream} found that a stream-stream collision can convert $\gtrsim 5 \%$ of the stream kinetic energy into radiation, for which the resulting luminosity is $\sim 10^{42} - 10^{44} \ \textrm{erg s}^{-1}$, and \cite{huang_2024_pre_peak_emission} found that a luminosity of $10^{44} \ \textrm{erg s}^{-1}$ can be sustained for several days.

The peak luminosity and duration of the ATLAS22kjn bump are both consistent with predictions for stream-stream collision. Reproducing the observed bump energy with $\sim 5 \%$ the kinetic energy of a stream with mass $M_{\textrm{stream}} \simeq 0.5 M_{\star}$ requires a stream velocity of $\sim 2000 \ \textrm{km s}^{-1}$ ($\sim 0.0067$c), similar to the lowest injected velocities, $0.0065$c, for the streams in the simulations from \cite{Huang_2023_stream_stream}, suggesting stream-stream collision is a sufficiently energetic scenario to produce the bump.

\subsection{Double TDE}
\label{sec:dtde}

Another form of stream-stream collisions are possible in a double TDE. In a double TDE, both members of a stellar binary are disrupted one after the other by a single SMBH, when approaching on an almost radial orbit \citep{Mandel2015}. This event could be identified by a light curve with two local maxima, but due to the short time interval separating the disruptions relative to the duration of each flare, these maxima are unlikely to be discernible observationally \citep{Mandel2015}. 
If the two stars or the mass lost by each star are significantly different, however, double TDE light curves may still be identifiable by a ``knee" after peak, and a decline steeper than the common TDE decline rate of $t^{-5/3}$ \citep{Mainetti_2016, Bonnerot_Rossi_2019}. 

We discuss the case of sequential tidal disruption occurring after separation of the binary, producing two debris streams that collide before they have returned to pericentre, resulting in a precursor when the most bound elements of the two streams collide \citep{Bonnerot_Rossi_2019}. This occurs at a distance  
\begin{equation}
    R^{\textrm{mb}}_{\textrm{col}} \approx 2 a_{\textrm{min}}
\end{equation}
from the SMBH (near apocentre), where 
\begin{equation}
    a_{\textrm{min}} = 23 \ M_{6}^{2/3} m_{\star}^{-2/3} r_{\star} \ \textrm{au} \ .
\end{equation}

For the streams to collide at all, the time between the first and second star passing through pericentre, $\Delta t$, and the angle measuring the relative shift in the pericentre location of the two stars, $\Delta \theta$, have a limited number of possible combinations. The time $\Delta t$ depends on the radii of the stars and the binary separation, as well as 
the time taken for the most bound debris to return to pericentre,
\begin{equation} \label{eq:t_min}
    t_{\textrm{min}} = 41 \ M_{6}^{1/2} \ m_{\star}^{-1} \ r_{\star}^{3/2} \ \textrm{days} \ .
\end{equation}
\cite{Bonnerot_Rossi_2019} predict the probability of collision between streams is maximised to $44 \%$ under certain ideal conditions. These conditions are that: 
both disrupted stars have $\beta=1$, 
the widths of the debris streams expand homologously, and that both stars follow Keplerian orbits.

In their numerical simulations, \cite{Bonnerot_Rossi_2019} begin with two stars on parabolic orbits that both have $M_{\star} = 1 M_{\odot}$, $R_{\star} = 1 R_{\odot}$, a binary separation $a=1000 R_{\odot}$, $\beta=2$, and $M_{BH} = 10^6 M_{\odot}$. In their strongest collision model, \cite{Bonnerot_Rossi_2019} determine that, assuming the gas internal energy of $\sim 10^{48} \ \textrm{erg}$ is rapidly radiated, a flare with luminosity $\sim 10^{43} \ \textrm{erg s}^{-1}$ results, and is likely observable as optical emission. 
\cite{Bonnerot_Rossi_2019} predict that the precursor is likely to be found in the weeks to months before the detection of the TDE. The duration of the precursor may be a few days to a few months, depending on the ratio $\Delta t / \Delta \theta$.
This ratio sets
\begin{equation} \label{eq:t_col_mu_orig}
    \frac{t_{\textrm{col}}^{\textrm{mu}}}{t_{\textrm{min}}} \approx 6.4 \ \beta^{-1/2} M_{6}^{-1/6} m_{\star}^{1/6} \left( \frac{\Delta \theta}{10^{-2} \pi} \right)^{-1} \frac{\Delta t}{t_{\textrm{min}}} \ ,
\end{equation}
the time at which the most unbound (``mu'') debris of the second stream collides with an element of the first stream. This in turn sets the duration of the collision. 
 
The peak luminosity of the ATLAS22kjn bump
is similar to the double TDE stream-stream collision luminosity predicted by \cite{Bonnerot_Rossi_2019} for their strongest collision model. Additionally, the ATLAS22kjn bump begins approximately one week before the discovery of ATLAS22kjn, consistent with shorter time of occurrence predicted by \cite{Bonnerot_Rossi_2019} for a double TDE precursor. Provided there exists a combination of $\Delta \theta$ and $\Delta t$ that allows for streams collision, the ATLAS22kjn bump duration is consistent with a precursor from this scenario.

To compare with the allowed space of $\Delta \theta$-$\Delta t$ values for streams collision determined by \cite{Bonnerot_Rossi_2019}, we compute possible $\Delta t$ and $\Delta \theta$ values. Under the same assumptions about stellar mass and radius as \cite{Bonnerot_Rossi_2019}, and with our SMBH mass estimate (Section \ref{sec:host}), we first determine the fallback time: $t_{\textrm{min}} = 68$ days. 
Since our observations do not provide constraints on the allowed values of the binary separation $a$ and the orbital angles, we evaluate $\Delta t$ and $\Delta \theta$ across the range of allowed values specified in \cite{Bonnerot_Rossi_2019}. We find that $\Delta t$ can be in the range $0$ to $\sim 260$ days.
We compare this range with the calculations of likely $\Delta t$ values by \cite{Yu_Dong_2024}. For a binary system in which both stars have mass $M_{\star}=1 M_{\odot}$, the binary semimajor axis is $a_{\textrm{min}} = 45$ AU as computed above, $\beta \simeq 1.6$, and our SMBH mass estimate, $\Delta t \simeq 110$ days \citep[see][]{Yu_Dong_2024}. The range of $\Delta t$ we compute with the assumptions in \cite{Bonnerot_Rossi_2019} is therefore plausible.
For $t_{\textrm{min}} = 68 \ \textrm{days}$, a range of combinations of $\Delta \theta$ and $\Delta t$ for which the two streams will collide exist, suggesting the bump duration we observe is also plausible in this scenario.

The ATLAS22kjn bump begins $\sim 3.3$ days prior to $t_1$, the start of the rise to the main UV/optical peak. Assuming this is the dynamical time from the collision point of the streams to the SMBH (approximately pericentre for the highly eccentric orbits of bound material), then the dynamical time is half the orbital timescale, so $t_{\textrm{orb}} \simeq 6.6 \ \textrm{days}$. With our SMBH mass estimate of $M_{BH} \simeq 10^{6.44} M_{\odot}$, our observations imply $a \simeq 10 \ \textrm{au}$, and therefore an apocentre distance of $R \simeq 20 \ \textrm{au}$. For our SMBH mass estimate and the same stellar mass and radius assumptions as \cite{Bonnerot_Rossi_2019}, $R_{\textrm{col}} \simeq 90 \ \textrm{au}$. Our observations therefore imply a collision radius that is significantly closer to the SMBH than the collision radius predicted by \cite{Bonnerot_Rossi_2019} for the most bound debris. While this suggests our observations are less consistent with the predictions for streams collision in a double TDE, a closer collision is not directly ruled out by the model.

Comparisons between an observed precursor and double TDE predictions were also made by \cite{Huang_2024} for the TDE AT 2023lli, which, like ATLAS22kjn, showed delayed X-ray emission.
\cite{Huang_2024} suggested that this delayed X-ray emission may also be a result of colliding streams in a double TDE. In this scenario, a collision-induced outflow is produced when the two streams collide \citep[e.g.,][]{lu20}. As the outflow expands with time, its optical depth may decrease, allowing X-rays to escape. Though we do not detect X-rays at early times, we observe CLs that require energies in the EUV/soft X-ray range, consistent with the presence of reprocessing material in ATLAS22kjn. Reprocessing, however, is not sufficient to explain why X-rays are again not detected at late times.

While some aspects of the double TDE explanation are consistent with our observations of ATLAS22kjn, important caveats exist. The ideal conditions that maximise the probability of collision between the two streams pairs $\beta =1$ with the condition that the streams expand homologously, expected when tidal forces set the expansion \citep[e.g.,][]{kochanek94} and typically only true for deeper encounters \citep[$\beta\gtrsim3$;][]{Bonnerot_Rossi_2019}. The actual collision probability is therefore most likely $\le 44\%$.
Furthermore, in their numerical simulations of double TDEs, \cite{Mandel2015} found that sequential disruption of both stars resulted from only 18\% of the simulated binary systems, but estimate that double TDEs comprise 10\% of all TDEs. As such, the likelihood of a double TDE with stream collision is uncertain across the broader TDE population. The decline of ATLAS22kjn is a mismatch for the predicted steeper or knee-like decline of double TDEs; the decline of ATLAS22kjn is shallower compared to many other TDEs, rather than steeper. A double TDE and subsequent streams collision is therefore unlikely to explain the ATLAS22kjn bump.

\begin{table*}
    \centering
    \caption{\normalfont This table summarises our discussion of possible explanations for the ATLAS22kjn bump. The first column lists the characteristics of the bump (top rows) and the main flare (bottom two rows). The subsequent columns indicate whether the theoretical models, identified in the top row, predict the characteristics observed for ATLAS22kjn. A green `{\color{Green} \checkmark}' denotes agreement between the model prediction and what we observe, an orange `{\color{orange} O}' indicates the model may match under certain conditions, a red `{\color{red} \text{\sffamily X}}' denotes disagreement, and a `?' indicates the model does not make a prediction.}
    \vspace{1mm}
    \begin{tabular}{c | c c c c c c c} \toprule
    \\[-0.5ex] 
         & \makecell{ATLAS22kjn \\ Observations} &  Nozzle Shock &  \makecell{Cooling of \\ Unbound \\ Stellar Debris}
        &  \makecell{Stream-Disk \\ Collision \\ in rpTDE} &  \makecell{Stream-Stream \\ Collision} & \makecell{Stream-Stream \\ Collision in \\ Double TDE} & \makecell{Wind- \\ Stream \\ Collision} \\[-0.5ex] \\ \midrule \\[-0.5ex]
         \multirow{7}{*}[-0.2em]{\rotatebox{90}{Bump}} 
         & $9 \substack{+4 \\ -2}$ -day duration & ? & {\color{orange} O} & {\color{orange} O} & {\color{orange} O} & {\color{orange} O} & {\color{Green} \checkmark} \\ \cline{2-8} \\[-0.5ex]
         & $\left(125 \substack{+5 \\ -3}\right)$ -day separation (rest-frame) & ? & ? & {\color{orange} O} & ? & {\color{orange} O} & ? \\ \cline{2-8} \\[-0.5ex]
         & $L = (3.0 \substack{+0.2 \\ -0.4})\times 10^{42} \ \textrm{erg s}^{-1}$  & {\color{red} \sffamily X} & {\color{orange} O} & {\color{Green} \checkmark} & {\color{Green} \checkmark} & {\color{Green} \checkmark} & {\color{orange} O} \\[1.0ex] \cline{2-8} \\[-0.5ex]
         & $E = (1.5  \substack{+0.9 \\ -0.5})\times 10^{48} \ \textrm{erg}$ & ? & {\color{red} \text{\sffamily X}} & {\color{orange} O} & {\color{Green} \checkmark} & {\color{orange} O} & {\color{orange} O} \\ \midrule \\
         \multirow{2}{*}[0.8em]{\rotatebox{90}{Main Flare}} & Slow rise ($125.4 \substack{+0.3 \\ -0.9}$ rest-frame days) & ? & ? & ? & ? & ? & {\color{Green} \checkmark} \\[1.0ex] \cline{2-8} \\[1.0ex]
         & Slow decline (see $\Delta L_{40}$ in Figure \protect\ref{fig:d_l40}) & ? & ? & ? & ? & {\color{red} \text{\sffamily X}} & ? \\[1.0ex] 
         \bottomrule
    \end{tabular}
    \label{tab:bump_explanations}
\end{table*}

\subsection{Wind-Stream Collision}
\label{sec:wind-stream collision}

The initial collision of a disk wind with the tidal debris stream can heat dense material in the innermost edge of the stream and potentially produce a pre-peak light curve feature \citep{Calderon2024}. In the simulations of \cite{Calderon2024}, the wind-stream interaction can produce a fast and sharp precursor, occurring in the first tens of days of the event, with a duration of $< 30$ days. When the wind first collides with the debris stream, the light curves of \cite{Calderon2024} show a $\lesssim 20 \%$ increase in luminosity, relative to light curves without a contribution from wind-stream collision.
For $M_{BH} \sim 10^6 M_{\odot}$ and low accretion efficiency $\eta \sim 0.01$, the amplitude of the precursor is comparable to the amplitude of the main flare. 
The precursor is likely indistinguishable from the main peak for events with high accretion efficiency $\eta \sim 0.1$, however, if $M_{BH} \gtrsim 10^7 M_{\odot}$, the simulated light curves show a rise to peak delayed by $50-100$ days.

To compare with the $\lesssim 20 \%$ increase in luminosity in the simulations of \cite{Calderon2024}, we compare our measured bump luminosity to the rise model (equation \ref{eq:power law}) fitted in Section \ref{sec:Early-Time Rise Modelling}.
The bump luminosity is a $\sim 1000\%$ increase relative to the rise model, in contrast with the $\lesssim 20 \%$ increase \cite{Calderon2024} found relative to their fiducial light curves, though we caution our rise model is not a one-to-one comparison with the fiducial models of \cite{Calderon2024}.
The $\sim 9$-day duration of the ATLAS22kjn bump is consistent with the $< 30$-day duration predicted for $M_{BH} \sim 10^6 M_{\odot}$. 
The slow rise of ATLAS22kjn to its main peak is also seen in the simulations of \citep{Calderon2024}, specifically for $M_{BH} \gtrsim 10^7 M_{\odot}$ and $\eta \sim 0.1$. Moreover, since our estimated SMBH mass lies between the two extremes explored by \cite{Calderon2024}, and ATLAS22kjn exhibits characteristics of both scenarios, it is plausible that wind–stream collisions in TDEs give rise to a continuum of observational signatures. The example of ATLAS22kjn is qualitatively consistent with such a scenario. Finally, the ATLAS22kjn bump began to rise 
$3.3 \substack{+1.0 \\ -1.1}$ days before the start of the rise to main peak $t_1$ (Section \ref{sec:BB modelling}), 
and therefore might reasonably occur in the ``first tens of days'' as predicted by \cite{Calderon2024}. 

Considering the ZTF $i$-band, since ATLAS22kjn was not in the TESS field for most of the main flare, the bump is not comparable to the amplitude of the main UV/optical peak, instead amounting to only $\sim 30 \%$ of the main peak.
If we crudely approximate a wind-stream bump as triangular, using a peak luminosity equivalent to the main flare and a duration of $\sim9$ days, the total radiated energy is $2 \times 10^{49} \ \textrm{erg}$, $\sim 20$ times greater than bump energy we measure.
We emphasise that this is only an order-of-magnitude estimate, however, and that \cite{Calderon2024} note the predicted bump amplitude depends on the relative kinetic energy injected into the debris stream, so the bump amplitude we observe may still be compatible with the this model. 

In this first exploration by \cite{Calderon2024}, simulations are performed in two dimensions, but the precessed tidal debris stream is inherently a three-dimensional process. Additionally, their simulations are performed under a grey approximation, but more detailed radiative transfer is likely needed in these simulations to construct a better estimate of the expected SED. Finally, as an important first exploration, the study by \cite{Calderon2024} necessarily focuses on a limited set of parameters. Future work that builds upon this study with three-dimensional simulations, improved radiative transfer, and additional parameters will be essential for refining theoretical predictions of the wind-stream process and enabling more robust comparisons with observations.

\vspace{10mm}

\subsection{Summary of Possible Bump Origins}
\label{sec:bump summary}

The observed characteristics of the ATLAS22kjn bump are most consistent with the possible explanations of stream-stream collision (Section \ref{sec:stream-stream collision}) and wind-stream collision (Section \ref{sec:wind-stream collision}).
We summarise our discussion of potential bump explanations in Table \ref{tab:bump_explanations}. We emphasise the need for more theoretical work in all these scenarios, and hope that this work will serve as a benchmark for future theoretical studies.

\section{Comparison of ATLAS22kjn With Other CLEs}
\label{sec:CLE discussion}

\begin{figure*}[t]
    \centering
    \includegraphics[width=1\textwidth]{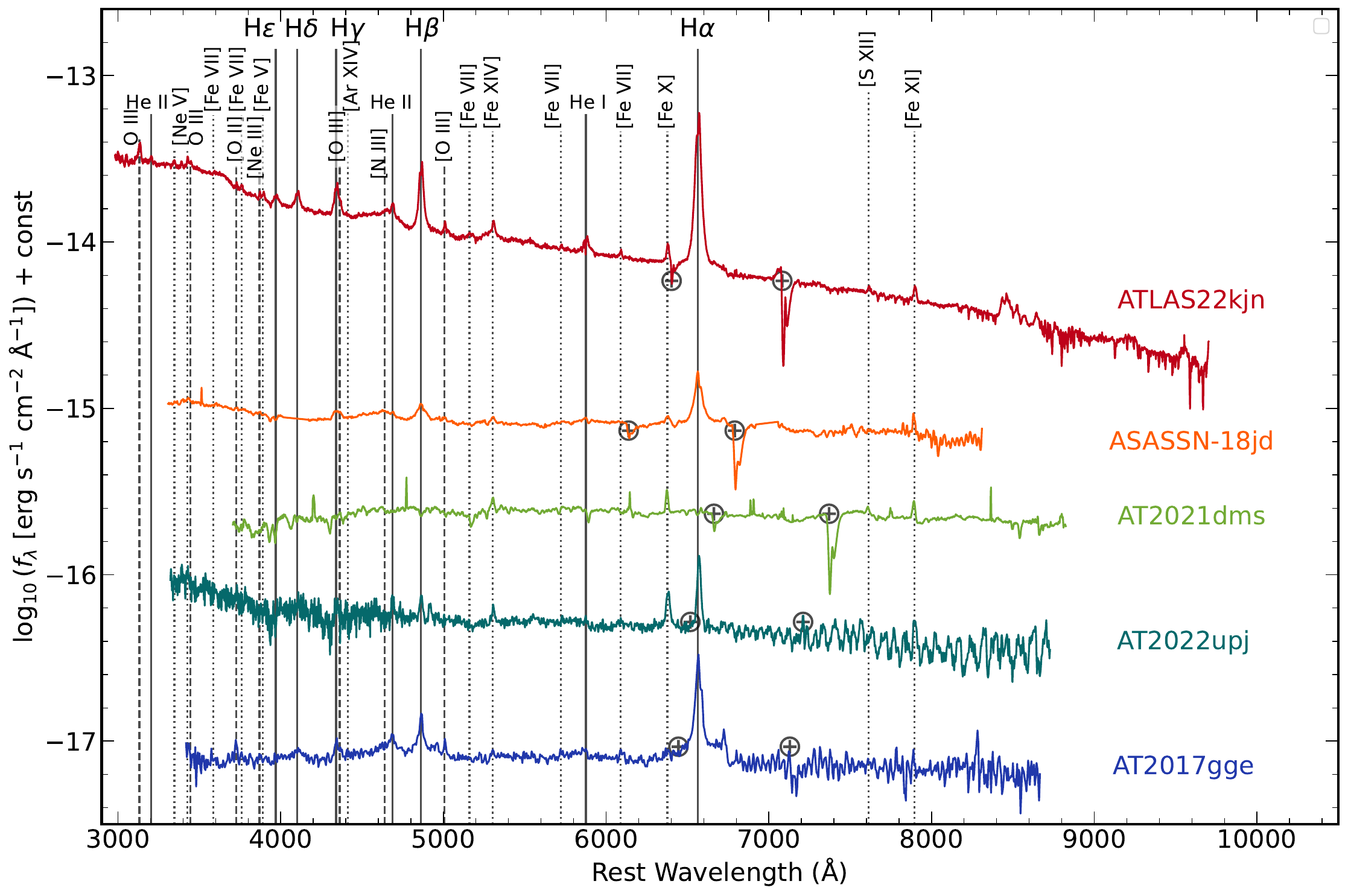}
    \caption{Milky Way extinction-corrected optical spectra for ATLAS22kjn and a comparison sample of CLEs, offset by a constant. The spectrum used for ATLAS22kjn is the LRIS $+8\textrm{d}$ spectrum (also presented in Figure \protect\ref{fig:uv opt spectrum}). Broad emission lines are represented by black solid lines, CLs by black dotted lines, and other narrow lines by black dashed lines. The Earth symbols $\oplus$ indicate telluric features, corrected for each CLE's host-galaxy redshift.}
    \label{fig:CLE spectra}
\end{figure*}

To contextualise ATLAS22kjn, we compare its emission with other CLEs. The CLs of ATLAS22kjn are similar to what has been observed for other CLEs, including ASASSN-18jd \citep{neustadt20}, AT 2019qiz \citep{nicholl20}, AT 2021dms \citep{forster21, Hinkle2024}, AT 2022upj \citep{fulton22, newsome22, Newsome_2024}, AT 2017gge/ATLAS17jrp \citep{wang22b, onori22}, and ASASSN-18ap \citep{Wang2024}. In Figure \ref{fig:CLE spectra}, we highlight the resemblance between early-time optical spectra for these CLEs and the $\textrm{LRIS} + 8\textrm{d}$ spectrum for ATLAS22kjn. With the exception of AT 2021dms, these objects all show blue continuum emission, strong Balmer lines, and [\ion{Fe}{vii}]. 
The absence of [\ion{Fe}{vii}] for AT 2021dms suggests the observed CL emission originates from gas of densities $\gtrsim 10^7 \ \textrm{cm}^{-3}$, for which [\ion{Fe}{vii}] is collisionally de-excited \citep{wang12}.

Iron CLs are most common for these CLEs, but neon, sulfur, and argon lines have also been observed in some cases. ASASSN-18ap, AT 2019qiz, possibly ASASSN-18jd, and ATLAS22kjn show [\ion{Ne}{v}] emission. AT 2022upj and ATLAS22kjn show [\ion{S}{XII}] $\lambda 7612$ as well. ATLAS22kjn is the only CLE that shows [\ion{Ar}{xiv}] $\lambda 4414$, which we observe only at late times. The $3\sigma$ luminosity upper limits we estimated for [\ion{Ar}{xiv}] (Section \ref{sec:CL evolution}) suggest [\ion{Ar}{xiv}] is likely produced but undetectable against the strong blue continuum present in the early spectra, as may be true for other CLEs too. 

These CLEs also share characteristics in the IR. 
In MIR photometry, AT 2017gge, ASASSN-18jd, ASASSN-18ap, AT 2022upj, and ATLAS22kjn show dust reprocessing echoes, suggesting the presence of a dusty torus in their nuclear environments. 
The dust covering fractions $f_c > 0.17$ for AT 2017gge and $f_c > 0.30$ for ASASSN-18jd \citep{hinkle22b} are similar to the covering fraction of $f_c = 0.40 \pm 0.03$ (Section \ref{sec:mir light curve}) we estimate for ATLAS22kjn. Only AT 2017gge \citep{onori22} and ATLAS22kjn have NIR spectra, but both show [\ion{Fe}{xiii}] $\lambda$10798 emission. 

In X-rays, as shown in Figure \ref{fig:uvot_xray_bb}, the CLEs show similar maximum blackbody luminosities in the $0.3-10 \ \textrm{keV}$ range, of order $\sim 10^{42} - 10^{43} \ \textrm{erg s}^{-1}$.   
The blackbody temperatures measured for AT 2017gge, ASASSN-18jd, and ATLAS22kjn are similar to those measured for typical TDEs \citep[$kT \sim 50-150 \ \textrm{eV}$; e.g.,][]{Guolo_2024}. 
For AT 2019qiz \citep[][]{nicholl20} and AT 2022upj \citep[][]{Newsome_2024}, the X-ray emission was better described by a power law, so we do not compare with the blackbody temperature estimates for these objects.
Lastly, ASASSN-18ap was not detected during observations taken both at early and late times during its evolution. As such, an X-ray blackbody temperature of $kT = 50 \ \textrm{eV}$ was assumed \citep{Wang2024}.

Many CLEs are initially not detected in X-rays. 
This includes AT 2017gge \citep{wang22b, onori22}, whose X-ray emission was delayed from UV/optical peak by $\sim 200$ days and preceded the appearance of CLs. AT 2019qiz \citep{nicholl20, Short2023} was also detected in the X-rays before the appearance of CLs, but showed X-ray emission even before UV/optical peak. 
AT 2019qiz has the longest delay between X-ray emission and the appearance of CLs, which may have appeared anytime between the last spectrum, at $+157$d, presented by \cite{nicholl20} and the first detection of CLs at $+428$d \citep{Short2023}. 
Like AT 2019qiz, ASASSN-18jd \citep{neustadt20} and AT 2022upj \citep{Newsome_2024} show early-time X-ray emission, but this is coincident with the appearance CLs.
Meanwhile, ASASSN-18ap shows CL emission around $+200$ days, though X-rays were not detected at either the early or late phases \citep{Wang2024}. Similarly, ATLAS22kjn shows CLs before the first X-ray detection, and CLs persist at late times when X-rays are again not detected. The presence of CLs without X-ray detections in these cases suggests the X-ray emission is obscured, either by a reprocessing layer or by the SMBH environment. 
We discuss this further for ATLAS22kjn in Section \ref{sec:implications for SMBH environments}.

\section{Implications for SMBH Environments}
\label{sec:implications for SMBH environments}

The environments of SMBHs provide insight into their accretion histories, and have been studied through the multi-wavelength emission of AGNs \citep[e.g.,][]{Riffel2006, lamperti17, Almeida_2017, Smith_2025}. The environments of quiescent SMBHs remain understudied due to their observational inaccessibility. The multi-wavelength emission of TDEs provides a probe of these environments, from the accretion disk out to any dusty torus.

\begin{figure*}[t]
    \centering    \includegraphics[width=1\linewidth]{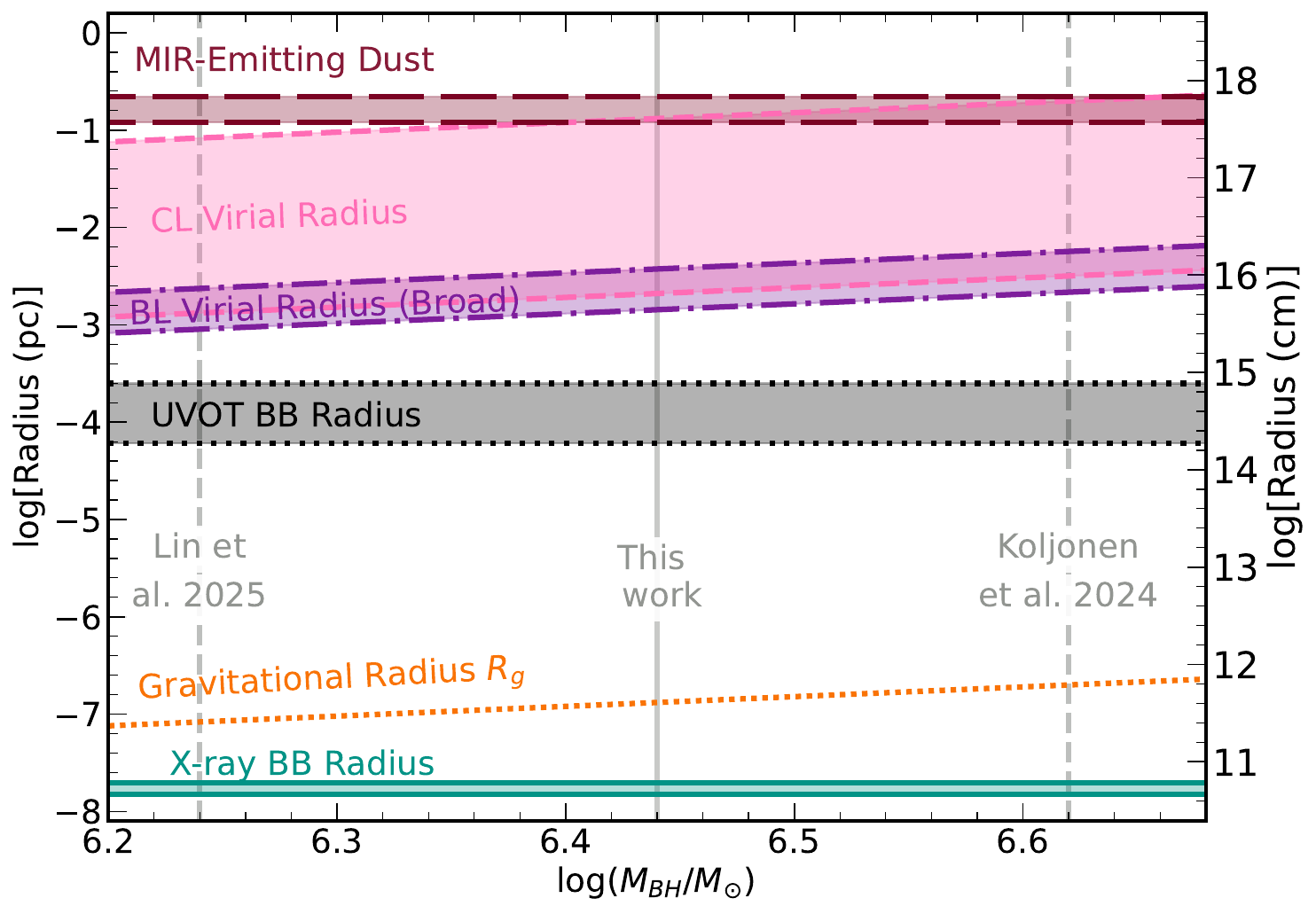}
    \caption{Radii of the emission regions surrounding the central SMBH of ATLAS22kjn, as a function of SMBH mass. The horizontal axis range is centered on our SMBH mass estimate, $\log{M_{BH}/M_{\odot}} \simeq 6.44$ \citep[using][see Section \ref{sec:host}]{reines15},
    and spans the intrinsic scatter of 0.24 dex. For completeness, we indicate the \cite{Koljonen2024} mass estimate of $\log{M_{BH}/M_{\odot}} \simeq 6.62$ with an intrinsic scatter of 0.58 dex, and the \cite{Lin_2025} mass estimate of $\log{M_{BH}/M_{\odot}} \simeq 6.24$ also using \cite{reines15} with an intrinsic scatter of 0.24 dex. We plot the gravitational radius $R_g$ (orange dotted line), as well as the range of measurements between the 5th and 95th percentile (90\% confidence interval) for: the X-ray blackbody (BB) radius (teal solid lines), the UVOT BB radius (black dotted lines), the broad-component (purple dash-dot lines) BL virial radii, CL virial radius (pink dashed lines), and the distance to the MIR-emitting dust (maroon long-dashed line), assuming lag-time is equivalent to light travel time. Our mass estimate, as well as that of \cite{Koljonen2024} and \cite{Lin_2025}, respectively, are reasonable for the expectation that the CL-emitting gas is within the torus and coincides with the inner edge of the torus, as probed by the MIR-emitting dust. The X-ray radius was derived under the assumption that the emission is well-fit by a blackbody (see Sections \ref{sec:Analysis-Xray} and \ref{sec:implications for SMBH environments}), however, this is a simplification and results in X-ray emission radius that is less than the gravitational radius.}
    \label{fig:dist_SMBH}
\end{figure*}

Figure \ref{fig:dist_SMBH} summarises the various scales we have inferred as a function of SMBH mass. From smallest to largest, we include the inferred size of the X-ray emitting region (Section \ref{sec:Analysis-Xray}), the UV/optical blackbody radius (Section \ref{sec:BB modelling}), virial estimates---simply estimated as $r \sim G M_{BH}/\textrm{FWHM}^2$---of the the broad lines (BLs) and the CLs, and the dust radius
(Section \ref{sec:mir light curve}).
As in Section \ref{sec:Analysis-Xray}, we emphasise that we obtained the X-ray radius under the assumption that the emission is well-fit by a blackbody, as results in an unphysical X-ray radius and suggests that detailed disk modelling would yield a better fit. We discussed the UVOT BB radius in Section \ref{sec:BB modelling}. The broad-component BL emission originates between the UVOT BB radius and the CL region, with some overlap into the CL region, and its radius is consistent with the BL emission radii of other TDEs \citep[e.g.,][]{Short2023, Newsome_2024, Earl_2025, Clark_2025}. The overlap of the BL region with the CL region is consistent with the results of \cite{Kynoch_2026}, who studied a sample of variable and non-variable CLEs and found no strong evidence for ionisation stratification in the line emission regions of ATLAS22kjn. We do not include an estimate of the narrow-component BL emission as the FWHMs we measure are mostly consistent with instrumental resolution limit.

\begin{figure*}[t]
    \centering    \includegraphics[width=1\linewidth]{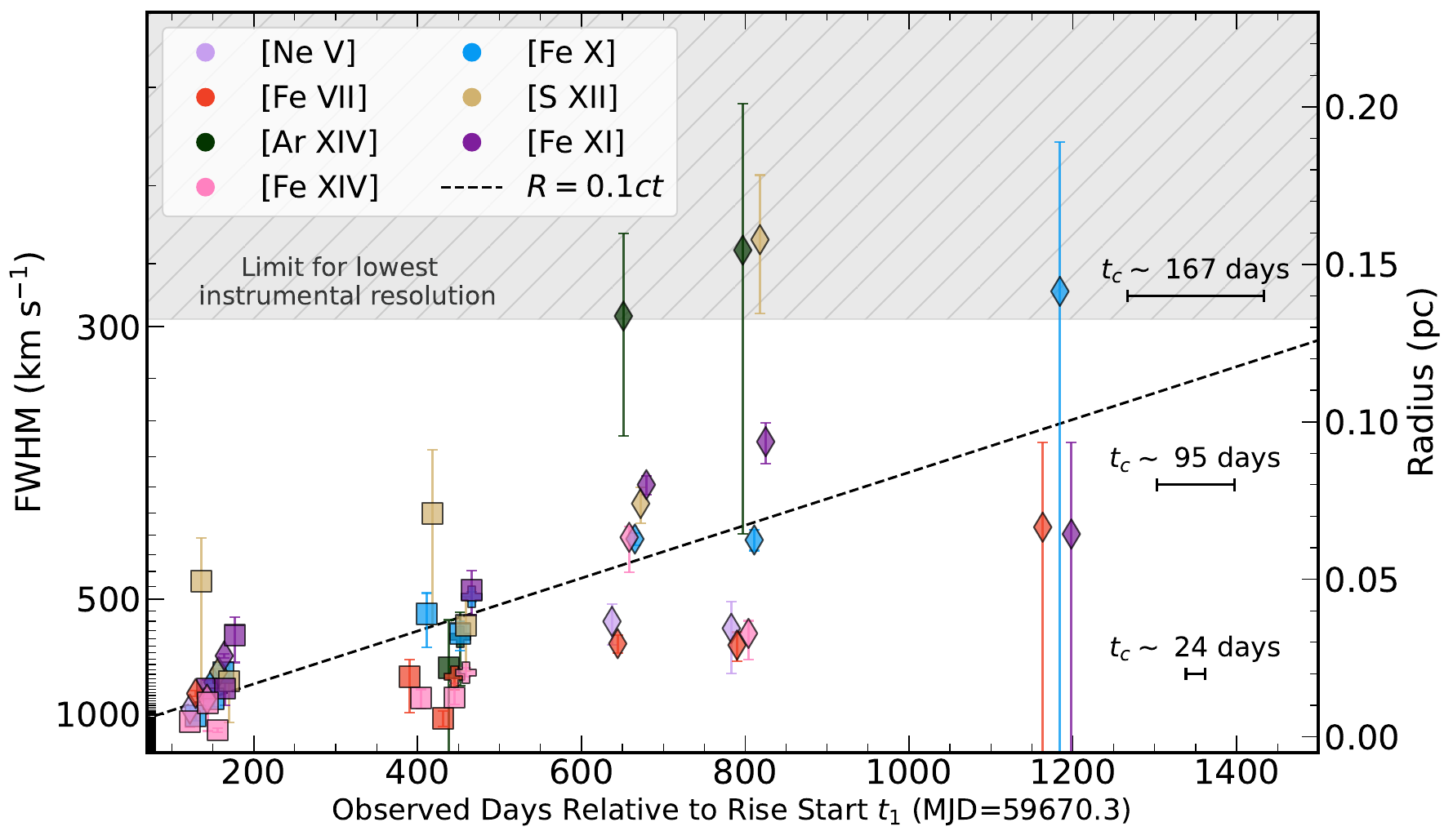}
    \caption{Widths of the optical CLs over time. The radius axis on the right assumes the velocities of the CLs are virial velocities. Squares indicate UH2.2m spectra, diamonds indicate Keck spectra, and plus signs indicate Gemini spectra. To aid in visualisation, we offset the different lines within an epoch by 7 days. [\ion{Fe}{XIV}] is not offset. The black dashed line indicates a characteristic velocity of $0.1c$. The grey hatched region indicates the limit for the lowest instrumental resolution (R=1000, blue side of the SNIFS instrument on UH2.2m). The black horizontal bars show the light travel time to the evenly-spaced, arbitrarily-chosen regions 0.02 pc, 0.08 pc, and 0.14 pc, respectively, and indicate the timescale over which variability in the ionising source is averaged out for the CLR. The individual CLs display a decrease in width over time.}
    \label{fig:CL_widths_time}
\end{figure*}

While the evolution of the BL widths is flat (see Section \ref{sec:opt spectra}), the CL widths decrease over time. This is shown in Figure \ref{fig:CL_widths_time}, where the CLs probe greater radii at later times under the assumption of virialised gas. In Figure \ref{fig:lum_CLs}, we illustrated that the CLs also decrease in luminosity with time, and this is the same line width-luminosity correlation typically seen in TDE BLs. For the iron lines, we observe [\ion{Fe}{XIV}] to be the broadest CL in most of the spectra, suggesting it is produced at smaller radii. Our findings are similar to those of \cite{Kynoch_2026}. Under the same assumption of virialised gas, \cite{Kynoch_2026} determined the emission radii of broad lines and CLs, including [\ion{Fe}{x}], [\ion{Fe}{xi}], and [\ion{Fe}{xiv}] in a DESI spectrum taken on MJD$=60093$ ($\sim 288$ days post- our measured UV/optical peak). Accounting for the difference in our SMBH mass estimates, we compare with our measurements at $+270 \textrm{d}$ and find that the line widths we measure and those measured by \cite{Kynoch_2026} are consistent within $1 \sigma$ for [\ion{Fe}{x}], and within $1.5 \sigma$ for [\ion{Fe}{xi}] and [\ion{Fe}{xiv}]. Compared with trends observed by \cite{Kynoch_2026} across their sample, ATLAS22kjn is unique in that [\ion{Fe}{vii}] is present before higher-ionisation lines fade, detected as early as $+8\textrm{d}$. 
Still, ATLAS22kjn shows the trend that high-ionisation lines both appear and disappear earlier than lower-ionisation lines, as is evident from the disappearance of [\ion{Fe}{xiv}]  (IP$=361.00$ eV), [\ion{S}{xii}] (IP$=504.78$ eV), and [\ion{Ar}{xiv}] (IP$=686.09$ eV) from the most recent spectrum, $\textrm{KCWI} \ +1042 \textrm{d}$, while [\ion{Fe}{vii}] (IP$=99.00$ eV), [\ion{Fe}{x}] (IP$=235.04$ eV), and [\ion{Fe}{xi}] (IP$=262.10$ eV) are still present. Though we do not detect [\ion{Ne}{v}] $\lambda 3347$ in the most recent spectrum of ATLAS22kjn, we note that [\ion{Ne}{v}] $\lambda 3427$ is still present, suggesting [\ion{Ne}{v}] (IP$=97.11$ eV) reflects the trend observed by \cite{Kynoch_2026} as well.

Our observations of the CL emission radii are also similar to theoretical predictions for CL emission in TDEs. \cite{Mummery_2025} modelled CL emission in TDEs using the TDE disk evolution code FitTED \citep{Mummery_2024} coupled with the photoionisation simulation code \textsc{Cloudy} \citep{Ferland_1998}. 
\cite{Mummery_2025} studied [\ion{Fe}{vii}],  [\ion{Fe}{x}], [\ion{Fe}{xiv}], and [\ion{O}{iii}] $\lambda 5007$ emission, and found that reprocessing by high-density ($n_H \sim 10^6-10^8 \ \textrm{cm}^{-3}$) clouds near the SMBH is capable of producing high-ionisation lines such as [\ion{Fe}{xiv}], while lower-ionisation lines like [\ion{O}{iii}] and [\ion{Fe}{vii}] generally originate at greater distances. 
These predictions match what we observe for the iron lines of ATLAS22kjn (Figure \ref{fig:CL_widths_time}).
Like \cite{Mummery_2025}, we observe [\ion{Fe}{x}] to be produced at a range of radii, while [\ion{Fe}{xiv}] is only produced at smaller radii. In contrast with \cite{Mummery_2025}, who note that [\ion{Fe}{xiv}] is generally a faint line, however, we observe [\ion{Fe}{xiv}] to be the strongest CL in the ATLAS22kjn spectra. Our observations also differ from the predictions of \cite{Mummery_2025} in the early-time characteristics of the lower ionisation potential lines [\ion{O}{iii}] and [\ion{Fe}{vii}]. We detect [\ion{O}{iii}] and [\ion{Fe}{vii}] as early as the $\textrm{LRIS}+ 8 \textrm{d}$ spectrum,
as opposed to switching on $\sim 1$ year post-flare.
The early-time detection of [\ion{O}{iii}] could be attributed to the LLAGN we found evidence for in Section \ref{sec:host}, but the early-time detection of [\ion{Fe}{vii}] and its luminosity relative to [\ion{O}{iii}] ($\sim 0.7$) suggests that it does not originate from an AGN.

At late times, the luminosity trends we observe for the CLs of ATLAS22kjn are similar to trends identified by \cite{Mummery_2024} for CLs arising from nuclear gas clouds with densities $n_{\textrm{H}} = 10^8 \ \textrm{cm}^{-3}$. As predicted by \cite{Mummery_2024}, [\ion{Fe}{xiv}] fades the fastest, and [\ion{Fe}{vii}] emission is slightly weaker than [\ion{O}{iii}] at all times. Our observations differ from \cite{Mummery_2024} in that [\ion{Fe}{xiv}] is still detected $\sim 2$ years after the onset of the TDE ($t_1$, Section \ref{sec:Early-Time Rise Modelling}), while \cite{Mummery_2024} predict [\ion{Fe}{xiv}] to fade after $\sim 1$ year. Additionally, we observe [\ion{Fe}{x}] emission to be more luminous than [\ion{O}{iii}] throughout most of the TDE, consistent with \cite{Mummery_2024}, but less luminous than [\ion{O}{iii}] after $\sim 3$ years. The luminosities we observe for [\ion{Fe}{vii}] and [\ion{Fe}{x}] after $\sim 3$ years are of order $10^{39} \ \textrm{erg s}^{-1}$, also in contrast with the $\sim 10^{37} \ \textrm{erg s}^{-1}$ line luminosities that \cite{Mummery_2024} predict. \cite{Mummery_2024} do not study [\ion{Fe}{xi}] or non-iron CLs, but we note that [\ion{Fe}{xi}] shows similar evolution and emission radii to [\ion{Fe}{x}]---as may be expected given their close ionisation potentials and critical densities---and in general, [\ion{Ne}{v}] $\lambda$3347, [\ion{Ar}{xiv}], and [\ion{S}{xii}] show similar luminosity trends to the iron CLs.
We suggest that some, if not all, of the differences between predictions for iron CLs of \cite{Mummery_2024} and our observations can be attributed to nuances of ATLAS22kjn's nuclear environment.

The presence of CLs at early and late times, when X-rays are not detected, has further implications for the environment of ATLAS22kjn. While early-time X-ray obscuration has been observed for several TDEs, this is typically attributed to delayed accretion disk formation \citep[e.g.,][]{gezari17, Pasham_2017, Kajava_2020, Liu_2022, wang22b, onori22} and is inconsistent with the early-time CL emission in ATLAS22kjn. Furthermore, the non-detection of X-rays at late times suggests the X-ray emission of ATLAS22kjn is also inconsistent with a reprocessing picture in which X-rays are initially obscured by an optically thick outflow that expands and decreases in optical depth. Instead, the X-rays may be obscured by the CLR itself or by the dusty torus---as also suggested by \cite{Koljonen2024}.

As shown in Figure \ref{fig:dist_SMBH}, the far edge of the CL emission coincides with the inner edge of the dusty torus we estimate to be at $0.17 \pm 0.03$ pc. This compares favourably with the result of \cite{Somalwar_2022}, who, for a potential TDE/AGN, also estimated the CL-emitting gas and MIR-emitting dust to be located at similar distances from the SMBH. 
The coincidence of this region with the MIR-emitting dust is in keeping with the consistent detections of MIR flares for CLEs \citep{Hinkle2024, Wang2024}. 

\section{Conclusions} 
\label{sec:Conclusions}

We have presented an analysis of multi-wavelength photometric and spectroscopic observations of the nuclear transient ATLAS22kjn. For the main UV/optical flare of ATLAS22kjn, we determined a peak bolometric luminosity of $(5.4 \pm 0.2) \times 10^{43} \textrm{erg s}^{-1}$, which is low compared to many CLEs and TDEs. Similarly, we observed a relatively cool blackbody temperature of $\sim 16,000 \ \textrm{K}$. Although cool compared to most TDEs, the flat temperature evolution is typical of TDEs, and similar temperatures have been observed for other CLEs. 
The UV/optical light curves show minimal variability prior to peak, aside from a $9 \substack{+4 \\ -2}$-day luminosity bump coinciding with the start of the main flare, with a peak luminosity of $L \simeq (1.5 \pm 0.3)\times 10^{42} \ \textrm{erg s}^{-1}$. 

We examined six possible mechanisms for powering the bump, and found that the theoretical predictions of the stream-stream model and wind-stream collision model were best matched to our observations. However, all of the possible models would benefit from further development to enable more meaningful observational comparisons. We note the serendipity of Gaia observing ATLAS22kjn during a rarely-caught bump, as well as its rise to the main UV/optical peak; flux and wavelength-calibrated Gaia low resolution BP/RP spectra in DR4 are currently expected to be released December 2026\footnote{ID 18.09, \url{https://www.cosmos.esa.int/web/gaia/dr4}}.

ATLAS22kjn exhibits many high-ionisation ($\gtrsim$ 100 eV) CLs in both the optical and NIR, that have been observed previously in the spectra of both AGNs and TDEs. The strengths of these CLs are comparable to those of the [\ion{O}{iii}] $\lambda$5007 line, indicating that the CL emission of ATLAS22kjn is powered by the TDE flare, and not by an AGN. 
These CLs probe the otherwise unobservable EUV and ultrasoft X-ray emission from the transient. We observe the evolution of the CL line luminosities to be ionisation-potential dependent, suggestive of a 10\% decrease in the temperature of the ionising source relative to the weighted-average X-ray temperature.
We encourage future theoretical work that makes observational predictions for TDE CL emission including for line species other than iron, and follow-up of future discoveries of CLEs with a greater number of high-S/N spectra with broad wavelength coverage, obtained at high cadence.
Additionally, upcoming UV missions like ULTRASAT \citep{ULTRASAT_2024} and UVEX \citep{UVEX_2023} will independently provide constraints on TDE SEDs further into the UV, enabling us to better leverage CLs to probe the EUV/soft X-ray evolution of TDEs.

We conclude that ATLAS22kjn is a TDE occurring in a galaxy nucleus with a high density of gas and a dusty torus. ATLAS22kjn contributes to the growing sample of nuclear transients with early-time light curve features, as well as the to sample of CLEs, both of which provide important avenues toward understanding the UV/optical emission mechanisms of TDEs. 
Future discoveries of nuclear transients like ATLAS22kjn will allow us to refine theoretical predictions for the physical mechanisms that power TDEs. In turn, these TDEs will provide insight into the accretion processes that drive SMBH growth, as well as the broader dynamics of galaxy evolution, and deepen our understanding of the connection between these properties.

\section*{Acknowledgements}

A.C.E. acknowledges support from Research Experience for Undergraduates program at the Institute for Astronomy, University of Hawai`i-Manoa funded through NSF grant \#2050710. 
A.C.E. thanks M. Jessen for encouragement and useful comments on the manuscript.

We thank Jack Neustadt for sharing the spectral data for ASASSN-18jd. We thank Helena Treiber for contributing the background sample of galaxies from HyperLeda. We thank Fabio Bresolin for helpful discussions on the changing Balmer decrement.

J.T.H. acknowledges support from NASA through the NASA Hubble Fellowship grant HST-HF2-51577.001-A, awarded by STScI. STScI is operated by the Association of Universities for Research in Astronomy, Incorporated, under NASA contract NAS5-26555.
The Shappee group at the University of Hawai‘i is supported with funds from NSF (grants AST-2407205) and NASA (grants HST-GO-17087, 80NSSC24K0521, 80NSSC24K0490, 80NSSC23K1431). CSK and KZS are supported by NSF grants AST-2307385, and 2407206. 

Parts of this research were supported by the Australian Research Council Discovery Early Career Researcher Award (DECRA) through project number DE230101069.

A.D. is supported by the European Research Council (ERC) under the European Union’s Horizon 2020 research and innovation programme under Grant Agreement No. 101002652 (BayeSN; PI K. Mandel) and Marie Skłodowska-Curie Grant Agreement No. 873089 (ASTROSTAT-II).

This work is based on observations made by Keck I, Keck II, UH2.2m, Gemini North, and IRTF. We wish to extend our special thanks to those of Hawaiian ancestry on whose sacred mountain of Maunakea we are privileged to be guests. The observations presented herein would not have been possible without their generosity.

The ZTF forced-photometry service was funded under the Heising-Simons Foundation grant \#12540303 (PI: Graham).

The CSS survey is funded by the National Aeronautics and Space Administration under Grant No. NNG05GF22G issued through the Science Mission Directorate Near-Earth Objects Observations Program. The CRTS survey is supported by the U.S. National Science Foundation under
grants AST-0909182.

We acknowledge ESA Gaia, DPAC and the Photometric Science Alerts Team (http://gsaweb.ast.cam.ac.uk/alerts).

This publication makes use of data products from the Near-Earth Object Wide-field Infrared Survey Explorer (NEOWISE), which is a joint project of the Jet Propulsion Laboratory/California Institute of Technology and the University of California Los Angeles. NEOWISE is funded by the National Aeronautics and Space Administration.

This research has made use of the SVO Filter Profile Service ``Carlos Rodrigo", funded by MCIN/AEI/10.13039/501100011033/ through grant PID2020-112949GB-I00.

Funding for the Sloan Digital Sky Survey V has been provided by the Alfred P. Sloan Foundation, the Heising-Simons Foundation, the National Science Foundation, and the Participating Institutions. SDSS acknowledges support and resources from the Center for High-Performance Computing at the University of Utah. SDSS telescopes are located at Apache Point Observatory, funded by the Astrophysical Research Consortium and operated by New Mexico State University, and at Las Campanas Observatory, operated by the Carnegie Institution for Science. The SDSS web site is \url{www.sdss.org}.

SDSS is managed by the Astrophysical Research Consortium for the Participating Institutions of the SDSS Collaboration, including Caltech, The Carnegie Institution for Science, Chilean National Time Allocation Committee (CNTAC) ratified researchers, The Flatiron Institute, the Gotham Participation Group, Harvard University, Heidelberg University, The Johns Hopkins University, L’Ecole polytechnique f\'{e}d\'{e}rale de Lausanne (EPFL), Leibniz-Institut für Astrophysik Potsdam (AIP), Max-Planck-Institut für Astronomie (MPIA Heidelberg), Max-Planck-Institut für Extraterrestrische Physik (MPE), Nanjing University, National Astronomical Observatories of China (NAOC), New Mexico State University, The Ohio State University, Pennsylvania State University, Smithsonian Astrophysical Observatory, Space Telescope Science Institute (STScI), the Stellar Astrophysics Participation Group, Universidad Nacional Aut\'{o}noma de M\'{e}xico, University of Arizona, University of Colorado Boulder, University of Illinois at Urbana-Champaign, University of Toronto, University of Utah, University of Virginia, Yale University, and Yunnan University.

\bibliographystyle{aasjournal}

\bibliography{ATLAS22kjn_OJA}

@ARTICLE{burgarella05,
       author = {{Burgarella}, D. and {Buat}, V. and {Iglesias-P{\'a}ramo}, J.},
        title = "{Star formation and dust attenuation properties in galaxies from a statistical ultraviolet-to-far-infrared analysis}",
      journal = {\mnras},
         year = 2005,
        month = jul,
       volume = {360},
       number = {4},
        pages = {1413-1425},
          doi = {10.1111/j.1365-2966.2005.09131.x},
archivePrefix = {arXiv},
       eprint = {astro-ph/0504434},
 primaryClass = {astro-ph},
       adsurl = {https://ui.adsabs.harvard.edu/abs/2005MNRAS.360.1413B}
}

@ARTICLE{boquien19,
       author = {{Boquien}, M. and {Burgarella}, D. and {Roehlly}, Y. and {Buat}, V. and {Ciesla}, L. and {Corre}, D. and {Inoue}, A.~K. and {Salas}, H.},
        title = "{CIGALE: a python Code Investigating GALaxy Emission}",
      journal = {\aap},
         year = 2019,
        month = feb,
       volume = {622},
          eid = {A103},
        pages = {A103},
          doi = {10.1051/0004-6361/201834156},
archivePrefix = {arXiv},
       eprint = {1811.03094},
 primaryClass = {astro-ph.GA},
       adsurl = {https://ui.adsabs.harvard.edu/abs/2019A\&A...622A.103B}
}

@ARTICLE{makarov14,
       author = {{Makarov}, Dmitry and {Prugniel}, Philippe and {Terekhova}, Nataliya and {Courtois}, H{\'e}l{\`e}ne and {Vauglin}, Isabelle},
        title = "{HyperLEDA. III. The catalogue of extragalactic distances}",
      journal = {\aap},
         year = 2014,
        month = oct,
       volume = {570},
          eid = {A13},
        pages = {A13},
          doi = {10.1051/0004-6361/201423496},
archivePrefix = {arXiv},
       eprint = {1408.3476},
 primaryClass = {astro-ph.GA},
       adsurl = {https://ui.adsabs.harvard.edu/abs/2014A\&A...570A..13M}
}

@ARTICLE{hinkle22a,
       author = {{Hinkle}, Jason T. and {Holoien}, Thomas W. -S. and {Shappee}, Benjamin. J. and {Neustadt}, Jack M.~M. and {Auchettl}, Katie and {Vallely}, Patrick J. and {Shahbandeh}, Melissa and {Kluge}, Matthias and {Kochanek}, Christopher S. and {Stanek}, K.~Z. and {Huber}, Mark E. and {Post}, Richard S. and {Bersier}, David and {Ashall}, Christopher and {Tucker}, Michael A. and {Williams}, Jonathan P. and {de Jaeger}, Thomas and {Do}, Aaron and {Fausnaugh}, Michael and {Gruen}, Daniel and {Hopp}, Ulrich and {Myles}, Justin and {Obermeier}, Christian and {Payne}, Anna V. and {Thompson}, Todd A.},
        title = "{The Curious Case of ASASSN-20hx: A Slowly Evolving, UV- and X-Ray-Luminous, Ambiguous Nuclear Transient}",
      journal = {\apj},
         year = 2022,
        month = may,
       volume = {930},
       number = {1},
          eid = {12},
        pages = {12},
          doi = {10.3847/1538-4357/ac5f54},
archivePrefix = {arXiv},
       eprint = {2108.03245},
 primaryClass = {astro-ph.HE},
       adsurl = {https://ui.adsabs.harvard.edu/abs/2022ApJ...930...12H}
}

@article{hinkle22b,
    author = {Hinkle, Jason T},
    title = {Mid-infrared echoes of ambiguous nuclear transients reveal high dust covering fractions: evidence for dusty tori},
    journal = {Monthly Notices of the Royal Astronomical Society},
    volume = {531},
    number = {2},
    pages = {2603-2614},
    year = {2024},
    month = {05},
    issn = {0035-8711},
    doi = {10.1093/mnras/stae1229},
    url = {https://doi.org/10.1093/mnras/stae1229},
    eprint = {https://academic.oup.com/mnras/article-pdf/531/2/2603/58061262/stae1229.pdf},
}

@ARTICLE{jiang21b,
       author = {{Jiang}, Ning and {Wang}, Tinggui and {Hu}, Xueyang and {Sun}, Luming and {Dou}, Liming and {Xiao}, Lin},
        title = "{Infrared Echoes of Optical Tidal Disruption Events: {\ensuremath{\sim}}1\% Dust-covering Factor or Less at Subparsec Scale}",
      journal = {\apj},
         year = 2021,
        month = apr,
       volume = {911},
       number = {1},
          eid = {31},
        pages = {31},
          doi = {10.3847/1538-4357/abe772},
archivePrefix = {arXiv},
       eprint = {2102.08044},
 primaryClass = {astro-ph.GA},
       adsurl = {https://ui.adsabs.harvard.edu/abs/2021ApJ...911...31J}
}

@ARTICLE{jiang21a,
       author = {{Jiang}, Ning and {Wang}, Tinggui and {Dou}, Liming and {Shu}, Xinwen and {Hu}, Xueyang and {Liu}, Hui and {Wang}, Yibo and {Yan}, Lin and {Sheng}, Zhenfeng and {Yang}, Chenwei and {Sun}, Luming and {Zhou}, Hongyan},
        title = "{Mid-infrared Outbursts in Nearby Galaxies (MIRONG). I. Sample Selection and Characterization}",
      journal = {\apjs},
         year = 2021,
        month = feb,
       volume = {252},
       number = {2},
          eid = {32},
        pages = {32},
          doi = {10.3847/1538-4365/abd1dc},
archivePrefix = {arXiv},
       eprint = {2012.06806},
 primaryClass = {astro-ph.GA},
       adsurl = {https://ui.adsabs.harvard.edu/abs/2021ApJS..252...32J}
}

@ARTICLE{newsome22,
       author = {{Newsome}, M. and {Dgany}, Y. and {Arcavi}, I. and {Howell}, D.~A. and {McCully}, C. and {Gonzalez}, E.~P. and {Pellegrino}, C.},
        title = "{Global SN Project Transient Classification Report for 2022-11-05}",
      journal = {Transient Name Server Classification Report},
         year = 2022,
        month = nov,
       volume = {2022-3231},
        pages = {1},
       adsurl = {https://ui.adsabs.harvard.edu/abs/2022TNSCR3231....1N}
}

@ARTICLE{fulton22,
       author = {{Fulton}, M. and {Srivastav}, S. and {Smith}, K.~W. and {Young}, D.~R. and {Moore}, T. and {Weston}, J. and {Sim}, S.~A. and {Smartt}, S.~J. and {Shingles}, L.},
        title = "{QUB Transient Classification Report for 2022-11-07}",
      journal = {Transient Name Server Classification Report},
         year = 2022,
        month = nov,
       volume = {2022-3245},
        pages = {1},
       adsurl = {https://ui.adsabs.harvard.edu/abs/2022TNSCR3245....1F}
}

@ARTICLE{forster21,
       author = {{Forster}, F. and {Bauer}, F.~E. and {Pignata}, G. and {Munoz-Arancibia}, A. and {Hernandez-Garcia}, L. and {Galbany}, L. and {Camacho}, E. and {Silva-Farfan}, J. and {Mourao}, A. and {Arredondo}, J. and {Cabrera-Vives}, G. and {Carrasco-Davis}, R. and {Estevez}, P.~A. and {Huijse}, P. and {Reyes}, E. and {Reyes}, I. and {Sanchez-Saez}, P. and {Valenzuela}, C. and {Castillo}, E. and {Ruz-Mieres}, D. and {Rodriguez-Mancini}, D. and {Catelan}, M. and {Eyheramendy}, S. and {Graham}, M.~J.},
        title = "{ALeRCE/ZTF Transient Discovery Report for 2021-02-21}",
      journal = {Transient Name Server Discovery Report},
         year = 2021,
        month = feb,
       volume = {2021-542},
        pages = {1-542},
       adsurl = {https://ui.adsabs.harvard.edu/abs/2021TNSTR.542....1F}
}

@ARTICLE{wang22b,
       author = {{Wang}, Yibo and {Jiang}, Ning and {Wang}, Tinggui and {Zhu}, Jiazheng and {Dou}, Liming and {Lin}, Zheyu and {Sun}, Luming and {Liu}, Hui and {Sheng}, Zhenfeng},
        title = "{Discovery of ATLAS17jrp as an Optical-, X-Ray-, and Infrared-bright Tidal Disruption Event in a Star-forming Galaxy}",
      journal = {\apjl},
         year = 2022,
        month = may,
       volume = {930},
       number = {1},
          eid = {L4},
        pages = {L4},
          doi = {10.3847/2041-8213/ac6670},
archivePrefix = {arXiv},
       eprint = {2204.05461},
 primaryClass = {astro-ph.GA},
       adsurl = {https://ui.adsabs.harvard.edu/abs/2022ApJ...930L...4W}
}

@ARTICLE{wang22a,
       author = {{Wang}, Yibo and {Jiang}, Ning and {Wang}, Tinggui and {Yan}, Lin and {Sheng}, Zhenfeng and {Dou}, Liming and {Ding}, Jiani and {Cai}, Zheng and {Sun}, Luming and {Yang}, Chenwei and {Shu}, Xinwen},
        title = "{Mid-infrared Outbursts in Nearby Galaxies (MIRONG). II. Optical Spectroscopic Follow-up}",
      journal = {\apjs},
         year = 2022,
        month = jan,
       volume = {258},
       number = {1},
          eid = {21},
        pages = {21},
          doi = {10.3847/1538-4365/ac33a6},
archivePrefix = {arXiv},
       eprint = {2111.12729},
 primaryClass = {astro-ph.GA},
       adsurl = {https://ui.adsabs.harvard.edu/abs/2022ApJS..258...21W}
}

@ARTICLE{palaversa16,
       author = {{Palaversa}, Lovro and {Gezari}, Suvi and {Sesar}, Branimir and {Stuart}, J. Scott and {Wozniak}, Przemyslaw and {Holl}, Berry and {Ivezi{\'c}}, {\v{Z}}eljko},
        title = "{Revealing the Nature of Extreme Coronal-line Emitter SDSS J095209.56+214313.3}",
      journal = {\apj},
         year = 2016,
        month = mar,
       volume = {819},
       number = {2},
          eid = {151},
        pages = {151},
          doi = {10.3847/0004-637X/819/2/151},
archivePrefix = {arXiv},
       eprint = {1512.08614},
 primaryClass = {astro-ph.GA},
       adsurl = {https://ui.adsabs.harvard.edu/abs/2016ApJ...819..151P}
}

@ARTICLE{onori22,
       author = {{Onori}, F. and {Cannizzaro}, G. and {Jonker}, P.~G. and {Kim}, M. and {Nicholl}, M. and {Mattila}, S. and {Reynolds}, T.~M. and {Fraser}, M. and {Wevers}, T. and {Brocato}, E. and {Anderson}, J.~P. and {Carini}, R. and {Charalampopoulos}, P. and {Clark}, P. and {Gromadzki}, M. and {Guti{\'e}rrez}, C.~P. and {Ihanec}, N. and {Inserra}, C. and {Lawrence}, A. and {Leloudas}, G. and {Lundqvist}, P. and {M{\"u}ller-Bravo}, T.~E. and {Piranomonte}, S. and {Pursiainen}, M. and {Rybicki}, K.~A. and {Somero}, A. and {Young}, D.~R. and {Chambers}, K.~C. and {Gao}, H. and {de Boer}, T.~J.~L. and {Magnier}, E.~A.},
        title = "{The nuclear transient AT 2017gge: a tidal disruption event in a dusty and gas-rich environment and the awakening of a dormant SMBH}",
      journal = {\mnras},
         year = 2022,
        month = nov,
       volume = {517},
       number = {1},
        pages = {76-98},
          doi = {10.1093/mnras/stac2673},
archivePrefix = {arXiv},
       eprint = {2206.00049},
 primaryClass = {astro-ph.HE},
       adsurl = {https://ui.adsabs.harvard.edu/abs/2022MNRAS.517...76O}
}

@ARTICLE{komossa08,
       author = {{Komossa}, S. and {Zhou}, H. and {Wang}, T. and {Ajello}, M. and {Ge}, J. and {Greiner}, J. and {Lu}, H. and {Salvato}, M. and {Saxton}, R. and {Shan}, H. and {Xu}, D. and {Yuan}, W.},
        title = "{Discovery of Superstrong, Fading, Iron Line Emission and Double-peaked Balmer Lines of the Galaxy SDSS J095209.56+214313.3: The Light Echo of a Huge Flare}",
      journal = {\apjl},
         year = 2008,
        month = may,
       volume = {678},
       number = {1},
        pages = {L13},
          doi = {10.1086/588281},
archivePrefix = {arXiv},
       eprint = {0804.2670},
 primaryClass = {astro-ph},
       adsurl = {https://ui.adsabs.harvard.edu/abs/2008ApJ...678L..13K}
}

@ARTICLE{hammerstein23,
       author = {{Hammerstein}, Erica and {van Velzen}, Sjoert and {Gezari}, Suvi and {Cenko}, S. Bradley and {Yao}, Yuhan and {Ward}, Charlotte and {Frederick}, Sara and {Villanueva}, Natalia and {Somalwar}, Jean J. and {Graham}, Matthew J. and {Kulkarni}, Shrinivas R. and {Stern}, Daniel and {Andreoni}, Igor and {Bellm}, Eric C. and {Dekany}, Richard and {Dhawan}, Suhail and {Drake}, Andrew J. and {Fremling}, Christoffer and {Gatkine}, Pradip and {Groom}, Steven L. and {Ho}, Anna Y.~Q. and {Kasliwal}, Mansi M. and {Karambelkar}, Viraj and {Kool}, Erik C. and {Masci}, Frank J. and {Medford}, Michael S. and {Perley}, Daniel A. and {Purdum}, Josiah and {van Roestel}, Jan and {Sharma}, Yashvi and {Sollerman}, Jesper and {Taggart}, Kirsty and {Yan}, Lin},
        title = "{The Final Season Reimagined: 30 Tidal Disruption Events from the ZTF-I Survey}",
      journal = {\apj},
         year = 2023,
        month = jan,
       volume = {942},
       number = {1},
          eid = {9},
        pages = {9},
          doi = {10.3847/1538-4357/aca283},
archivePrefix = {arXiv},
       eprint = {2203.01461},
 primaryClass = {astro-ph.HE},
       adsurl = {https://ui.adsabs.harvard.edu/abs/2023ApJ...942....9H}
}

@ARTICLE{wright06,
       author = {{Wright}, E.~L.},
        title = "{A Cosmology Calculator for the World Wide Web}",
      journal = {\pasp},
         year = 2006,
        month = dec,
       volume = {118},
       number = {850},
        pages = {1711-1715},
          doi = {10.1086/510102},
archivePrefix = {arXiv},
       eprint = {astro-ph/0609593},
 primaryClass = {astro-ph},
       adsurl = {https://ui.adsabs.harvard.edu/abs/2006PASP..118.1711W}
}

@ARTICLE{bennett14,
       author = {{Bennett}, C.~L. and {Larson}, D. and {Weiland}, J.~L. and {Hinshaw}, G.},
        title = "{The 1\% Concordance Hubble Constant}",
      journal = {\apj},
         year = 2014,
        month = oct,
       volume = {794},
       number = {2},
          eid = {135},
        pages = {135},
          doi = {10.1088/0004-637X/794/2/135},
archivePrefix = {arXiv},
       eprint = {1406.1718},
 primaryClass = {astro-ph.CO},
       adsurl = {https://ui.adsabs.harvard.edu/abs/2014ApJ...794..135B}
}

@ARTICLE{lamperti17,
       author = {{Lamperti}, Isabella and {Koss}, Michael and {Trakhtenbrot}, Benny and {Schawinski}, Kevin and {Ricci}, Claudio and {Oh}, Kyuseok and {Landt}, Hermine and {Riffel}, Rog{\'e}rio and {Rodr{\'\i}guez-Ardila}, Alberto and {Gehrels}, Neil and {Harrison}, Fiona and {Masetti}, Nicola and {Mushotzky}, Richard and {Treister}, Ezequiel and {Ueda}, Yoshihiro and {Veilleux}, Sylvain},
        title = "{BAT AGN Spectroscopic Survey - IV: Near-Infrared Coronal Lines, Hidden Broad Lines, and Correlation with Hard X-ray Emission}",
      journal = {\mnras},
         year = 2017,
        month = may,
       volume = {467},
       number = {1},
        pages = {540-572},
          doi = {10.1093/mnras/stx055},
archivePrefix = {arXiv},
       eprint = {1701.02755},
 primaryClass = {astro-ph.GA},
       adsurl = {https://ui.adsabs.harvard.edu/abs/2017MNRAS.467..540L}
}

@article{cendes21,
	author = {Cendes, Y. and Alexander, K. D. and Berger, E. and Eftekhari, T. and Williams, P. K. G. and Chornock, R.},
	doi = {10.3847/1538-4357/ac110a},
	journal = {The Astrophysical Journal},
	month = {oct},
	number = {2},
	pages = {127},
	publisher = {The American Astronomical Society},
	title = {Radio Observations of an Ordinary Outflow from the Tidal Disruption Event AT2019dsg},
	url = {https://doi.org/10.3847/1538-4357/ac110a},
	volume = {919},
	year = {2021}}

@ARTICLE{vandenberk01,
       author = {{Vanden Berk}, Daniel E. and {Richards}, Gordon T. and {Bauer}, Amanda and {Strauss}, Michael A. and {Schneider}, Donald P. and {Heckman}, Timothy M. and {York}, Donald G. and {Hall}, Patrick B. and {Fan}, Xiaohui and {Knapp}, G.~R. and {Anderson}, Scott F. and {Annis}, James and {Bahcall}, Neta A. and {Bernardi}, Mariangela and {Briggs}, John W. and {Brinkmann}, J. and {Brunner}, Robert and {Burles}, Scott and {Carey}, Larry and {Castander}, Francisco J. and {Connolly}, A.~J. and {Crocker}, J.~H. and {Csabai}, Istv{\'a}n and {Doi}, Mamoru and {Finkbeiner}, Douglas and {Friedman}, Scott and {Frieman}, Joshua A. and {Fukugita}, Masataka and {Gunn}, James E. and {Hennessy}, G.~S. and {Ivezi{\'c}}, {\v{Z}}eljko and {Kent}, Stephen and {Kunszt}, Peter Z. and {Lamb}, D.~Q. and {Leger}, R. French and {Long}, Daniel C. and {Loveday}, Jon and {Lupton}, Robert H. and {Meiksin}, Avery and {Merelli}, Aronne and {Munn}, Jeffrey A. and {Newberg}, Heidi Jo and {Newcomb}, Matt and {Nichol}, R.~C. and {Owen}, Russell and {Pier}, Jeffrey R. and {Pope}, Adrian and {Rockosi}, Constance M. and {Schlegel}, David J. and {Siegmund}, Walter A. and {Smee}, Stephen and {Snir}, Yehuda and {Stoughton}, Chris and {Stubbs}, Christopher and {SubbaRao}, Mark and {Szalay}, Alexander S. and {Szokoly}, Gyula P. and {Tremonti}, Christy and {Uomoto}, Alan and {Waddell}, Patrick and {Yanny}, Brian and {Zheng}, Wei},
        title = "{Composite Quasar Spectra from the Sloan Digital Sky Survey}",
      journal = {\aj},
         year = 2001,
        month = aug,
       volume = {122},
       number = {2},
        pages = {549-564},
          doi = {10.1086/321167},
archivePrefix = {arXiv},
       eprint = {astro-ph/0105231},
 primaryClass = {astro-ph},
       adsurl = {https://ui.adsabs.harvard.edu/abs/2001AJ....122..549V}
}

@ARTICLE{bianchi05,
       author = {{Bianchi}, S. and {Guainazzi}, M. and {Matt}, G. and {Chiaberge}, M. and {Iwasawa}, K. and {Fiore}, F. and {Maiolino}, R.},
        title = "{A search for changing-look AGN in the Grossan catalog}",
      journal = {\aap},
         year = 2005,
        month = oct,
       volume = {442},
       number = {1},
        pages = {185-194},
          doi = {10.1051/0004-6361:20053389},
archivePrefix = {arXiv},
       eprint = {astro-ph/0507323},
 primaryClass = {astro-ph},
       adsurl = {https://ui.adsabs.harvard.edu/abs/2005A\&A...442..185B}
}

@INPROCEEDINGS{piascik14,
       author = {{Piascik}, A.~S. and {Steele}, Iain A. and {Bates}, Stuart D. and {Mottram}, Christopher J. and {Smith}, R.~J. and {Barnsley}, R.~M. and {Bolton}, B.},
        title = "{SPRAT: Spectrograph for the Rapid Acquisition of Transients}",
    booktitle = {Ground-based and Airborne Instrumentation for Astronomy V},
         year = 2014,
       editor = {{Ramsay}, Suzanne K. and {McLean}, Ian S. and {Takami}, Hideki},
       series = {Society of Photo-Optical Instrumentation Engineers (SPIE) Conference Series},
       volume = {9147},
        month = jul,
          eid = {91478H},
        pages = {91478H},
          doi = {10.1117/12.2055117},
       adsurl = {https://ui.adsabs.harvard.edu/abs/2014SPIE.9147E..8HP}
}

@ARTICLE{hinkle21b,
       author = {{Hinkle}, Jason T. and {Holoien}, Thomas W. -S. and {Shappee}, Benjamin. J. and {Auchettl}, Katie},
        title = "{A Swift Fix for Nuclear Outbursts}",
      journal = {\apj},
         year = 2021,
        month = apr,
       volume = {910},
       number = {2},
          eid = {83},
        pages = {83},
          doi = {10.3847/1538-4357/abe4d8},
archivePrefix = {arXiv},
       eprint = {2012.08521},
 primaryClass = {astro-ph.HE},
       adsurl = {https://ui.adsabs.harvard.edu/abs/2021ApJ...910...83H}
}

@ARTICLE{hinkle21a,
       author = {{Hinkle}, Jason T. and {Holoien}, T.~W. -S. and {Auchettl}, K. and {Shappee}, B.~J. and {Neustadt}, J.~M.~M. and {Payne}, A.~V. and {Brown}, J.~S. and {Kochanek}, C.~S. and {Stanek}, K.~Z. and {Graham}, M.~J. and {Tucker}, M.~A. and {Do}, A. and {Anderson}, J.~P. and {Bose}, S. and {Chen}, P. and {Coulter}, D.~A. and {Dimitriadis}, G. and {Dong}, Subo and {Foley}, R.~J. and {Huber}, M.~E. and {Hung}, T. and {Kilpatrick}, C.~D. and {Pignata}, G. and {Piro}, A.~L. and {Rojas-Bravo}, C. and {Siebert}, M.~R. and {Stalder}, B. and {Thompson}, Todd A. and {Tonry}, J.~L. and {Vallely}, P.~J. and {Wisniewski}, J.~P.},
        title = "{Discovery and follow-up of ASASSN-19dj: an X-ray and UV luminous TDE in an extreme post-starburst galaxy}",
      journal = {\mnras},
         year = 2021,
        month = jan,
       volume = {500},
       number = {2},
        pages = {1673-1696},
          doi = {10.1093/mnras/staa3170},
archivePrefix = {arXiv},
       eprint = {2006.06690},
 primaryClass = {astro-ph.HE},
       adsurl = {https://ui.adsabs.harvard.edu/abs/2021MNRAS.500.1673H}
}

@article{nicholl20,
    author = {Nicholl, M and Wevers, T and Oates, S R and Alexander, K D and Leloudas, G and Onori, F and Jerkstrand, A and Gomez, S and Campana, S and Arcavi, I and Charalampopoulos, P and Gromadzki, M and Ihanec, N and Jonker, P G and Lawrence, A and Mandel, I and Schulze, S and Short, P and Burke, J and McCully, C and Hiramatsu, D and Howell, D A and Pellegrino, C and Abbot, H and Anderson, J P and Berger, E and Blanchard, P K and Cannizzaro, G and Chen, T-W and Dennefeld, M and Galbany, L and González-Gaitán, S and Hosseinzadeh, G and Inserra, C and Irani, I and Kuin, P and Müller-Bravo, T and Pineda, J and Ross, N P and Roy, R and Smartt, S J and Smith, K W and Tucker, B and Wyrzykowski, Ł and Young, D R},
    title = {An outflow powers the optical rise of the nearby, fast-evolving tidal disruption event AT2019qiz},
    journal = {Monthly Notices of the Royal Astronomical Society},
    volume = {499},
    number = {1},
    pages = {482-504},
    year = {2020},
    month = {11},
    issn = {0035-8711},
    doi = {10.1093/mnras/staa2824},
    url = {https://doi.org/10.1093/mnras/staa2824},
    eprint = {https://academic.oup.com/mnras/article-pdf/499/1/482/33857159/staa2824.pdf},
}

@article{holoien20,
	author = {Holoien, Thomas W.-S. and Auchettl, Katie and Tucker, Michael A. and Shappee, Benjamin J. and Patel, Shannon G. and Miller-Jones, James C. A. and Mockler, Brenna and Groenewald, Dani{\`e}l N. and Hinkle, Jason T. and Brown, Jonathan S. and Kochanek, Christopher S. and Stanek, K. Z. and Chen, Ping and Dong, Subo and Prieto, Jose L. and Thompson, Todd A. and Beaton, Rachael L. and Connor, Thomas and Cowperthwaite, Philip S. and Dahmen, Linnea and French, K. Decker and Morrell, Nidia and Buckley, David A. H. and Gromadzki, Mariusz and Roy, Rupak and Coulter, David A. and Dimitriadis, Georgios and Foley, Ryan J. and Kilpatrick, Charles D. and Piro, Anthony L. and Rojas-Bravo, C{\'e}sar and Siebert, Matthew R. and Velzen, Sjoert van},
	doi = {10.3847/1538-4357/ab9f3d},
	journal = {The Astrophysical Journal},
	month = {aug},
	number = {2},
	pages = {161},
	publisher = {The American Astronomical Society},
	title = {The Rise and Fall of ASASSN-18pg: Following a TDE from Early to Late Times},
	url = {https://doi.org/10.3847/1538-4357/ab9f3d},
	volume = {898},
	year = {2020}}

@ARTICLE{vanvelzen21b,
       author = {{van Velzen}, Sjoert and {Pasham}, Dheeraj R. and {Komossa}, Stefanie and {Yan}, Lin and {Kara}, Erin A.},
        title = "{Reverberation in Tidal Disruption Events: Dust Echoes, Coronal Emission Lines, Multi-wavelength Cross-correlations, and QPOs}",
      journal = {\ssr},
         year = 2021,
        month = aug,
       volume = {217},
       number = {5},
          eid = {63},
        pages = {63},
          doi = {10.1007/s11214-021-00835-6},
archivePrefix = {arXiv},
       eprint = {2107.12268},
 primaryClass = {astro-ph.HE},
       adsurl = {https://ui.adsabs.harvard.edu/abs/2021SSRv..217...63V}
}

@ARTICLE{vanvelzen21,
       author = {{van Velzen}, Sjoert and {Gezari}, Suvi and {Hammerstein}, Erica and {Roth}, Nathaniel and {Frederick}, Sara and {Ward}, Charlotte and {Hung}, Tiara and {Cenko}, S. Bradley and {Stein}, Robert and {Perley}, Daniel A. and {Taggart}, Kirsty and {Foley}, Ryan J. and {Sollerman}, Jesper and {Blagorodnova}, Nadejda and {Andreoni}, Igor and {Bellm}, Eric C. and {Brinnel}, Valery and {De}, Kishalay and {Dekany}, Richard and {Feeney}, Michael and {Fremling}, Christoffer and {Giomi}, Matteo and {Golkhou}, V. Zach and {Graham}, Matthew J. and {Ho}, Anna. Y.~Q. and {Kasliwal}, Mansi M. and {Kilpatrick}, Charles D. and {Kulkarni}, Shrinivas R. and {Kupfer}, Thomas and {Laher}, Russ R. and {Mahabal}, Ashish and {Masci}, Frank J. and {Miller}, Adam A. and {Nordin}, Jakob and {Riddle}, Reed and {Rusholme}, Ben and {van Santen}, Jakob and {Sharma}, Yashvi and {Shupe}, David L. and {Soumagnac}, Maayane T.},
        title = "{Seventeen Tidal Disruption Events from the First Half of ZTF Survey Observations: Entering a New Era of Population Studies}",
      journal = {\apj},
         year = 2021,
        month = feb,
       volume = {908},
       number = {1},
          eid = {4},
        pages = {4},
          doi = {10.3847/1538-4357/abc258},
archivePrefix = {arXiv},
       eprint = {2001.01409},
 primaryClass = {astro-ph.HE},
       adsurl = {https://ui.adsabs.harvard.edu/abs/2021ApJ...908....4V}
}

@article{hinkle20a,
	author = {Hinkle, Jason T. and Holoien, Thomas W.-S. and Shappee, Benjamin. J. and Auchettl, Katie and Kochanek, Christopher S. and Stanek, K. Z. and Payne, Anna V. and Thompson, Todd A.},
	doi = {10.3847/2041-8213/ab89a2},
	journal = {The Astrophysical Journal Letters},
	month = {may},
	number = {1},
	pages = {L10},
	publisher = {The American Astronomical Society},
	title = {Examining a Peak-luminosity/Decline-rate Relationship for Tidal Disruption Events},
	url = {https://doi.org/10.3847/2041-8213/ab89a2},
	volume = {894},
	year = {2020}}

@ARTICLE{lu20,
       author = {{Lu}, Wenbin and {Bonnerot}, Cl{\'e}ment},
        title = "{Self-intersection of the fallback stream in tidal disruption events}",
      journal = {\mnras},
         year = 2020,
        month = feb,
       volume = {492},
       number = {1},
        pages = {686-707},
          doi = {10.1093/mnras/stz3405},
archivePrefix = {arXiv},
       eprint = {1904.12018},
 primaryClass = {astro-ph.HE},
       adsurl = {https://ui.adsabs.harvard.edu/abs/2020MNRAS.492..686L}
}

@ARTICLE{yalinewich19,
       author = {{Yalinewich}, A. and {Steinberg}, E. and {Piran}, T. and {Krolik}, J.~H.},
        title = "{Radio emission from the unbound debris of tidal disruption events}",
      journal = {\mnras},
         year = "2019",
        month = "Aug",
       volume = {487},
       number = {3},
        pages = {4083-4092},
          doi = {10.1093/mnras/stz1567},
archivePrefix = {arXiv},
       eprint = {1903.02575},
 primaryClass = {astro-ph.HE},
       adsurl = {https://ui.adsabs.harvard.edu/abs/2019MNRAS.487.4083Y}
}

@ARTICLE{masci19,
       author = {{Masci}, Frank J. and {Laher}, Russ R. and {Rusholme}, Ben and
         {Shupe}, David L. and {Groom}, Steven and {Surace}, Jason and
         {Jackson}, Edward and {Monkewitz}, Serge and {Beck}, Ron and
         {Flynn}, David and {Terek}, Scott and {Landry}, Walter and
         {Hacopians}, Eugean and {Desai}, Vandana and {Howell}, Justin and
         {Brooke}, Tim and {Imel}, David and {Wachter}, Stefanie and
         {Ye}, Quan-Zhi and {Lin}, Hsing-Wen and {Cenko}, S. Bradley and
         {Cunningham}, Virginia and {Rebbapragada}, Umaa and {Bue}, Brian and
         {Miller}, Adam A. and {Mahabal}, Ashish and {Bellm}, Eric C. and
         {Patterson}, Maria T. and {Juri{\'c}}, Mario and {Golkhou}, V. Zach and
         {Ofek}, Eran O. and {Walters}, Richard and {Graham}, Matthew and
         {Kasliwal}, Mansi M. and {Dekany}, Richard G. and {Kupfer}, Thomas and
         {Burdge}, Kevin and {Cannella}, Christopher B. and {Barlow}, Tom and
         {Van Sistine}, Angela and {Giomi}, Matteo and {Fremling}, Christoffer and
         {Blagorodnova}, Nadejda and {Levitan}, David and {Riddle}, Reed and
         {Smith}, Roger M. and {Helou}, George and {Prince}, Thomas A. and
         {Kulkarni}, Shrinivas R.},
        title = "{The Zwicky Transient Facility: Data Processing, Products, and Archive}",
      journal = {\pasp},
         year = "2019",
        month = "Jan",
       volume = {131},
       number = {995},
        pages = {018003},
          doi = {10.1088/1538-3873/aae8ac},
archivePrefix = {arXiv},
       eprint = {1902.01872},
 primaryClass = {astro-ph.IM},
       adsurl = {https://ui.adsabs.harvard.edu/abs/2019PASP..131a8003M}
}

@ARTICLE{alexander17,
       author = {{Alexander}, K.~D. and {Wieringa}, M.~H. and {Berger}, E. and
         {Saxton}, R.~D. and {Komossa}, S.},
        title = "{Radio Observations of the Tidal Disruption Event XMMSL1 J0740-85}",
      journal = {\apj},
         year = "2017",
        month = "Mar",
       volume = {837},
       number = {2},
          eid = {153},
        pages = {153},
          doi = {10.3847/1538-4357/aa6192},
archivePrefix = {arXiv},
       eprint = {1610.03861},
 primaryClass = {astro-ph.HE},
       adsurl = {https://ui.adsabs.harvard.edu/abs/2017ApJ...837..153A}
}

@ARTICLE{atkinson05,
       author = {{Atkinson}, J.~W. and {Collett}, J.~L. and {Marconi}, A. and
         {Axon}, D.~J. and {Alonso-Herrero}, A. and {Batcheldor}, D. and
         {Binney}, J.~J. and {Capetti}, A. and {Carollo}, C.~M. and
         {Dressel}, L. and {Ford}, H. and {Gerssen}, J. and {Hughes}, M.~A. and
         {Macchetto}, D. and {Maciejewski}, W. and {Merrifield}, M.~R. and
         {Scarlata}, C. and {Sparks}, W. and {Stiavelli}, M. and
         {Tsvetanov}, Z. and {van der Marel}, R.~P.},
        title = "{Supermassive black hole mass measurements for NGC 1300 and 2748 based on Hubble Space Telescope emission-line gas kinematics}",
      journal = {\mnras},
         year = "2005",
        month = "May",
       volume = {359},
       number = {2},
        pages = {504-520},
          doi = {10.1111/j.1365-2966.2005.08904.x},
archivePrefix = {arXiv},
       eprint = {astro-ph/0502573},
 primaryClass = {astro-ph},
       adsurl = {https://ui.adsabs.harvard.edu/abs/2005MNRAS.359..504A}
}

@ARTICLE{ford94,
       author = {{Ford}, Holland C. and {Harms}, Richard J. and {Tsvetanov}, Zlatan I. and
         {Hartig}, George F. and {Dressel}, Linda L. and {Kriss}, Gerard A. and
         {Bohlin}, Ralph C. and {Davidsen}, Arthur F. and {Margon}, Bruce and
         {Kochhar}, Ajay K.},
        title = "{Narrowband HST Images of M87: Evidence for a Disk of Ionized Gas around a Massive Black Hole}",
      journal = {\apjl},
         year = "1994",
        month = "Nov",
       volume = {435},
        pages = {L27},
          doi = {10.1086/187586},
       adsurl = {https://ui.adsabs.harvard.edu/abs/1994ApJ...435L..27F}
}

@ARTICLE{gebhardt11,
       author = {{Gebhardt}, Karl and {Adams}, Joshua and {Richstone}, Douglas and
         {Lauer}, Tod R. and {Faber}, S.~M. and {G{\"u}ltekin}, Kayhan and
         {Murphy}, Jeremy and {Tremaine}, Scott},
        title = "{The Black Hole Mass in M87 from Gemini/NIFS Adaptive Optics Observations}",
      journal = {\apj},
         year = "2011",
        month = "Mar",
       volume = {729},
       number = {2},
          eid = {119},
        pages = {119},
          doi = {10.1088/0004-637X/729/2/119},
archivePrefix = {arXiv},
       eprint = {1101.1954},
 primaryClass = {astro-ph.CO},
       adsurl = {https://ui.adsabs.harvard.edu/abs/2011ApJ...729..119G}
}

@ARTICLE{Alard2000,
       author = {{Alard}, C.},
        title = "{Image subtraction using a space-varying kernel}",
      journal = {\aaps},
         year = "2000",
        month = "Jun",
       volume = {144},
        pages = {363-370},
          doi = {10.1051/aas:2000214},
       adsurl = {https://ui.adsabs.harvard.edu/abs/2000A\&AS..144..363A}
}

@INPROCEEDINGS{Henden2015,
       author = {{Henden}, Arne A. and {Levine}, Stephen and {Terrell}, Dirk and
         {Welch}, Douglas L.},
        title = "{APASS - The Latest Data Release}",
    booktitle = {American Astronomical Society Meeting Abstracts \#225},
         year = "2015",
       series = {American Astronomical Society Meeting Abstracts},
       volume = {225},
        month = "Jan",
          eid = {336.16},
        pages = {336.16},
       adsurl = {https://ui.adsabs.harvard.edu/abs/2015AAS...22533616H}
}

@ARTICLE{martin05,
       author = {{Martin}, D. Christopher and {Fanson}, James and {Schiminovich}, David and
         {Morrissey}, Patrick and {Friedman}, Peter G. and {Barlow}, Tom A. and
         {Conrow}, Tim and {Grange}, Robert and {Jelinsky}, Patrick N. and
         {Milliard}, Bruno and {Siegmund}, Oswald H.~W. and {Bianchi}, Luciana and
         {Byun}, Yong-Ik and {Donas}, Jose and {Forster}, Karl and
         {Heckman}, Timothy M. and {Lee}, Young-Wook and {Madore}, Barry F. and
         {Malina}, Roger F. and {Neff}, Susan G. and {Rich}, R. Michael and
         {Small}, Todd and {Surber}, Frank and {Szalay}, Alex S. and
         {Welsh}, Barry and {Wyder}, Ted K.},
        title = "{The Galaxy Evolution Explorer: A Space Ultraviolet Survey Mission}",
      journal = {\apjl},
         year = "2005",
        month = "Jan",
       volume = {619},
       number = {1},
        pages = {L1-L6},
          doi = {10.1086/426387},
archivePrefix = {arXiv},
       eprint = {astro-ph/0411302},
 primaryClass = {astro-ph},
       adsurl = {https://ui.adsabs.harvard.edu/abs/2005ApJ...619L...1M}
}

@ARTICLE{ho08,
       author = {{Ho}, L.~C.},
        title = "{Nuclear activity in nearby galaxies.}",
      journal = {\araa},
         year = "2008",
        month = "Sep",
       volume = {46},
        pages = {475-539},
          doi = {10.1146/annurev.astro.45.051806.110546},
archivePrefix = {arXiv},
       eprint = {0803.2268},
 primaryClass = {astro-ph},
       adsurl = {https://ui.adsabs.harvard.edu/abs/2008ARA\&A..46..475H}
}

@ARTICLE{holoien19b,
       author = {{Holoien}, T.~W. -S. and {Huber}, M.~E. and {Shappee}, B.~J. and
         {Eracleous}, M. and {Auchettl}, K. and {Brown}, J.~S. and
         {Tucker}, M.~A. and {Chambers}, K.~C. and {Kochanek}, C.~S. and
         {Stanek}, K.~Z. and {Rest}, A. and {Bersier}, D. and {Post}, R.~S. and
         {Aldering}, G. and {Ponder}, K.~A. and {Simon}, J.~D. and
         {Kankare}, E. and {Dong}, D. and {Hallinan}, G. and {Reddy}, N.~A. and
         {Sanders}, R.~L. and {Topping}, M.~W. and {Pan-STARRS} and
         {Bulger}, J. and {Lowe}, T.~B. and {Magnier}, E.~A. and
         {Schultz}, A.~S.~B. and {Waters}, C.~Z. and {Willman}, M. and
         {Wright}, D. and {Young}, D.~R. and {ASAS-SN} and {Dong}, Subo and
         {Prieto}, J.~L. and {Thompson}, Todd A. and {ATLAS} and {Denneau}, L. and
         {Flewelling}, H. and {Heinze}, A.~N. and {Smartt}, S.~J. and
         {Smith}, K.~W. and {Stalder}, B. and {Tonry}, J.~L. and {Weiland}, H.},
        title = "{PS18kh: A New Tidal Disruption Event with a Non-axisymmetric Accretion Disk}",
      journal = {\apj},
         year = "2019",
        month = "Aug",
       volume = {880},
       number = {2},
          eid = {120},
        pages = {120},
          doi = {10.3847/1538-4357/ab2ae1},
archivePrefix = {arXiv},
       eprint = {1808.02890},
 primaryClass = {astro-ph.HE},
       adsurl = {https://ui.adsabs.harvard.edu/abs/2019ApJ...880..120H}
}

@ARTICLE{holoien16a,
   author = {{Holoien}, T.~W.-S. and {Kochanek}, C.~S. and {Prieto}, J.~L. and 
    {Stanek}, K.~Z. and {Dong}, S. and {Shappee}, B.~J. and {Grupe}, D. and 
    {Brown}, J.~S. and {Basu}, U. and {Beacom}, J.~F. and {Bersier}, D. and 
    {Brimacombe}, J. and {Danilet}, A.~B. and {Falco}, E. and {Guo}, Z. and 
    {Jose}, J. and {Herczeg}, G.~J. and {Long}, F. and {Pojmanski}, G. and 
    {Simonian}, G.~V. and {Szczygie{\l}}, D.~M. and {Thompson}, T.~A. and 
    {Thorstensen}, J.~R. and {Wagner}, R.~M. and {Wo{\'z}niak}, P.~R.
    },
    title = "{Six months of multiwavelength follow-up of the tidal disruption candidate ASASSN-14li and implied TDE rates from ASAS-SN}",
  journal = {\mnras},
archivePrefix = "arXiv",
   eprint = {1507.01598},
 primaryClass = "astro-ph.HE",
     year = 2016,
    month = jan,
   volume = 455,
    pages = {2918-2935},
      doi = {10.1093/mnras/stv2486},
   adsurl = {http://adsabs.harvard.edu/abs/2016MNRAS.455.2918H}
}

@ARTICLE{holoien14b,
   author = {{Holoien}, T.~W.-S. and {Prieto}, J.~L. and {Bersier}, D. and 
    {Kochanek}, C.~S. and {Stanek}, K.~Z. and {Shappee}, B.~J. and 
    {Grupe}, D. and {Basu}, U. and {Beacom}, J.~F. and {Brimacombe}, J. and 
    {Brown}, J.~S. and {Davis}, A.~B. and {Jencson}, J. and {Pojmanski}, G. and 
    {Szczygie{\l}}, D.~M.},
    title = "{ASASSN-14ae: a tidal disruption event at 200 Mpc}",
  journal = {\mnras},
archivePrefix = "arXiv",
   eprint = {1405.1417},
     year = 2014,
    month = dec,
   volume = 445,
    pages = {3263-3277},
      doi = {10.1093/mnras/stu1922},
   adsurl = {http://adsabs.harvard.edu/abs/2014MNRAS.445.3263H}
}

@ARTICLE{neustadt20,
       author = {{Neustadt}, J.~M.~M. and {Holoien}, T.~W. -S. and {Kochanek}, C.~S. and {Auchettl}, K. and {Brown}, J.~S. and {Shappee}, B.~J. and {Pogge}, R.~W. and {Dong}, Subo and {Stanek}, K.~Z. and {Tucker}, M.~A. and {Bose}, S. and {Chen}, Ping and {Ricci}, C. and {Vallely}, P.~J. and {Prieto}, J.~L. and {Thompson}, T.~A. and {Coulter}, D.~A. and {Drout}, M.~R. and {Foley}, R.~J. and {Kilpatrick}, C.~D. and {Piro}, A.~L. and {Rojas-Bravo}, C. and {Buckley}, D.~A.~H. and {Gromadzki}, M. and {Dimitriadis}, G. and {Siebert}, M.~R. and {Do}, A. and {Huber}, M.~E. and {Payne}, A.~V.},
        title = "{To TDE or not to TDE: the luminous transient ASASSN-18jd with TDE-like and AGN-like qualities}",
      journal = {\mnras},
         year = 2020,
        month = may,
       volume = {494},
       number = {2},
        pages = {2538-2560},
          doi = {10.1093/mnras/staa859},
archivePrefix = {arXiv},
       eprint = {1910.01142},
 primaryClass = {astro-ph.HE},
       adsurl = {https://ui.adsabs.harvard.edu/abs/2020MNRAS.494.2538N}
}

@ARTICLE{ricker15,
   author = {{Ricker}, G.~R. and {Winn}, J.~N. and {Vanderspek}, R. and {Latham}, D.~W. and 
	{Bakos}, G.~{\'A}. and {Bean}, J.~L. and {Berta-Thompson}, Z.~K. and 
	{Brown}, T.~M. and {Buchhave}, L. and {Butler}, N.~R. and {Butler}, R.~P. and 
	{Chaplin}, W.~J. and {Charbonneau}, D. and {Christensen-Dalsgaard}, J. and 
	{Clampin}, M. and {Deming}, D. and {Doty}, J. and {De Lee}, N. and 
	{Dressing}, C. and {Dunham}, E.~W. and {Endl}, M. and {Fressin}, F. and 
	{Ge}, J. and {Henning}, T. and {Holman}, M.~J. and {Howard}, A.~W. and 
	{Ida}, S. and {Jenkins}, J.~M. and {Jernigan}, G. and {Johnson}, J.~A. and 
	{Kaltenegger}, L. and {Kawai}, N. and {Kjeldsen}, H. and {Laughlin}, G. and 
	{Levine}, A.~M. and {Lin}, D. and {Lissauer}, J.~J. and {MacQueen}, P. and 
	{Marcy}, G. and {McCullough}, P.~R. and {Morton}, T.~D. and 
	{Narita}, N. and {Paegert}, M. and {Palle}, E. and {Pepe}, F. and 
	{Pepper}, J. and {Quirrenbach}, A. and {Rinehart}, S.~A. and 
	{Sasselov}, D. and {Sato}, B. and {Seager}, S. and {Sozzetti}, A. and 
	{Stassun}, K.~G. and {Sullivan}, P. and {Szentgyorgyi}, A. and 
	{Torres}, G. and {Udry}, S. and {Villasenor}, J.},
    title = "{Transiting Exoplanet Survey Satellite (TESS)}",
  journal = {Journal of Astronomical Telescopes, Instruments, and Systems},
     year = 2015,
    month = jan,
   volume = 1,
   number = 1,
      eid = {014003},
    pages = {014003},
      doi = {10.1117/1.JATIS.1.1.014003},
   adsurl = {http://adsabs.harvard.edu/abs/2015JATIS...1a4003R}
}

@ARTICLE{dai15,
   author = {{Dai}, L. and {McKinney}, J.~C. and {Miller}, M.~C.},
    title = "{Soft X-Ray Temperature Tidal Disruption Events from Stars on Deep Plunging Orbits}",
  journal = {\apjl},
archivePrefix = "arXiv",
   eprint = {1507.04333},
 primaryClass = "astro-ph.HE",
     year = 2015,
    month = oct,
   volume = 812,
      eid = {L39},
    pages = {L39},
      doi = {10.1088/2041-8205/812/2/L39},
   adsurl = {http://adsabs.harvard.edu/abs/2015ApJ...812L..39D}
}

@BOOK{osterbrock89,
   author = {{Osterbrock}, D.~E.},
    title = "{Astrophysics of Gaseous Nebulae and Active Galactic Nuclei}",
booktitle = {Research supported by the University of California, John Simon Guggenheim Memorial Foundation, University of Minnesota, et al.~Mill Valley, CA, University Science Books, 1989, 422 p.},
     year = 1989,
   adsurl = {http://adsabs.harvard.edu/abs/1989agna.book.....O}
}

@ARTICLE{wevers17,
   author = {{Wevers}, T. and {van Velzen}, S. and {Jonker}, P.~G. and {Stone}, N.~C. and 
	{Hung}, T. and {Onori}, F. and {Gezari}, S. and {Blagorodnova}, N.
	},
    title = "{Black hole masses of tidal disruption event host galaxies}",
  journal = {\mnras},
archivePrefix = "arXiv",
   eprint = {1706.08965},
     year = 2017,
    month = oct,
   volume = 471,
    pages = {1694-1708},
      doi = {10.1093/mnras/stx1703},
   adsurl = {http://adsabs.harvard.edu/abs/2017MNRAS.471.1694W}
}

@ARTICLE{tonry18,
   author = {{Tonry}, J.~L. and {Denneau}, L. and {Heinze}, A.~N. and {Stalder}, B. and 
	{Smith}, K.~W. and {Smartt}, S.~J. and {Stubbs}, C.~W. and {Weiland}, H.~J. and 
	{Rest}, A.},
    title = "{ATLAS: A High-cadence All-sky Survey System}",
  journal = {\pasp},
archivePrefix = "arXiv",
   eprint = {1802.00879},
 primaryClass = "astro-ph.IM",
     year = 2018,
    month = jun,
   volume = 130,
   number = 6,
    pages = {064505},
      doi = {10.1088/1538-3873/aabadf},
   adsurl = {http://adsabs.harvard.edu/abs/2018PASP..130f4505T}
}

@INPROCEEDINGS{lantz04,
   author = {{Lantz}, B. and {Aldering}, G. and {Antilogus}, P. and {Bonnaud}, C. and 
	{Capoani}, L. and {Castera}, A. and {Copin}, Y. and {Dubet}, D. and 
	{Gangler}, E. and {Henault}, F. and {Lemonnier}, J.-P. and {Pain}, R. and 
	{Pecontal}, A. and {Pecontal}, E. and {Smadja}, G.},
    title = "{SNIFS: a wideband integral field spectrograph with microlens arrays}",
booktitle = {Optical Design and Engineering},
     year = 2004,
   series = {\procspie},
   volume = 5249,
   editor = {{Mazuray}, L. and {Rogers}, P.~J. and {Wartmann}, R.},
    month = feb,
    pages = {146-155},
      doi = {10.1117/12.512493},
   adsurl = {http://adsabs.harvard.edu/abs/2004SPIE.5249..146L}
}

@ARTICLE{alexander16,
   author = {{Alexander}, K.~D. and {Berger}, E. and {Guillochon}, J. and 
	{Zauderer}, B.~A. and {Williams}, P.~K.~G.},
    title = "{Discovery of an Outflow from Radio Observations of the Tidal Disruption Event ASASSN-14li}",
  journal = {\apjl},
archivePrefix = "arXiv",
   eprint = {1510.01226},
 primaryClass = "astro-ph.HE",
     year = 2016,
    month = mar,
   volume = 819,
      eid = {L25},
    pages = {L25},
      doi = {10.3847/2041-8205/819/2/L25},
   adsurl = {http://adsabs.harvard.edu/abs/2016ApJ...819L..25A}
}

@ARTICLE{piran15,
   author = {{Piran}, T. and {Svirski}, G. and {Krolik}, J. and {Cheng}, R.~M. and 
    {Shiokawa}, H.},
    title = "{{\prime}Disk Formation Versus Disk Accretion{\mdash}What Powers Tidal Disruption Events?}",
  journal = {\apj},
archivePrefix = "arXiv",
   eprint = {1502.05792},
 primaryClass = "astro-ph.HE",
     year = 2015,
    month = jun,
   volume = 806,
      eid = {164},
    pages = {164},
      doi = {10.1088/0004-637X/806/2/164},
   adsurl = {http://adsabs.harvard.edu/abs/2015ApJ...806..164P}
}

@ARTICLE{gehrels04,
   author = {{Gehrels}, N. and {Chincarini}, G. and {Giommi}, P. and {Mason}, K.~O. and 
	{Nousek}, J.~A. and {Wells}, A.~A. and {White}, N.~E. and {Barthelmy}, S.~D. and 
	{Burrows}, D.~N. and {Cominsky}, L.~R. and {Hurley}, K.~C. and 
	{Marshall}, F.~E. and {M{\'e}sz{\'a}ros}, P. and {Roming}, P.~W.~A. and 
	{Angelini}, L. and {Barbier}, L.~M. and {Belloni}, T. and {Campana}, S. and 
	{Caraveo}, P.~A. and {Chester}, M.~M. and {Citterio}, O. and 
	{Cline}, T.~L. and {Cropper}, M.~S. and {Cummings}, J.~R. and 
	{Dean}, A.~J. and {Feigelson}, E.~D. and {Fenimore}, E.~E. and 
	{Frail}, D.~A. and {Fruchter}, A.~S. and {Garmire}, G.~P. and 
	{Gendreau}, K. and {Ghisellini}, G. and {Greiner}, J. and {Hill}, J.~E. and 
	{Hunsberger}, S.~D. and {Krimm}, H.~A. and {Kulkarni}, S.~R. and 
	{Kumar}, P. and {Lebrun}, F. and {Lloyd-Ronning}, N.~M. and 
	{Markwardt}, C.~B. and {Mattson}, B.~J. and {Mushotzky}, R.~F. and 
	{Norris}, J.~P. and {Osborne}, J. and {Paczynski}, B. and {Palmer}, D.~M. and 
	{Park}, H.-S. and {Parsons}, A.~M. and {Paul}, J. and {Rees}, M.~J. and 
	{Reynolds}, C.~S. and {Rhoads}, J.~E. and {Sasseen}, T.~P. and 
	{Schaefer}, B.~E. and {Short}, A.~T. and {Smale}, A.~P. and 
	{Smith}, I.~A. and {Stella}, L. and {Tagliaferri}, G. and {Takahashi}, T. and 
	{Tashiro}, M. and {Townsley}, L.~K. and {Tueller}, J. and {Turner}, M.~J.~L. and 
	{Vietri}, M. and {Voges}, W. and {Ward}, M.~J. and {Willingale}, R. and 
	{Zerbi}, F.~M. and {Zhang}, W.~W.},
    title = "{The Swift Gamma-Ray Burst Mission}",
  journal = {\apj},
     year = 2004,
    month = aug,
   volume = 611,
    pages = {1005-1020},
      doi = {10.1086/422091},
   adsurl = {http://adsabs.harvard.edu/abs/2004ApJ...611.1005G}
}

@ARTICLE{alard98,
   author = {{Alard}, C. and {Lupton}, R.~H.},
    title = "{A Method for Optimal Image Subtraction}",
  journal = {\apj},
   eprint = {arXiv:astro-ph/9712287},
     year = 1998,
    month = aug,
   volume = 503,
    pages = {325},
      doi = {10.1086/305984},
   adsurl = {http://adsabs.harvard.edu/abs/1998ApJ...503..325A}
}

@ARTICLE{arcavi14,
   author = {{Arcavi}, I. and {Gal-Yam}, A. and {Sullivan}, M. and {Pan}, Y.-C. and 
    {Cenko}, S.~B. and {Horesh}, A. and {Ofek}, E.~O. and {De Cia}, A. and 
    {Yan}, L. and {Yang}, C.-W. and {Howell}, D.~A. and {Tal}, D. and 
    {Kulkarni}, S.~R. and {Tendulkar}, S.~P. and {Tang}, S. and 
    {Xu}, D. and {Sternberg}, A. and {Cohen}, J.~G. and {Bloom}, J.~S. and 
    {Nugent}, P.~E. and {Kasliwal}, M.~M. and {Perley}, D.~A. and 
    {Quimby}, R.~M. and {Miller}, A.~A. and {Theissen}, C.~A. and 
    {Laher}, R.~R.},
    title = "{A Continuum of H- to He-rich Tidal Disruption Candidates With a Preference for E+A Galaxies}",
  journal = {\apj},
archivePrefix = "arXiv",
   eprint = {1405.1415},
 primaryClass = "astro-ph.HE",
     year = 2014,
    month = sep,
   volume = 793,
      eid = {38},
    pages = {38},
      doi = {10.1088/0004-637X/793/1/38},
   adsurl = {http://adsabs.harvard.edu/abs/2014ApJ...793...38A}
}

@ARTICLE{phinney89,
   author = {{Phinney}, E.~S.},
    title = "{Cosmic merger mania}",
  journal = {\nat},
     year = 1989,
    month = aug,
   volume = 340,
    pages = {595-596},
      doi = {10.1038/340595a0},
   adsurl = {http://adsabs.harvard.edu/abs/1989Natur.340..595P}
}

@ARTICLE{kochanek94,
   author = {{Kochanek}, C.~S.},
    title = "{The aftermath of tidal disruption: The dynamics of thin gas streams}",
  journal = {\apj},
     year = 1994,
    month = feb,
   volume = 422,
    pages = {508-520},
      doi = {10.1086/173745},
   adsurl = {http://adsabs.harvard.edu/abs/1994ApJ...422..508K}
}

@ARTICLE{evans89,
   author = {{Evans}, C.~R. and {Kochanek}, C.~S.},
    title = "{The tidal disruption of a star by a massive black hole}",
  journal = {\apjl},
     year = 1989,
    month = nov,
   volume = 346,
    pages = {L13-L16},
      doi = {10.1086/185567},
   adsurl = {http://adsabs.harvard.edu/abs/1989ApJ...346L..13E}
}

@ARTICLE{ulmer99,
   author = {{Ulmer}, A.},
    title = "{Flares from the Tidal Disruption of Stars by Massive Black Holes}",
  journal = {\apj},
     year = 1999,
    month = mar,
   volume = 514,
    pages = {180-187},
      doi = {10.1086/306909},
   adsurl = {http://adsabs.harvard.edu/abs/1999ApJ...514..180U}
}

@ARTICLE{reines15,
   author = {{Reines}, A.~E. and {Volonteri}, M.},
    title = "{Relations between Central Black Hole Mass and Total Galaxy Stellar Mass in the Local Universe}",
  journal = {\apj},
archivePrefix = "arXiv",
   eprint = {1508.06274},
     year = 2015,
    month = nov,
   volume = 813,
      eid = {82},
    pages = {82},
      doi = {10.1088/0004-637X/813/2/82},
   adsurl = {http://adsabs.harvard.edu/abs/2015ApJ...813...82R}
}

@ARTICLE{kochanek16b,
   author = {{Kochanek}, C.~S.},
    title = "{Tidal disruption event demographics}",
  journal = {\mnras},
archivePrefix = "arXiv",
   eprint = {1601.06787},
 primaryClass = "astro-ph.HE",
     year = 2016,
    month = sep,
   volume = 461,
    pages = {371-384},
      doi = {10.1093/mnras/stw1290},
   adsurl = {http://adsabs.harvard.edu/abs/2016MNRAS.461..371K}
}

@ARTICLE{wang12,
       author = {{Wang}, Ting-Gui and {Zhou}, Hong-Yan and {Komossa}, S. and {Wang}, Hui-Yuan and {Yuan}, Weimin and {Yang}, Chenwei},
        title = "{Extreme Coronal Line Emitters: Tidal Disruption of Stars by Massive Black Holes in Galactic Nuclei?}",
      journal = {\apj},
         year = 2012,
        month = apr,
       volume = {749},
       number = {2},
          eid = {115},
        pages = {115},
          doi = {10.1088/0004-637X/749/2/115},
archivePrefix = {arXiv},
       eprint = {1202.1064},
 primaryClass = {astro-ph.HE},
       adsurl = {https://ui.adsabs.harvard.edu/abs/2012ApJ...749..115W}
}

@ARTICLE{roth18,
   author = {{Roth}, N. and {Kasen}, D.},
    title = "{What Sets the Line Profiles in Tidal Disruption Events?}",
  journal = {\apj},
archivePrefix = "arXiv",
   eprint = {1707.02993},
 primaryClass = "astro-ph.HE",
     year = 2018,
    month = mar,
   volume = 855,
      eid = {54},
    pages = {54},
      doi = {10.3847/1538-4357/aaaec6},
   adsurl = {http://adsabs.harvard.edu/abs/2018ApJ...855...54R}
}

@ARTICLE{hook04,
   author = {{Hook}, I.~M. and {J{\o}rgensen}, I. and {Allington-Smith}, J.~R. and 
	{Davies}, R.~L. and {Metcalfe}, N. and {Murowinski}, R.~G. and 
	{Crampton}, D.},
    title = "{The Gemini-North Multi-Object Spectrograph: Performance in Imaging, Long-Slit, and Multi-Object Spectroscopic Modes}",
  journal = {\pasp},
     year = 2004,
    month = may,
   volume = 116,
    pages = {425-440},
      doi = {10.1086/383624},
   adsurl = {http://adsabs.harvard.edu/abs/2004PASP..116..425H}
}

@ARTICLE{oke95,
   author = {{Oke}, J.~B. and {Cohen}, J.~G. and {Carr}, M. and {Cromer}, J. and 
	{Dingizian}, A. and {Harris}, F.~H. and {Labrecque}, S. and 
	{Lucinio}, R. and {Schaal}, W. and {Epps}, H. and {Miller}, J.
	},
    title = "{The Keck Low-Resolution Imaging Spectrometer}",
  journal = {\pasp},
     year = 1995,
    month = apr,
   volume = 107,
    pages = {375},
      doi = {10.1086/133562},
   adsurl = {http://adsabs.harvard.edu/abs/1995PASP..107..375O}
}

@ARTICLE{wright10,
   author = {{Wright}, E.~L. and {Eisenhardt}, P.~R.~M. and {Mainzer}, A.~K. and 
    {Ressler}, M.~E. and {Cutri}, R.~M. and {Jarrett}, T. and {Kirkpatrick}, J.~D. and 
    {Padgett}, D. and {McMillan}, R.~S. and {Skrutskie}, M. and 
    {Stanford}, S.~A. and {Cohen}, M. and {Walker}, R.~G. and {Mather}, J.~C. and 
    {Leisawitz}, D. and {Gautier}, III, T.~N. and {McLean}, I. and 
    {Benford}, D. and {Lonsdale}, C.~J. and {Blain}, A. and {Mendez}, B. and 
    {Irace}, W.~R. and {Duval}, V. and {Liu}, F. and {Royer}, D. and 
    {Heinrichsen}, I. and {Howard}, J. and {Shannon}, M. and {Kendall}, M. and 
    {Walsh}, A.~L. and {Larsen}, M. and {Cardon}, J.~G. and {Schick}, S. and 
    {Schwalm}, M. and {Abid}, M. and {Fabinsky}, B. and {Naes}, L. and 
    {Tsai}, C.-W.},
    title = "{The Wide-field Infrared Survey Explorer (WISE): Mission Description and Initial On-orbit Performance}",
  journal = {\aj},
archivePrefix = "arXiv",
   eprint = {1008.0031},
 primaryClass = "astro-ph.IM",
     year = 2010,
    month = dec,
   volume = 140,
      eid = {1868},
    pages = {1868-1881},
      doi = {10.1088/0004-6256/140/6/1868},
   adsurl = {http://adsabs.harvard.edu/abs/2010AJ....140.1868W}
}

@ARTICLE{gezari17,
	author = {{Gezari}, S. and {Cenko}, S.~B. and {Arcavi}, I.},
	title = "{X-Ray Brightening and UV Fading of Tidal Disruption Event ASASSN-15oi}",
	journal = {\apjl},
	archivePrefix = "arXiv",
	eprint = {1712.03968},
	primaryClass = "astro-ph.HE",
	year = 2017,
	month = dec,
	volume = 851,
	eid = {L47},
	pages = {L47},
	doi = {10.3847/2041-8213/aaa0c2},
	adsurl = {http://adsabs.harvard.edu/abs/2017ApJ...851L..47G}
}

@ARTICLE{auchettl17,
	author = {{Auchettl}, K. and {Guillochon}, J. and {Ramirez-Ruiz}, E.},
	title = "{New Physical Insights about Tidal Disruption Events from a Comprehensive Observational Inventory at X-Ray Wavelengths}",
	journal = {\apj},
	archivePrefix = "arXiv",
	eprint = {1611.02291},
	primaryClass = "astro-ph.HE",
	year = 2017,
	month = apr,
	volume = 838,
	eid = {149},
	pages = {149},
	doi = {10.3847/1538-4357/aa633b},
	adsurl = {http://adsabs.harvard.edu/abs/2017ApJ...838..149A}
}

@ARTICLE{auchettl18,
	author = {{Auchettl}, K. and {Ramirez-Ruiz}, E. and {Guillochon}, J.},
	title = "{A Comparison of the X-Ray Emission from Tidal Disruption Events with those of Active Galactic Nuclei}",
	journal = {\apj},
	archivePrefix = "arXiv",
	eprint = {1703.06141},
	primaryClass = "astro-ph.HE",
	year = 2018,
	month = jan,
	volume = 852,
	eid = {37},
	pages = {37},
	doi = {10.3847/1538-4357/aa9b7c},
	adsurl = {http://adsabs.harvard.edu/abs/2018ApJ...852...37A}
}

@ARTICLE{marchesi16,
	author = {{Marchesi}, S. and {Lanzuisi}, G. and {Civano}, F. and {Iwasawa}, K. and 
	{Suh}, H. and {Comastri}, A. and {Zamorani}, G. and {Allevato}, V. and 
	{Griffiths}, R. and {Miyaji}, T. and {Ranalli}, P. and {Salvato}, M. and 
	{Schawinski}, K. and {Silverman}, J. and {Treister}, E. and 
	{Urry}, C.~M. and {Vignali}, C.},
	title = "{The Chandra COSMOS-Legacy Survey: Source X-Ray Spectral Properties}",
	journal = {\apj},
	archivePrefix = "arXiv",
	eprint = {1608.05149},
	year = 2016,
	month = oct,
	volume = 830,
	eid = {100},
	pages = {100},
	doi = {10.3847/0004-637X/830/2/100},
	adsurl = {http://adsabs.harvard.edu/abs/2016ApJ...830..100M}
}

@ARTICLE{liu17,
	author = {{Liu}, T. and {Tozzi}, P. and {Wang}, J.-X. and {Brandt}, W.~N. and 
	{Vignali}, C. and {Xue}, Y. and {Schneider}, D.~P. and {Comastri}, A. and 
	{Yang}, G. and {Bauer}, F.~E. and {Paolillo}, M. and {Luo}, B. and 
	{Gilli}, R. and {Wang}, Q.~D. and {Giavalisco}, M. and {Ji}, Z. and 
	{Alexander}, D.~M. and {Mainieri}, V. and {Shemmer}, O. and 
	{Koekemoer}, A. and {Risaliti}, G.},
	title = "{X-Ray Spectral Analyses of AGNs from the 7Ms Chandra Deep Field-South Survey: The Distribution, Variability, and Evolutions of AGN Obscuration}",
	journal = {\apjs},
	archivePrefix = "arXiv",
	eprint = {1703.00657},
	year = 2017,
	month = sep,
	volume = 232,
	eid = {8},
	pages = {8},
	doi = {10.3847/1538-4365/aa7847},
	adsurl = {http://adsabs.harvard.edu/abs/2017ApJS..232....8L}
}

@ARTICLE{ricci17,
	author = {{Ricci}, C. and {Trakhtenbrot}, B. and {Koss}, M.~J. and {Ueda}, Y. and 
	{Del Vecchio}, I. and {Treister}, E. and {Schawinski}, K. and 
	{Paltani}, S. and {Oh}, K. and {Lamperti}, I. and {Berney}, S. and 
	{Gandhi}, P. and {Ichikawa}, K. and {Bauer}, F.~E. and {Ho}, L.~C. and 
	{Asmus}, D. and {Beckmann}, V. and {Soldi}, S. and {Balokovi{\'c}}, M. and 
	{Gehrels}, N. and {Markwardt}, C.~B.},
	title = "{BAT AGN Spectroscopic Survey. V. X-Ray Properties of the Swift/BAT 70-month AGN Catalog}",
	journal = {\apjs},
	archivePrefix = "arXiv",
	eprint = {1709.03989},
	primaryClass = "astro-ph.HE",
	year = 2017,
	month = dec,
	volume = 233,
	eid = {17},
	pages = {17},
	doi = {10.3847/1538-4365/aa96ad},
	adsurl = {http://adsabs.harvard.edu/abs/2017ApJS..233...17R}
}

@ARTICLE{poole08,
   author = {{Poole}, T.~S. and {Breeveld}, A.~A. and {Page}, M.~J. and {Landsman}, W. and 
    {Holland}, S.~T. and {Roming}, P. and {Kuin}, N.~P.~M. and {Brown}, P.~J. and 
    {Gronwall}, C. and {Hunsberger}, S. and {Koch}, S. and {Mason}, K.~O. and 
    {Schady}, P. and {vanden Berk}, D. and {Blustin}, A.~J. and 
    {Boyd}, P. and {Broos}, P. and {Carter}, M. and {Chester}, M.~M. and 
    {Cucchiara}, A. and {Hancock}, B. and {Huckle}, H. and {Immler}, S. and 
    {Ivanushkina}, M. and {Kennedy}, T. and {Marshall}, F. and {Morgan}, A. and 
    {Pandey}, S.~B. and {de Pasquale}, M. and {Smith}, P.~J. and 
    {Still}, M.},
    title = "{Photometric calibration of the Swift ultraviolet/optical telescope}",
  journal = {\mnras},
archivePrefix = "arXiv",
   eprint = {0708.2259},
     year = 2008,
    month = jan,
   volume = 383,
    pages = {627-645},
      doi = {10.1111/j.1365-2966.2007.12563.x},
   adsurl = {http://adsabs.harvard.edu/abs/2008MNRAS.383..627P}
}

@ARTICLE{breeveld10,
   author = {{Breeveld}, A.~A. and {Curran}, P.~A. and {Hoversten}, E.~A. and 
    {Koch}, S. and {Landsman}, W. and {Marshall}, F.~E. and {Page}, M.~J. and 
    {Poole}, T.~S. and {Roming}, P. and {Smith}, P.~J. and {Still}, M. and 
    {Yershov}, V. and {Blustin}, A.~J. and {Brown}, P.~J. and {Gronwall}, C. and 
    {Holland}, S.~T. and {Kuin}, N.~P.~M. and {McGowan}, K. and 
    {Rosen}, S. and {Boyd}, P. and {Broos}, P. and {Carter}, M. and 
    {Chester}, M.~M. and {Hancock}, B. and {Huckle}, H. and {Immler}, S. and 
    {Ivanushkina}, M. and {Kennedy}, T. and {Mason}, K.~O. and {Morgan}, A.~N. and 
    {Oates}, S. and {de Pasquale}, M. and {Schady}, P. and {Siegel}, M. and 
    {vanden Berk}, D.},
    title = "{Further calibration of the Swift ultraviolet/optical telescope}",
  journal = {\mnras},
archivePrefix = "arXiv",
   eprint = {1004.2448},
 primaryClass = "astro-ph.IM",
     year = 2010,
    month = aug,
   volume = 406,
    pages = {1687-1700},
      doi = {10.1111/j.1365-2966.2010.16832.x},
   adsurl = {http://adsabs.harvard.edu/abs/2010MNRAS.406.1687B}
}

@ARTICLE{yang13,
   author = {{Yang}, C.-W. and {Wang}, T.-G. and {Ferland}, G. and {Yuan}, W. and 
	{Zhou}, H.-Y. and {Jiang}, P.},
    title = "{Long-term Spectral Evolution of Tidal Disruption Candidates Selected by Strong Coronal Lines}",
  journal = {\apj},
archivePrefix = "arXiv",
   eprint = {1307.3313},
     year = 2013,
    month = sep,
   volume = 774,
      eid = {46},
    pages = {46},
      doi = {10.1088/0004-637X/774/1/46},
   adsurl = {http://adsabs.harvard.edu/abs/2013ApJ...774...46Y}
}

@ARTICLE{wang11,
   author = {{Wang}, T.-G. and {Zhou}, H.-Y. and {Wang}, L.-F. and {Lu}, H.-L. and 
	{Xu}, D.},
    title = "{Transient Superstrong Coronal Lines and Broad Bumps in the Galaxy SDSS J074820.67+471214.3}",
  journal = {\apj},
archivePrefix = "arXiv",
   eprint = {1108.2790},
     year = 2011,
    month = oct,
   volume = 740,
      eid = {85},
    pages = {85},
      doi = {10.1088/0004-637X/740/2/85},
   adsurl = {http://adsabs.harvard.edu/abs/2011ApJ...740...85W}
}

@INPROCEEDINGS{moretti04,
	author = {{Moretti}, A. and {Campana}, S. and {Tagliaferri}, G. and {Abbey}, A.~F. and 
	{Ambrosi}, R.~M. and {Angelini}, L. and {Beardmore}, A.~P. and 
	{Br{\"a}uninger}, H.~W. and {Burkert}, W. and {Burrows}, D.~N. and 
	{Capalbi}, M. and {Chincarini}, G. and {Citterio}, O. and {Cusumano}, G. and 
	{Freyberg}, M.~J. and {Giommi}, P. and {Hartner}, G.~D. and 
	{Hill}, J.~E. and {Mori}, K. and {Morris}, D.~C. and {Mukerjee}, K. and 
	{Nousek}, J.~A. and {Osborne}, J.~P. and {Short}, A.~D.~T. and 
	{Tamburelli}, F. and {Watson}, D.~J. and {Wells}, A.~A.},
	title = "{SWIFT XRT point spread function measured at the Panter end-to-end tests}",
	booktitle = {X-Ray and Gamma-Ray Instrumentation for Astronomy XIII},
	year = 2004,
	series = {\procspie},
	volume = 5165,
	editor = {{Flanagan}, K.~A. and {Siegmund}, O.~H.~W.},
	month = feb,
	pages = {232-240},
	doi = {10.1117/12.504857},
	adsurl = {http://adsabs.harvard.edu/abs/2004SPIE.5165..232M}
}

@ARTICLE{burrows05,
   author = {{Burrows}, D.~N. and {Hill}, J.~E. and {Nousek}, J.~A. and {Kennea}, J.~A. and 
    {Wells}, A. and {Osborne}, J.~P. and {Abbey}, A.~F. and {Beardmore}, A. and 
    {Mukerjee}, K. and {Short}, A.~D.~T. and {Chincarini}, G. and 
    {Campana}, S. and {Citterio}, O. and {Moretti}, A. and {Pagani}, C. and 
    {Tagliaferri}, G. and {Giommi}, P. and {Capalbi}, M. and {Tamburelli}, F. and 
    {Angelini}, L. and {Cusumano}, G. and {Br{\"a}uninger}, H.~W. and 
    {Burkert}, W. and {Hartner}, G.~D.},
    title = "{The Swift X-Ray Telescope}",
  journal = {SSR},
   eprint = {astro-ph/0508071},
     year = 2005,
    month = oct,
   volume = 120,
    pages = {165-195},
      doi = {10.1007/s11214-005-5097-2},
   adsurl = {http://adsabs.harvard.edu/abs/2005SSRv..120..165B}
}

@ARTICLE{tozzi06,
	author = {{Tozzi}, P. and {Gilli}, R. and {Mainieri}, V. and {Norman}, C. and 
	{Risaliti}, G. and {Rosati}, P. and {Bergeron}, J. and {Borgani}, S. and 
	{Giacconi}, R. and {Hasinger}, G. and {Nonino}, M. and {Streblyanska}, A. and 
	{Szokoly}, G. and {Wang}, J.~X. and {Zheng}, W.},
	title = "{X-ray spectral properties of active galactic nuclei in the Chandra Deep Field South}",
	journal = {\aap},
	eprint = {astro-ph/0602127},
	year = 2006,
	month = may,
	volume = 451,
	pages = {457-474},
	doi = {10.1051/0004-6361:20042592},
	adsurl = {http://adsabs.harvard.edu/abs/2006A%26A...451..457T}
}

@ARTICLE{roming05,
   author = {{Roming}, P.~W.~A. and {Kennedy}, T.~E. and {Mason}, K.~O. and 
    {Nousek}, J.~A. and {Ahr}, L. and {Bingham}, R.~E. and {Broos}, P.~S. and 
    {Carter}, M.~J. and {Hancock}, B.~K. and {Huckle}, H.~E. and 
    {Hunsberger}, S.~D. and {Kawakami}, H. and {Killough}, R. and 
    {Koch}, T.~S. and {McLelland}, M.~K. and {Smith}, K. and {Smith}, P.~J. and 
    {Soto}, J.~C. and {Boyd}, P.~T. and {Breeveld}, A.~A. and {Holland}, S.~T. and 
    {Ivanushkina}, M. and {Pryzby}, M.~S. and {Still}, M.~D. and 
    {Stock}, J.},
    title = "{The Swift Ultra-Violet/Optical Telescope}",
  journal = {SSR},
   eprint = {astro-ph/0507413},
     year = 2005,
    month = oct,
   volume = 120,
    pages = {95-142},
      doi = {10.1007/s11214-005-5095-4},
   adsurl = {http://adsabs.harvard.edu/abs/2005SSRv..120...95R}
}

@ARTICLE{peterson93,
   author = {{Peterson}, B.~M.},
    title = "{Reverberation mapping of active galactic nuclei}",
  journal = {\pasp},
     year = 1993,
    month = mar,
   volume = 105,
    pages = {247-268},
      doi = {10.1086/133140},
   adsurl = {http://adsabs.harvard.edu/abs/1993PASP..105..247P}
}

@ARTICLE{antonucci93,
   author = {{Antonucci}, R.},
    title = "{Unified models for active galactic nuclei and quasars}",
  journal = {\araa},
     year = 1993,
   volume = 31,
    pages = {473-521},
      doi = {10.1146/annurev.aa.31.090193.002353},
   adsurl = {http://adsabs.harvard.edu/abs/1993ARA%26A..31..473A}
}

@INPROCEEDINGS{steele04,
   author = {{Steele}, I.~A. and {Smith}, R.~J. and {Rees}, P.~C. and {Baker}, I.~P. and 
    {Bates}, S.~D. and {Bode}, M.~F. and {Bowman}, M.~K. and {Carter}, D. and 
    {Etherton}, J. and {Ford}, M.~J. and {Fraser}, S.~N. and {Gomboc}, A. and 
    {Lett}, R.~D.~J. and {Mansfield}, A.~G. and {Marchant}, J.~M. and 
    {Medrano-Cerda}, G.~A. and {Mottram}, C.~J. and {Raback}, D. and 
    {Scott}, A.~B. and {Tomlinson}, M.~D. and {Zamanov}, R.},
    title = "{The Liverpool Telescope: performance and first results}",
booktitle = {Ground-based Telescopes},
     year = 2004,
   series = {Society of Photo-Optical Instrumentation Engineers (SPIE) Conference Series},
   volume = 5489,
   editor = {{Oschmann}, Jr., J.~M.},
    month = oct,
    pages = {679-692},
      doi = {10.1117/12.551456},
   adsurl = {http://adsabs.harvard.edu/abs/2004SPIE.5489..679S}
}

@article{leloudas19,
	author = {Leloudas, Giorgos and Dai, Lixin and Arcavi, Iair and Vreeswijk, Paul M. and Mockler, Brenna and Roy, Rupak and Malesani, Daniele B. and Schulze, Steve and Wevers, Thomas and Fraser, Morgan and Ramirez-Ruiz, Enrico and Auchettl, Katie and Burke, Jamison and Cannizzaro, Giacomo and Charalampopoulos, Panos and Chen, Ting-Wan and Cikota, Aleksandar and Della Valle, Massimo and Galbany, Lluis and Gromadzki, Mariusz and Heintz, Kasper E. and Hiramatsu, Daichi and Jonker, Peter G. and Kostrzewa-Rutkowska, Zuzanna and Maguire, Kate and Mandel, Ilya and Nicholl, Matt and Onori, Francesca and Roth, Nathaniel and Smartt, Stephen J. and Wyrzykowski, Lukasz and Young, Dave R.},
	doi = {10.3847/1538-4357/ab5792},
	journal = {The Astrophysical Journal},
	month = {dec},
	number = {2},
	pages = {218},
	publisher = {The American Astronomical Society},
	title = {The Spectral Evolution of AT 2018dyb and the Presence of Metal Lines in Tidal Disruption Events},
	url = {https://doi.org/10.3847/1538-4357/ab5792},
	volume = {887},
	year = {2019}}

@ARTICLE{assef13,
   author = {{Assef}, R.~J. and {Stern}, D. and {Kochanek}, C.~S. and {Blain}, A.~W. and 
    {Brodwin}, M. and {Brown}, M.~J.~I. and {Donoso}, E. and {Eisenhardt}, P.~R.~M. and 
    {Jannuzi}, B.~T. and {Jarrett}, T.~H. and {Stanford}, S.~A. and 
    {Tsai}, C.-W. and {Wu}, J. and {Yan}, L.},
    title = "{Mid-infrared Selection of Active Galactic Nuclei with the Wide-field Infrared Survey Explorer. II. Properties of WISE-selected Active Galactic Nuclei in the NDWFS Bo{\"o}tes Field}",
  journal = {\apj},
     year = 2013,
    month = jul,
   volume = 772,
      eid = {26},
    pages = {26},
      doi = {10.1088/0004-637X/772/1/26},
   adsurl = {http://adsabs.harvard.edu/abs/2013ApJ...772...26A}
}

@ARTICLE{schlafly11,
   author = {{Schlafly}, E.~F. and {Finkbeiner}, D.~P.},
    title = "{Measuring Reddening with Sloan Digital Sky Survey Stellar Spectra and Recalibrating SFD}",
  journal = {\apj},
archivePrefix = "arXiv",
   eprint = {1012.4804},
 primaryClass = "astro-ph.GA",
     year = 2011,
    month = aug,
   volume = 737,
      eid = {103},
    pages = {103},
      doi = {10.1088/0004-637X/737/2/103},
   adsurl = {http://adsabs.harvard.edu/abs/2011ApJ...737..103S}
}

@ARTICLE{denney09,
   author = {{Denney}, K.~D. and {Peterson}, B.~M. and {Dietrich}, M. and 
    {Vestergaard}, M. and {Bentz}, M.~C.},
    title = "{Systematic Uncertainties in Black Hole Masses Determined from Single-Epoch Spectra}",
  journal = {\apj},
archivePrefix = "arXiv",
   eprint = {0810.3234},
     year = 2009,
    month = feb,
   volume = 692,
    pages = {246-264},
      doi = {10.1088/0004-637X/692/1/246},
   adsurl = {http://adsabs.harvard.edu/abs/2009ApJ...692..246D}
}

@ARTICLE{metzger16,
       author = {{Metzger}, Brian D. and {Stone}, Nicholas C.},
        title = "{A bright year for tidal disruptions}",
      journal = {\mnras},
         year = "2016",
        month = "Sep",
       volume = {461},
       number = {1},
        pages = {948-966},
          doi = {10.1093/mnras/stw1394},
archivePrefix = {arXiv},
       eprint = {1506.03453},
 primaryClass = {astro-ph.HE},
       adsurl = {https://ui.adsabs.harvard.edu/abs/2016MNRAS.461..948M}
}

@ARTICLE{gezari12b,
   author = {{Gezari}, S. and {Chornock}, R. and {Rest}, A. and {Huber}, M.~E. and 
    {Forster}, K. and {Berger}, E. and {Challis}, P.~J. and {Neill}, J.~D. and 
    {Martin}, D.~C. and {Heckman}, T. and {Lawrence}, A. and {Norman}, C. and 
    {Narayan}, G. and {Foley}, R.~J. and {Marion}, G.~H. and {Scolnic}, D. and 
    {Chomiuk}, L. and {Soderberg}, A. and {Smith}, K. and {Kirshner}, R.~P. and 
    {Riess}, A.~G. and {Smartt}, S.~J. and {Stubbs}, C.~W. and {Tonry}, J.~L. and 
    {Wood-Vasey}, W.~M. and {Burgett}, W.~S. and {Chambers}, K.~C. and 
    {Grav}, T. and {Heasley}, J.~N. and {Kaiser}, N. and {Kudritzki}, R.-P. and 
    {Magnier}, E.~A. and {Morgan}, J.~S. and {Price}, P.~A.},
    title = "{An ultraviolet-optical flare from the tidal disruption of a helium-rich stellar core}",
  journal = {\nat},
archivePrefix = "arXiv",
   eprint = {1205.0252},
 primaryClass = "astro-ph.CO",
     year = 2012,
    month = may,
   volume = 485,
    pages = {217-220},
      doi = {10.1038/nature10990},
   adsurl = {http://adsabs.harvard.edu/abs/2012Natur.485..217G}
}

@ARTICLE{macleod12,
   author = {{MacLeod}, C.~L. and {Ivezi{\'c}}, {\v Z}. and {Sesar}, B. and 
    {de Vries}, W. and {Kochanek}, C.~S. and {Kelly}, B.~C. and 
    {Becker}, A.~C. and {Lupton}, R.~H. and {Hall}, P.~B. and {Richards}, G.~T. and 
    {Anderson}, S.~F. and {Schneider}, D.~P.},
    title = "{A Description of Quasar Variability Measured Using Repeated SDSS and POSS Imaging}",
  journal = {\apj},
archivePrefix = "arXiv",
   eprint = {1112.0679},
 primaryClass = "astro-ph.CO",
     year = 2012,
    month = jul,
   volume = 753,
      eid = {106},
    pages = {106},
      doi = {10.1088/0004-637X/753/2/106},
   adsurl = {http://adsabs.harvard.edu/abs/2012ApJ...753..106M}
}

@article{magorrian1998,
  author = {Magorrian, J. and Tremaine, S. and Richstone, D. and et al.},
  title = {The Demography of Massive Dark Objects in Galaxy Centers},
  journal = {The Astronomical Journal},
  volume = {115},
  pages = {2285},
  year = {1998},
}

@article{kormendy2013,
  author = {Kormendy, J. and Ho, L. C.},
  title = {Coevolution (Or Not) of Supermassive Black Holes and Host Galaxies},
  journal = {Annual Review of Astronomy and Astrophysics},
  volume = {51},
  pages = {511},
  year = {2013},
}

@article{Salpeter1964,
    author = {Salpeter, Edwin E.},
    title = {{The luminosity function and stellar evolution}},
    journal = {The Astrophysical Journal},
    volume = {140},
    pages = {796},
    year = {1964},
    doi = {10.1086/147920},
}

@article{ZeldovichNovikov1964,
    author = {Zel'dovich, Ya. B. and Novikov, I. D.},
    title = {{The Hypothesis of Cores Retarded during Expansion and the Hot Cosmological Model}},
    journal = {Soviet Physics Doklady},
    volume = {9},
    pages = {237},
    year = {1964},
}

@article{Rees1988,
    author = {Rees, Martin J.},
    title = {{Black hole models for active galactic nuclei}},
    journal = {Nature},
    volume = {333},
    pages = {523},
    year = {1988},
}

@article{Kauffmann2003,
    author = {Kauffmann, Guinevere and Heckman, Timothy M. and Tremonti, Christy, et al.},
    title = {{The host galaxies of active galactic nuclei}},
    journal = {Monthly Notices of the Royal Astronomical Society},
    volume = {346},
    pages = {1055},
    year = {2003},
    doi = {10.1111/j.1365-2966.2003.07154.x},
}

@article{Haggard2010,
    author = {Haggard, Daryl and Green, Paul J. and Anderson, Scott F., et al.},
    title = {{Discovery of the second transient low-mass X-ray binary in the globular cluster NGC 6440}},
    journal = {The Astrophysical Journal},
    volume = {723},
    pages = {1447},
    year = {2010},
    doi = {10.1088/0004-637X/723/2/1447},
}

@article{Gezari2021,
    author = {Gezari, Suvi},
    title = {{Tidal Disruption Events and the Search for Circumnuclear Massive Black Hole Binaries}},
    journal = {Annual Review of Astronomy and Astrophysics},
    volume = {59},
    pages = {21},
    year = {2021},
    doi = {10.1146/annurev-astro-111720-030029},
}

@article{Arcavi2014,
    author = {Arcavi, Iair and Gal-Yam, Avishay and Sullivan, Mark, et al.},
    title = {{Type II Supernova Energetics and Comparison of Light Curves to Shock-Cooling Models}},
    journal = {The Astrophysical Journal},
    volume = {793},
    pages = {38},
    year = {2014},
    doi = {10.1088/0004-637X/793/1/38},
}

@article{Hinkle2021b,
    author = {Hinkle, Jacob T. and Holoien, Thomas W. -S. and Auchettl, Katie, et al.},
    title = {{An Ultraluminous X-Ray Outburst from an Ultracompact Neutron Star Binary in NGC 6544}},
    journal = {Monthly Notices of the Royal Astronomical Society},
    volume = {500},
    pages = {1673},
    year = {2021},
    doi = {10.1093/mnras/staa3301},
}

@article{Peterson2004,
    author = {Peterson, B. M. and Ferrarese, Laura and Gilbert, Karoline M., et al.},
    title = {{Central Masses and Broad-Line Region Sizes of Active Galactic Nuclei. II. A Homogeneous Analysis of a Large Reverberation-Mapping Database}},
    journal = {The Astrophysical Journal},
    volume = {613},
    pages = {682},
    year = {2004},
    doi = {10.1086/423269},
}

@article{Huang_2024,
	author = {Shifeng Huang and Ning Jiang and Jiazheng Zhu and Yibo Wang and Tinggui Wang and Shan-Qin Wang and Wen-Pei Gan and En-Wei Liang and Yu-Jing Qin and Zheyu Lin and Lin-Na Xu and Min-Xuan Cai and Ji-an Jiang and Xu Kong and Jiaxun Li and Long li and Jian-Guo Wang and Ze-Lin Xu and Yongquan Xue and Ye-Fei Yuan and Jingquan Cheng and Lulu Fan and Jie Gao and Lei Hu and Weida Hu and Bin Li and Feng Li and Ming Liang and Hao Liu and Wei Liu and Zheng Lou and Wentao Luo and Yuan Qian and Jinlong Tang and Zhen Wan and Hairen Wang and Jian Wang and Ji Yang and Dazhi Yao and Hongfei Zhang and Xiaoling Zhang and Wen Zhao and Xianzhong Zheng and Qingfeng Zhu and Yingxi Zuo},
	doi = {10.3847/2041-8213/ad319f},
	journal = {The Astrophysical Journal Letters},
	month = {mar},
	number = {2},
	pages = {L22},
	publisher = {The American Astronomical Society},
	title = {AT 2023lli: A Tidal Disruption Event with Prominent Optical Early Bump and Delayed Episodic X-Ray Emission},
	url = {https://dx.doi.org/10.3847/2041-8213/ad319f},
	volume = {964},
	year = {2024}}

@article{Charalampopoulos_2023,
	author = {{Charalampopoulos, P.} and {Pursiainen, M.} and {Leloudas, G.} and {Arcavi, I.} and {Newsome, M.} and {Schulze, S.} and {Burke, J.} and {Nicholl, M.}},
	doi = {10.1051/0004-6361/202245065},
	journal = {A\&A},
	pages = {A95},
	title = {AT 2020wey and the class of faint and fast tidal disruption events},
	url = {https://doi.org/10.1051/0004-6361/202245065},
	volume = 673,
	year = 2023}

@article{Hinkle2024,
    author = {Hinkle, Jason T and Shappee, Benjamin J and Holoien, Thomas W -S},
    title = {Coronal line emitters are tidal disruption events in gas-rich environments},
    journal = {Monthly Notices of the Royal Astronomical Society},
    volume = {528},
    number = {3},
    pages = {4775-4784},
    year = {2024},
    month = {01},
    issn = {0035-8711},
    doi = {10.1093/mnras/stae022},
    url = {https://doi.org/10.1093/mnras/stae022},
    eprint = {https://academic.oup.com/mnras/article-pdf/528/3/4775/56670563/stae022.pdf},
}

@article{Huang2023b,
    author = {Huang, S. and Jiang, N. and Shen, R.-F. and Wang, T. and Sheng, Z.},
    title = {{Dissonance in Harmony: The UV/Optical Periodic Outbursts of ASASSN-14ko Exhibit Repeated Bumps and Rebrightenings}},
    journal = {The Astrophysical Journal Letters},
    volume = {956},
    pages = {L46},
    year = {2023},
    doi = {10.3847/2041-8213/acffc5},
}

@article{Short2023,
    author = {Short, P and Lawrence, A and Nicholl, M and Ward, M and Reynolds, T M and Mattila, S and Yin, C and Arcavi, I and Carnall, A and Charalampopoulos, P and Gromadzki, M and Jonker, P G and Kim, S and Leloudas, G and Mandel, I and Onori, F and Pursiainen, M and Schulze, S and Villforth, C and Wevers, T},
    title = "{Delayed appearance and evolution of coronal lines in the TDE AT2019qiz}",
    journal = {Monthly Notices of the Royal Astronomical Society},
    volume = {525},
    number = {1},
    pages = {1568-1587},
    year = {2023},
    month = {07},
    issn = {0035-8711},
    doi = {10.1093/mnras/stad2270},
    url = {https://doi.org/10.1093/mnras/stad2270},
}

@article{KasenRamirez2010,
    author = {Kasen, Daniel and Ramirez-Ruiz, Enrico},
    title = {{Optical Transients from the Unbound Debris of Tidal Disruption}},
    journal = {The Astrophysical Journal},
    volume = {714},
    number = {1},
    pages = {155},
    year = {2010},
    doi = {10.1088/0004-637X/714/1/155},
}

@article{callow2024,
   title={The rate of extreme coronal line emitting galaxies in the Sloan Digital Sky Survey and their relation to tidal disruption events},
   volume={535},
   ISSN={1365-2966},
   url={http://dx.doi.org/10.1093/mnras/stae2384},
   DOI={10.1093/mnras/stae2384},
   number={1},
   journal={Monthly Notices of the Royal Astronomical Society},
   publisher={Oxford University Press (OUP)},
   author={Callow, J and Graur, O and Clark, P and Palmese, A and Aguilar, J and Ahlen, S and BenZvi, S and Brooks, D and Claybaugh, T and de la Macorra, A and Doel, P and Forero-Romero, J E and Gaztañaga, E and Gontcho A Gontcho, S and Lambert, A and Landriau, M and Manera, M and Meisner, A and Miquel, R and Moustakas, J and Nie, J and Poppett, C and Prada, F and Rezaie, M and Rossi, G and Sanchez, E and Silber, J and Tarlé, G and Weaver, B A and Zhou, Z},
   year={2024},
   month=oct, pages={1095–1122} }

@article{Clark2024,
   title={Long-term follow-up observations of extreme coronal line emitting galaxies},
   volume={528},
   ISSN={1365-2966},
   url={http://dx.doi.org/10.1093/mnras/stae460},
   DOI={10.1093/mnras/stae460},
   number={4},
   journal={Monthly Notices of the Royal Astronomical Society},
   publisher={Oxford University Press (OUP)},
   author={Clark, Peter and Graur, Or and Callow, Joseph and Aguilar, Jessica and Ahlen, Steven and Anderson, Joseph P and Berger, Edo and Müller-Bravo, Tomás E and Brink, Thomas G and Brooks, David and Chen, Ting-Wan and Claybaugh, Todd and de la Macorra, Axel and Doel, Peter and Filippenko, Alexei V and Forero-Romero, Jamie E and Gomez, Sebastian and Gromadzki, Mariusz and Honscheid, Klaus and Inserra, Cosimo and Kisner, Theodore and Landriau, Martin and Makrygianni, Lydia and Manera, Marc and Meisner, Aaron and Miquel, Ramon and Moustakas, John and Nicholl, Matt and Nie, Jundan and Onori, Francesca and Palmese, Antonella and Poppett, Claire and Reynolds, Thomas and Rezaie, Mehdi and Rossi, Graziano and Sanchez, Eusebio and Schubnell, Michael and Tarlé, Gregory and Weaver, Benjamin A and Wevers, Thomas and Young, David R and Zheng, WeiKang and Zhou, Zhimin},
   year={2024},
   month=feb, pages={7076–7102} }

@ARTICLE{Noll2009,
       author = {{Noll}, S. and {Burgarella}, D. and {Giovannoli}, E. and {Buat}, V. and {Marcillac}, D. and {Mu{\~n}oz-Mateos}, J.~C.},
        title = "{Analysis of galaxy spectral energy distributions from far-UV to far-IR with CIGALE: studying a SINGS test sample}",
      journal = {\aap},
         year = 2009,
        month = dec,
       volume = {507},
       number = {3},
        pages = {1793-1813},
          doi = {10.1051/0004-6361/200912497},
archivePrefix = {arXiv},
       eprint = {0909.5439},
 primaryClass = {astro-ph.CO}
}

@article{Vallely2020,
   title={High-cadence, early-time observations of core-collapse supernovae from the TESS prime mission},
   volume={500},
   ISSN={1365-2966},
   url={http://dx.doi.org/10.1093/mnras/staa3675},
   DOI={10.1093/mnras/staa3675},
   number={4},
   journal={Monthly Notices of the Royal Astronomical Society},
   publisher={Oxford University Press (OUP)},
   author={Vallely, P J and Kochanek, C S and Stanek, K Z and Fausnaugh, M and Shappee, B J},
   year={2020},}

@article{ForemanMackey2013,
   title={<tt>emcee</tt>: The MCMC Hammer},
   volume={125},
   ISSN={1538-3873},
   url={http://dx.doi.org/10.1086/670067},
   DOI={10.1086/670067},
   number={925},
   journal={Publications of the Astronomical Society of the Pacific},
   publisher={IOP Publishing},
   author={Foreman-Mackey, Daniel and Hogg, David W. and Lang, Dustin and Goodman, Jonathan},
   year={2013},
   month=mar, pages={306–312} }

@ARTICLE{Perez-Fournon2022,
       author = {{Perez-Fournon}, I. and {Poidevin}, F. and {Angel}, C.~J. and {Delgado-Gonzalez}, Z. and {Shirley}, R. and {Marques-Chaves}, R. and {Geier}, S. and {Shu}, Y. and {Rodney}, S. and {Roberts-Pierel}, J. and {Bolton}, A. and {Chakrabarti}, S. and {Craig}, P. and {Alamiri}, B.},
        title = "{SGLF Transient Classification Report for 2022-06-25}",
      journal = {Transient Name Server Classification Report},
         year = 2022,
        month = jun,
       volume = {2022-1771},
        pages = {1},
       adsurl = {https://ui.adsabs.harvard.edu/abs/2022TNSCR1771....1P}
}

@article{Million2016,
   title={gPhoton: THE GALEX PHOTON DATA ARCHIVE},
   volume={833},
   ISSN={1538-4357},
   url={http://dx.doi.org/10.3847/1538-4357/833/2/292},
   DOI={10.3847/1538-4357/833/2/292},
   number={2},
   journal={The Astrophysical Journal},
   publisher={American Astronomical Society},
   author={Million, Chase and Fleming, Scott W. and Shiao, Bernie and Seibert, Mark and Loyd, Parke and Tucker, Michael and Smith, Myron and Thompson, Randy and White, Richard L.},
   year={2016},
   month=dec, pages={292} }

@misc{Cutri2021,
    author = {Cutri, R. M. and Wright, E. L. and Conrow, T., et al.},
    title = {{WISE All-Sky Data Release}},
    year = {2021},
    howpublished = {VizieR Online Data Catalog, II/328},
    note = {Originally published in 2012},
}

@misc{Wright2019,
    author = {Wright, Edward L. and Eisenhardt, Peter R. M. and Mainzer, Amy K., et al.},
    title = {{AllWISE Source Catalog}},
    year = {2019},
    publisher = {IPAC},
    doi = {10.26131/IRSA1},
}

@ARTICLE{pypeit:joss_arXiv,
       author = {{Prochaska}, J. and {Hennawi}, Joseph and {Westfall}, Kyle and {Cooke}, Ryan and {Wang}, Feige and {Hsyu}, Tiffany and {Davies}, Frederick and {Farina}, Emanuele and {Pelliccia}, Debora},
        title = "{PypeIt: The Python Spectroscopic Data Reduction Pipeline}",
      journal = {The Journal of Open Source Software},
         year = 2020,
        month = dec,
       volume = {5},
       number = {56},
          eid = {2308},
        pages = {2308},
          doi = {10.21105/joss.02308},
archivePrefix = {arXiv},
       eprint = {2005.06505},
 primaryClass = {astro-ph.IM},
       adsurl = {https://ui.adsabs.harvard.edu/abs/2020JOSS....5.2308P}
}

@article{pypeit:joss_pub,
    doi = {10.21105/joss.02308},
    url = {https://doi.org/10.21105/joss.02308},
    year = {2020},
    publisher = {The Open Journal},
    volume = {5},
    number = {56},
    pages = {2308},
    author = {J. Xavier Prochaska and Joseph F. Hennawi and Kyle B. Westfall and Ryan J. Cooke and Feige Wang and Tiffany Hsyu and Frederick B. Davies and Emanuele Paolo Farina and Debora Pelliccia},
    title = {PypeIt: The Python Spectroscopic Data Reduction Pipeline},
    journal = {Journal of Open Source Software}
}

@MISC{pypeit:zenodo,
       author = {{Prochaska}, J. Xavier and {Hennawi}, Joseph and {Cooke}, Ryan and
         {Westfall}, Kyle and {Wang}, Feige and {EmAstro} and {Tiffanyhsyu} and
         {Wasserman}, Asher and {Villaume}, Alexa and {Marijana777} and
         {Schindler}, JT and {Young}, David and {Simha}, Sunil and
         {Wilde}, Matt and {Tejos}, Nicolas and {Isbell}, Jacob and
         {Fl{\"o}rs}, Andreas and {Sandford}, Nathan and {Vasovi{\'c}}, Zlatan and
         {Betts}, Edward and {Holden}, Brad},
        title = "{pypeit/PypeIt: Release 1.0.0}",
         year = 2020,
        month = apr,
          eid = {10.5281/zenodo.3743493},
          doi = {10.5281/zenodo.3743493},
      version = {v1.0.0},
    publisher = {Zenodo},
       adsurl = {https://ui.adsabs.harvard.edu/abs/2020zndo...3743493P}
}

@article{Rayner2003,
	author = {J. T. Rayner and D. W. Toomey and P. M. Onaka and A. J. Denault and W. E. Stahlberger and W. D. Vacca and M. C. Cushing and S. Wang},
	journal = {Publications of the Astronomical Society of the Pacific},
	month = {mar},
	number = {805},
	pages = {362},
	title = {SpeX: A Medium‐Resolution 0.8--5.5 Micron Spectrograph and Imager for the NASA Infrared Telescope Facility},
	volume = {115},
	year = {2003}}

@ARTICLE{Tucker2022,
       author = {{Tucker}, M.~A. and {Shappee}, B.~J. and {Huber}, M.~E. and {Payne}, A.~V. and {Do}, A. and {Hinkle}, J.~T. and {de Jaeger}, T. and {Ashall}, C. and {Desai}, D.~D. and {Hoogendam}, W.~B. and {Aldering}, G. and {Auchettl}, K. and {Baranec}, C. and {Bulger}, J. and {Chambers}, K. and {Chun}, M. and {Hodapp}, K.~W. and {Lowe}, T.~B. and {McKay}, L. and {Rampy}, R. and {Rubin}, D. and {Tonry}, J.~L.},
        title = "{The Spectroscopic Classification of Astronomical Transients (SCAT) Survey: Overview, Pipeline Description, Initial Results, and Future Plans}",
      journal = {\pasp},
         year = 2022,
        month = dec,
       volume = {134},
       number = {1042},
          eid = {124502},
        pages = {124502},
          doi = {10.1088/1538-3873/aca719},
archivePrefix = {arXiv},
       eprint = {2210.09322},
 primaryClass = {astro-ph.IM},
       adsurl = {https://ui.adsabs.harvard.edu/abs/2022PASP..134l4502T}
}

@ARTICLE{Labrie2023,
       author = {{Labrie}, K. and {Simpson}, C. and {Cardenes}, R. and {Turner}, J. and {Soraisam}, M. and {Quint}, B. and {Oberdorf}, O. and {Placco}, V.~M. and {Berke}, D. and {Smirnova}, O. and {Conseil}, S. and {Vacca}, W.~D. and {Thomas-Osip}, J.},
        title = "{DRAGONS-A Quick Overview}",
      journal = {Research Notes of the American Astronomical Society},
         year = 2023,
        month = oct,
       volume = {7},
       number = {10},
          eid = {214},
        pages = {214},
          doi = {10.3847/2515-5172/ad0044},
archivePrefix = {arXiv},
       eprint = {2310.03048},
 primaryClass = {astro-ph.IM},
       adsurl = {https://ui.adsabs.harvard.edu/abs/2023RNAAS...7..214L}
}

@article{Morrissey2018,
	author = {Patrick Morrissey and Matuesz Matuszewski and D. Christopher Martin and James D. Neill and Harland Epps and Jason Fucik and Bob Weber and Behnam Darvish and Sean Adkins and Steve Allen and Randy Bartos and Justin Belicki and Jerry Cabak and Shawn Callahan and Dave Cowley and Marty Crabill and Willian Deich and Alex Delecroix and Greg Doppman and David Hilyard and Ean James and Steve Kaye and Michael Kokorowski and Shui Kwok and Kyle Lanclos and Steve Milner and Anna Moore and Donal O'Sullivan and Prachi Parihar and Sam Park and Andrew Phillips and Luca Rizzi and Constance Rockosi and Hector Rodriguez and Yves Salaun and Kirk Seaman and David Sheikh and Jason Weiss and Ray Zarzaca},
	journal = {The Astrophysical Journal},
	month = {sep},
	number = {1},
	pages = {93},
	title = {The Keck Cosmic Web Imager Integral Field Spectrograph},
	volume = {864},
	year = {2018}}

@INPROCEEDINGS{Kollmeier2019,
       author = {{Kollmeier}, Juna and {Anderson}, S.~F. and {Blanc}, G.~A. and {Blanton}, M.~R. and {Covey}, K.~R. and {Crane}, J. and {Drory}, N. and {Frinchaboy}, P.~M. and {Froning}, C.~S. and {Johnson}, J.~A. and {Kneib}, J. -P. and {Kreckel}, K. and {Merloni}, A. and {Pellegrini}, E.~W. and {Pogge}, R.~W. and {Ramirez}, S.~V. and {Rix}, H.~W. and {Sayres}, C. and {S{\'a}nchez-Gallego}, Jos{\'e} and {Shen}, Yue and {Tkachenko}, A. and {Trump}, J.~R. and {Tuttle}, S.~E. and {Weijmans}, A. and {Zasowski}, G. and {Barbuy}, B. and {Beaton}, R.~L. and {Bergemann}, M. and {Bochanski}, J.~J. and {Brandt}, W.~N. and {Casey}, A.~R. and {Cherinka}, B.~A. and {Eracleous}, M. and {Fan}, X. and {Garc{\'\i}a}, R.~A. and {Green}, P.~J. and {Hekker}, S. and {Lane}, R.~R. and {Longa-Pe{\~n}a}, P. and {Mathur}, S. and {Meza}, A. and {Minchev}, I. and {Myers}, A.~D. and {Nidever}, D.~L. and {Nitschelm}, C. and {O'Connell}, J.~E. and {Price-Whelan}, A.~M. and {Raddick}, M.~J. and {Rossi}, G. and {Sankrit}, R. and {Simon}, J.~D. and {Stutz}, A.~M. and {Ting}, Y. -S. and {Trakhtenbrot}, B. and {Weaver}, B.~A. and {Willmer}, C.~N.~A. and {Weinberg}, D.~H.},
        title = "{SDSS-V Pioneering Panoptic Spectroscopy}",
    booktitle = {Bulletin of the American Astronomical Society},
         year = 2019,
       volume = {51},
        month = sep,
          eid = {274},
        pages = {274},
       adsurl = {https://ui.adsabs.harvard.edu/abs/2019BAAS...51g.274K}
}

@ARTICLE{Wang2024,
       author = {{Wang}, Yibo and {Wang}, Tinggui and {Jiang}, Ning and {Zhang}, Xiaer and {Zhu}, Jiazheng and {Shu}, Xinwen and {Huang}, Shifeng and {Zhang}, FaBao and {Sheng}, Zhenfeng and {Lin}, Zheyu},
        title = "{ASASSN-18ap: A Dusty Tidal Disruption Event Candidate with an Early Bump in the Light Curve}",
      journal = {\apj},
         year = 2024,
        month = may,
       volume = {966},
       number = {1},
          eid = {136},
        pages = {136},
          doi = {10.3847/1538-4357/ad2ae4},
archivePrefix = {arXiv},
       eprint = {2312.12015},
 primaryClass = {astro-ph.HE},
       adsurl = {https://ui.adsabs.harvard.edu/abs/2024ApJ...966..136W}
}

@article{Riffel2006,
   title={A 0.8–2.4μm spectral atlas of active galactic nuclei},
   volume={457},
   ISSN={1432-0746},
   url={http://dx.doi.org/10.1051/0004-6361:20065291},
   DOI={10.1051/0004-6361:20065291},
   number={1},
   journal={Astronomy \& Astrophysics},
   publisher={EDP Sciences},
   author={Riffel, R. and Rodríguez-Ardila, A. and Pastoriza, M. G.},
   year={2006},
   month=sep, pages={61–70} }

@article{Calderon2024,
    author = {Calderón, Diego and Pejcha, Ondřej and Metzger, Brian D and Duffell, Paul C},
    title = "{The effect of relativistic precession on light curves of tidal disruption events}",
    journal = {Monthly Notices of the Royal Astronomical Society},
    volume = {528},
    number = {2},
    pages = {2568-2587},
    year = {2024},
    month = {01},
    issn = {0035-8711},
    doi = {10.1093/mnras/stae194},
    url = {https://doi.org/10.1093/mnras/stae194},
    eprint = {https://academic.oup.com/mnras/article-pdf/528/2/2568/56485544/stae194.pdf},
}

@article{Bonnerot2022,
    author = {Bonnerot, Clément and Lu, Wenbin},
    title = {The nozzle shock in tidal disruption events},
    journal = {Monthly Notices of the Royal Astronomical Society},
    volume = {511},
    number = {2},
    pages = {2147-2169},
    year = {2022},
    month = {01},
    issn = {0035-8711},
    doi = {10.1093/mnras/stac146},
    url = {https://doi.org/10.1093/mnras/stac146},
    eprint = {https://academic.oup.com/mnras/article-pdf/511/2/2147/42497386/stac146.pdf},
}

@ARTICLE{Cushing2004,
       author = {{Cushing}, Michael C. and {Vacca}, William D. and {Rayner}, John T.},
        title = "{Spextool: A Spectral Extraction Package for SpeX, a 0.8-5.5 Micron Cross-Dispersed Spectrograph}",
      journal = {\pasp},
         year = 2004,
        month = apr,
       volume = {116},
       number = {818},
        pages = {362-376},
          doi = {10.1086/382907},
       adsurl = {https://ui.adsabs.harvard.edu/abs/2004PASP..116..362C}
}

@article{Koljonen2024,
    author = {Koljonen, Karri I I and Liodakis, Ioannis and Lindfors, Elina and Nilsson, Kari and Reynolds, Thomas M and Charalampopoulos, Panos and Kouroumpatzakis, Konstantinos and McCall, Callum and Jermak, Helen E and Steele, Iain A and Carbajo-Hijarrubia, Juan},
    title = {The extreme coronal line emitter AT 2022fpx: varying optical polarization properties and late-time X-ray flare},
    journal = {Monthly Notices of the Royal Astronomical Society},
    volume = {532},
    number = {1},
    pages = {112-125},
    year = {2024},
    month = {06},
    issn = {0035-8711},
    doi = {10.1093/mnras/stae1466},
    url = {https://doi.org/10.1093/mnras/stae1466},
    eprint = {https://academic.oup.com/mnras/article-pdf/532/1/112/58341271/stae1466.pdf},
}

@article{Greene_2020,
	author = {Greene, Jenny E. and Strader, Jay and Ho, Luis C.},
	journal = {Annual Review of Astronomy and Astrophysics},
	number = {Volume 58, 2020},
	pages = {257-312},
	title = {Intermediate-Mass Black Holes},
	volume = {58},
	year = {2020}}

@article{Mummery_2023,
    author = {Mummery, Andrew and van Velzen, Sjoert and Nathan, Edward and Ingram, Adam and Hammerstein, Erica and Fraser-Taliente, Ludovic and Balbus, Steven},
    title = {Fundamental scaling relationships revealed in the optical light curves of tidal disruption events},
    journal = {Monthly Notices of the Royal Astronomical Society},
    volume = {527},
    number = {2},
    pages = {2452-2489},
    year = {2023},
    month = {10},
    issn = {0035-8711},
    doi = {10.1093/mnras/stad3001},
    url = {https://doi.org/10.1093/mnras/stad3001},
    eprint = {https://academic.oup.com/mnras/article-pdf/527/2/2452/53452929/stad3001.pdf},
}

@ARTICLE{mainzer23,
       author = {{Mainzer}, A.~K. and {Masiero}, J.~R. and {Abell}, Paul A. and {Bauer}, J.~M. and {Bottke}, William and {Buratti}, Bonnie J. and {Carey}, Sean J. and {Cotto-Figueroa}, D. and {Cutri}, R.~M. and {Dahlen}, D. and {Eisenhardt}, Peter R.~M. and {Fernandez}, Y.~R. and {Furfaro}, Roberto and {Grav}, Tommy and {Hoffman}, T.~L. and {Kelley}, Michael S. and {Kim}, Yoonyoung and {Kirkpatrick}, J. Davy and {Lawler}, Christopher R. and {Lilly}, Eva and {Liu}, X. and {Marocco}, Federico and {Marsh}, K.~A. and {Masci}, Frank J. and {McMurtry}, Craig W. and {Pourrahmani}, Milad and {Reinhart}, Lennon and {Ressler}, Michael E. and {Satpathy}, Akash and {Schambeau}, C.~A. and {Sonnett}, S. and {Spahr}, Timothy B. and {Surace}, Jason A. and {Vaquero}, Mar and {Wright}, E.~L. and {Zengilowski}, Gregory R. and {NEO Surveyor Mission Team}},
        title = "{The Near-Earth Object Surveyor Mission}",
      journal = {PSJ},
         year = 2023,
        month = dec,
       volume = {4},
       number = {12},
          eid = {224},
        pages = {224},
          doi = {10.3847/PSJ/ad0468},
archivePrefix = {arXiv},
       eprint = {2310.12918},
 primaryClass = {astro-ph.EP},
       adsurl = {https://ui.adsabs.harvard.edu/abs/2023PSJ.....4..224M}
}

@article{Mazzalay_2010,
    author = {Mazzalay, Ximena and Rodríguez-Ardila, Alberto and Komossa, S.},
    title = "{Demystifying the coronal-line region of active galactic nuclei: spatially resolved spectroscopy with the Hubble Space Telescope}",
    journal = {Monthly Notices of the Royal Astronomical Society},
    volume = {405},
    number = {2},
    pages = {1315-1338},
    year = {2010},
    month = {06},
    issn = {0035-8711},
    doi = {10.1111/j.1365-2966.2010.16533.x},
    url = {https://doi.org/10.1111/j.1365-2966.2010.16533.x},
    eprint = {https://academic.oup.com/mnras/article-pdf/405/2/1315/4011855/mnras0405-1315.pdf},
}

@article{Negus_2023,
   title={A Catalog of 71 Coronal Line Galaxies in MaNGA: [Ne v] Is an Effective AGN Tracer},
   volume={945},
   ISSN={1538-4357},
   url={http://dx.doi.org/10.3847/1538-4357/acb772},
   DOI={10.3847/1538-4357/acb772},
   number={2},
   journal={The Astrophysical Journal},
   publisher={American Astronomical Society},
   author={Negus, James and Comerford, Julia M. and Sánchez, Francisco Müller and Revalski, Mitchell and Riffel, Rogemar A. and Bundy, Kevin and Nevin, Rebecca and Rembold, Sandro B.},
   year={2023},
   month=mar, pages={127} }

@article{huang_2024_pre_peak_emission,
	author = {Huang, Xiaoshan and Davis, Shane W. and Jiang, Yan-fei},
	journal = {The Astrophysical Journal},
	month = {oct},
	number = {2},
	pages = {165},
	title = {Pre-peak Emission in Tidal Disruption Events},
	volume = {974},
	year = {2024}}

@article{Mandel2015,
	author = {Ilya Mandel and Yuri Levin},
	journal = {The Astrophysical Journal Letters},
	month = {may},
	number = {1},
	pages = {L4},
	title = {DOUBLE TIDAL DISRUPTIONS IN GALACTIC NUCLEI},
	volume = {805},
	year = {2015},
        doi = {10.1088/2041-8205/805/1/L4}}

@article{Wu_Yuan_2018,
    author = {Wu, Xiao-Jun and Yuan, Ye-Fei},
    title = {Double tidal disruption events with massive black hole binaries},
    journal = {Monthly Notices of the Royal Astronomical Society},
    volume = {479},
    number = {2},
    pages = {1569-1578},
    year = {2018},
    month = {07},
    issn = {0035-8711},
    doi = {10.1093/mnras/sty1423},
    url = {https://doi.org/10.1093/mnras/sty1423},
    eprint = {https://academic.oup.com/mnras/article-pdf/479/2/1569/25135055/sty1423.pdf},
}

@article{Ferguson1997,
   title={Physical Conditions of the Coronal Line Region in Seyfert Galaxies},
   volume={110},
   ISSN={1538-4365},
   url={http://dx.doi.org/10.1086/312998},
   DOI={10.1086/312998},
   number={2},
   journal={The Astrophysical Journal Supplement Series},
   publisher={American Astronomical Society},
   author={Ferguson, Jason W. and Korista, Kirk T. and Ferland, Gary J.},
   year={1997},
   month=jun, pages={287–297} }

@INPROCEEDINGS{Oliva1997,
       author = {{Oliva}, E.},
        title = "{Coronal Lines in Active Galactic Nuclei}",
    booktitle = {IAU Colloq. 159: Emission Lines in Active Galaxies: New Methods and Techniques},
         year = 1997,
       editor = {{Peterson}, Bradley M. and {Cheng}, Fu-Zhen and {Wilson}, Andrew S.},
       series = {Astronomical Society of the Pacific Conference Series},
       volume = {113},
        month = jan,
        pages = {288},
       adsurl = {https://ui.adsabs.harvard.edu/abs/1997ASPC..113..288O}
}

@article{Rodriguez_Ardila_2002,
   title={Near‐Infrared Coronal Lines in Narrow‐Line Seyfert 1 Galaxies},
   volume={579},
   ISSN={1538-4357},
   url={http://dx.doi.org/10.1086/342840},
   DOI={10.1086/342840},
   number={1},
   journal={The Astrophysical Journal},
   publisher={American Astronomical Society},
   author={Rodriguez‐Ardila, A. and Viegas, S. M. and Pastoriza, M. G. and Prato, L.},
   year={2002},
   month=nov, pages={214–226} }

@article{Almudena_Prieto_2005,
   title={Morphology of the coronal-line region in active galactic nuclei},
   volume={364},
   ISSN={1745-3925},
   url={http://dx.doi.org/10.1111/j.1745-3933.2005.00099.x},
   DOI={10.1111/j.1745-3933.2005.00099.x},
   number={1},
   journal={Monthly Notices of the Royal Astronomical Society: Letters},
   publisher={Oxford University Press (OUP)},
   author={Almudena Prieto, M. and Marco, Olivier and Gallimore, Jack},
   year={2005},
   month=nov, pages={L28–L32} }

@article{Gelbord_2009,
    author = {Gelbord, Jonathan M. and Mullaney, James R. and Ward, Martin J.},
    title = {AGN with strong forbidden high-ionization lines selected from the Sloan Digital Sky Survey},
    journal = {Monthly Notices of the Royal Astronomical Society},
    volume = {397},
    number = {1},
    pages = {172-189},
    year = {2009},
    month = {07},
    issn = {0035-8711},
    doi = {10.1111/j.1365-2966.2009.14961.x},
    url = {https://doi.org/10.1111/j.1365-2966.2009.14961.x},
    eprint = {https://academic.oup.com/mnras/article-pdf/397/1/172/18744751/mnras0397-0172.pdf},
}

@article{Mullaney_2009,
   title={The location and kinematics of the coronal-line emitting regions in active galactic nuclei},
   volume={394},
   ISSN={1745-3925},
   url={http://dx.doi.org/10.1111/j.1745-3933.2008.00599.x},
   DOI={10.1111/j.1745-3933.2008.00599.x},
   number={1},
   journal={Monthly Notices of the Royal Astronomical Society: Letters},
   publisher={Oxford University Press (OUP)},
   author={Mullaney, J. R. and Ward, M. J. and Done, C. and Ferland, G. J. and Schurch, N.},
   year={2009},
   month=mar, pages={L16–L20} }

@ARTICLE{MullerSanchez_2011,
       author = {{M{\"u}ller-S{\'a}nchez}, F. and {Prieto}, M.~A. and {Hicks}, E.~K.~S. and {Vives-Arias}, H. and {Davies}, R.~I. and {Malkan}, M. and {Tacconi}, L.~J. and {Genzel}, R.},
        title = "{Outflows from Active Galactic Nuclei: Kinematics of the Narrow-line and Coronal-line Regions in Seyfert Galaxies}",
      journal = {\apj},
         year = 2011,
        month = oct,
       volume = {739},
       number = {2},
          eid = {69},
        pages = {69},
          doi = {10.1088/0004-637X/739/2/69},
archivePrefix = {arXiv},
       eprint = {1107.3140},
 primaryClass = {astro-ph.CO},
       adsurl = {https://ui.adsabs.harvard.edu/abs/2011ApJ...739...69M}
}

@article{Glidden_2016,
   title={A MODEL FOR TYPE 2 CORONAL LINE FOREST (CLiF) AGNs},
   volume={824},
   ISSN={1538-4357},
   url={http://dx.doi.org/10.3847/0004-637X/824/1/34},
   DOI={10.3847/0004-637x/824/1/34},
   number={1},
   journal={The Astrophysical Journal},
   publisher={American Astronomical Society},
   author={Glidden, Ana and Rose, Marvin and Elvis, Martin and McDowell, Jonathan},
   year={2016},
   month=jun, pages={34} 
}

@article{Riffel_2021,
    author = {Riffel, Rogemar A and Bianchin, Marina and Riffel, Rogério and Storchi-Bergmann, Thaisa and Schönell, Astor J and Dahmer-Hahn, Luis Gabriel and Dametto, Natacha Z and Diniz, Marlon R},
    title = {Gemini NIFS survey of feeding and feedback in nearby active galaxies – IV. Excitation},
    journal = {Monthly Notices of the Royal Astronomical Society},
    volume = {503},
    number = {4},
    pages = {5161-5178},
    year = {2021},
    month = {03},
    issn = {0035-8711},
    doi = {10.1093/mnras/stab788},
    url = {https://doi.org/10.1093/mnras/stab788},
    eprint = {https://academic.oup.com/mnras/article-pdf/503/4/5161/37018383/stab788.pdf},
}

@article{TrindadeFalcao_2022,
    author = {Trindade Falcão, Anna and Kraemer, S B and Crenshaw, D M and Melendez, M and Revalski, M and Fischer, T C and Schmitt, H R and Turner, T J},
    title = {Tracking X-ray outflows with optical/infrared footprint lines},
    journal = {Monthly Notices of the Royal Astronomical Society},
    volume = {511},
    number = {1},
    pages = {1420-1430},
    year = {2022},
    month = {01},
    issn = {0035-8711},
    doi = {10.1093/mnras/stac173},
    url = {https://doi.org/10.1093/mnras/stac173},
    eprint = {https://academic.oup.com/mnras/article-pdf/511/1/1420/42425289/stac173.pdf},
}

@article{Treiber_2023,
    author = {Treiber, Helena P and Hinkle, Jason T and Fausnaugh, Michael M and Shappee, Benjamin J and Kochanek, Christopher S and Vallely, Patrick J and Auchettl, Katie and Holoien, Thomas W-S and Payne, Anna V and Dai, Xinyu},
    title = {Revealing AGNs through TESS variability},
    journal = {Monthly Notices of the Royal Astronomical Society},
    volume = {525},
    number = {4},
    pages = {5795-5812},
    year = {2023},
    month = {09},
    issn = {0035-8711},
    doi = {10.1093/mnras/stad2530},
    url = {https://doi.org/10.1093/mnras/stad2530},
    eprint = {https://academic.oup.com/mnras/article-pdf/525/4/5795/51754993/stad2530.pdf},
}

@ARTICLE{Masci_2023,
       author = {{Masci}, Frank J. and {Laher}, Russ R. and {Rusholme}, Benjamin and {Shupe}, David and {Paladini}, Roberta and {Groom}, Steve and {Wold}, Avery and {Miller}, Adam A. and {Drake}, Andrew},
        title = "{A New Forced Photometry Service for the Zwicky Transient Facility}",
      journal = {arXiv e-prints},
         year = 2023,
        month = may,
          eid = {arXiv:2305.16279},
        pages = {arXiv:2305.16279},
          doi = {10.48550/arXiv.2305.16279},
archivePrefix = {arXiv},
       eprint = {2305.16279},
 primaryClass = {astro-ph.IM},
       adsurl = {https://ui.adsabs.harvard.edu/abs/2023arXiv230516279M}
}

@ARTICLE{Somalwar_2022,
       author = {{Somalwar}, Jean J. and {Ravi}, Vikram and {Dong}, Dillon and {Graham}, Matthew and {Hallinan}, Gregg and {Law}, Casey and {Lu}, Wenbin and {Myers}, Steven T.},
        title = "{The Nascent Milliquasar VT J154843.06+220812.6: Tidal Disruption Event or Extreme Accretion State Change?}",
      journal = {\apj},
         year = 2022,
        month = apr,
       volume = {929},
       number = {2},
          eid = {184},
        pages = {184},
          doi = {10.3847/1538-4357/ac5e29},
archivePrefix = {arXiv},
       eprint = {2108.12431},
 primaryClass = {astro-ph.HE},
       adsurl = {https://ui.adsabs.harvard.edu/abs/2022ApJ...929..184S}
}

@article{Liu_2022,
	author = {Liu, Xiao-Long and Dou, Li-Ming and Chen, Jin-Hong and Shen, Rong-Feng},
	doi = {10.3847/1538-4357/ac33a9},
	journal = {The Astrophysical Journal},
	month = {jan},
	number = {1},
	pages = {67},
	publisher = {The American Astronomical Society},
	title = {The UV/Optical Peak and X-Ray Brightening in TDE Candidate AT 2019azh: A Case of Stream--Stream Collision and Delayed Accretion},
	url = {https://dx.doi.org/10.3847/1538-4357/ac33a9},
	volume = {925},
	year = {2022}}

@article{Charalampopoulos_2022,
	author = {{Charalampopoulos} and {Leloudas, G.} and {Malesani, D. B.} and {Wevers, T.} and {Arcavi, I.} and {Nicholl, M.} and {Pursiainen, M.} and {Lawrence, A.} and {Anderson, J. P.} and {Benetti, S.} and {Cannizzaro, G.} and {Chen, T.-W.} and {Galbany, L.} and {Gromadzki, M.} and {Guti{\'e}rrez, C. P.} and {Inserra, C.} and {Jonker, P. G.} and {M{\"u}ller-Bravo, T. E.} and {Onori, F.} and {Short, P.} and {Sollerman, J.} and {Young, D. R.}},
	journal = {A\&A},
	pages = {A34},
	title = {A detailed spectroscopic study of tidal disruption events},
	volume = 659,
	year = 2022}

@ARTICLE{Tarrant_2025,
       author = {{Tarrant}, Ashley and {Hinkle}, Jason and {Shappee}, Benjamin and {Kochanek}, Christopher and {Hey}, Daniel and {Auge}, Connor and {Payne}, Anna and {Bolish}, Michael and {Yuk}, Heechan and {Dai}, Xinyu and {Auchettl}, Katie and {Thompson}, Todd and {Treiber}, Helena},
        title = "{The AGN Optical Variability Fundamental Plane}",
      journal = {arXiv e-prints},
         year = 2025,
        month = jan,
          eid = {arXiv:2501.12444},
        pages = {arXiv:2501.12444},
          doi = {10.48550/arXiv.2501.12444},
archivePrefix = {arXiv},
       eprint = {2501.12444},
 primaryClass = {astro-ph.GA},
       adsurl = {https://ui.adsabs.harvard.edu/abs/2025arXiv250112444T}
}

@ARTICLE{ATLAS22kjn_classification,
       author = {{Perez-Fournon}, I. and {Poidevin}, F. and {Angel}, C.~J. and {Delgado-Gonzalez}, Z. and {Shirley}, R. and {Marques-Chaves}, R. and {Geier}, S. and {Shu}, Y. and {Rodney}, S. and {Roberts-Pierel}, J. and {Bolton}, A. and {Chakrabarti}, S. and {Craig}, P. and {Alamiri}, B.},
        title = "{SGLF Transient Classification Report for 2022-06-25}",
      journal = {Transient Name Server Classification Report},
         year = 2022,
        month = jun,
       volume = {2022-1771},
        pages = {1},
       adsurl = {https://ui.adsabs.harvard.edu/abs/2022TNSCR1771....1P}
}

@article{Faris2024,
	author = {Faris, Sara and Arcavi, Iair and Makrygianni, Lydia and Hiramatsu, Daichi and Terreran, Giacomo and Farah, Joseph and Howell, D. Andrew and McCully, Curtis and Newsome, Megan and Padilla Gonzalez, Estefania and Pellegrino, Craig and Bostroem, K. Azalee and Abojanb, Wiam and Lam, Marco C. and Tomasella, Lina and Brink, Thomas G. and Filippenko, Alexei V. and French, K. Decker and Clark, Peter and Graur, Or and Leloudas, Giorgos and Gromadzki, Mariusz and Anderson, Joseph P. and Nicholl, Matt and Guti{\'e}rrez, Claudia P. and Kankare, Erkki and Inserra, Cosimo and Galbany, Llu{\'\i}s and Reynolds, Thomas and Mattila, Seppo and Heikkil{\"a}, Teppo and Wang, Yanan and Onori, Francesca and Wevers, Thomas and Coughlin, Eric R. and Charalampopoulos, Panos and Johansson, Joel},
	journal = {The Astrophysical Journal},
	month = {jul},
	number = {2},
	pages = {104},
	title = {Light-curve Structure and Hα Line Formation in the Tidal Disruption Event AT 2019azh},
	volume = {969},
	year = {2024}}

@article{Li_Shen_2023,
	author = {Li, Junyao and Shen, Yue},
	journal = {The Astrophysical Journal},
	month = {jun},
	number = {2},
	pages = {122},
	title = {Constraining AGN Torus Sizes with Optical and Mid-infrared Ensemble Structure Functions},
	volume = {950},
	year = {2023}}

@article{Lyu_2019,
	author = {Lyu, Jianwei and Rieke, George H. and Smith, Paul S.},
	journal = {The Astrophysical Journal},
	month = {nov},
	number = {1},
	pages = {33},
	title = {Mid-IR Variability and Dust Reverberation Mapping of Low-z Quasars. I. Data, Methods, and Basic Results},
	volume = {886},
	year = {2019}}

@article{Bonnerot_Rossi_2019,
    author = {Bonnerot, Clément and Rossi, Elena M},
    title = {Streams collision as possible precursor of double tidal disruption events},
    journal = {Monthly Notices of the Royal Astronomical Society},
    volume = {484},
    number = {1},
    pages = {1301-1316},
    year = {2019},
    month = {01},
    issn = {0035-8711},
    doi = {10.1093/mnras/stz062},
    url = {https://doi.org/10.1093/mnras/stz062},
    eprint = {https://academic.oup.com/mnras/article-pdf/484/1/1301/27579705/stz062.pdf},
}

@ARTICLE{Steinberg_2024,
       author = {{Steinberg}, Elad and {Stone}, Nicholas C.},
        title = "{Stream-disk shocks as the origins of peak light in tidal disruption events}",
      journal = {\nat},
         year = 2024,
        month = jan,
       volume = {625},
       number = {7995},
        pages = {463-467},
          doi = {10.1038/s41586-023-06875-y},
archivePrefix = {arXiv},
       eprint = {2206.10641},
 primaryClass = {astro-ph.HE},
       adsurl = {https://ui.adsabs.harvard.edu/abs/2024Natur.625..463S}
}

@article{Ryu_2023,
	author = {Ryu, Taeho and Krolik, Julian and Piran, Tsvi and Noble, Scott C. and Avara, Mark},
	journal = {The Astrophysical Journal},
	month = {oct},
	number = {1},
	pages = {12},
	title = {Shocks Power Tidal Disruption Events},
	volume = {957},
	year = {2023}}

@article{Huang_2023_stream_stream,
	author = {Huang, Xiaoshan and Davis, Shane W. and Jiang, Yan-fei},
	journal = {The Astrophysical Journal},
	month = {aug},
	number = {1},
	pages = {117},
	title = {A Bright First Day for Tidal Disruption Events},
	volume = {953},
	year = {2023}}

@article{Pasham_2017,
	author = {Pasham, Dheeraj R. and Cenko, S. Bradley and Sadowski, Aleksander and Guillochon, James and Stone, Nicholas C. and Velzen, Sjoert van and Cannizzo, John K.},
	journal = {The Astrophysical Journal Letters},
	month = {mar},
	number = {2},
	pages = {L30},
	title = {Optical/UV-to-X-Ray Echoes from the Tidal Disruption Flare ASASSN-14li},
	volume = {837},
	year = {2017}}

@article{Kajava_2020,
	author = {Kajava, Jari J. E. and Giustini, Margherita and Saxton, Richard D. and Miniutti, Giovanni},
	journal = {A\&A},
	pages = {A100},
	title = {Rapid late-time X-ray brightening of the tidal disruption event OGLE16aaa},
	volume = 639,
	year = 2020}

@ARTICLE{Mummery_2024,
       author = {{Mummery}, Andrew and {Nathan}, Edward and {Ingram}, Adam and {Gardner}, M.},
        title = "{Fitting transients with discs (FITTED): a public light curve and spectral fitting package based on evolving relativistic discs}",
      journal = {\mnras},
         year = 2025,
        month = dec,
       volume = {544},
       number = {2},
        pages = {2225-2240},
          doi = {10.1093/mnras/staf1565},
archivePrefix = {arXiv},
       eprint = {2408.15048},
 primaryClass = {astro-ph.HE},
       adsurl = {https://ui.adsabs.harvard.edu/abs/2025MNRAS.544.2225M}
}

@article{Mummery_2025,
    author = {Mummery, Andrew and Guolo, Muryel and Matthews, James and Newsome, Megan and Lintott, Chris and Keel, William},
    title = {Galaxy-scale consequences of tidal disruption events: extended emission-line regions, extreme coronal lines, and infrared-to-optical light echoes},
    journal = {Monthly Notices of the Royal Astronomical Society},
    volume = {544},
    number = {2},
    pages = {2262-2295},
    year = {2025},
    month = {12},
    issn = {0035-8711},
    doi = {10.1093/mnras/staf1649},
    url = {https://doi.org/10.1093/mnras/staf1649},
    eprint = {https://academic.oup.com/mnras/article-pdf/544/2/2262/64417999/staf1649.pdf},
}

@article{Ferland_1998,
	author = {Ferland, G. J. and Korista, K. T. and Verner, D. A. and Ferguson, J. W. and Kingdon, J. B. and Verner, E. M.},
	journal = {Publications of the Astronomical Society of the Pacific},
	month = {jul},
	number = {749},
	pages = {761},
	title = {CLOUDY 90: Numerical Simulation of Plasmas and Their Spectra},
	volume = {110},
	year = {1998}}

@article{Newsome_2024,
	author = {Newsome, Megan and Arcavi, Iair and Howell, D. Andrew and McCully, Curtis and Terreran, Giacomo and Hosseinzadeh, Griffin and Bostroem, K. Azalee and Dgany, Yael and Farah, Joseph and Faris, Sara and Padilla-Gonzalez, Estefania and Pellegrino, Craig and Andrews, Moira},
	journal = {The Astrophysical Journal},
	month = {dec},
	number = {2},
	pages = {258},
	title = {Mapping the Inner 0.1 pc of a Supermassive Black Hole Environment with the Tidal Disruption Event and Extreme Coronal-line Emitter AT 2022upj},
	volume = {977},
	year = {2024}}

@article{Frederick_2019,
	author = {Frederick, Sara and Gezari, Suvi and Graham, Matthew J. and Cenko, S. Bradley and van Velzen, Sjoert and Stern, Daniel and Blagorodnova, Nadejda and Kulkarni, Shrinivas R. and Yan, Lin and De, Kishalay and Fremling, U. Christoffer and Hung, Tiara and Kara, Erin and Shupe, David L. and Ward, Charlotte and Bellm, Eric C. and Dekany, Richard and Duev, Dmitry A. and Feindt, Ulrich and Giomi, Matteo and Kupfer, Thomas and Laher, Russ R. and Masci, Frank J. and Miller, Adam A. and Neill, James D. and Ngeow, Chow-Choong and Patterson, Maria T. and Porter, Michael and Rusholme, Ben and Sollerman, Jesper and Walters, Richard},
	doi = {10.3847/1538-4357/ab3a38},
	journal = {The Astrophysical Journal},
	month = {sep},
	number = {1},
	pages = {31},
	publisher = {The American Astronomical Society},
	title = {A New Class of Changing-look LINERs},
	url = {https://dx.doi.org/10.3847/1538-4357/ab3a38},
	volume = {883},
	year = {2019}}

@ARTICLE{Smith_2020,
       author = {{Smith}, K.~W. and {Smartt}, S.~J. and {Young}, D.~R. and {Tonry}, J.~L. and {Denneau}, L. and {Flewelling}, H. and {Heinze}, A.~N. and {Weiland}, H.~J. and {Stalder}, B. and {Rest}, A. and {Stubbs}, C.~W. and {Anderson}, J.~P. and {Chen}, T. -W. and {Clark}, P. and {Do}, A. and {F{\"o}rster}, F. and {Fulton}, M. and {Gillanders}, J. and {McBrien}, O.~R. and {O'Neill}, D. and {Srivastav}, S. and {Wright}, D.~E.},
        title = "{Design and Operation of the ATLAS Transient Science Server}",
      journal = {\pasp},
         year = 2020,
        month = aug,
       volume = {132},
       number = {1014},
          eid = {085002},
        pages = {085002},
          doi = {10.1088/1538-3873/ab936e},
archivePrefix = {arXiv},
       eprint = {2003.09052},
 primaryClass = {astro-ph.IM},
       adsurl = {https://ui.adsabs.harvard.edu/abs/2020PASP..132h5002S}
}

@article{Mainetti_2016,
	author = {Mainetti, Deborah and Lupi, Alessandro and Campana, Sergio and Colpi, Monica},
	doi = {10.1093/mnras/stw197},
	eprint = {https://academic.oup.com/mnras/article-pdf/457/3/2516/8002243/stw197.pdf},
	issn = {0035-8711},
	journal = {Monthly Notices of the Royal Astronomical Society},
	month = {02},
	number = {3},
	pages = {2516-2529},
	title = {Hydrodynamical simulations of the tidal stripping of binary stars by massive black holes},
	url = {https://doi.org/10.1093/mnras/stw197},
	volume = {457},
	year = {2016}}

@ARTICLE{ULTRASAT_2024,
       author = {{Shvartzvald}, Y. and {Waxman}, E. and {Gal-Yam}, A. and {Ofek}, E.~O. and {Ben-Ami}, S. and {Berge}, D. and {Kowalski}, M. and {B{\"u}hler}, R. and {Worm}, S. and {Rhoads}, J.~E. and {Arcavi}, I. and {Maoz}, D. and {Polishook}, D. and {Stone}, N. and {Trakhtenbrot}, B. and {Ackermann}, M. and {Aharonson}, O. and {Birnholtz}, O. and {Chelouche}, D. and {Guetta}, D. and {Hallakoun}, N. and {Horesh}, A. and {Kushnir}, D. and {Mazeh}, T. and {Nordin}, J. and {Ofir}, A. and {Ohm}, S. and {Parsons}, D. and {Pe'er}, A. and {Perets}, H.~B. and {Perdelwitz}, V. and {Poznanski}, D. and {Sadeh}, I. and {Sagiv}, I. and {Shahaf}, S. and {Soumagnac}, M. and {Tal-Or}, L. and {Santen}, J. Van and {Zackay}, B. and {Guttman}, O. and {Rekhi}, P. and {Townsend}, A. and {Weinstein}, A. and {Wold}, I.},
        title = "{ULTRASAT: A Wide-field Time-domain UV Space Telescope}",
      journal = {\apj},
         year = 2024,
        month = mar,
       volume = {964},
       number = {1},
          eid = {74},
        pages = {74},
          doi = {10.3847/1538-4357/ad2704},
archivePrefix = {arXiv},
       eprint = {2304.14482},
 primaryClass = {astro-ph.IM},
       adsurl = {https://ui.adsabs.harvard.edu/abs/2024ApJ...964...74S}
}

@misc{UVEX_2023,
      title={Science with the Ultraviolet Explorer (UVEX)}, 
      author={S. R. Kulkarni and Fiona A. Harrison and Brian W. Grefenstette and Hannah P. Earnshaw and Igor Andreoni and Danielle A. Berg and Joshua S. Bloom and S. Bradley Cenko and Ryan Chornock and Jessie L. Christiansen and Michael W. Coughlin and Alexander Wuollet Criswell and Behnam Darvish and Kaustav K. Das and Kishalay De and Luc Dessart and Don Dixon and Bas Dorsman and Kareem El-Badry and Christopher Evans and K. E. Saavik Ford and Christoffer Fremling and Boris T. Gansicke and Suvi Gezari and Y. Goetberg and Gregory M. Green and Matthew J. Graham and Marianne Heida and Anna Y. Q. Ho and Amruta D. Jaodand and Christopher M. Johns-Krull and Mansi M. Kasliwal and Margaret Lazzarini and Wenbin Lu and Raffaella Margutti and D. Christopher Martin and Daniel Charles Masters and Barry McKernan and Yael Naze and Samaya M. Nissanke and B. Parazin and Daniel A. Perley and E. Sterl Phinney and Anthony L. Piro and G. Raaijmakers and Gregor Rauw and Antonio C. Rodriguez and Hugues Sana and Peter Senchyna and Leo P. Singer and Jessica J. Spake and Keivan G. Stassun and Daniel Stern and Harry I. Teplitz and Daniel R. Weisz and Yuhan Yao},
      year={2023},
      eprint={2111.15608},
      archivePrefix={arXiv},
      primaryClass={astro-ph.GA},
      url={https://arxiv.org/abs/2111.15608}, 
}

@article{Pedregosa11,
  author  = {Fabian Pedregosa and Ga{{\"e}}l Varoquaux and Alexandre Gramfort and Vincent Michel and Bertrand Thirion and Olivier Grisel and Mathieu Blondel and Peter Prettenhofer and Ron Weiss and Vincent Dubourg and Jake Vanderplas and Alexandre Passos and David Cournapeau and Matthieu Brucher and Matthieu Perrot and {{\'E}}douard Duchesnay},
  title   = {Scikit-learn: Machine Learning in Python},
  journal = {Journal of Machine Learning Research},
  year    = {2011},
  volume  = {12},
  number  = {85},
  pages   = {2825--2830},
  url     = {http://jmlr.org/papers/v12/pedregosa11a.html}
}

@BOOK{RasmussenWilliams06,
       author = {{Rasmussen}, Carl Edward and {Williams}, Christopher K.~I.},
        title = "{Gaussian Processes for Machine Learning}",
         year = 2006,
       adsurl = {https://ui.adsabs.harvard.edu/abs/2006gpml.book.....R}
}

@phdthesis{Duvenaud14,
    title={Automatic model construction with Gaussian processes},
    url={https://www.repository.cam.ac.uk/handle/1810/247281},
    DOI={10.17863/CAM.14087},
    school={Apollo - University of Cambridge Repository},
    author={Duvenaud, David},
    year={2014}
}

@article{Sun_2025,
	author = {Sun, Jingbo and Guo, Hengxiao and Gu, Minfeng and Li, Ya-Ping and Chen, Yongjun and Gonz{\'a}lez-Buitrago, D. and Wang, Jian-Guo and Li, Sha-Sha and Feng, Hai-Cheng and Xiong, Dingrong and Wang, Yanan and Yuan, Qi and Jin, Jun-jie and Zhang, Wenda and Deng, Hongping and Zhang, Minghao},
	doi = {10.3847/1538-4357/adb724},
	journal = {The Astrophysical Journal},
	month = {mar},
	number = {2},
	pages = {150},
	publisher = {The American Astronomical Society},
	title = {AT2021aeuk: A Repeating Partial Tidal Disruption Event Candidate in a Narrow-line Seyfert 1 Galaxy},
	url = {https://dx.doi.org/10.3847/1538-4357/adb724},
	volume = {982},
	year = {2025}}

@article{Miniutti_2019,
	author = {Miniutti, G. and Saxton, R. D. and Giustini, M. and Alexander, K. D. and Fender, R. P. and Heywood, I. and Monageng, I. and Coriat, M. and Tzioumis, A. K. and Read, A. M. and Knigge, C. and Gandhi, P. and Pretorius, M. L. and Ag{\'\i}s-Gonz{\'a}lez, B.},
	date = {2019/09/01},
	doi = {10.1038/s41586-019-1556-x},
	id = {Miniutti2019},
	isbn = {1476-4687},
	journal = {Nature},
	number = {7774},
	pages = {381--384},
	title = {Nine-hour X-ray quasi-periodic eruptions from a low-mass black hole galactic nucleus},
	url = {https://doi.org/10.1038/s41586-019-1556-x},
	volume = {573},
	year = {2019}}

@ARTICLE{Giustini_2020,
       author = {{Giustini}, Margherita and {Miniutti}, Giovanni and {Saxton}, Richard D.},
        title = "{X-ray quasi-periodic eruptions from the galactic nucleus of RX J1301.9+2747}",
      journal = {\aap},
         year = 2020,
        month = apr,
       volume = {636},
          eid = {L2},
        pages = {L2},
          doi = {10.1051/0004-6361/202037610},
archivePrefix = {arXiv},
       eprint = {2002.08967},
 primaryClass = {astro-ph.HE},
       adsurl = {https://ui.adsabs.harvard.edu/abs/2020A\&A...636L...2G}
}

@article{Arcodia_2021,
	author = {Arcodia, R. and Merloni, A. and Nandra, K. and Buchner, J. and Salvato, M. and Pasham, D. and Remillard, R. and Comparat, J. and Lamer, G. and Ponti, G. and Malyali, A. and Wolf, J. and Arzoumanian, Z. and Bogensberger, D. and Buckley, D. A. H. and Gendreau, K. and Gromadzki, M. and Kara, E. and Krumpe, M. and Markwardt, C. and Ramos-Ceja, M. E. and Rau, A. and Schramm, M. and Schwope, A.},
	date = {2021/04/01},
	doi = {10.1038/s41586-021-03394-6},
	id = {Arcodia2021},
	isbn = {1476-4687},
	journal = {Nature},
	number = {7856},
	pages = {704--707},
	title = {X-ray quasi-periodic eruptions from two previously quiescent galaxies},
	url = {https://doi.org/10.1038/s41586-021-03394-6},
	volume = {592},
	year = {2021}}

@article{Arcodia_2024,
	author = {{Arcodia, R.} and {Linial, I.} and {Miniutti, G.} and {Franchini, A.} and {Giustini, M.} and {Bonetti, M.} and {Sesana, A.} and {Soria, R.} and {Chakraborty, J.} and {Dotti, M.} and {Kara, E.} and {Merloni, A.} and {Ponti, G.} and {Vincentelli, F.}},
	doi = {10.1051/0004-6361/202451218},
	journal = {A\&A},
	pages = {A80},
	title = {Ticking away: The long-term X-ray timing and spectral evolution of eRO-QPE2},
	url = {https://doi.org/10.1051/0004-6361/202451218},
	volume = 690,
	year = 2024}

@article{Nicholl_2024,
	author = {Nicholl, M. and Pasham, D. R. and Mummery, A. and Guolo, M. and Gendreau, K. and Dewangan, G. C. and Ferrara, E. C. and Remillard, R. and Bonnerot, C. and Chakraborty, J. and Hajela, A. and Dhillon, V. S. and Gillan, A. F. and Greenwood, J. and Huber, M. E. and Janiuk, A. and Salvesen, G. and van Velzen, S. and Aamer, A. and Alexander, K. D. and Angus, C. R. and Arzoumanian, Z. and Auchettl, K. and Berger, E. and de Boer, T. and Cendes, Y. and Chambers, K. C. and Chen, T. -W. and Chornock, R. and Fulton, M. D. and Gao, H. and Gillanders, J. H. and Gomez, S. and Gompertz, B. P. and Fabian, A. C. and Herman, J. and Ingram, A. and Kara, E. and Laskar, T. and Lawrence, A. and Lin, C. -C. and Lowe, T. B. and Magnier, E. A. and Margutti, R. and McGee, S. L. and Minguez, P. and Moore, T. and Nathan, E. and Oates, S. R. and Patra, K. C. and Ramsden, P. and Ravi, V. and Ridley, E. J. and Sheng, X. and Smartt, S. J. and Smith, K. W. and Srivastav, S. and Stein, R. and Stevance, H. F. and Turner, S. G. D. and Wainscoat, R. J. and Weston, J. and Wevers, T. and Young, D. R.},
	date = {2024/10/01},
	doi = {10.1038/s41586-024-08023-6},
	id = {Nicholl2024},
	isbn = {1476-4687},
	journal = {Nature},
	number = {8035},
	pages = {804--808},
	title = {Quasi-periodic X-ray eruptions years after a nearby tidal disruption event},
	url = {https://doi.org/10.1038/s41586-024-08023-6},
	volume = {634},
	year = {2024}}

@article{Chakraborty_2025,
	author = {Chakraborty, Joheen and Kara, Erin and Arcodia, Riccardo and Buchner, Johannes and Giustini, Margherita and Hern{\'a}ndez-Garc{\'\i}a, Lorena and Linial, Itai and Masterson, Megan and Miniutti, Giovanni and Mummery, Andrew and Panagiotou, Christos and Quintin, Erwan and S{\'a}nchez-S{\'a}ez, Paula},
	doi = {10.3847/2041-8213/adc2f8},
	journal = {The Astrophysical Journal Letters},
	month = {apr},
	number = {2},
	pages = {L39},
	publisher = {The American Astronomical Society},
	title = {Discovery of Quasiperiodic Eruptions in the Tidal Disruption Event and Extreme Coronal Line Emitter AT2022upj: Implications for the QPE/TDE Fraction and a Connection to ECLEs},
	url = {https://doi.org/10.3847/2041-8213/adc2f8},
	volume = {983},
	year = {2025}}

@ARTICLE{Hernandez_Garcia_2025,
       author = {{Hern{\'a}ndez-Garc{\'\i}a}, Lorena and {Chakraborty}, Joheen and {S{\'a}nchez-S{\'a}ez}, Paula and {Ricci}, Claudio and {Cuadra}, Jorge and {McKernan}, Barry and {Ford}, K.~E. Saavik and {Ar{\'e}valo}, Patricia and {Rau}, Arne and {Arcodia}, Riccardo and {Kara}, Erin and {Liu}, Zhu and {Merloni}, Andrea and {Bruni}, Gabriele and {Goodwin}, Adelle and {Arzoumanian}, Zaven and {Assef}, Roberto J. and {Baldini}, Pietro and {Bayo}, Amelia and {Bauer}, Franz E. and {Bernal}, Santiago and {Brightman}, Murray and {Calistro Rivera}, Gabriela and {Gendreau}, Keith and {Homan}, David and {Krumpe}, Mirko and {Lira}, Paulina and {Mart{\'\i}nez-Aldama}, Mary Loli and {Salvato}, Mara and {Sotomayor}, Bel{\'e}n},
        title = "{Discovery of extreme quasi-periodic eruptions in a newly accreting massive black hole}",
      journal = {Nature Astronomy},
         year = 2025,
        month = jun,
       volume = {9},
        pages = {895-906},
          doi = {10.1038/s41550-025-02523-9},
archivePrefix = {arXiv},
       eprint = {2504.07169},
 primaryClass = {astro-ph.HE},
       adsurl = {https://ui.adsabs.harvard.edu/abs/2025NatAs...9..895H}
}

@article{Linial_2025,
	author = {Linial, Itai and Metzger, Brian D. and Quataert, Eliot},
	doi = {10.3847/1538-4357/adfa0e},
	journal = {The Astrophysical Journal},
	month = {sep},
	number = {2},
	pages = {147},
	publisher = {The American Astronomical Society},
	title = {QPEs from EMRI Debris Streams Impacting Accretion Disks in Galactic Nuclei},
	url = {https://doi.org/10.3847/1538-4357/adfa0e},
	volume = {991},
	year = {2025}}

@ARTICLE{Jiang_2016,
       author = {{Jiang}, Yan-Fei and {Guillochon}, James and {Loeb}, Abraham},
        title = "{Prompt Radiation and Mass Outflows from the Stream-Stream Collisions of Tidal Disruption Events}",
      journal = {\apj},
         year = 2016,
        month = oct,
       volume = {830},
       number = {2},
          eid = {125},
        pages = {125},
          doi = {10.3847/0004-637X/830/2/125},
archivePrefix = {arXiv},
       eprint = {1603.07733},
 primaryClass = {astro-ph.HE},
       adsurl = {https://ui.adsabs.harvard.edu/abs/2016ApJ...830..125J}
}

@article{Savitzky_Golay_1964,
	author = {Savitzky, Abraham. and Golay, M. J. E.},
	date = {1964/07/01},
	doi = {10.1021/ac60214a047},
	isbn = {0003-2700},
	journal = {Analytical Chemistry},
	journal1 = {Analytical Chemistry},
	journal2 = {Anal. Chem.},
	month = {07},
	number = {8},
	pages = {1627--1639},
	publisher = {American Chemical Society},
	title = {Smoothing and Differentiation of Data by Simplified Least Squares Procedures.},
	type = {doi: 10.1021/ac60214a047},
	url = {https://doi.org/10.1021/ac60214a047},
	volume = {36},
	year = {1964},
	year1 = {1964}}

@article{Lin_2025,
	author = {Lin, Zheyu and Jiang, Ning and Wang, Yibo and Kong, Xu and Huang, Shifeng and Lin, Zesen and Qin, Chen and Xia, Tianyu},
	doi = {10.3847/1538-4357/adef10},
	journal = {The Astrophysical Journal},
	month = {aug},
	number = {1},
	pages = {22},
	publisher = {The American Astronomical Society},
	title = {Insights from the ``Red Devil'' AT 2022fpx: A Dust-reddened Family of Tidal Disruption Events Excluded by Their Apparent Red Color?},
	url = {https://doi.org/10.3847/1538-4357/adef10},
	volume = {990},
	year = {2025}}

@ARTICLE{Guolo_2024,
       author = {{Guolo}, Muryel and {Gezari}, Suvi and {Yao}, Yuhan and {van Velzen}, Sjoert and {Hammerstein}, Erica and {Cenko}, S. Bradley and {Tokayer}, Yarone M.},
        title = "{A Systematic Analysis of the X-Ray Emission in Optically Selected Tidal Disruption Events: Observational Evidence for the Unification of the Optically and X-Ray-selected Populations}",
      journal = {\apj},
         year = 2024,
        month = may,
       volume = {966},
       number = {2},
          eid = {160},
        pages = {160},
          doi = {10.3847/1538-4357/ad2f9f},
archivePrefix = {arXiv},
       eprint = {2308.13019},
 primaryClass = {astro-ph.HE},
       adsurl = {https://ui.adsabs.harvard.edu/abs/2024ApJ...966..160G}
}

@ARTICLE{Ferland_Osterbrock_1987,
       author = {{Ferland}, Gary J. and {Osterbrock}, Donald E.},
        title = "{The Ultraviolet and Optical Emission-Line Spectrum of III ZW 77}",
      journal = {\apj},
         year = 1987,
        month = jul,
       volume = {318},
        pages = {145},
          doi = {10.1086/165357},
       adsurl = {https://ui.adsabs.harvard.edu/abs/1987ApJ...318..145F}
}

@ARTICLE{Rose_2011,
       author = {{Rose}, M. and {Tadhunter}, C.~N. and {Holt}, J. and {Ramos Almeida}, C. and {Littlefair}, S.~P.},
        title = "{The forbidden high-ionization-line region of the type 2 quasar SDSS J11311.05+162739.5: a clear view of the inner face of the torus?}",
      journal = {\mnras},
         year = 2011,
        month = jul,
       volume = {414},
       number = {4},
        pages = {3360-3380},
          doi = {10.1111/j.1365-2966.2011.18639.x},
archivePrefix = {arXiv},
       eprint = {1103.0660},
 primaryClass = {astro-ph.CO},
       adsurl = {https://ui.adsabs.harvard.edu/abs/2011MNRAS.414.3360R}
}

@article{Hauschild-Roier_2025,
	author = {Hauschild-Roier, Gabriel R and Storchi-Bergmann, Thaisa and Riffel, Rog{\'e}rio and Mainieri, Vincenzo},
	doi = {10.1093/mnras/staf1388},
	eprint = {https://academic.oup.com/mnras/article-pdf/542/3/2525/64110984/staf1388.pdf},
	issn = {0035-8711},
	journal = {Monthly Notices of the Royal Astronomical Society},
	month = {08},
	number = {3},
	pages = {2525-2541},
	title = {The differences in the narrow-line region of nearby QSOs 1 and 2 -- I. Higher excitation and contribution of shocks in type 1s},
	url = {https://doi.org/10.1093/mnras/staf1388},
	volume = {542},
	year = {2025}}

@ARTICLE{Shingles_2021,
       author = {{Shingles}, L. and {Smith}, K.~W. and {Young}, D.~R. and {Smartt}, S.~J. and {Tonry}, J. and {Denneau}, L. and {Heinze}, A. and {Weiland}, H. and {Flewelling}, H. and {Stalder}, B. and {Clocchiatti}, A. and {F{\"o}rster}, F. and {Pignata}, G. and {Rest}, A. and {Anderson}, J. and {Stubbs}, C. and {Erasmus}, N.},
        title = "{Release of the ATLAS Forced Photometry server for public use}",
      journal = {Transient Name Server AstroNote},
         year = 2021,
        month = jan,
       volume = {7},
        pages = {1-7},
       adsurl = {https://ui.adsabs.harvard.edu/abs/2021TNSAN...7....1S}
}

@ARTICLE{Rodriguez_Ardila_2020,
       author = {{Rodr{\'\i}guez-Ardila}, Alberto and {Fonseca-Faria}, Marcos A.},
        title = "{A 700 pc Extended Coronal Gas Emission in the Circinus Galaxy}",
      journal = {\apjl},
         year = 2020,
        month = may,
       volume = {895},
       number = {1},
          eid = {L9},
        pages = {L9},
          doi = {10.3847/2041-8213/ab901b},
archivePrefix = {arXiv},
       eprint = {2005.03113},
 primaryClass = {astro-ph.GA},
       adsurl = {https://ui.adsabs.harvard.edu/abs/2020ApJ...895L...9R}
}

@ARTICLE{Yu_Dong_2024,
       author = {{Yu}, Fangyuan and {Lai}, Dong},
        title = "{Binary Stars Approaching Supermassive Black Holes: Tidal Breakup, Double Stellar Disruptions, and Stellar Collision}",
      journal = {\apj},
         year = 2024,
        month = dec,
       volume = {977},
       number = {2},
          eid = {268},
        pages = {268},
          doi = {10.3847/1538-4357/ad93a6},
archivePrefix = {arXiv},
       eprint = {2409.09597},
 primaryClass = {astro-ph.HE},
       adsurl = {https://ui.adsabs.harvard.edu/abs/2024ApJ...977..268Y}
}

@ARTICLE{Coughlin_2023,
       author = {{Coughlin}, Eric R.},
        title = "{The dynamics of debris streams from tidal disruption events: exact solutions, critical stream density, and hydrogen recombination}",
      journal = {\mnras},
         year = 2023,
        month = jul,
       volume = {522},
       number = {4},
        pages = {5500-5516},
          doi = {10.1093/mnras/stad1347},
archivePrefix = {arXiv},
       eprint = {2305.01677},
 primaryClass = {astro-ph.HE},
       adsurl = {https://ui.adsabs.harvard.edu/abs/2023MNRAS.522.5500C}
}

@misc{Hinkle_2025,
      title={On the Double: Two Luminous Flares from the Nearby Tidal Disruption Event ASASSN-22ci (AT2022dbl) and Connections to Repeating TDE Candidates}, 
      author={Jason T. Hinkle and Katie Auchettl and Willem B. Hoogendam and Anna V. Payne and Thomas W. -S. Holoien and Benjamin J. Shappee and Michael A. Tucker and Christopher S. Kochanek and K. Z. Stanek and Patrick J. Vallely and Charlotte R. Angus and Chris Ashall and Thomas de Jaeger and Dhvanil D. Desai and Aaron Do and Michael M. Fausnaugh and Mark E. Huber and Ryan J. Rickards Vaught and Jennifer Shi},
      year={2025},
      eprint={2412.15326},
      archivePrefix={arXiv},
      primaryClass={astro-ph.HE},
      url={https://arxiv.org/abs/2412.15326}, 
}

@ARTICLE{Almeida_2017,
       author = {{Ramos Almeida}, Cristina and {Ricci}, Claudio},
        title = "{Nuclear obscuration in active galactic nuclei}",
      journal = {Nature Astronomy},
         year = 2017,
        month = oct,
       volume = {1},
        pages = {679-689},
          doi = {10.1038/s41550-017-0232-z},
archivePrefix = {arXiv},
       eprint = {1709.00019},
 primaryClass = {astro-ph.GA},
       adsurl = {https://ui.adsabs.harvard.edu/abs/2017NatAs...1..679R}
}

@ARTICLE{Smith_2025,
       author = {{Smith}, Theodore B. and {Fries}, Logan B. and {Trump}, Jonathan R. and {Grier}, Catherine J. and {Shen}, Yue and {Anderson}, Scott F. and {Brandt}, W.~N. and {Davis}, Megan C. and {Dwelly}, Tom and {Hall}, P.~B. and {Horne}, Keith and {Homayouni}, Y. and {McKaig}, J. and {Morrison}, Sean and {Sharp}, Hugh W. and {Assef}, Roberto J. and {Bauer}, Franz E. and {Koekemoer}, Anton M. and {Schneider}, Donald P. and {Trakhtenbrot}, Benny and {Ibarra-Medel}, Hector Javier and {Pe{\~n}aloza}, Castalia Alenka Negrete},
        title = "{The SDSS-V Black Hole Mapper Reverberation Mapping Project: Light Echoes of the Coronal-line Region in a Luminous Quasar}",
      journal = {\apj},
         year = 2025,
        month = dec,
       volume = {995},
       number = {2},
          eid = {185},
        pages = {185},
          doi = {10.3847/1538-4357/ae1f18},
archivePrefix = {arXiv},
       eprint = {2510.16099},
 primaryClass = {astro-ph.GA},
       adsurl = {https://ui.adsabs.harvard.edu/abs/2025ApJ...995..185S}
}

@misc{Kynoch_2026,
      title={Mapping the nuclear environments of extreme coronal line emitting galaxies}, 
      author={Daniel Kynoch and Or Graur and Peter Clark and J. N. Aguilar and S. Ahlen and D. Bianchi and D. Brooks and T. Claybaugh and A. de la Macorra and P. Doel and J. E. Forero-Romero and S. Gontcho A Gontcho and G. Gutierrez and R. Joyce and S. Juneau and M. Landriau and L. Le Guillou and A. Meisner and R. Miquel and J. Moustakas and S. Panda and W. J. Percival and F. Prada and I. Pérez-Ràfols and G. Rossi and E. Sanchez and D. Schlegel and J. Silber and D. Sprayberry and G. Tarlé and B. A. Weaver and R. Zhou and H. Zou},
      year={2026},
      eprint={2606.04090},
      archivePrefix={arXiv},
      primaryClass={astro-ph.GA},
      url={https://arxiv.org/abs/2606.04090}, 
}

@article{Yao_2026,
    author = {Yao, Yuhan and Chornock, Ryan and Mummery, Andrew and Margutti, Raffaella and Gilfanov, Marat and Guolo, Muryel and Coughlin, Eric R and Lu, Wenbin and Chakraborty, Joheen and Pasham, Dheeraj R and Alexander, Kate D and Aspegren, Olivia and Angus, Charlotte R and Guo, Xinze and Hall, Xander J and Hammerstein, Erica and Hinds, K -Ryan and Ho, Anna Y Q and Huang, Xiaoshan and Kammoun, Elias and LeBaron, Natalie and Lucchini, Matteo and McGrath, Zoë and Nicholl, Matt and Perley, Daniel A and Rich, R Michael and Schroeder, Genevieve and Sheng, Xinyue and Sollerman, Jesper and Somalwar, Jean and Wise, Jacob L and Coughlin, Michael W and Drake, Andrew and Graham, Matthew J and Helou, George and Jaimes, Joahan C and Kasliwal, Mansi M and Mahabal, Ashish A and Medvedev, Pavel and Purdum, Josiah and Rusholme, Ben and Sunyaev, Rashid},
    title = {AT2024lhc and AT2024kmq in the landscape of featureless tidal disruption events},
    journal = {Monthly Notices of the Royal Astronomical Society},
    volume = {549},
    number = {2},
    pages = {stag920},
    year = {2026},
    month = {06},
    issn = {0035-8711},
    doi = {10.1093/mnras/stag920},
    url = {https://doi.org/10.1093/mnras/stag920},
    eprint = {https://academic.oup.com/mnras/article-pdf/549/2/stag920/68386097/stag920.pdf},
}

@misc{Tucker_2026,
      title={SCAT Data Release 1: 1810 optical spectra of 1330 transients}, 
      author={Michael A. Tucker and Mark E. Huber and Benjamin J. Shappee and Jason T. Hinkle and Willem B. Hoogendam and Charlotte R. Angus and Chris Ashall and Katie Auchettl and Kenneth C. Chambers and Dhvanil D. Desai and Aaron Do and Joseph Ghammashi and Catherine J. Grier and Joanna Herman and Thomas de Jaeger and Jodie Kiyokawa and Thomas B. Lowe and Eugene A. Magnier and Anna V. Payne and Sara Romagnoli and David Rubin},
      year={2026},
      eprint={2604.23794},
      archivePrefix={arXiv},
      primaryClass={astro-ph.HE},
      url={https://arxiv.org/abs/2604.23794}, 
}

@ARTICLE{tucker18_scat,
       author = {{Tucker}, Michael A. and {Huber}, Mark and {Shappee}, Benjamin J. and {Prieto}, Jose L. and {Holoien}, T.~W.-S. and {Dong}, Subo and {Bose}, S. and {Chen}, Ping and {Falco}, E. and {Calkins}, M. and {Chambers}, K.~C. and {Flewelling}, H. and {Magnier}, T. Lowe. E. and {Schultz}, A. and {Waters}, C. and {Wainscoat}, R.~J. and {Wilman}, M. and {Smith}, K.~W. and {Smartt}, S.~J. and {Young}, D.~R. and {Wright}, D.~E.},
        title = "{SCAT Classification of PS18kh as a potential TDE}",
      journal = {The Astronomer's Telegram},
         year = 2018,
        month = mar,
       volume = {11473},
        pages = {1},
       adsurl = {https://ui.adsabs.harvard.edu/abs/2018ATel11473....1T}
}

@ARTICLE{holoien18b,
       author = {{Holoien}, T.~W.-S. and {Huber}, M.~E. and {Shappee}, B.~J. and {Eracleous}, M. and {Auchettl}, K. and {Brown}, J.~S. and {Tucker}, M.~A. and {Chambers}, K.~C. and {Kochanek}, C.~S. and {Stanek}, K.~Z. and {Rest}, A. and {Bersier}, D. and {Post}, R.~S. and {Aldering}, G. and {Ponder}, K.~A. and {Simon}, J.~D. and {Kankare}, E. and {Dong}, D. and {Hallinan}, G. and {Reddy}, N.~A. and {Sanders}, R.~L. and {Topping}, M.~W. and {Pan-STARRS} and {Bulger}, J. and {Lowe}, T.~B. and {Magnier}, E.~A. and {Schultz}, A.~S.~B. and {Waters}, C.~Z. and {Willman}, M. and {Wright}, D. and {Young}, D.~R. and {ASAS-SN} and {Dong}, Subo and {Prieto}, J.~L. and {Thompson}, Todd A. and {ATLAS} and {Denneau}, L. and {Flewelling}, H. and {Heinze}, A.~N. and {Smartt}, S.~J. and {Smith}, K.~W. and {Stalder}, B. and {Tonry}, J.~L. and {Weiland}, H.},
        title = "{PS18kh: A New Tidal Disruption Event with a Non-axisymmetric Accretion Disk}",
      journal = {\apj},
         year = 2019,
        month = aug,
       volume = {880},
       number = {2},
          eid = {120},
        pages = {120},
          doi = {10.3847/1538-4357/ab2ae1},
archivePrefix = {arXiv},
       eprint = {1808.02890},
 primaryClass = {astro-ph.HE},
       adsurl = {https://ui.adsabs.harvard.edu/abs/2019ApJ...880..120H}
}

@ARTICLE{vanvelzen19b,
       author = {{van Velzen}, Sjoert and {Gezari}, Suvi and {Cenko}, S. Bradley and {Kara}, Erin and {Miller-Jones}, James C.~A. and {Hung}, Tiara and {Bright}, Joe and {Roth}, Nathaniel and {Blagorodnova}, Nadejda and {Huppenkothen}, Daniela and {Yan}, Lin and {Ofek}, Eran and {Sollerman}, Jesper and {Frederick}, Sara and {Ward}, Charlotte and {Graham}, Matthew J. and {Fender}, Rob and {Kasliwal}, Mansi M. and {Canella}, Chris and {Stein}, Robert and {Giomi}, Matteo and {Brinnel}, Valery and {van Santen}, Jakob and {Nordin}, Jakob and {Bellm}, Eric C. and {Dekany}, Richard and {Fremling}, Christoffer and {Golkhou}, V. Zach and {Kupfer}, Thomas and {Kulkarni}, Shrinivas R. and {Laher}, Russ R. and {Mahabal}, Ashish and {Masci}, Frank J. and {Miller}, Adam A. and {Neill}, James D. and {Riddle}, Reed and {Rigault}, Mickael and {Rusholme}, Ben and {Soumagnac}, Maayane T. and {Tachibana}, Yutaro},
        title = "{The First Tidal Disruption Flare in ZTF: From Photometric Selection to Multi-wavelength Characterization}",
      journal = {\apj},
         year = 2019,
        month = feb,
       volume = {872},
       number = {2},
          eid = {198},
        pages = {198},
          doi = {10.3847/1538-4357/aafe0c},
archivePrefix = {arXiv},
       eprint = {1809.02608},
 primaryClass = {astro-ph.HE},
       adsurl = {https://ui.adsabs.harvard.edu/abs/2019ApJ...872..198V}
}

@article{Cendes_2022,
	author = {Cendes, Y. and Berger, E. and Alexander, K. D. and Gomez, S. and Hajela, A. and Chornock, R. and Laskar, T. and Margutti, R. and Metzger, B. and Bietenholz, M. F. and Brethauer, D. and Wieringa, M. H.},
	doi = {10.3847/1538-4357/ac88d0},
	journal = {The Astrophysical Journal},
	month = {oct},
	number = {1},
	pages = {28},
	publisher = {The American Astronomical Society},
	title = {A Mildly Relativistic Outflow Launched Two Years after Disruption in Tidal Disruption Event AT2018hyz},
	url = {https://doi.org/10.3847/1538-4357/ac88d0},
	volume = {938},
	year = {2022}}

@ARTICLE{Cendes_2026,
       author = {{Cendes}, Yvette and {Berger}, Edo and {Beniamini}, Paz and {Gill}, Ramandeep and {Matsumoto}, Tatsuya and {Alexander}, Kate D. and {Bietenholz}, Michael F. and {Hajela}, Aprajita and {Christy}, Collin T. and {Chornock}, Ryan and {Gomez}, Sebastian and {Gurwell}, Mark A. and {Keating}, Garrett K. and {Laskar}, Tanmoy and {Margutti}, Raffaella and {Rao}, Ramprasad and {Velez}, Natalie and {Wieringa}, Mark H.},
        title = "{Continued Rapid Radio Brightening of the Tidal Disruption Event AT2018hyz}",
      journal = {\apj},
         year = 2026,
        month = feb,
       volume = {998},
       number = {1},
          eid = {111},
        pages = {111},
          doi = {10.3847/1538-4357/ae286d},
archivePrefix = {arXiv},
       eprint = {2507.08998},
 primaryClass = {astro-ph.HE},
       adsurl = {https://ui.adsabs.harvard.edu/abs/2026ApJ...998..111C}
}

@article{Leloudas_2022,
	author = {Leloudas, Giorgos and Bulla, Mattia and Cikota, Aleksandar and Dai, Lixin and Thomsen, Lars L. and Maund, Justyn R. and Charalampopoulos, Panos and Roth, Nathaniel and Arcavi, Iair and Auchettl, Katie and Malesani, Daniele B. and Nicholl, Matt and Ramirez-Ruiz, Enrico},
	date = {2022/10/01},
	doi = {10.1038/s41550-022-01767-z},
	id = {Leloudas2022},
	isbn = {2397-3366},
	journal = {Nature Astronomy},
	number = {10},
	pages = {1193--1202},
	title = {An asymmetric electron-scattering photosphere around optical tidal disruption events},
	url = {https://doi.org/10.1038/s41550-022-01767-z},
	volume = {6},
	year = {2022}}

@ARTICLE{scipy,
  author  = {Virtanen, Pauli and Gommers, Ralf and Oliphant, Travis E. and
            Haberland, Matt and Reddy, Tyler and Cournapeau, David and
            Burovski, Evgeni and Peterson, Pearu and Weckesser, Warren and
            Bright, Jonathan and {van der Walt}, St{\'e}fan J. and
            Brett, Matthew and Wilson, Joshua and Millman, K. Jarrod and
            Mayorov, Nikolay and Nelson, Andrew R. J. and Jones, Eric and
            Kern, Robert and Larson, Eric and Carey, C J and
            Polat, {\.I}lhan and Feng, Yu and Moore, Eric W. and
            {VanderPlas}, Jake and Laxalde, Denis and Perktold, Josef and
            Cimrman, Robert and Henriksen, Ian and Quintero, E. A. and
            Harris, Charles R. and Archibald, Anne M. and
            Ribeiro, Ant{\^o}nio H. and Pedregosa, Fabian and
            {van Mulbregt}, Paul and {SciPy 1.0 Contributors}},
  title   = {{{SciPy} 1.0: Fundamental Algorithms for Scientific
            Computing in Python}},
  journal = {Nature Methods},
  year    = {2020},
  volume  = {17},
  pages   = {261--272},
  adsurl  = {https://rdcu.be/b08Wh},
  doi     = {10.1038/s41592-019-0686-2},
}

@article{Mondal_2026,
	author = {Mondal, Samaresh and French, K. Decker and Hinkle, Jason T.},
	doi = {10.3847/1538-4357/ae7dec},
	journal = {The Astrophysical Journal},
	month = {jul},
	number = {1},
	pages = {3},
	publisher = {The American Astronomical Society},
	title = {Comparison of X-Ray Emission Properties of TDEs and Soft Flares from AGNs},
	url = {https://doi.org/10.3847/1538-4357/ae7dec},
	volume = {1006},
	year = {2026}}

@article{Hu_2026,
	author = {Hu, Fangyi (Fitz) and Mandel, Ilya and Nealon, Rebecca and Price, Daniel J.},
	doi = {10.3847/2041-8213/ae27cc},
	journal = {The Astrophysical Journal Letters},
	month = {dec},
	number = {2},
	pages = {L21},
	publisher = {The American Astronomical Society},
	title = {Converged Simulations of the Nozzle Shock in Tidal Disruption Events},
	url = {https://doi.org/10.3847/2041-8213/ae27cc},
	volume = {996},
	year = {2025}}

@misc{Sharma_2026,
      title={Are most detected tidal disruption events partial?}, 
      author={Megha Sharma and Daniel J. Price and Alexander Heger and Katie Auchettle},
      year={2026},
      eprint={2606.22919},
      archivePrefix={arXiv},
      primaryClass={astro-ph.HE},
      url={https://arxiv.org/abs/2606.22919}, 
}

@article{Oh_2022,
	author = {Oh, Kyuseok and Koss, Michael J. and Ueda, Yoshihiro and Stern, Daniel and Ricci, Claudio and Trakhtenbrot, Benny and Powell, Meredith C. and den Brok, Jakob S. and Lamperti, Isabella and Mushotzky, Richard and Ricci, Federica and B{\"a}r, Rudolf E. and Rojas, Alejandra F. and Ichikawa, Kohei and Riffel, Rog{\'e}rio and Treister, Ezequiel and Harrison, Fiona and Urry, C. Megan and Bauer, Franz E. and Schawinski, Kevin},
	doi = {10.3847/1538-4365/ac5b68},
	journal = {The Astrophysical Journal Supplement Series},
	month = {jul},
	number = {1},
	pages = {4},
	publisher = {The American Astronomical Society},
	title = {BASS. XXIV. The BASS DR2 Spectroscopic Line Measurements and AGN Demographics},
	url = {https://doi.org/10.3847/1538-4365/ac5b68},
	volume = {261},
	year = {2022}}

@ARTICLE{Ho_1997,
       author = {{Ho}, Luis C. and {Filippenko}, Alexei V. and {Sargent}, Wallace L.~W. and {Peng}, Chien Y.},
        title = "{A Search for ``Dwarf'' Seyfert Nuclei. IV. Nuclei with Broad H{\ensuremath{\alpha}} Emission}",
      journal = {\apjs},
         year = 1997,
        month = oct,
       volume = {112},
       number = {2},
        pages = {391-414},
          doi = {10.1086/313042},
archivePrefix = {arXiv},
       eprint = {astro-ph/9704099},
 primaryClass = {astro-ph},
       adsurl = {https://ui.adsabs.harvard.edu/abs/1997ApJS..112..391H}
}

@ARTICLE{Ho_2003,
       author = {{Ho}, Luis C. and {Filippenko}, Alexei V. and {Sargent}, Wallace L.~W.},
        title = "{A Search for ``Dwarf'' Seyfert Nuclei. VI. Properties of Emission-Line Nuclei in Nearby Galaxies}",
      journal = {\apj},
         year = 2003,
        month = jan,
       volume = {583},
       number = {1},
        pages = {159-177},
          doi = {10.1086/345354},
archivePrefix = {arXiv},
       eprint = {astro-ph/0210048},
 primaryClass = {astro-ph},
       adsurl = {https://ui.adsabs.harvard.edu/abs/2003ApJ...583..159H}
}

@ARTICLE{Koss_2017,
       author = {{Koss}, Michael and {Trakhtenbrot}, Benny and {Ricci}, Claudio and {Lamperti}, Isabella and {Oh}, Kyuseok and {Berney}, Simon and {Schawinski}, Kevin and {Balokovi{\'c}}, Mislav and {Baronchelli}, Linda and {Crenshaw}, D. Michael and {Fischer}, Travis and {Gehrels}, Neil and {Harrison}, Fiona and {Hashimoto}, Yasuhiro and {Hogg}, Drew and {Ichikawa}, Kohei and {Masetti}, Nicola and {Mushotzky}, Richard and {Sartori}, Lia and {Stern}, Daniel and {Treister}, Ezequiel and {Ueda}, Yoshihiro and {Veilleux}, Sylvain and {Winter}, Lisa},
        title = "{BAT AGN Spectroscopic Survey. I. Spectral Measurements, Derived Quantities, and AGN Demographics}",
      journal = {\apj},
         year = 2017,
        month = nov,
       volume = {850},
       number = {1},
          eid = {74},
        pages = {74},
          doi = {10.3847/1538-4357/aa8ec9},
archivePrefix = {arXiv},
       eprint = {1707.08123},
 primaryClass = {astro-ph.HE},
       adsurl = {https://ui.adsabs.harvard.edu/abs/2017ApJ...850...74K}
}

@ARTICLE{Hobbs_1974,
       author = {{Hobbs}, L.~M.},
        title = "{A comparison of interstellar Na I, Ca II, and K I absorption.}",
      journal = {\apj},
         year = 1974,
        month = jul,
       volume = {191},
        pages = {381-393},
          doi = {10.1086/152976},
       adsurl = {https://ui.adsabs.harvard.edu/abs/1974ApJ...191..381H}
}

@ARTICLE{Poznanski_2012,
       author = {{Poznanski}, Dovi and {Prochaska}, J. Xavier and {Bloom}, Joshua S.},
        title = "{An empirical relation between sodium absorption and dust extinction}",
      journal = {\mnras},
         year = 2012,
        month = oct,
       volume = {426},
       number = {2},
        pages = {1465-1474},
          doi = {10.1111/j.1365-2966.2012.21796.x},
archivePrefix = {arXiv},
       eprint = {1206.6107},
 primaryClass = {astro-ph.IM},
       adsurl = {https://ui.adsabs.harvard.edu/abs/2012MNRAS.426.1465P}
}

@misc{Franz_2026,
      title={The Radio Properties of Extreme Coronal Line Emitters: Constraints on the Sub-parsec Environment}, 
      author={Noah Franz and Kate D. Alexander and Collin T. Christy and Tanmoy Laskar and Stefanie Komossa and Enrico Ramirez-Ruiz and Jean Somalwar and Edo Berger and Ryan Chornock and Fabio De Colle and Gavin Farley and Megan Newsome and B. Ashley VanderLey},
      year={2026},
      eprint={2607.09871},
      archivePrefix={arXiv},
      primaryClass={astro-ph.HE},
      url={https://arxiv.org/abs/2607.09871}, 
}

@ARTICLE{Anderson_2020,
       author = {{Anderson}, M.~M. and {Mooley}, K.~P. and {Hallinan}, G. and {Dong}, D. and {Phinney}, E.~S. and {Horesh}, A. and {Bourke}, S. and {Cenko}, S.~B. and {Frail}, D. and {Kulkarni}, S.~R. and {Myers}, S.},
        title = "{Caltech-NRAO Stripe 82 Survey (CNSS). III. The First Radio-discovered Tidal Disruption Event, CNSS J0019+00}",
      journal = {\apj},
         year = 2020,
        month = nov,
       volume = {903},
       number = {2},
          eid = {116},
        pages = {116},
          doi = {10.3847/1538-4357/abb94b},
archivePrefix = {arXiv},
       eprint = {1910.11912},
 primaryClass = {astro-ph.HE},
       adsurl = {https://ui.adsabs.harvard.edu/abs/2020ApJ...903..116A}
}

@article{Yuan_2014,
	author = {Yuan, Feng and Narayan, Ramesh},
	doi = {https://doi.org/10.1146/annurev-astro-082812-141003},
	issn = {1545-4282},
	journal = {Annual Review of Astronomy and Astrophysics},
	number = {Volume 52, 2014},
	pages = {529-588},
	publisher = {Annual Reviews},
	title = {Hot Accretion Flows Around Black Holes},
	type = {Journal Article},
	url = {https://www.annualreviews.org/content/journals/10.1146/annurev-astro-082812-141003},
	volume = {52},
	year = {2014}}

@article{Earl_2025,
	author = {Earl, Nicholas and French, K. Decker and Ramirez-Ruiz, Enrico and Auchettl, Katie and Raimundo, Sandra I. and Davis, Kyle W. and Masterson, Megan and Arcavi, Iair and Lu, Wenbin and Baldassare, Vivienne F. and Coulter, David A. and de Boer, Thomas and Drout, Maria R. and Dykaar, Hannah and Foley, Ryan J. and Gall, Christa and Gao, Hua and Huber, Mark E. and Jones, David O. and Langeroodi, Danial and Lin, Chien-Cheng and Magnier, Eugene A. and Mockler, Brenna and Shepherd, Margaret and Verrico, Margaret E.},
	doi = {10.3847/1538-4357/adb974},
	journal = {The Astrophysical Journal},
	month = {apr},
	number = {1},
	pages = {28},
	publisher = {The American Astronomical Society},
	title = {AT 2020nov: Evidence for Disk Reprocessing in a Rare Tidal Disruption Event},
	url = {https://doi.org/10.3847/1538-4357/adb974},
	volume = {983},
	year = {2025}}

@article{Clark_2025,
    author = {Clark, Peter and Callow, Joseph and Graur, Or and Greenwell, Claire and Hu, Lei and Aguilar, Jessica and Ahlen, Steven and Bianchi, Davide and Brooks, David and Claybaugh, Todd and Dawson, Kyle and de la Macorra, Axel and Doel, Peter and Gontcho A Gontcho, Satya and Gutierrez, Gaston and Honscheid, Klaus and Juneau, Stephanie and Kehoe, Robert and Kisner, Theodore and Kremin, Anthony and Landriau, Martin and Le Guillou, Laurent and Meisner, Aaron and Miquel, Ramon and Moustakas, John and Pérez-Ràfols, Ignasi and Sanchez, Eusebio and Schubnell, Michael and Sprayberry, David and Tarlé, Gregory and Weaver, Benjamin A and Zou, Hu},
    title = {AT 2018dyk: tidal disruption event or active galactic nucleus? Follow-up observations of an extreme coronal line emitter with the Dark Energy Spectroscopic Instrument},
    journal = {Monthly Notices of the Royal Astronomical Society},
    volume = {540},
    number = {1},
    pages = {871-906},
    year = {2025},
    month = {06},
    issn = {0035-8711},
    doi = {10.1093/mnras/staf724},
    url = {https://doi.org/10.1093/mnras/staf724},
    eprint = {https://academic.oup.com/mnras/article-pdf/540/1/871/63052714/staf724.pdf},
}

@article{Wu_2023,
	author = {Wu, Jiancheng and Wu, Qingwen and Xue, Hanrui and Lei, Weihua and Lyu, Bing},
	doi = {10.3847/1538-4357/acce9e},
	journal = {The Astrophysical Journal},
	month = {jun},
	number = {2},
	pages = {106},
	publisher = {The American Astronomical Society},
	title = {Steep Balmer Decrement in Weak AGNs May Not Be Caused by Dust Extinction: Clues from Low-luminosity AGNs and Changing-look AGNs},
	url = {https://doi.org/10.3847/1538-4357/acce9e},
	volume = {950},
	year = {2023}}

@INPROCEEDINGS{Mockler_2026,
       author = {{Mockler}, Brenna and {Hammerstein}, Erica and {Coughlin}, Eric R. and {Nicholl}, Matt},
        title = "{Tidal disruption events}",
    booktitle = {Encyclopedia of Astrophysics, Volume 3},
         year = 2026,
       volume = {3},
        month = jan,
        pages = {423-457},
          doi = {10.1016/B978-0-443-21439-4.00102-4},
archivePrefix = {arXiv},
       eprint = {2511.14911},
 primaryClass = {astro-ph.HE},
       adsurl = {https://ui.adsabs.harvard.edu/abs/2026enap....3..423M}
}

@ARTICLE{Tombesi_2011,
       author = {{Tombesi}, F. and {Cappi}, M. and {Reeves}, J.~N. and {Palumbo}, G.~G.~C. and {Braito}, V. and {Dadina}, M.},
        title = "{Evidence for Ultra-fast Outflows in Radio-quiet Active Galactic Nuclei. II. Detailed Photoionization Modeling of Fe K-shell Absorption Lines}",
      journal = {\apj},
         year = 2011,
        month = nov,
       volume = {742},
       number = {1},
          eid = {44},
        pages = {44},
          doi = {10.1088/0004-637X/742/1/44},
archivePrefix = {arXiv},
       eprint = {1109.2882},
 primaryClass = {astro-ph.HE},
       adsurl = {https://ui.adsabs.harvard.edu/abs/2011ApJ...742...44T}
}

@ARTICLE{Gofford_2013,
       author = {{Gofford}, Jason and {Reeves}, James N. and {Tombesi}, Francesco and {Braito}, Valentina and {Turner}, T. Jane and {Miller}, Lance and {Cappi}, Massimo},
        title = "{The Suzaku view of highly ionized outflows in AGN - I. Statistical detection and global absorber properties}",
      journal = {\mnras},
         year = 2013,
        month = mar,
       volume = {430},
       number = {1},
        pages = {60-80},
          doi = {10.1093/mnras/sts481},
archivePrefix = {arXiv},
       eprint = {1211.5810},
 primaryClass = {astro-ph.HE},
       adsurl = {https://ui.adsabs.harvard.edu/abs/2013MNRAS.430...60G}
}

@article{King_Pounds_2015,
	author = {King, Andrew and Pounds, Ken},
	doi = {https://doi.org/10.1146/annurev-astro-082214-122316},
	issn = {1545-4282},
	journal = {Annual Review of Astronomy and Astrophysics},
	number = {Volume 53, 2015},
	pages = {115-154},
	publisher = {Annual Reviews},
	title = {Powerful Outflows and Feedback from Active Galactic Nuclei},
	type = {Journal Article},
	url = {https://www.annualreviews.org/content/journals/10.1146/annurev-astro-082214-122316},
	volume = {53},
	year = {2015}}

@article{Igo_2020,
    author = {Igo, Z and Parker, M L and Matzeu, G A and Alston, W and Alvarez Crespo, N and Fürst, F and Buisson, D J K and Lobban, A and Joyce, A M and Mallick, L and Schartel, N and Santos-Lleó, M},
    title = {Searching for ultra-fast outflows in AGN using variability spectra},
    journal = {Monthly Notices of the Royal Astronomical Society},
    volume = {493},
    number = {1},
    pages = {1088-1108},
    year = {2020},
    month = {03},
    issn = {0035-8711},
    doi = {10.1093/mnras/staa265},
    url = {https://doi.org/10.1093/mnras/staa265},
    eprint = {https://academic.oup.com/mnras/article-pdf/493/1/1088/32533917/staa265.pdf},
}

@article{SFR_Kennicutt_2012,
	author = {Kennicutt, Robert C. and Evans, Neal J.},
	doi = {https://doi.org/10.1146/annurev-astro-081811-125610},
	issn = {1545-4282},
	journal = {Annual Review of Astronomy and Astrophysics},
	number = {Volume 50, 2012},
	pages = {531-608},
	publisher = {Annual Reviews},
	title = {Star Formation in the Milky Way and Nearby Galaxies},
	type = {Journal Article},
	url = {https://www.annualreviews.org/content/journals/10.1146/annurev-astro-081811-125610},
	volume = {50},
	year = {2012}}

\begin{appendix}

\section{A. Rise Model Parameters}
\label{ap:rise_model_params}

\begin{table*}[ht]
\centering
\caption{\normalfont Median model parameters from fitting the pre-peak light curves of ATLAS22kjn in Section \ref{sec:Early-Time Rise Modelling}, as shown in Figure \ref{fig:rise_modelling}. The filters are listed in order of increasing central wavelength.}
\vspace{1mm}
\begin{tabular}{c@{\hspace{5.0pt}}c@{\hspace{5.0pt}}c@{\hspace{5.0pt}}c@{\hspace{6.0pt}}c}
\hline \\[-1.0ex]
Filter & \makecell{$h$ \\[1mm] $\left(10^{-6} \ \textrm{mJy}\right)$} & $a_1$ & \makecell{$a_2$ \\[1mm] $\left(10^{-3} \ \textrm{day}^{-1} \right)$} & \makecell{$f_0$ \\[1mm] $\left(10^{-3} \ \textrm{mJy} \right)$} \\[3mm]
\hline & \\[-1.5ex]
ASAS-SN $g$ & $1.7\substack{+0.8\\ -0.5}$ & $2.87\substack{+0.12\\ -0.11}$ & $-1.12\substack{+0.13\\ -0.11}$ & $-4\pm 2$ \\[1mm]
ZTF $g$ & $0.81\substack{+0.19\\ -0.16}$ & $3.00\substack{+0.05\\ -0.06}$ & $-0.92\substack{+0.03\\ -0.02}$ & $-0.9\pm 0.2$ \\[2mm]
ATLAS $c$ & $0.9\substack{+0.4\\ -0.3}$ & $3.04\substack{+0.09\\ -0.10}$ & $-1.01\pm 0.05$ & $-1.4\pm 0.7$ \\[2mm]
Gaia $G$ & $2.4\substack{+0.6\\ -0.5}$ & $2.76\pm 0.08$ & $-0.76\substack{+0.10\\ -0.07}$ & $-0.3\substack{+0.5\\ -0.6}$ \\[2mm]
ZTF $r$ & $1.5\substack{+0.3\\ -0.2}$ & $2.94\substack{+0.03\\ -0.04}$ & $-1.03\substack{+0.03\\ -0.02}$ & $-0.6\pm 0.3$ \\[2mm]
ATLAS $o$ & $2.1\substack{+0.6\\ -0.5}$ & $2.90\pm 0.08$ & $-1.07\pm 0.5$ & $-0.5\pm 0.7$ \\[2mm]
ZTF $i$ & $1.0\substack{+0.3\\ -0.2}$ & $3.04\pm 0.07$ & $-1.15\pm 0.4$ & $0.2\pm 0.7$ \\[2mm]
TESS & $3.3\substack{+0.7\\ -0.6}$ & $2.78\substack{+0.10\\ -0.11}$ & $-1.7\substack{+0.4\\ -0.3}$ & $4.4\pm 0.3$ \\[2mm]
\hline
\end{tabular}
\label{tab:MCMC params}
\end{table*}

\clearpage
\section{B. Spectroscopy Tables}
\label{ap:spectroscopy_tables}

\begin{table}[ht]
        \caption{\normalfont Spectroscopic observations of ATLAS22kjn taken on the Liverpool Telescope, UH2.2m, Keck I, Keck II, NASA IRTF, and Gemini. The SPRAT spectrum and the first UH2.2m spectrum were taken prior to the UV/optical peak of ATLAS22kjn. All other spectra were taken post-peak.}
        \vspace{1mm}
    \centering
    \begin{tabular}{cccccc}
    \hline \\[-1.5ex]
        MJD & UTC Date & Telescope & Instrument & Rest Wavelength Range (Å) & Exposure Time (s) \\[1mm] \hline \\[-1.5ex]
        59753.9 & 2022 June 23 & Liverpool 2 m Telescope & SPRAT & 3747--7450 & 1500.0 \\
        59792.3 & 2022 August 1.3 & University of Hawai`i 2.2m Telescope & SNIFS & 3172--8480 & 1800.0 \\ 
        59813.3 & 2022 August 22.3 & Keck I 10m Telescope & LRIS & 2983--9703 & 3600.0 \\ 
        59814.3 & 2022 August 23.3 & University of Hawai`i 2.2m Telescope & SNIFS & 3172--8481 & 1800.0 \\ 
        59826.3 & 2022 September 4.3 & University of Hawai`i 2.2m Telescope & SNIFS & 3172--8481 & 2100.0 \\ 
        59990.6 & 2023 February 15.6 & NASA Infrared Telescope Facility & SpeX & 6661--23953 & 6000.0 \\ 
        60074.5 & 2023 May 10.5 & University of Hawai`i 2.2m Telescope & SNIFS & 3170--8479 & 3600.0 \\ 
        60115.4 & 2023 June 20.4 & University of Hawai`i 2.2m Telescope & SNIFS & 3170--8479 & 3600.0 \\ 
        60129.3 & 2023 July 4.3 & Gemini North & GMOS & 3409--7285 & 3600.0 \\ 
        60328.6 & 2024 Jan 19.6 & Keck II 10m Telescope & KCWI & 3188--8293 & 3600.0 \\ 
        60401.5 & 2024 Apr 1.6 & NASA Infrared Telescope Facility & SpeX & 6661--23953 & 6000.0 \\
        60432.6 & 2024 May 2.6 & NASA Infrared Telescope Facility & SpeX & 6661--23953 & 6000.0 \\ 
        60523.3 & 2024 Aug 1.3 & NASA Infrared Telescope Facility & SpeX & 6661--23953 & 6000.0 \\ 
        60474.3 & 2024 June 13.3 & Keck II 10m Telescope & KCWI & 3188--8293 & 4500.0 \\ 
        60847.5 & 2025 June 21.5 & Keck II 10m Telescope & KCWI & 3188--8293 & 4500.0 \\ \hline
    \end{tabular}
    \label{spectra table}
\end{table}
\FloatBarrier

\begin{longtable}{lcccccc}
        \caption{\normalfont Balmer line measurements, listed in order of wavelength. We report $1 \sigma$ uncertainties for the luminosities and the broad FWHMs. We report $3 \sigma$ uncertainties for the narrow FWHMs.} 
        \vspace{-1mm}
    \label{tab:balmer_lines} \\
    \hline \\[-1.5ex]
    Line & Instrument & MJD & \makecell{Broad \\ $\log(L/\mathrm{erg\,s^{-1}})$} & \makecell{Narrow \\ $\log(L/\mathrm{erg\,s^{-1}})$} & \makecell{Broad FWHM \\ ($\mathrm{km\,s^{-1}}$)} & \makecell{Narrow FWHM \\ ($\mathrm{km\,s^{-1}}$)} \\[1.5ex]
    \hline \\[-1.5ex]
    \endfirsthead

    \multicolumn{7}{c}{\tablename\ \thetable{} -- continued from previous page} \\
    \hline \\[-1.5ex]
    Line & Instrument & MJD & \makecell{Broad \\ $\log(L/\mathrm{erg\,s^{-1}})$} & \makecell{Narrow \\ $\log(L/\mathrm{erg\,s^{-1}})$} & \makecell{Broad FWHM \\ ($\mathrm{km\,s^{-1}}$)} & \makecell{Narrow FWHM \\ ($\mathrm{km\,s^{-1}}$)} \\
    \hline \\[-1.5ex]
    \endhead

    \hline
    \multicolumn{7}{r}{\textit{Continued on next page}} \\
    \endfoot

    \endlastfoot

    H$\beta$ $\lambda4861$ & SPRAT & 59754.0 & $40.84^{+0.07}_{-0.01}$ & $40.3^{+0.1}_{-0.4}$ & $2260^{+180}_{-20}$ & $800^{+10}_{-700}$ \\[1mm]
     & SNIFS & 59792.0 & $41.22^{+0.02}_{-0.03}$ & $39.7^{+0.6}_{-0.2}$ & $2080^{+70}_{-100}$ & $400^{+400}_{-400}$ \\[1mm]
     & LRIS & 59813.3 & $41.224^{+0.017}_{-0.012}$ & $39.71^{+0.03}_{-0.03}$ & $1967^{+4}_{-5}$ & $339^{+18}_{-18}$ \\[1mm]
     & SNIFS & 59814.0 & $41.27^{+0.02}_{-0.02}$ & $40.0^{+0.2}_{-0.4}$ & $1960^{+80}_{-100}$ & $400^{+400}_{-300}$ \\[1mm]
     & SNIFS & 59826.0 & $41.36^{+0.03}_{-0.01}$ & $40.40^{+0.03}_{-0.30}$ & $2900^{+50}_{-200}$ & $800^{+20}_{-700}$ \\[1mm]
     & SNIFS & 60074.0 & $41.00^{+0.03}_{-0.02}$ & $39.6^{+0.2}_{-0.4}$ & $2900^{+200}_{-100}$ & $500^{+300}_{-400}$ \\[1mm]
     & SNIFS & 60115.0 & $40.91^{+0.03}_{-0.03}$ & $39.8^{+0.4}_{-0.4}$ & $2200^{+100}_{-200}$ & $500^{+300}_{-500}$ \\[1mm]
     & GMOS & 60129.3 & $40.751^{+0.017}_{-0.013}$ & $39.34^{+0.06}_{-0.04}$ & $1737^{+18}_{-13}$ & $360^{+70}_{-60}$ \\[1mm]
     & KCWI & 60328.0 & $40.658^{+0.017}_{-0.013}$ & $39.74^{+0.04}_{-0.02}$ & $2040^{+30}_{-20}$ & $490^{+60}_{-50}$ \\[1mm]
     & KCWI & 60474.3 & $40.364^{+0.019}_{-0.017}$ & $39.62^{+0.05}_{-0.04}$ & $2250^{+70}_{-60}$ & $500^{+90}_{-70}$ \\[1mm]
     & KCWI & 60847.5 & $40.12^{+0.03}_{-0.04}$ & $39.42^{+0.08}_{-0.06}$ & $2800^{+300}_{-200}$ & $390^{+200}_{-130}$ \\[1mm]
    \hline \\[-1.5ex]
    H$\alpha$ $\lambda6563$ & SPRAT & 59754.0 & $41.595^{+0.015}_{-0.010}$ & $40.0^{+0.3}_{-0.4}$ & $2300^{+50}_{-70}$ & $300^{+400}_{-200}$ \\[1mm]
     & SNIFS & 59792.0 & $41.583^{+0.003}_{-0.003}$ & $40.910^{+0.014}_{-0.013}$ & $2793^{+18}_{-13}$ & $490^{+30}_{-30}$ \\[1mm]
     & LRIS & 59813.3 & $41.761^{+0.002}_{-0.002}$ & $40.839^{+0.011}_{-0.012}$ & $1917.1^{+1.0}_{-1.5}$ & $373^{+5}_{-5}$ \\[1mm]
     & SNIFS & 59814.0 & $41.640^{+0.003}_{-0.004}$ & $40.78^{+0.03}_{-0.02}$ & $2250^{+20}_{-10}$ & $590^{+130}_{-70}$ \\[1mm]
     & SNIFS & 59826.0 & $41.666^{+0.003}_{-0.002}$ & $40.22^{+0.03}_{-0.04}$ & $2070^{+8}_{-7}$ & $500^{+60}_{-60}$ \\[1mm]
     & SNIFS & 60074.0 & $41.441^{+0.003}_{-0.002}$ & $40.727^{+0.014}_{-0.016}$ & $1965^{+11}_{-8}$ & $753.0^{+0.1}_{-15.0}$ \\[1mm]
     & SNIFS & 60115.0 & $41.441^{+0.003}_{-0.003}$ & $40.738^{+0.015}_{-0.015}$ & $1981^{+10}_{-8}$ & $753.00^{+0.04}_{-10.00}$ \\[1mm]
     & GMOS & 60129.3 & $41.623^{+0.002}_{-0.002}$ & $40.18^{+0.02}_{-0.02}$ & $1772^{+2}_{-3}$ & $321^{+12}_{-12}$ \\[1mm]
     & KCWI & 60328.6 & $41.279^{+0.002}_{-0.002}$ & $40.863^{+0.011}_{-0.012}$ & $2086^{+3}_{-6}$ & $753.00^{+0.02}_{-3.00}$ \\[1mm]
     & KCWI & 60474.3 & $41.026^{+0.003}_{-0.002}$ & $40.506^{+0.012}_{-0.015}$ & $2068^{+9}_{-11}$ & $590^{+20}_{-20}$ \\[1mm]
     & KCWI & 60847.5 & $40.805^{+0.010}_{-0.010}$ & $40.07^{+0.05}_{-0.04}$ & $2320^{+70}_{-60}$ & $430^{+80}_{-70}$ \\[1mm]
    \hline \\[-1.5ex]
\end{longtable}

\clearpage

\begin{longtable}{lcccc}
    \caption{[\normalfont \ion{O}{iii}] $\lambda$5007 measurements and $1 \sigma$ uncertainties. The asterisk * indicates a $3\sigma$ upper limit, determined by assuming a fixed-width in the Gaussian (see \protect Section \ref{sec:CL evolution}); FWHM is therefore not available for this line.}
    \vspace{-1mm}
    \label{tab:oiii} \\
    \hline \\[-1.5ex]
    Line & Instrument & MJD & $\log_{10}[L (\mathrm{erg\,s^{-1}}])$ & FWHM ($\mathrm{km\,s^{-1}}$) \\[0.5ex]
    \hline \\[-1.5ex]
    \endfirsthead

    \hline \\[-1.5ex]
    Line & Instrument & MJD & $\log_{10}[L (\mathrm{erg\,s^{-1}}])$ & FWHM ($\mathrm{km\,s^{-1}}$) \\
    \hline \\[-1.5ex]
    \endhead

    \hline
    
    \endfoot

    \endlastfoot

    [O III] $\lambda5007$ (IP = 35.12~eV) & SNIFS$^{*}$ & 59792.3 & $< 40.03$ & N/A \\[1mm]
     & LRIS & 59813.3 & $39.951^{+0.010}_{-0.009}$ & $850^{+30}_{-20}$ \\[1mm]
     & SNIFS & 59814.3 & $40.2^{+0.2}_{-0.1}$ & $1000^{+800}_{-100}$ \\[1mm]
     & SNIFS & 59826.3 & $40.0^{+0.8}_{-0.1}$ & $1000^{+4000}_{-50}$ \\[1mm]
     & SNIFS & 60074.5 & $39.98^{+0.04}_{-0.04}$ & $630^{+100}_{-60}$ \\[1mm]
     & SNIFS & 60115.4 & $39.83^{+0.08}_{-0.05}$ & $700^{+200}_{-100}$ \\[1mm]
     & GMOS & 60129.3 & $39.857^{+0.009}_{-0.009}$ & $427^{+12}_{-12}$ \\[1mm]
     & KCWI & 60328.6 & $39.886^{+0.008}_{-0.009}$ & $382^{+10}_{-10}$ \\[1mm]
     & KCWI & 60474.3 & $39.791^{+0.013}_{-0.015}$ & $318^{+13}_{-12}$ \\[1mm]
     & KCWI & 60847.5 & $39.88^{+0.02}_{-0.02}$ & $309^{+19}_{-16}$ \\[1mm]
    \hline \\[-1.5ex]
\end{longtable}

\vspace{8mm}

\begin{longtable}{lcccc}
    \caption{Measurements of coronal lines (CLs) and $1 \sigma$ uncertainties, in order of wavelengths. ``IP'' denotes the ionisation potential of a given line. Wavelengths for the first four optical CLs were obtained from \protect\cite{Negus_2023}, and the rest from \protect\cite{wang12}. The ionisation potentials for optical CLs were obtained from \cite{Ferland_Osterbrock_1987}, \cite{komossa08}, \cite{Rose_2011}, \cite{wang11}, and \cite{Hauschild-Roier_2025}. Central wavelengths and ionisation potentials for NIR CLs were obtained from \protect\cite{lamperti17}. Asterisks * indicate $3\sigma$ upper limits, determined by assuming a fixed-width in the Gaussian (see \protect Section \ref{sec:CL evolution}); FWHM is therefore not available for these instances.}
    \vspace{-1mm}
    \label{tab:CL_measurements} \\
    \hline \\[-1.5ex]
    Line & Instrument & MJD & $\log_{10}[L (\mathrm{erg\,s^{-1}}])$ & FWHM ($\mathrm{km\,s^{-1}}$) \\[0.5ex]
    \hline \\[-1.5ex]
    \endfirsthead
    
    \multicolumn{5}{l}{\textsc{Table \thetable.} \ {\normalfont Measurements of coronal lines (CLs), continued from the previous page.}} \\[1mm]

    \hline \\[-1.5ex]
    Line & Instrument & MJD & $\log_{10}[L (\mathrm{erg\,s^{-1}}])$ & FWHM ($\mathrm{km\,s^{-1}}$) \\[0.5ex]
    \endhead
    
    \\

    \endfoot
    
    \endlastfoot
    
    [Ne V] $\lambda3347$ (IP = 97.11~eV) & SNIFS$^{*}$ & 59792.3 & $< 40.05$ & N/A \\[1mm]
     & LRIS & 59813.3 & $39.95^{+0.02}_{-0.02}$ & $1160^{+70}_{-60}$ \\[1mm]
     & SNIFS$^{*}$ & 59814.0 & $< 40.22$ & N/A \\[1mm]
     & SNIFS$^{*}$ & 59826.3 & $< 39.76$ & N/A \\[1mm]
     & SNIFS$^{*}$ & 60074.5 & $< 40.03$ & N/A \\[1mm]
     & SNIFS$^{*}$ & 60115.4 & $< 39.67$ & N/A \\[1mm]
     & KCWI & 60328.6 & $39.90^{+0.02}_{-0.03}$ & $570^{+60}_{-40}$ \\[1mm]
     & KCWI & 60474.3 & $39.81^{+0.04}_{-0.05}$ & $590^{+120}_{-70}$ \\[1mm]
     & KCWI$^{*}$ & 60847.5 & $< 39.85$ & N/A \\[1mm]
    \hline \\[-1.5ex]
    [Fe VII] $\lambda3760$ (IP = 99.00~eV) & SNIFS$^{*}$ & 59792.3 & $< 39.95$ & N/A \\[1mm]
     & LRIS & 59813.3 & $39.825^{+0.013}_{-0.013}$ & $930^{+30}_{-30}$ \\[1mm]
     & SNIFS$^{*}$ & 59814.3 & $< 39.63$ & N/A \\[1mm]
     & SNIFS$^{*}$ & 59826.3 & $< 40.11$ & N/A \\[1mm]
     & SNIFS & 60074.5 & $39.89^{+0.08}_{-0.08}$ & $800^{+200}_{-100}$ \\[1mm]
     & SNIFS & 60115.4 & $40.07^{+0.09}_{-0.09}$ & $1400^{+300}_{-300}$ \\[1mm]
     & GMOS & 60129.3 & $39.78^{+0.03}_{-0.03}$ & $790^{+90}_{-70}$ \\[1mm]
     & KCWI & 60328.6 & $39.734^{+0.014}_{-0.017}$ & $630^{+30}_{-30}$ \\[1mm]
     & KCWI & 60474.3 & $39.69^{+0.02}_{-0.03}$ & $640^{+60}_{-40}$ \\[1mm]
     & KCWI & 60847.5 & $39.34^{+0.07}_{-0.14}$ & $400^{+300}_{-100}$ \\[1mm]
    \hline \\[-1.5ex]
    [Ar XIV] $\lambda4414$ (IP = 686.09~eV) & SNIFS$^{*}$ & 59792.3 & $< 39.83$ & N/A \\[1mm]
     & LRIS$^{*}$ & 59813.3 & $< 38.95$ & N/A \\[1mm]
     & SNIFS$^{*}$ & 59814.3 & $< 39.69$ & N/A \\[1mm]
     & SNIFS$^{*}$ & 59826.3 & $< 39.68$ & N/A \\[1mm]
     & SNIFS$^{*}$ & 60074.5 & $< 39.36$ & N/A \\[1mm]
     & SNIFS & 60115.4 & $39.3^{+0.3}_{-0.2}$ & $700^{+1200}_{-300}$ \\[1mm]
     & GMOS & 60129.3 & $39.23^{+0.05}_{-0.06}$ & $610^{+150}_{-70}$ \\[1mm]
     & KCWI & 60328.6 & $39.16^{+0.05}_{-0.06}$ & $300^{+40}_{-30}$ \\[1mm]
     & KCWI & 60474.3 & $39.03^{+0.10}_{-0.12}$ & $280^{+80}_{-40}$ \\[1mm]
     & KCWI$^{*}$ & 60847.5 & $< 39.12$ & N/A \\[1mm]
    \hline \\[-1.5ex]
    [Fe XIV] $\lambda5304$ (IP = 361.00~eV) & SNIFS & 59792.3 & $40.31^{+0.03}_{-0.13}$ & $1600^{+300}_{-300}$ \\[1mm]
     & LRIS & 59813.3 & $40.091^{+0.015}_{-0.004}$ & $1000^{+40}_{-10}$ \\[1mm]
     & SNIFS & 59814.3 & $40.15^{+0.10}_{-0.06}$ & $1000^{+400}_{-100}$ \\[1mm]
     & SNIFS & 59826.3 & $40.42^{+0.03}_{-0.10}$ & $2400^{+300}_{-400}$ \\[1mm]
     & SNIFS & 60074.5 & $39.94^{+0.05}_{-0.05}$ & $980^{+130}_{-110}$ \\[1mm]
     & SNIFS & 60115.4 & $40.02^{+0.03}_{-0.07}$ & $980^{+80}_{-180}$ \\[1mm]
     & GMOS & 60129.3 & $39.875^{+0.011}_{-0.016}$ & $760^{+20}_{-30}$ \\[1mm]
     & KCWI & 60328.6 & $39.61^{+0.03}_{-0.01}$ & $430^{+40}_{-10}$ \\[1mm]
     & KCWI & 60474.3 & $39.57^{+0.06}_{-0.03}$ & $600^{+80}_{-40}$ \\[1mm]
     & KCWI$^{*}$ & 60847.5 & $< 39.78$ & N/A \\[1mm]
    \hline \\[-1.5ex]
    [Fe X] $\lambda6376$ (IP = 235.04~eV) & SNIFS & 59792.3 & $40.10^{+0.03}_{-0.03}$ & $1340^{+140}_{-110}$ \\[1mm]
     & LRIS & 59813.3 & $40.142^{+0.004}_{-0.004}$ & $828^{+8}_{-8}$ \\[1mm]
     & SNIFS & 59814.3 & $40.08^{+0.04}_{-0.03}$ & $1000^{+110}_{-70}$ \\[1mm]
     & SNIFS & 59826.3 & $40.04^{+0.03}_{-0.03}$ & $760^{+60}_{-40}$ \\[1mm]
     & SNIFS & 60074.5 & $39.64^{+0.04}_{-0.03}$ & $550^{+80}_{-50}$ \\[1mm]
     & SNIFS & 60115.4 & $39.80^{+0.03}_{-0.02}$ & $600^{+50}_{-40}$ \\[1mm]
     & GMOS & 60129.3 & $39.982^{+0.005}_{-0.005}$ & $517^{+9}_{-8}$ \\[1mm]
     & KCWI & 60328.6 & $39.910^{+0.004}_{-0.005}$ & $435^{+7}_{-6}$ \\[1mm]
     & KCWI & 60474.3 & $39.722^{+0.008}_{-0.007}$ & $437^{+12}_{-11}$ \\[1mm]
     & KCWI & 60847.5 & $39.31^{+0.08}_{-0.13}$ & $300^{+300}_{-50}$ \\[1mm]
    \hline \\[-1.5ex]
    [S XII] $\lambda7612$ (IP = 504.78~eV) & SNIFS & 59792.3 & $39.41^{+0.11}_{-0.05}$ & $490^{+190}_{-70}$ \\[1mm]
     & LRIS & 59813.3 & $39.516^{+0.011}_{-0.011}$ & $760^{+20}_{-20}$ \\[1mm]
     & SNIFS$^{*}$ & 59814.3 & $< 39.19$ & N/A \\[1mm]
     & SNIFS & 59826.3 & $39.43^{+0.10}_{-0.07}$ & $800^{+300}_{-100}$ \\[1mm]
     & SNIFS & 60074.5 & $39.03^{+0.07}_{-0.07}$ & $410^{+100}_{-60}$ \\[1mm]
     & SNIFS & 60115.4 & $39.17^{+0.08}_{-0.06}$ & $580^{+120}_{-70}$ \\[1mm]
     & KCWI & 60328.6 & $39.332^{+0.011}_{-0.011}$ & $401^{+17}_{-14}$ \\[1mm]
     & KCWI & 60474.3 & $38.93^{+0.02}_{-0.03}$ & $270^{+20}_{-20}$ \\[1mm]
     & KCWI$^{*}$ & 60847.5 & $< 39.01$ & N/A \\[1mm]
    \hline \\[-1.5ex]
    [Fe XI] $\lambda7894$ (IP = 262.10~eV) & SNIFS & 59792.3 & $39.93^{+0.04}_{-0.04}$ & $880^{+110}_{-80}$ \\[1mm]
     & LRIS & 59813.3 & $39.799^{+0.006}_{-0.006}$ & $681^{+9}_{-8}$ \\[1mm]
     & SNIFS & 59814.3 & $39.90^{+0.05}_{-0.04}$ & $880^{+150}_{-70}$ \\[1mm]
     & SNIFS & 59826.3 & $39.64^{+0.05}_{-0.04}$ & $610^{+80}_{-50}$ \\[1mm]
     & SNIFS$^{*}$ & 60074.5 & $< 39.76$ & N/A \\[1mm]
     & SNIFS & 60115.4 & $39.56^{+0.04}_{-0.03}$ & $510^{+40}_{-30}$ \\[1mm]
     & KCWI & 60328.6 & $39.724^{+0.005}_{-0.005}$ & $386^{+8}_{-7}$ \\[1mm]
     & KCWI & 60474.3 & $39.432^{+0.008}_{-0.009}$ & $357^{+13}_{-11}$ \\[1mm]
     & KCWI & 60847.5 & $39.2^{+0.2}_{-0.1}$ & $400^{+400}_{-100}$ \\[1mm]
    \hline \\[-1.5ex]
    [Si X] $\lambda14300$ (IP = 351.10~eV) & IRTF$^{*}$ & 59990.6 & $< 34.93$ & N/A \\[1mm]
     & IRTF & 60401.6 & $35.73^{+0.14}_{-0.12}$ & $2400^{+1500}_{-500}$ \\[1mm]
     & IRTF & 60432.6 & $35.5^{+0.6}_{-0.3}$ & $1000^{+7000}_{-1000}$ \\[1mm]
     & IRTF$^{*}$ & 60523.3 & $< 35.98$ & N/A \\[1mm]
    \hline \\[-1.5ex]
    [Si VI] $\lambda19620$ (IP = 166.80~eV) & IRTF & 59990.6 & $35.76^{+0.05}_{-0.09}$ & $980^{+150}_{-160}$ \\[1mm]
     & IRTF & 60401.6 & $35.69^{+0.07}_{-0.10}$ & $2500^{+600}_{-600}$ \\[1mm]
     & IRTF & 60432.6 & $35.3^{+0.3}_{-0.3}$ & $800^{+1200}_{-400}$ \\[1mm]
     & IRTF & 60523.3 & $35.82^{+0.06}_{-0.11}$ & $2900^{+400}_{-700}$ \\[1mm]
    \hline \\[-1.5ex]
\end{longtable}

\FloatBarrier

\section{C. Alternative Analysis of the Coronal Line Evolution}
\label{ap:alt_CL_analysis}

For this analysis, we adopted the same two assumptions as in Section \ref{sec:CLs_lum_trend_analysis}, that: (1) the source of ionisation for the CLR and the source of the X-rays is the same, and (2) the SED of this ionising source is well-fit by a blackbody. 
We also assume that the density and fundamental geometries of the gas that makes up the CL-emitting region do not change over the 1-2 years for which we have spectral coverage, since the dynamical timescale $t=(R^3/GM_{BH})^{1/2}$ of the CLR is $\sim 1$ to $\sim 500$ years.
While we assume that the structure of the CLR itself is stable, the distribution of any material between the ionising source and the CLR may vary.
As inferred from the 5th to 95th percentile of virial radii from broad-line velocities, the characteristic timescale of any reprocessing material must be much shorter, ranging from $\sim 10$ days to $\sim 3$ months. In this analysis, however, we aimed to probe only long-term changes in the illumination of the CLR 
because the CLR responds to changes in the ionising continuum only after a delay set by the light travel time, which ranges from $\sim 3$ days to $\sim 5$ months. As a result, rapid fluctuations in the EUV/X-ray source are effectively smoothed out in the observed CLs, 
and we are sensitive only to long-term trends in the accretion rate. We emphasise that we do not attempt to probe short-term X-ray variability with this method.

We adopt a toy model where the CLs are probes of blackbody emission that is allowed to change between epochs. This model consists of coupled non-linear equations describing the line luminosity for each CL ($l$) in each spectrum ($s$):
\begin{equation} \label{eq:toy model}
L_{l,s} = f_l \cdot R_s^2 \int_{E_{l}}^{\infty} \frac{ 2 \pi B_{E}(E, T_s)}{E} dE \ ,
\end{equation}
in which 
$B_{E}(E, T_s)$ is the Planck function.
The integral yields a number density of photons at or above the line energy (i.e., the ionisation potential) $E_{l}$ for a given temperature and line energy. 
In this model, the blackbody temperature $T_s$ and radius $R_s$ are allowed to be completely independent in each spectrum. The term $f_l$ encapsulates the unknown physical conditions producing the CLs and is fit for each CL $l$, but held constant across the spectral epochs. We therefore have a total of 25 free parameters. We employ the same nine optical spectra and seven CLs as in Section \ref{sec:CLs_lum_trend_analysis}, but since the wavelength range of GMOS does not extend to three of the CLs, we obtain 60 total constraints. This means we have an overdetermined set of coupled nonlinear equations. We explored the parameter space and find a best-fit solution with uncertainties using Markov Chain Monte Carlo, again employing the ensemble sampler in the Python package \texttt{emcee} \citep{ForemanMackey2013}.

We fit the luminosity of the CLs in all nine spectra simultaneously. 
Before fitting, to account for systematics in the line fits that were not captured by our formal uncertainties, we inflated the uncertainties on the data with the aim of achieving a post-fit reduced chi-squared $\chi^2/ \nu \sim 1$. For each data point, we added the 50th percentile uncertainty in quadrature with the formal uncertainty, then multiplied by $\sim 3.11$.
We use MCMC processes and fit our parameters in log space to increase efficiency and prevent issues with numerical instability, with priors
$4 < \log_{10}\left(T_s \ [\textrm{K}]\right) < 8$ and $8 < \log_{10}\left(R_s \ [\textrm{cm}]\right) < 11$ per spectrum $s$, and 
$-70 < \log_{10}\left(f_l \ [ \textrm{eV} \ \textrm{cm}^{-2}]\right) < -30$ per line $l$. Our temperature priors are centred on the order of magnitude peak X-ray temperature (Section \ref{sec:Analysis-Xray}), 
as the CLs and X-rays probe similar energies.  
Based on our priors for $T_s$ and $R_s$, we centred the priors for $f_l$ on $10^{-50} \ \textrm{eV} \ \textrm{cm}^{-2}$, the order of magnitude necessary for our model to reproduce the observed CL luminosities. 

We encountered strong degeneracies between the $f_l$ parameters that prevented our MCMC runs from converging to a solution. 
This degeneracy likely persists because the change in the ionising source over the duration of spectral coverage is too small for the fit to gain leverage. This is evidenced by our analysis in \ref{sec:CLs_lum_trend_analysis}.
In an attempt to break these degeneracies, we experimented with fixing the $R_s$ value for the $\textrm{SNIFS} + 270 \textrm{d}$ spectrum to the \textit{XMM-Newton} (+267.0d) X-ray radius measurement, fixing the $T_s$ value for the same spectrum to be the \textit{XMM-Newton} X-ray temperature measurement, or fixing both. We further attempted to fix $R_s$ and/or $T_s$ for the $\textrm{SNIFS} + 310\textrm{d}$ spectrum to the first \textit{Swift} (+299.5d) X-ray measurement instead, since \textit{Swift} does not suffer from the same periods of high background as \textit{XMM-Newton}. We also separately tested fixing $R_s$ and/or $T_s$ for the KCWI + 532.7d spectrum to the second \textit{Swift} (+484.1d) X-ray measurements since the Keck spectra are higher S/N than the UH2.2m spectra. Throughout these tests, however, the $f_l$ degeneracies remained unbroken.

We therefore explored a model with a more limited set of free parameters by fitting a linear trend for temperature and a linear trend for radius evolution. We fit to the same data as before, but with free parameters for only the temperature slope $s_T$, the temperature intercept $b_T$, and the radius slope $s_R$, in addition to the $f_l$ parameters. We fixed the radius intercept $b_R$ to be weighted average $R$ from the X-ray measurements, such that $b_R = \bar{R}_{\textrm{X-ray}} \simeq 5.2 \times 10^{10} \ \textrm{cm}$. Additionally, since $T$ and $R$ are now a functions of time, we normalised the temperature and radius trends to be near the middle of the spectral coverage ($\sim -295$ observed days relative to UV/optical peak, or $\textrm{MJD} \sim 60100$) to reduce degeneracies between slopes and intercepts. Our model is still equation \ref{eq:toy model}, but now,
\begin{align}
    T(t) &= (\textrm{MJD} - 60100) \ s_T + b_T \\
    R(t) &= (\textrm{MJD} - 60100) \ s_R + b_R \ .
\end{align}
These modifications reduce our free parameters from 25 to 11. Where possible, we fit our parameters in log space, with priors
$-300 \ \textrm{K day}^{-1} \le s_T \le 300 \ \textrm{K day}^{-1}$, $4 \le \log_{10}\left(b_T \ [\textrm{K}]\right) \le 8$, $-8 \times 10^7 \ \textrm{cm day}^{-1} \le s_R \le 8 \times 10^7 \ \textrm{cm day}^{-1}$, and 
$-70 \le \log_{10}\left(f_l \ [ \textrm{eV} \ \textrm{cm}^{-2}]\right) \le -30$ per line $l$. 

We again encountered strong degeneracies, both between different $f_l$ parameters and between $f_l$ parameters and $b_T$. We were unable to break these degeneracies by normalising the $f_l$ values by $[T(t)]^4$. In addition to this normalisation, since [\ion{Fe}{XIV}] exhibits a more extreme decrease in luminosity over time than do the other CLs (see Figure \ref{fig:lum_CLs}), we tested the effect of removing this line from the data fitted. This resulted in better constraints on some $f_l$ parameters, but did not improve degenerate behaviour between the $f_l$ parameters and $b_T$. Separately, because [\ion{Ar}{XIV}] was not detected in the first of the four spectra used in this analysis, we tested removing [\ion{Ar}{XIV}] from the data fitted in combination with normalisation. This adjustment had similar effects on the $f_l$ parameters as removing [\ion{Fe}{XIV}], with no other significant changes. We note that though our MCMC did not fully converge, the $s_T$ value obtained from every version of our linear fits was of order $10 \ \textrm{K day}^{-1}$ ($0.005 \ \bar{T}_{\textrm{X}}/500 \ \textrm{days}$) and negative, and consistent with zero to within $1\sigma$. 
This change is slightly less than the decrease of $\lesssim 0.10 \ \bar{T}_{\textrm{X}}/500 \ \textrm{days}$ presented in Section \ref{sec:CLs_lum_trend_analysis} and Figure \ref{fig:lum_IP_CLs_w_lines}.

The method presented here may still enable future studies to infer the temperature of the unobservable portion of the SED using CL emission. In a more physically-motivated model that places better constraints on $f_l$ values, this method would also yield tighter constraints on any change in the ionising source.
Furthermore, our model may benefit from an alternative assumption about the SED of the ionising source, such as by replacing the blackbody with a disk model. Finally, if the source of ionisation persists to very late times, studies of CLs will also become easier as UV/optical blackbody emission fades. We therefore encourage others to pursue spectroscopic follow-up of CLEs out to several years post-UV/optical peak.
In any case, ATLAS22kjn illustrates the need for high cadence, high S/N spectroscopic follow-up of TDEs with CLs to probe the EUV/soft X-ray emission from TDEs, as this will allow us to investigate the ``missing energy" problem that TDEs have long faced.

\end{appendix}

\end{document}